\documentclass[%
 reprint,
 amsmath,amssymb,
 aps,
 prl,superscriptaddress
]{revtex4-2}

\usepackage{dcolumn}% Align table columns on decimal point
\usepackage{bm}% bold math

\usepackage{xcolor}
\usepackage{comment}
\usepackage{nccmath}

\usepackage[unicode=true,
 bookmarks=true,bookmarksnumbered=true,bookmarksopen=false,
 breaklinks=true,pdfborder={0 0 0},backref=false,colorlinks=true]
 {hyperref}
\hypersetup{
citecolor={blue},
urlcolor={blue}
}

\usepackage[normalem]{ulem}

\newcommand{\beq}{\begin{equation}}
\newcommand{\eeq}{\end{equation}}
\newcommand{\bea}{\begin{eqnarray}}
\newcommand{\eea}{\end{eqnarray}}
\newcommand{\bwt}{\begin{widetext}}
\newcommand{\ewt}{\end{widetext}}

\newcommand{\bk}{\mathbf{k}}
\newcommand{\ba}{\mathbf{a}}
\newcommand{\bA}{\mathbf{A}}

\newcommand{\bB}{\mathbf{B}}

\newcommand{\bK}{\mathbf{K}}
\newcommand{\bL}{\mathbf{L}}

\newcommand{\bR}{\mathbf{R}}

\newcommand{\bq}{\mathbf{q}}
\newcommand{\bp}{\mathbf{p}}
\newcommand{\bg}{\mathbf{g}}
\newcommand{\bG}{\mathbf{G}}
\newcommand{\br}{\mathbf{r}}
\newcommand{\bs}{\mathbf{s}}

\newcommand{\ii}{\mathrm{i}}

\usepackage{babel}
\usepackage{physics}
\usepackage{xr}
\usepackage{stmaryrd}
\usepackage{graphicx}
\usepackage{esint}
\usepackage{subfigure}
\usepackage{multirow}

\newcommand{\rr}{{\bf{r}}}

\global\long\def\ket#1{\left| #1\right\rangle }%

\global\long\def\bra#1{\left\langle #1 \right|}%

\global\long\def\bs#1{\boldsymbol{#1}}%

\def\ba#1\ea{\begin{align}#1\end{align}}  

\definecolor{colorhhy}{rgb}{0.9, 0.17, 0.31}

\definecolor{colorlmr}{rgb}{0.1, 0.2, 0.7}

\newcommand{\ie}{{\it i.e.}}

\def\rr{\mathbf{r}}

\makeatother

\begin{document}

\preprint{APS/LL spectroscopy}
\title{Distinguishing Mott and strain-induced anisotropic semimetals in magic-angle twisted bilayer graphene using Landau level spectroscopy}

\author{Keshav Singh}
\affiliation{National High Magnetic Field Laboratory, Tallahassee, Florida, 32310}
\email{ksingh6@fsu.edu}
\affiliation{Department of Physics, Florida State University, Tallahassee, Florida, 32306}
\author{Juan Felipe Mendez-Valderrama}
\affiliation{Department of Physics, Princeton University, Princeton, New Jersey 08544, USA}
\author{Erez Berg}
\affiliation{Department of Condensed Matter Physics, Weizmann Institute of Science, Rehovot 76100, Israel}
\affiliation{Materials Department, University of California Santa Barbara, Santa Barbara 93106 USA}
\affiliation{Department of Electrical and Computer Engineering, University of California, Santa Barbara, CA 93106,
USA}
\author{B. Andrei Bernevig}
\affiliation{Department of Physics, Princeton University, Princeton, New Jersey 08544, USA}
\affiliation{Donostia International Physics Center, P. Manuel de Lardizabal 4, 20018 Donostia-San Sebastian, Spain}
\affiliation{IKERBASQUE, Basque Foundation for Science, Bilbao, Spain}

\author{Oskar Vafek}
\affiliation{W.~I.~Fine Theoretical Physics Institute and School of Physics and Astronomy,
University of Minnesota, Minneapolis, Minnesota 55455, USA}
\email{vafek@umn.edu}

% !TEX root = ./main.tex

% \maketitle
% \paperAuthors

\begin{abstract}
Recent quantum twisting microscopy experiments \cite{inbar2023quantum} have provided unprecedented momentum- and energy-resolved imaging of magic-angle twisted bilayer graphene, revealing a stark dichotomy between the heavy and light electronic characters of its narrow bands. In this work, we provide a sharp distinguishing criterion between the two most likely candidate states that exhibit spectroscopic features consistent with the experimental results. Focusing on charge neutrality, these states are 1) the `Mott semimetal', governed by strong dynamical correlations, and 2) the `strain-induced anisotropic semimetal', characterized by weak Hartree-Fock effects. We demonstrate that an out-of-plane magnetic field serves as a sharp discriminating probe based on the expected 
%gap hierarchy of the magnetic field induced Chern insulators. 
Landau level gap size hierarchy. We find that the strain-induced anisotropic semimetal has a robust gap hierarchy where the Chern $C$ gaps, $\Delta_C$, decrease in order $\Delta_{\pm4}>\Delta_{\pm8}>\Delta_{\pm12}$. In contrast, for the Mott semimetal,  $\Delta_{\pm4}$ remains the largest gap, while the relative magnitude of the $C=\pm 8 ,\pm 12$ gaps depends sensitively on the applied magnetic field. We test the robustness of these results against strain orientation and lattice relaxation finding consistency between the gap hierarchy of the strained semimetal and previous incompressibility measurements at finite magnetic field \cite{yu2022correlated}. 
\end{abstract}
\maketitle

The initial discovery of  non-trivial topological narrow bands \cite{bistritzer2011moire,ahn2019failure,po2019faithful,song2019all,PhysRevB.103.205412,cao2018correlated,cao2018unconventional} in magic-angle twisted bilayer graphene (MATBG) has spurred intense research \cite{PhysRevX.8.031088,PhysRevX.8.031087,PhysRevX.8.031089,PhysRevX.8.041041,PhysRevLett.121.257001,ahn2019failure,PhysRevLett.122.246401,PhysRevB.104.075143,PhysRevB.98.085435,PhysRevLett.122.106405,PhysRevLett.122.257002,PhysRevLett.124.046403,
PhysRevLett.124.097601,PhysRevB.102.035161,PhysRevLett.124.166601,
PhysRevX.10.031034,vafek2020renormalization,PhysRevB.103.205411,PhysRevB.103.205412,PhysRevB.103.205413,PhysRevB.103.205414,PhysRevB.103.205415,PhysRevB.103.205416,khalaf2021charged,parker2021strain,PhysRevResearch.3.013033,PhysRevLett.127.266402,PhysRevX.11.041063,PhysRevLett.129.117602,ghawri2022breakdown,hofmann2022fermionic,2023PhRvB.107g5156X,wang2024theory} into its intricate phase diagram, which hosts strongly correlated states \cite{kerelsky2019maximized,xie2019spectroscopic,sharpe2019emergent,jiang2019charge,choi2019electronic},  superconductivity, \cite{yankowitz2019tuning,lu2019superconductors,stepanov2020untying,saito2020independent,kim2022evidence} and a plethora of topological phases \cite{tomarken2019electronic,cao2020strange,lu2021multiple,lyu2021strange,hesp2021observation,rozen2021entropic,saito2021isospin,hubmann2022nonlinear,paul2022interaction,jaoui2022quantum,diez2023symmetry,wang2023unusual}.  The topological heavy fermion (THF) framework  \cite{song2022magic,cualuguaru2023twisted,singh2024topological,herzog2025topological,herzog2025kekule} reconciles the signatures of local moment physics \cite{2019Natur.572..101X,benschop2021measuring,nuckolls2023quantum,rozen2021entropic,saito2021isospin,zhang2025heavy} and delocalized topological itineracy \cite{sharpe2019emergent,tomarken2019electronic,yankowitz2019tuning,lu2019superconductors,zondiner2020cascade,stepanov2020untying,saito2020independent,kim2022evidence} in MATBG, through a unified principle. In analogy to heavy fermion materials, the THF model recasts the low energy degrees of freedom in MATBG in terms of a set of heavy $p_x\pm ip_y$-like Wannier states, localized at the AA stacking sites, and topological conduction fermions, denoted by $f$ and $c$, respectively. The narrow topological bands of MATBG are spanned by the $f$-electron orbitals throughout most regions in momentum space, except in the vicinity of the $\Gamma$ point of the moir\'e Brillouin zone (mBZ), where they acquire a sharp itinerant $c-$fermion character. 
%This dual fermionic nature of the carriers in moir\'e graphene is consistent with the experimental reports in \cite{zhang2025heavy, merino2025interplay,batlle2025photovoltage,kim2026resolving}. 

\begin{figure}
    \centering
    \includegraphics[width=1\linewidth]{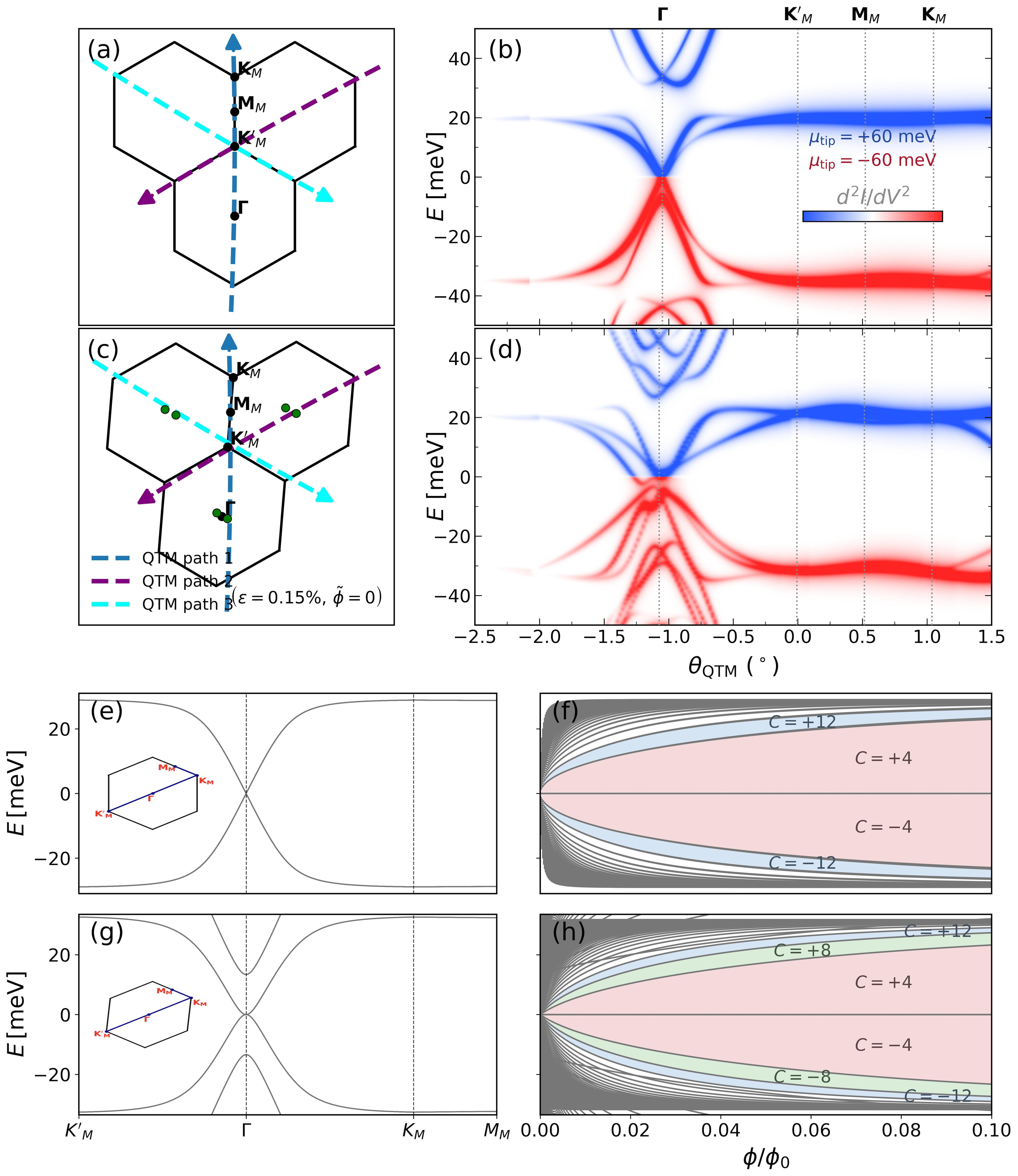}
    \caption{The paths in reciprocal space scanned by the quantum twisting microscope
(QTM) and the expected $d^2I/dV^2$ maps for the (a,b) relaxed Mott and the (c,d) relaxed strain-induced anisotropic semimetals. The $d^2I/dV^2$ map is plotted as a function of band energy $E$ and the relative angle between the top graphene layer of the MATBG sample and the probing monolayer graphene tip, labeled by $\theta_{QTM}(^\circ)$, where we use the result obtained in  \cite{wei2025dirac} (also see Appendix(\ref{Appx:Robust_Strain_Semimetal})). We set the sample chemical potential $\mu_{\rm{sample}}=0$, such that the corresponding bias between the tip and sample is given by $-eV = E+\mu_{\rm tip}$\cite{wei2025dirac} where $\mu_{\rm{tip}}$ is the tip chemical potential.  We mark the high symmetry points that are closest to the QTM path 1 as $\theta_{QTM}$ varies (also marked in b an d). The $d^2I/dV^2>0$  (in red) for $\mu_{\rm tip}=-60$ meV while $d^2I/dV^2<0$  (in blue) for $\mu_{\rm tip}=+60$ meV, respectively. The location of Dirac points for the strain-induced anisotropic semimetal are shown in green in (c). In contrast to the $d^2I/dV^2$ response for the Mott semimetal, the three paths in reciprocal space shown in (b) do not contribute identically for the strain-induced anisotropic semimetal, as $C_{3z}$ symmetry is broken by heterostrain\cite{wei2025dirac}. The excitation spectrum and its continuation in $\bB$ for the unrelaxed flat-chiral limit of (e, f) Mott and (g, h)strain-induced anisotropic semimetals. Here $\phi/\phi_0$ represents the total number of magnetic flux quanta ($hc/e$) piercing the moir\'e unit cell.}
    \label{fig:d^2I/dV^2_main_text}
\end{figure}

While this dual fermionic character of the carriers in moir\'e graphene is consistent across multiple experimental reports \cite{zhang2025heavy, merino2025interplay,batlle2025photovoltage,kim2026resolving}, we specifically focus on its manifestation in recent studies using the quantum twisting microscope (QTM) \cite{inbar2023quantum}. The unprecedented momentum- and energy-resolved spectroscopy of the interaction-renormalized narrow bands of MATBG using QTM directly confirms the momentum-dependent light and heavy fermionic nature of these bands. In this work, we focus on the semimetallic state reported at charge-neutrality point (CNP) at $T=4$K. In QTM, this
state features a set of narrow bands, separated by a large gap of $\sim$ 35 meV, over a wide region of the mBZ except in the vicinity of $\Gamma$ where the gap closes \cite{xiao2026imaging}. Although the spectroscopic features clearly identify the dichotomy between the $f$ and $c$ fermions, the nature of the ground state - and of its parent state at non-zero temperature - is not completely determined by the spectroscopic signatures in QTM \cite{xiao2026imaging}. The calculated $d^2 I/dV^2$ QTM response \cite{wei2025dirac} for the two leading energetically competitive and distinct candidate phases, namely the Mott and strain-induced anisotropic semimetals, are shown in Figs.(\ref{fig:d^2I/dV^2_main_text}b) and (\ref{fig:d^2I/dV^2_main_text}d). The calculated $d^2 I/dV^2$ QTM response for both semimetals features flat bands in most regions of momentum space except in vicinity of $\Gamma$ or $\theta_{QTM}=-1^\circ$ where they dip toward neutrality, which is consistent with the QTM report \cite{inbar2023quantum}. By contrast, we find that the two states can be distinguished in the presence of an out-of-plane magnetic field $\mathbf{B}=B\hat{z}$  based on the magnitudes of the expected Chern gaps revealed by Landau level (LL) spectroscopy. Although the technique in \cite{singh2024topological} directly applies to the analysis of Landau quantization of symmetry broken states of the THF model in the absence of dynamical correlations, we find that it naturally admits an extension enabling similar analysis of generic heavy fermion systems with local self-energies \cite{aux_map_B2026}. 
We thus employ it to examine the Landau quantization of the Mott semimetal.  

\begin{figure}
    \centering
    \includegraphics[width=0.99\linewidth]{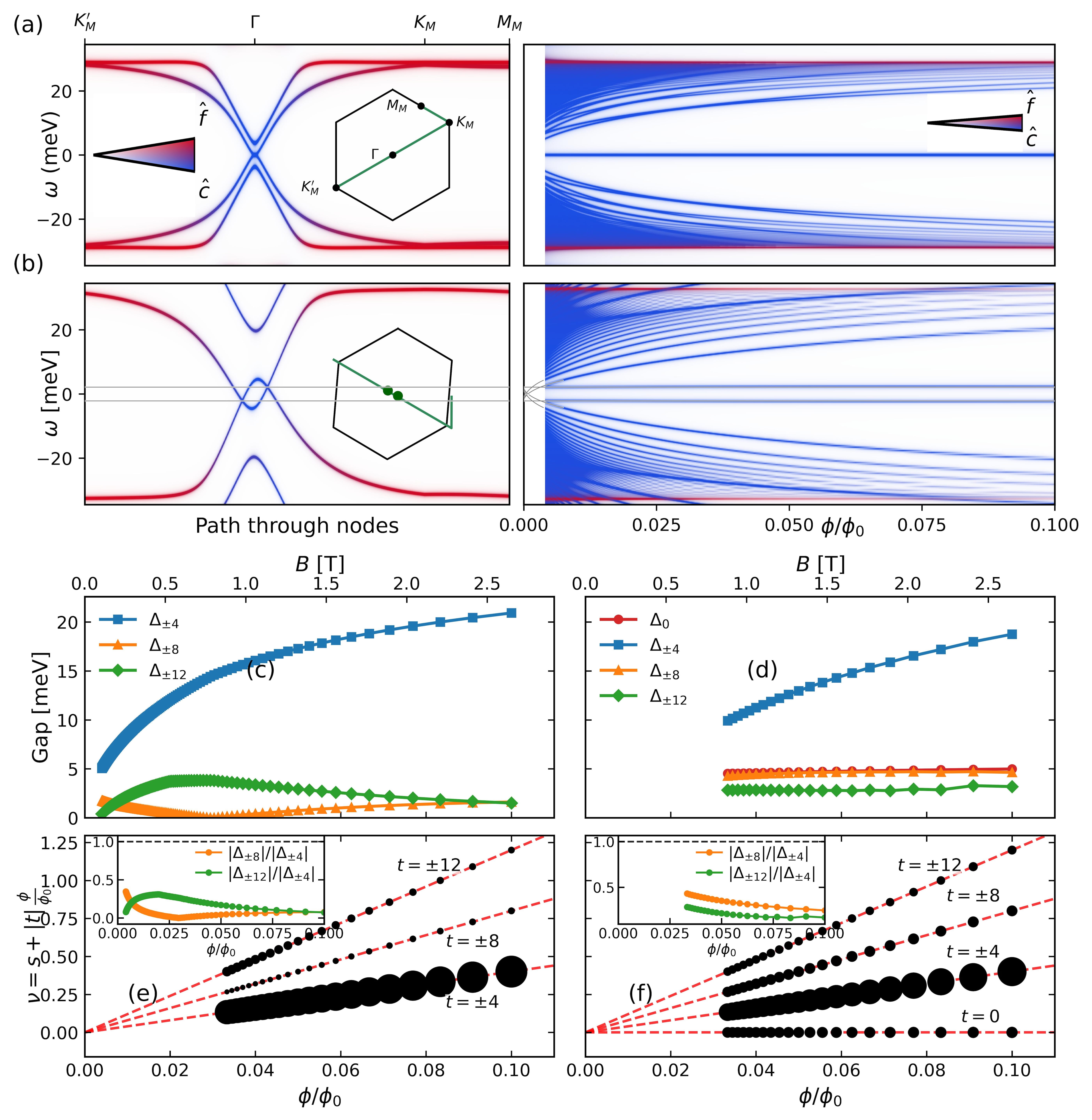}
    \caption{The orbital-resolved spectral function and its continuation in $\bB$ for (a) Mott and (b) strain-induced anisotropic semimetals in absence of the Zeeman and lattice relaxation effects. At a given $\bk$ (for $B=0$) or $\phi/\phi_0$ (for $B\neq0$), the spectral function is given by $A(\omega)=\sum_{n}P_{\rm{phy}}|u_n\rangle\langle u_n|P_{\rm{phy}}\delta(\omega-\epsilon_n)$, where $n$ labels the normalized eigenstates $|u_n\rangle$ with eigenvalues $\epsilon_n$ obtained at that $\bk$ or $\phi/\phi_0$, and $P_{\rm{phy}}$ is the projector onto the (non-auxiliary) physical fermions. In the used color scheme, we encode the orbital character via $A_{c(f)}=\rm{Tr}[P_{c(f)}A(\omega)]$, where $P_{c(f)}$ is the projector onto $c(f)$ orbitals. 
    A large $A_c$ ($A_f$) shifts the pixel color toward blue (red). Comparable weight in both orbitals produces intermediate hues, while regions with negligible spectral weight remain close to white (see Appendix(\ref{Appx:Spectral_map_color})).
    Only the excitations from valley $\bK$ are included in the left panel of (b) and the two grey solid lines mark the energies of the two Dirac nodes. The grey solid lines in the right panel of (b) show the zero modes and the $\sqrt{B}$ LLs\cite{vafek2014dirac} obtained from semiclassical Onsager quantization near the Dirac nodes and continuously merge with the numerically obtained spectrum (details in Appendix(\ref{Appx:Semiclassical:Prescription})). The magnitudes $\Delta_{t}$ of resulting Chern $t=\pm 4,\pm 8,\pm 12$ gaps and $t=0,\pm 4,\pm 8,\pm 12$ gaps are shown in (c) and (d) for the Mott and strain-induced anisotropic semimetals respectively. The ratio of the gaps $\Delta_{\pm 8}/\Delta_{\pm 4}$ and $\Delta_{\pm 12}/\Delta_{\pm 4}$ are shown as insets in (e) and (f) for the Mott and strain-induced anisotropic semimetals respectively. Single-electron excitation gap as a function of electron filling $\nu =s + t\frac{\phi}{\phi_0}$ along (e) $t=\pm 4, \pm8,\pm12$ and (f)  $t=0,\pm 4, \pm8,\pm12$ at $s=0$ (CNP) for Mott and strain-induced anisotropic semimetals respectively. The radius of the dots scales with magnitude of the gap. The gap magnitudes in (d) and (f) were obtained using self-consistent Hartree-Fock (details in Appendix.(\ref{appx:One_shot_in_B_strain})). While $\Delta_{\pm4}\gg\Delta_{\pm 8}>\Delta_{\pm 12}$ for the strain-induced anisotropic semimetal, we find $\Delta_{\pm4}\gg\Delta_{\pm 12}\gtrsim \Delta_{\pm 8}$ for the Mott semimetal.}
    \label{fig:main:fig2}
\end{figure}

\emph{Leading Candidates and spectral function at zero magnetic field ---} 
The first candidate is the Mott semimetal, i.e., the symmetric, strain-free, state \cite{hofmann2022fermionic,rai2024dynamical,cualuguaru2025obtaining,hu2025projectedsolvabletopologicalheavy,nonlocalmoments2025} of the THF model at CNP. 
This candidate 
describes the fluctuating local moment regime, i.e. when the temperature is much below the leading charge excitation energy scale $ U_1\gg k_B T$, but still above any ordering temperature, leaving behind fluctuating local spin, valley and orbital moments. 
We examine this state within the Hubbard-I approximation, wherein the local dynamical self-energy of the $f$-modes is approximated by its atomic limit, which captures the physics of Hubbard-band formation. Despite this, the spectrum is gapless at the $\Gamma$ 
point due to itinerant $c$-fermions whose double $\Gamma_3$ representation remains unbroken in the symmetric state.
Since the Hubbard-I self-energy is purely local,  the Green's function zeros do not generate additional winding \cite{GurarieGreens2011} and the spectrum can efficiently be characterized through an effective non-interacting auxiliary-fermion construction \cite{cualuguaru2025obtaining}. At valley $\bK$ the corresponding auxiliary Hamiltonian is
\begin{eqnarray}
    H^\bK_{\mathrm{aux}}(\bk) = \left(\begin{array}{ccc}
        h^{cc}(\bk) & h^{cf}(\bk)e^{-\frac{\bk^2\lambda^2}{2}}  &0_{4\times 2}\\
      e^{-\frac{\bk^2\lambda^2}{2}} h^{cf\dagger}(\bk)  & 0_{2\times 2} &\frac{U_1}{2}\sigma_0\\
      0_{2\times 4}  & \frac{U_1}{2}\sigma_0 &0_{2\times 2}
\end{array}\right)\label{Eq:main:Symmetric_state_} 
\end{eqnarray}
where $\lambda=0.3L_m$ is the width of the Wannier states where $L_m$ is the moir\'e lattice period and $\sigma_0$ is the identity matrix $1_{2\times 2}$. The upper $4\times 4$ block acts on the four $c$ orbitals 
%labeled by  $a=\{1,\dots,4\}$. Here $a=\{1,2\}$ and $a=\{3,4\}$ $c$-fermions
which form the $\Gamma_3$ and $\Gamma_1\oplus\Gamma_2$ irreps at $\Gamma$\cite{song2022magic}, respectively. The middle $2\times2$ block acts on the two $p_x\pm i p_y$-like  $f$-Wannier states. %labelled by $b=\{1,2\}$. 
The lowest $2\times2$ block acts on the auxiliary fermion degrees of freedom. The single-particle THF couplings are given by \cite{song2022magic} 
\begin{equation}
    h^{cc} =\left(\begin{array}{cc}
        0_{2\times 2} &v_\ast\bk\cdot(\sigma_0,i\sigma_z)  \\
      h.c.   & M\sigma_x
    \end{array}\right),~
    h^{cf} = \left(\begin{array}{cc}
       \gamma\sigma_0 + v_\ast'\bk \cdot\sigma   \\0_{2\times2}
   \end{array}\right),\label{Eq:main:THF:coupling}
\end{equation}
where $h.c.$ represents Hermitian conjugate. The model at valley $\bK'$ is related by time reversal.
The resulting orbital-resolved spectral function is shown in the left panel of Fig.(\ref{fig:main:fig2}a).

The second candidate we consider is the weakly dynamic Hartree-Fock and strain-induced anisotropic semimetal at $T=0$ \cite{kang2020non,liu2021nematic, parker2021strain, PhysRevX.11.041063,herzog2025kekule,crippa2025dynamical,keshet2026controlledloopexpansionstrained}. We focus on the effects of uniaxial heterostrain, as it produces the most significant distortion of the moir\'e lattice \cite{Kazmierczak2021,ZhenStrain2019}. We start by noting that the single-particle strained THF is given by \cite{herzog2025topological}
\begin{eqnarray}
    H^\bK_{strain} =\left(\begin{array}{cc}
        h_\epsilon^{cc}(\bk) & h^{cf}_\epsilon(\bk)e^{-\lambda^2\bk^2/2} \\
      h.c.   & h^{ff}_\epsilon
    \end{array}\right)\label{Eq:Main:Strain}
\end{eqnarray}
where $h_\epsilon^{cc}=h^{cc}+\tilde{h}_\epsilon^{cc}$, $h_\epsilon^{cf}=h^{cf} +\tilde{h}_\epsilon^{cf}$ with 
\begin{equation}
    \tilde{h}^{cc}_\epsilon =\left(\begin{array}{cc}
      c\boldsymbol{\epsilon}\cdot\boldsymbol{\sigma}  &  c'\boldsymbol{\epsilon}\cdot\boldsymbol{\sigma^\ast}\\
       h.c. & M'\epsilon_+\sigma_y
   \end{array}\right),\, \tilde{h}^{cf}_\epsilon = \left(\begin{array}{cc}
        i\gamma'\epsilon_+\sigma_z\\
       c''\boldsymbol{\epsilon}\cdot(\sigma_0,-i\sigma_z)
   \end{array}\right)
\end{equation} 
and $h^{ff}_\epsilon =M_f\boldsymbol{\epsilon}\cdot\boldsymbol{\sigma}$
at valley $\bK$. The model at $\bK'$ is related by time reversal. Here $\boldsymbol{\epsilon}=(\epsilon_{xy},\epsilon_-)=(\sin2\tilde{\phi}, -\cos2\tilde{\phi})\epsilon_a$ and $\epsilon_+= \epsilon(\nu_G-1)/2$, where $\epsilon_a=\epsilon(\nu_G+1)/2$,  $(\epsilon,\tilde{\phi})$ parametrizes the magnitude and orientation of the heterostrain axis, such that $\tilde{\phi}=0$ is the x-axis and $\nu_G=0.16$ is the Poisson ratio of graphene. 
For generic directions of heterostrain, both $C_{2x}$ and $C_{3z}$ symmetries are broken explicitly while all other symmetries of MATBG are preserved \cite{herzog2025topological}. 
The breaking of $C_{3z}$ by $h^{ff}_\epsilon$ 
lifts the degeneracy of the underlying $p_x \pm ip_y$-like Wannier states and splits the $f$-modes\cite{herzog2025topological}. 
At $T=0$, at CNP, half of the $f$-electrons are occupied \cite{herzog2025kekule} while the rest remain above the chemical potential. On-site interactions then enhance the gap via Fock exchange\cite{herzog2025kekule}. In particular, the on-site mean-field (MF) interaction preserves both the valley $U(1)$ and spin $SU(2)$ symmetries and takes the form
$H^{\mathrm{strain}}_{U_1}
=\frac{U_1}{2}\left(\sin 2\tilde{\phi}\,\sigma_x-\cos 2\tilde{\phi}\,\sigma_y\right)$, at CNP in valley $\bK$, where $\sigma$ acts in the subspace spanned by $p_x\pm ip_y$ Wannier states. In other words, the MF interactions are incorporated by adding  $H^{\mathrm{strain}}_{U_1}$ to $h^{ff}_\epsilon$ in Eq.(\ref{Eq:Main:Strain}). Thus, away from the $\Gamma$ point, where the band structure is dominated by $f$ fermions, we find the high-energy flat bands $\sim \pm 35$ meV (gap of $M_f\epsilon(\nu_G+1)+U_1\sim 70$ meV).
By contrast, the bands acquire a sharp $c$-fermionic nature in the vicinity of $\Gamma$ which hosts the gapless Dirac nodes. These Dirac nodes, previously pinned at the corner of the moir\'e Brillouin zone (mBZ) by $C_{3z}$ in the strainless and non-interacting THF, now migrate towards the $\Gamma$ point and remain protected by $C_{2z}\mathcal{T}$ which is preserved by uniaxial heterostrain \cite{herzog2025topological}. Consequently, the resulting bands are characterized by a large bandwidth everywhere in the mBZ except in the vicinity of $\Gamma$, where the symmetry-protected gapless Dirac nodes reside\cite{herzog2025topological}. The resulting orbital-resolved spectral function is shown in the left panel of Fig.(\ref{fig:main:fig2}b). 

\emph{Landau level spectrum and hierarchy of Chern-gap magnitudes---}
The sharp distinction between the Chern-gap hierarchies of the two candidates admits the simplest interpretation in the flat-chiral limit $v_\ast'=M=0$ of the THF model. 
In this limit, the spectrum of the Mott semimetal is exactly solvable and doubly degenerate (per valley per spin) at $\bB=0$. The lowest energy single-particle charge $\pm 1$ excitations take the form 
 \begin{eqnarray}\label{Eq:main:E_mott_chiral}
    && E^{\rm{flat-chiral}}_{\rm{Mott}}(\bk)=\nonumber\\&&\pm
   \frac{1}{\sqrt{2}} \sqrt{v_\ast^2\bk^2+\frac{U_1^2\mathcal{A}_\bk}{4}
-\sqrt{\left(v_\ast^2\bk^2+\frac{U_1^2\mathcal{A}_\bk}{4}\right)^2-v_\ast^2\bk^2U_1^2}}\nonumber\\
 \end{eqnarray}
 where $\mathcal{A}_\bk=1+4\gamma^2e^{-\bk^2\lambda^2}/U_1^2$, and asymptotes to  $\pm |v_\ast\bk|/\sqrt{\mathcal{A}_\mathbf{0}}$ as $\bk\rightarrow0$. In other words, the spectrum near Fermi level is that of a doubly degenerate massless Dirac particle with velocity $|v_\ast|/\sqrt{\mathcal{A}_\mathbf{0}}$ (see Fig.(\ref{fig:d^2I/dV^2_main_text}e)). To obtain the single-particle excitation spectrum at $\bB\neq 0$ we promote\cite{singh2024topological}
 $v_\ast(k_x-ik_y)$ and $e^{-\bk^2\lambda^2/2}$ to $ i\Delta_B\hat{a}^\dagger$ and $\Sigma(\hat{a}^\dagger\hat{a})$, respectively, in Eq.(\ref{Eq:main:Symmetric_state_}).  Here $\hat{a}$ is simple harmonic oscillator (h.o.) annihilation operator with simple h.o. eigenstate $|m\rangle$, $\Delta_B=\sqrt{2}\frac{v_\ast}{\ell}$, $\Sigma(m)\equiv\Sigma_m\approx e^{-(m+\frac{1}{2})\frac{\lambda^2}{\ell^2}}$ and $\ell=\sqrt{\frac{\hbar c}{eB}}$ is the magnetic length. The resulting operator is block diagonalized in the ansatz 
\begin{eqnarray}
    \Theta^{(1)}_{n}= [c^{n}_1|n-1\rangle,c^{n}_2|n-2\rangle,c^{n}_3|n\rangle,c^{n}_4|n-3\rangle,\nonumber\\c^{n}_5|n-1\rangle,c^{n}_6|n-2\rangle,c^{n}_7|n-1\rangle,c^{n}_8|n-2\rangle]^T
\end{eqnarray}
where $n\geq0$ and $c^{n<1}_1,c^{n<2}_2, c^{n<3}_4,c^{n<1}_{5,7},c^{n<2}_{6,8}=0$. The coefficients $c^{n}_{7,8}$ represent the weight on auxiliary fermions. The two Dirac nodes yield two zero modes with non-zero coefficients $\{c^{0}_3=1\}$ and $\{c^{2}_2=-U_1, c^{2}_8=2\gamma\Sigma_0\}$. The non-zero modes arising from Landau quantization of the single-particle charge $\pm 1$ excitations are almost degenerate and approximately given by Eq.(\ref{Eq:main:E_mott_chiral}) with $\mathcal{A}_\bk$ and  $v_\ast^2\bk^2$ replaced by $\mathcal{A}_\mathbf{0}$ and $\Delta_B^2n$ ($n\geq 1$), respectively. Their splitting is extremely small for the range $B$ analyzed in this work and is set by $\frac{\lambda^2}{\ell^2}\ll1$. Their energy vanishes as $\pm \sqrt{2n}|v_\ast|/\ell\sqrt{\mathcal{A}_{\mathbf{0}}}$ at low $\bB$, which is expected from Landau quantization of a massless Dirac particle. Including the spin-valley degeneracy of each of these modes, the leading gaps thus have Chern number $\pm4,\pm 12$ with $\Delta_{\pm4}\gg\Delta_{\pm 12}$, as can be seen in Fig.(\ref{fig:d^2I/dV^2_main_text}f). 

Having illustrated the flat-chiral limit for the Mott semimetal we now analyze the flat-chiral limit for the strain-induced anisotropic semimetal. For simplicity we only consider the $C_{3z}$ (and generically $C_{2x}$) breaking coupling $M_f$ and the $C_{2x}$ breaking coupling $\gamma'$.
The resulting single-particle excitation spectrum is independent of $\tilde{\phi}$. At $\bK$ it is  obtained by computing eigenvalues of  $\left(\begin{array}{cc}
   0_{3\times 3}  & \mathcal{D}(\bk) \\
   \mathcal{D}^\dagger(\bk)   &  0_{3\times 3}
\end{array}\right)$ where
\begin{figure}
    \centering
    \includegraphics[width=0.99\linewidth]{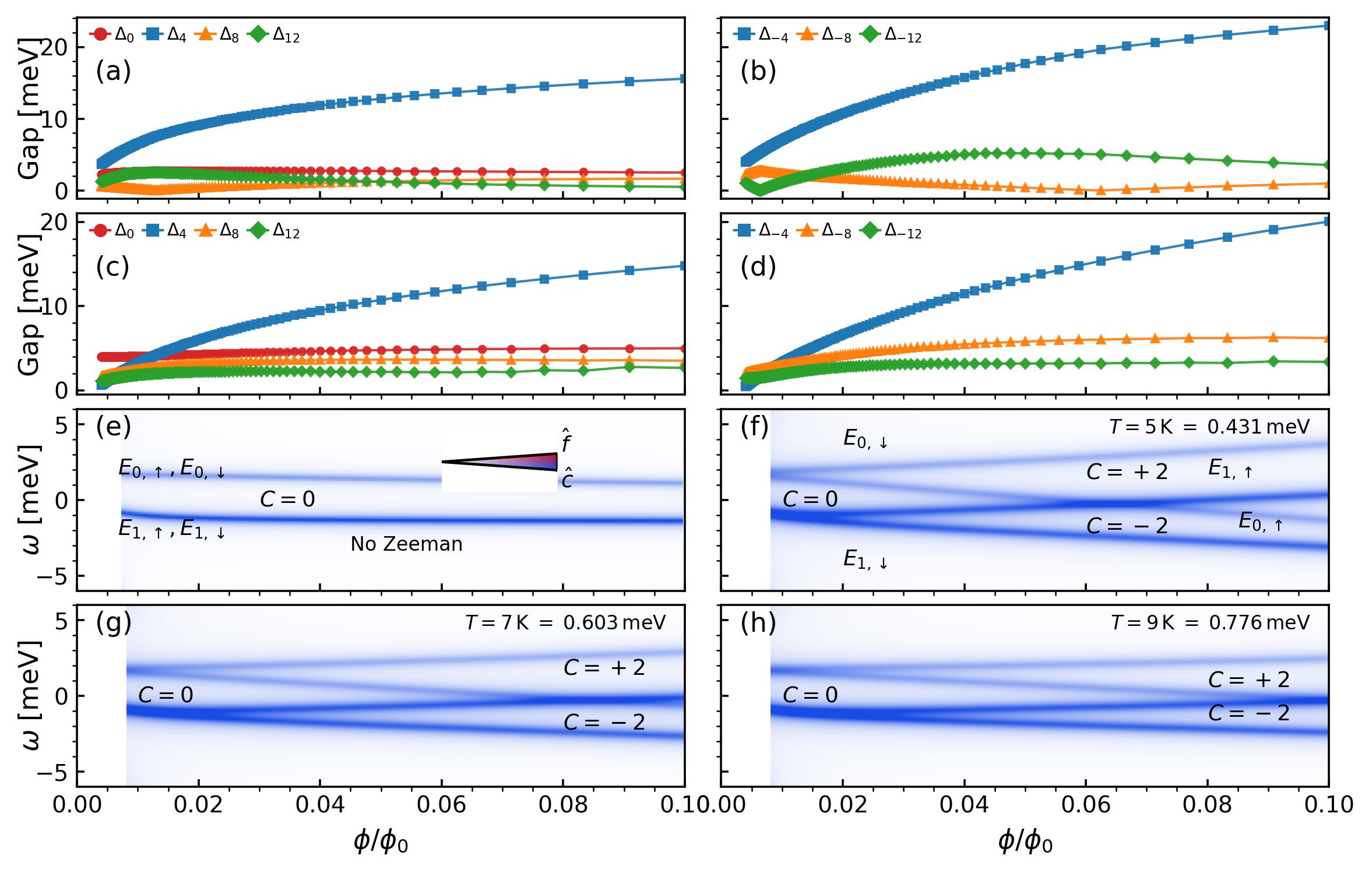}
    \caption{(a-d) The magnitudes of Chern gaps for the (a,b) Mott and (c,d) strain-induced anisotropic semimetal in the presence of lattice relaxation effects. (e-h) The spectral function of the valley degenerate anomalous modes $\{E_{0,\uparrow\downarrow},{E_{1,\uparrow\downarrow}}\}$ of the Mott
    semimetal including lattice relaxation effects (e) without Zeeman and (f-h) with Zeeman
    for  $T=5K,7K$ and $9K$ as a function of $\phi/\phi_0$ and energy (see Appendix(\ref{sec:LLzeemann})). The $C=\pm 2$ gaps formed via Zeeman induced splitting marked. The spectral weight of the modes at $E_{1,\uparrow\downarrow}$ is $\sim (1+\mathcal{A}_\ell)/(\mathcal{A}_\mathbf{0}+\mathcal{A}_\ell)$, which is larger than the spectral weight of the modes at $E_{0,\uparrow\downarrow}$ which is $\sim \mathcal{A}^{-1}_{\mathbf{0}}$, where $\mathcal{A}_\ell = 2 (\frac{v_\ast}{M\ell})^2$ (see Appendix(\ref{sec:anomalous})). Thus modes $E_{1,\uparrow\downarrow}$ appear with enhanced brightness.}
    \label{fig:main:Relaxation:Zeeman}
\end{figure}
\begin{eqnarray}
    \mathcal{D}(\bk) =\left(\begin{array}{ccc}
        v_\ast(k_x-ik_y) & 0 &0\\
        \tilde{\gamma}e^{-\bk^2\lambda^2/2} & i\tilde{U}_f/2 &0\\
       0  & \tilde{\gamma}e^{-\bk^2\lambda^2/2} &v_\ast(k_x-ik_y)
    \end{array}\right)
\end{eqnarray}
with $\tilde{\gamma}=\gamma-i\gamma'\epsilon_+$ and $\tilde{U}_f=\left(2M_{f}\epsilon_a+U_1\right)$ (see Appendix(\ref{Appx:Simple:model:Strain})). In contrast to the doubly degenerate Dirac crossing for the Mott semimetal, the lowest energy charge $\pm 1$ excitations here asymptote to $\pm |v_\ast\bk|^2 \tilde{U}_f/4|\tilde{\gamma}|^2$ as $\bk\rightarrow 0$, and form a quadratic band touching at $\Gamma$ (see Fig.(\ref{fig:d^2I/dV^2_main_text}g)). To obtain the single-particle excitation spectrum at $\bB\neq 0$
we again substitute \cite{singh2024topological} 
$v_\ast(k_x-ik_y),e^{-\bk^2\lambda^2/2}\rightarrow i\Delta_B\hat{a}^\dagger,\Sigma(\hat{a}^\dagger\hat{a})$ in $\mathcal{D}(\bk)$ (see Appendix(\ref{Appx:Simple:model:Strain})). The resulting operator is block diagonalized in the ansatz 
\begin{eqnarray}
    \Theta^{(2)}_{n}= [c^{n}_1|n\rangle,c^{n}_2|n-1\rangle,c^{n}_3|n-1\rangle,c^{n}_4|n-1\rangle,\nonumber\\c^{n}_5|n-1\rangle,c^{n}_6|n-2\rangle]^T
\end{eqnarray}
where $n\geq 0$ and $c^{n<1}_{2,3,4,5},c^{n<2}_6=0$. We obtain two zero modes with non-zero coefficients $\{c^{0}_1=1\}$ and $\{c^{1}_1=-2i\tilde{\gamma}^{\ast 2}\Sigma^2_0, c^{1}_2=2\tilde{\gamma}^{\ast}\Sigma_0\Delta_B, c^{1}_3=i\tilde{U}_f\Delta_B\}$, respectively. The LLs emerging from the Landau quantization of the charge $\pm 1$ excitation bands approach  $\pm  \sqrt{n(n+1)}\Delta_B^2\tilde{U}_f/2|\tilde{\gamma}|^2$ as $\bB\rightarrow 0$ ($n\geq 1$), as anticipated from Landau quantization of a quadratic band touching. For $\phi/\phi_0\gtrsim 0.06$ where $|\tilde{\gamma}|\ll\Delta_B$, the energies of these singly degenerate non-zero modes is approximately given by Eq.(\ref{Eq:main:E_mott_chiral}) with replacements $\mathcal{A}_\bk,v_\ast^2\bk^2\rightarrow\mathcal{A}_{\mathbf{0}},n\Delta^2_B$ and $U_1,\gamma^2\rightarrow\tilde{U}_f,|\tilde{\gamma}|^2$ (see Appendix(\ref{Appx:Simple:model:Strain})). Including the spin and valley degeneracy, we thus note that the leading gaps have Chern number $\pm4,\pm 8,\pm 12$ with $\Delta_{\pm4}\gg\Delta_{\pm 8}>\Delta_{\pm 12}$, as can be seen in Fig.(\ref{fig:d^2I/dV^2_main_text}h).

We now briefly examine the effects of reinstating the terms omitted in the Eqs.(\ref{Eq:main:Symmetric_state_}) and (\ref{Eq:Main:Strain}) on the hierarchy of leading Chern gaps of the two semimetals. Recall that, for $M,v_\ast'=0$ the single-particle charge $\pm 1$ excitations of the Mott semimetal are formed by (almost) degenerate LL pairs. Their energy is approximately given by Eq.(\ref{Eq:main:E_mott_chiral}) with the substitution $\mathcal{A}_\bk,v_\ast^2\bk^2\rightarrow \mathcal{A}_\mathbf{0},n\Delta_B^2$ $(n\geq 1)$. We denote these energies by $\pm \mathcal{E}_{n}$ in the remainder of this paragraph. 
A finite $M,v_\ast'\neq 0$ splits the double degeneracy at $\pm \mathcal{E}_{n}$. In particular, the splitting at $\pm\mathcal{E}_{1}$  gives rise to the  $C=\pm 8$ gap 
(see right panel of Fig.(\ref{fig:main:fig2}a)). The modified energies upto $\mathcal{O}(v_\ast'^2)$ and $\mathcal{O}(M^2)$ are given by $\pm(\mathcal{E}_{1}+E^{A}_{corr,LL})$ and $\pm(\mathcal{E}_{1}+E^{B}_{corr,LL})$ where (see Appendix(\ref{sec:m-one-low-LL-reduction})) 
\begin{align}
    E^{A}_{corr,LL}&= \frac{2(M\mathcal{B}_1\mathcal{B}_2 )^2\mathcal{E}_{1}}{(\mathcal{E}_{1}^2-\mathcal{E}_{2}^2)}\\
    E^{B}_{corr,LL}&= \frac{M^2\mathcal{B}_1^2}{\mathcal{E}_1} + \frac{2(\Delta_B^\prime\gamma \mathcal{E}_3\mathcal{E}_1)^2}{\tilde{N}^2_1\tilde{N}^2_3}\sum_{\zeta=\pm}\frac{(\mathcal{D}_3\mathcal{E}_1+\zeta \mathcal{D}_1\mathcal{E}_3)^2}{\mathcal{E}_1-\zeta\mathcal{E}_3}
\end{align}
with  $\Delta_B^\prime= \frac{\sqrt{2}v_\ast'}{\ell}$, $\mathcal{B}_n=\frac{\sqrt{n}\Delta_B\mathcal{D}_n}{N_n}$, $\mathcal{D}_n=\mathcal{E}_n^2-\frac{U_1^2}{4}$ and $\tilde{N}_n=\left(\mathcal{D}_n^2(\mathcal{E}_n^2+n\Delta_B^2)+\gamma^2\mathcal{E}_n^2(\mathcal{E}_n^2+\frac{U_1^2}{4})\right)^{1/2}$. At $\phi/\phi_0\sim 0.03$, where $E^{A}_{corr,LL}= E^{B}_{corr,LL}$, the $C=\pm 8$ gap vanishes (see Fig.(\ref{fig:main:fig2}c)).
 More generally, $\Delta_{C=\pm 8}\lesssim \Delta_{C=\pm 12}$ except for very small fluxes $\phi/\phi_0<0.01$, and the hierarchy $\Delta_{C=\pm 4}\gg\Delta_{C=\pm 12}$ is preserved. 
 
We now discuss the effects of reinstating the terms omitted in Eq.(\ref{Eq:Main:Strain}) for the strain-induced anisotropic semimetal. While the Chern gap hierarchy $\Delta_{\pm4}\gg\Delta_{\pm 8}>\Delta_{\pm 12}$ remains robust, we find that a $C=0$ gap emerges (see Fig.(\ref{fig:main:fig2} b,d,f)). To examine the origin of this gap, we consider the following $\bk\cdot\bp$ expansion of the lowest energy charge $\pm 1$ excitation bands in vicinity of $\Gamma$ at $\bB=0$:  $H^{\bK,\tilde{\phi}}_{eff}(\bk)=\left(\alpha_{\tilde{\phi} }k+\alpha^\ast_{\tilde{\phi}}\bar{k}\right)\sigma_0 + M\sigma_x-M'\epsilon_+\sigma_y+\left(\begin{array}{cc}
 0    &  g_{\tilde{\phi}}(\bk)\\
  g_{\tilde{\phi}}^\ast(\bk)   & 0
\end{array}\right)$ (derivation in Appendix(\ref{Appx:Effective_model_CNP_oneshot})). Here $k=k_x+ik_y$, $\bar{k}=k^\ast$,  Pauli $\sigma$ acts in the subspace spanned by $\Gamma_1\oplus\Gamma_2$ $c$-fermions and
\begin{eqnarray}
    \alpha_{\tilde{\phi}} &=& -\frac{\epsilon_a v_\ast\tilde{\gamma}}{2|\tilde{\gamma}|^4}\left(\tilde{U}_fc'\tilde{\gamma}e^{+4i\tilde{\phi}}+2ic''|\tilde{\gamma}|^2 e^{-2i\tilde{\phi}}\right)\\
    g_{\tilde{\phi}}(\bk) &=& -\frac{\tilde{U}_f}{2}\left(\frac{v_\ast^2e^{2i\tilde{\phi}}k^2}{\tilde{\gamma}^{\ast 2}}+\frac{c'^2\epsilon_a^2 e^{-6i\tilde{\phi}}}{\tilde{\gamma}^2}\right) - \frac{2c'c''\epsilon_a^2}{\tilde{\gamma}}.
\end{eqnarray}
The reinstated terms split the quadratic band touching at $\Gamma$ into two Dirac nodes whose location is determined by $g_{\tilde{\phi}}(\pm \bk^{\tilde{\phi}}_D)+(M+iM'\epsilon_+)=0$, such that $|k^{\tilde{\phi}}_D|\approx \left(2|\tilde{\gamma}|^2(|M+iM'\epsilon_+- 2c'c''\epsilon_a^2/\tilde{\gamma}|)/\tilde{U}_fv_\ast^2\right)^{1/2}\sim 0.05|\bg_m|>0$, where $|\bg_m|=2\pi/L_m$. Since $k_D^{\tilde{\phi}+\frac{\pi}{3}}=e^{i2\pi/3}k_D^{\tilde{\phi}}$ we note that  $E^{\tilde{\phi}+\pi/3}_D=-E^{\tilde{\phi}}_D$, where $\pm E^{\tilde{\phi}}_D = \pm(\alpha_{\tilde{\phi}} k_D^{\tilde{\phi}}+\alpha^\ast_{\tilde{\phi}} \bar{k}_D^{\tilde{\phi}})$ are the Dirac node energies. Thus $\exists \tilde{\phi}^\ast$ for which $E_D^{\tilde{\phi}^\ast}=0$. For realistic values of $\epsilon$ we find that $\tilde{\phi}^\ast\sim \pi/4$. The two nodes provide two zero-LLs emanating out of them, giving rise to a Chern $0$ gap of magnitude $2|E^{\tilde{\phi}}_D|$. These modes and resulting $C=0$ gap $\Delta_0\sim 5$ meV at $(\epsilon,\tilde{\phi})=(0.15\%,0)$ are shown in Fig.(\ref{fig:main:fig2} b,d,f). 

\emph{Lattice relaxation and Zeeman Coupling---} The effects of lattice relaxation are accounted by adding the correction $h^{rel}$ to $h^{cc}$ (defined in Eq.(\ref{Eq:main:THF:coupling})) where 
\begin{equation}
    h^{rel}(\bk)=\left(\begin{array}{cc}
   \mu_1\sigma_0+v_2\bk\cdot\sigma  & v_1\bk\cdot\sigma^\ast  \\
   h.c.  &  \mu_2\sigma_0
\end{array}\right)
\end{equation}
at valley $\bK$, $\mu_{1,2}$ and $v_{1,2}\ll v_\ast',v_\ast$ are the particle-hole symmetry breaking orbital potentials and velocities, respectively\cite{herzog2025topological}. The corresponding finite $\bB$ contribution is obtained via promoting \cite{singh2024topological} $(k_x-ik_y)\rightarrow i\frac{\sqrt{2}}{\ell}\hat{a}^\dagger$. For the Mott semimetal, breaking of the particle-hole symmetry sensitively affects the hierarchy between $\{\Delta_{8},\Delta_{12}\}$, while leaving behind the hierarchy $\Delta_{-8}\lesssim \Delta_{- 12}$ robust. In particular, while $\Delta_{- 8}\lesssim \Delta_{- 12}$ for $\phi/\phi_0\gtrsim 0.01$,
we find that $\Delta_{+8}\lesssim \Delta_{+12}$ only for $\phi/\phi_0\lesssim 0.05$(see Fig(\ref{fig:main:Relaxation:Zeeman}a,b)). We further note that the inclusion of lattice relaxation will, in addition to breaking particle-hole symmetry, open a $C=0$ gap at  finite $B$ (see Appendix(\ref{sec:anomalous})). This gap arises from the lifting of the degeneracy between the two zero modes and has a magnitude $\mu_1/\mathcal{A}_{\bf{0}}-\mu_2\sim 3$ meV for $\phi/\phi_0\gtrsim 0.005$ (see Fig(\ref{fig:main:Relaxation:Zeeman}a,e)). In contrast, for the strain-induced anisotropic semimetal, we find that the hierarchy $\Delta_{\pm 4}\gg\Delta_{\pm 8}\gtrsim\Delta_{\pm 12}$ stays robust for most regions of the fluxes that we analyzed (see Appendix(\ref{Appx:Robust_Strain_Semimetal})). The obtained gaps for $(\epsilon,\tilde{\phi})=(0.15\%,0)$ are shown in Fig(\ref{fig:main:Relaxation:Zeeman}c,d)). In addition to relaxation, we also examine the effects of spin Zeeman coupling $h\sim 0.058B[\rm{T}]$ meV for the Mott semimetal within the Hubbard-I approximation. Besides the conventional Zeeman splitting of the bare LLs, the coupling between the $f$ and auxiliary fermions in Eq.(\ref{Eq:main:Symmetric_state_}) is promoted to $ \frac{U_1}{2}\sigma_0\rightarrow U_1\sqrt{n_\uparrow n_\downarrow}\sigma_0$, and the previously vanishing auxiliary fermion self-coupling in Eq.(\ref{Eq:main:Symmetric_state_}) now becomes $-U_1(n_\sigma-\frac{1}{2})\sigma_0$. Here  $\sigma=\uparrow\downarrow$ represents the spin sector and   $n_\sigma$ represents the $B$-dependent thermally averaged occupation of the $f$-fermions with spin $\sigma$ relative to charge neutrality in the atomic-limit, with $n_\downarrow+n_\uparrow=1$ (see Appendix(\ref{sec:LLzeemann})). In addition, we also include symmetry-breaking mean-field terms in our analysis, which previously were forbidden by spin-SU(2) symmetry (see Appendix(\ref{sec:LLzeemann})). 
For temperatures comparable to previous spectroscopic measurement $\sim 5K-6K$ ($\sim 0.5$ meV) \cite{2020Natur.582..198W}  and $B<3$T, the above discussed Chern gap hierarchy between $\{\Delta_{\pm 4},\Delta_{\pm 8},\Delta_{\pm 12}\}$ stays fairly robust (see Appendix(\ref{sec:LLzeemann})). We also note that spin splitting of the two zero modes by Zeeman (see Fig(\ref{fig:main:Relaxation:Zeeman}f-h)) can give rise to $C=\pm 2$ gaps at CNP (also see Appendix.(\ref{sec:LLzeemann})).

\emph{Discussion---} In this work, we examine the two leading candidates whose spectroscopic signatures are consistent with the report in Ref~\cite{xiao2026imaging} at CNP: Mott semimetal and strain-induced anisotropic semimetal. We demonstrate that these two states yield a distinct LL gap size hierarchy,
enabling LL spectroscopy to distinguish them. We also note that the hierarchy $\Delta_{\pm4}\gg\Delta_{\pm8}>\Delta_{\pm12}$ obtained for the strain-induced anisotropic semimetal is in agreement with previous incompressibility measurements at finite magnetic field \cite{yu2022correlated}. We expect the sharp single-particle excitation gaps reported in this work to show as broadened incompressible features in spectroscopy due to finite lifetime effects. Recent works suggest lifetime effects to be more acute for the unstrained Mott semimetal state \cite{wei2026lifetimespectralfunctiontopological,hu2026twistedbilayergraphenelifetimes,vituri2026controlledloopexpansiontopological,nosov2026controlledexpansioncorrelatedelectrons} compared to that for the strain-induced anisotropic semimetal \cite{keshet2026controlledloopexpansionstrained}. We also note that in local moment regime of the Mott semimetal, $U\gg k_B T \gg h$, each anomalous mode exhibits a $1/T$ dependent energy dispersion as a function of $B$. This behaviour is governed by $\gamma ^2 \chi_0 / U_1 $ and $J \chi_0$ where $\chi_0$ is the Curie susceptibily of the local moments and $J$ is the $f$-$c$ exchange interaction strength (see Appendix \ref{sec:LLzeemann}). This temperature dependence is reminiscent of the Curie gap reported in \cite{nonlocalmoments2025}. Concurrent with the preparation of this work, Ref\cite{hofmann2026montecarlostudiestwisted} proposed alternate experimental discriminants for distinguishing the two semimetals based on the response to temperature and an in-plane Zeeman field.

\section*{acknowledgments}
We wish to thank Prof. Ben Feldman and Dr. Jiachen Yu for helpful correspondence.
JFMV was supported by the Gordon and Betty Moore Foundation EPiQS Initiative Grant No. GBMF8685. BAB was supported by DOE Grant No. DE-SC001623. OV was funded in part by the Gordon and Betty Moore
Foundation’s EPiQS Initiative Grant GBMF11070. OV and EB acknowledge support from the Simons Foundation Collaboration on New Frontiers in Superconductivity (Grant no. SFI-MPS-NFS00006741-09 and SFI-MPS-NFS00006741-03, respectively). 

\clearpage

\bibliography{Paper2}

\clearpage
\onecolumngrid

%\appendix
%\setcounter{secnumdepth}{1}
%\setcounter{section}{0}

\appendix
\tableofcontents
\clearpage
\setcounter{secnumdepth}{3}
\setcounter{tocdepth}{3}

\section{Introduction to the appendices}\label{sec:Introduction-to-the}

\begin{table}[t]
\centering
\small
\setlength{\tabcolsep}{4pt}        % tighter columns
\renewcommand{\arraystretch}{1.15} % a bit taller rows
\setlength{\arrayrulewidth}{0.4pt} % thinner lines 

\resizebox{\linewidth}{!}{%
\begin{tabular}{|cccc|cccc|ccc|ccc|ccccc|}
\hline\hline
$\gamma$ & $M$ & $v_\ast$ & $v'_\ast$
& $\mu_1$ & $\mu_2$ & $v_1$ & $v_2$
& $c$ & $c'$ & $c''$
& $M_f$ & $\gamma'$ & $M'$
& $U_1$ & $U_2$ & $W_1$ & $W_3$ & $J$
\\
\hline
$-24.8$ & $3.7$ & $-4.3$ & $1.6$
& $14.4$ & $4.5$ & $0.2$ & $-0.4$
& $-8750$ & $2050$ & $-3362$
& $4380$ & $-3352$ & $-4580$
& $57.95$ & $2.33$ & $44.03$ & $50.20$ & $16.38$
\\
\hline\hline
\end{tabular}%
}

\caption{Parameters of THFM determined using first principles for BM model at $\theta=1.05^\circ$ at $w_0/w_1= 0.8$\cite{herzog2025topological}.  All values reported are in meV except the velocities $v_\ast,v_\ast', v_1$ and $v_2$ are in $eV\mathring{A}$.}
\label{tab:model_params}
\end{table}

In these appendices, we start by reviewing the Topological Heavy Fermion (THF) model of magic-angle twisted bilayer
graphene (MATBG) in presence of an out of plane magnetic field $\mathbf{B}$. We then discuss the self-consistent Hartree-Fock (SCHF) calculations at finite $\mathbf{B}$ used
to quantify the gaps and determine the sequence of Chern numbers.
shown in Fig.(\ref{fig:main:fig2}d,f). Next, we derive the effective model for the light modes (near $\Gamma$) in the presence of heterostrain.  The second half of the appendix focuses on the relaxed unstrained symmetric state, where we discuss details of its Landau quantization within the Hubbard I approximation to the THFM used to obtain Fig.(\ref{fig:main:fig2}c,e) and provide analytical results. 
\section{Review of Topological heavy fermions in magnetic field}\label{appx:Section_THFM_in_B}

\subsection{Bistritzer-Macdonald model and the magnetic translations }
\label{apdx:BM-model and MTG}
The Bistritzer-Macdonald (BM) model of magic-angle twisted bilayer graphene  \cite{bistritzer2011moire} at valley $\bK(\tau=+1)$ is given by ($\bB=0$)
\begin{eqnarray}
H_{BM}^{\bK}(\bp) = \left(\begin{array}{cc} v_F \sigma \cdot \bp & T(\br)e^{i\bq_1 \cdot\br} \\
e^{-i\bq_1 \cdot\br}T^{\dagger}(\br) & v_F \sigma \cdot (\bp+ \hbar\bq_1)
\end{array}\right)\label{appxEq:BM model},
\end{eqnarray}
 where $\sigma$ acts in sublattice space and $\bp=0$ denotes the moir\'e Dirac point $K_M$. The interlayer hopping functions are 
 \begin{eqnarray}
T(\br)=\sum_{j=1}^{3}T_{j}e^{-i\bq_1 \cdot \br}, \label{Eq:Appx:moire_q}
 \end{eqnarray}
 where $\bq_1 = k_{\theta}(0,-1)$, $\bq_{2,3}=k_{\theta}(\pm \frac{\sqrt{3}}{2},\frac{1}{2})$, $k_{\theta}=\frac{8\pi}{3a_0}\sin\frac{\theta}{2}=\frac{4\pi}{3L_m}$, $a_0 \approx 0.246$nm, $L_m= \frac{a_0}{2\sin{\theta/2}}$ is the period of the moir\'e lattice. $\theta=1.05^{\circ}$ in this work.
 \begin{eqnarray}
T_{j+1}=w_0I_2+w_1\left(\cos\left(\frac{2\pi}{3}j\right)\sigma_x + \sin\left(\frac{2\pi}{3}j\right)\sigma_y\right) 
 \end{eqnarray}
where $I_n$ is the $n \times n$ unit matrix. We will present results for the case $w_0/w_1=0.8$. The BM model is invariant under translation by integer multiples of moir\'e lattice vectors $\bL_1=L_m(\frac{\sqrt{3}}{2},\frac{1}{2})$ and $\bL_2=L_m(0,1)$. The corresponding reciprocal lattice vectors are $\tilde{\bg}_1=\frac{4\pi}{\sqrt{3}L_m}(1,0)$ and $\tilde{\bg}_2=\frac{4\pi}{\sqrt{3}L_m}(-\frac{1}{2},\frac{\sqrt{3}}{2})$ respectively.
Thus if $f(\br)$ is an eigenstate, then so is
\begin{eqnarray}
\hat{T}_{\bL_{1,2}}f(\br) = f(\br-\bL_{1,2})  
\end{eqnarray}
where $\hat{T}_{\bL_{1,2}}$ are the usual discrete translation operators. In the presence of an out-of-plane magnetic field $\bB=(0,B\hat{z})$, using the Landau gauge $\bA=(0,Bx)$, the BM model is promoted via a minimal substitution, 
$H^{\bK}_{BM}\left(\bp-\frac{e}{c}\bA\right)=H^{\bK}_{BM}\left(p_x,p_y-\frac{eB}{c}x\right)$. The finite $\bB$ BM model at $\bK'$$(\tau=-1$) can be obtained similarly by applying time reversal to Eq.(\ref{appxEq:BM model}) followed by the minimal substitution. Although the $\bB\neq 0$ BM model is still invariant under the translation by $\bL_2$, a translation by $\bL_1$ needs to be accompanied by a gauge transformation as
\begin{eqnarray}
&&e^{i\frac{eB}{\hbar c}L_{1x}y}H^{\tau}_{BM}\left(p_x,p_y-\frac{eB}{c}x+\frac{eB}{c}L_{1x}\right)e^{-i\frac{eB}{\hbar c}L_{1x}y}=
H^{\tau}_{BM}\left(p_x,p_y-\frac{eB}{c}x\right).
\end{eqnarray}
In other words if $\tilde{f}(\br)$ is an eigenstate of $H^{\tau}_{BM}\left(p_x,p_y-\frac{eB}{c}x\right)$, then so is
\begin{eqnarray}
\tilde{f}(\br) \rightarrow \hat{t}_{\bL_1}\tilde{f}(\br)=e^{i\frac{eB}{\hbar c}L_{1x}y}\tilde{f}(\br-\bL_{1})
\end{eqnarray}
Thus $\hat{t}_{\bL_1}=e^{i\frac{eB}{\hbar c}L_{1x}y} \hat{T}_{\bL_1}= e^{2\pi i \frac{\phi}{\phi_0}\frac{y}{L_2}}\hat{T}_{\bL_1}$ is the generator of magnetic translations by $\bL_1$, where $L_2=|\bL_2|= L_m$,  $\phi=BL_{1x}L_m$ is magnetic flux per moir\'e unit cell and $\phi_0=hc/e$ . Magnetic translation by $\bL_2$ are generated by $\hat{t}_{\bL_2}\tilde{f}(\br)=\tilde{f}(\br-\bL_2)$. We thus have
\begin{eqnarray}
\hat{t}_{\bL_2}\hat{t}_{\bL_1}\tilde{f}(\br)&=&e^{i\frac{eB}{\hbar c}L_{1x}(y-L_2)}\tilde{f}(\br-\bL_1-\bL_2)\\
\hat{t}_{\bL_1}\hat{t}_{\bL_2}\tilde{f}(\br)&=&e^{i\frac{eB}{\hbar c}L_{1x}y}\tilde{f}(\br-\bL_1-\bL_2)\\
\Rightarrow \hat{t}_{\bL_2}\hat{t}_{\bL_1}&=&e^{-2\pi i\frac{\phi}{\phi_0}}\hat{t}_{\bL_1}\hat{t}_{\bL_2}
\end{eqnarray}
If $\phi/\phi_0=p/q$, with $p$ and $q$ relatively prime integers,
\begin{eqnarray}
 \left[\hat{t}^q_{\bL_2},\hat{t}_{\bL_1}\right]=0
\end{eqnarray}
and 
\begin{eqnarray}
\left[\hat{t}_{\bL_{1,2}},H^{\tau}_{BM}\left(p_x,p_y-\frac{eB}{c}x\right)\right]=0.   
\end{eqnarray}

%---------------------Section change----

\subsection{Basis states in presence of magnetic field}
We begin by reviewing the basis states for $c$ and $f$ fermions in finite magnetic field, which were systematically constructed in Ref\cite{singh2024topological} to be irreps of the magnetic translation group (MTG). In other words, orthonormal simultaneous eigenstates of the magnetic translation operators $\hat{t}_{\bL_1}$ and $\hat{t}^{q}_{\bL_2}$ were constructed to solve the $\bB\neq 0$ BM model. For the $f$ fermions, following the Hybrid Wannier method introduced in Ref\cite{wang2022narrow}, 
one first builds the hybrid-Wannier states (delocalized along $y$ axis) out of zero $\bB$ Wannier states, and then project them onto the irreps of MTG. On the other hand, the $\bB=0$ $\bk\cdot\bp$ $c$ fermions at $\Gamma$ are promoted to $\bB$ by replacing the slowly varying phase $e^{ikr}$ by Landau gauge Landau level (LL) wavefunctions  , followed by projecting it onto irreps of MTG\cite{singh2024topological}. The low energy field can be expanded in terms of these single particle degrees of freedom within a single spin-valley flavor $s=\uparrow,\downarrow$ and valley  $\tau=\pm 1$ 
as \cite{singh2024topological}
\begin{eqnarray}
 \hat{\psi}_{\tau s}(\br)= \sum_{k\in[0,1)\otimes [0,\frac{1}{q})}\left(\sum_{b=1}^2\sum_{r'= 0}^{q-1} \eta_{b\tau k r'}(\br) f_{b\tau k r's} + \sum_{a=1}^4 \sum_{m=0}^{m_{a,\tau}}\sum_{r = 0}^{p-1} \Psi_{a\tau}(\br)\chi_{k r m}(\br)
c_{a\tau k r m s} \right)\label{Eq:appx:Low_energy_expansion}. 
\end{eqnarray} 
where the single-particle wavefunctions read
 \begin{eqnarray}
\eta_{b\tau k r'}(\br)&=&\frac{1}{\sqrt{\mathcal{N}}}
\sum_{s,n\in \mathbb{Z}} e^{2\pi i \left(s k_1+n\left(k_2+\frac{r'}{q}\right)\right)}\hat{t}^s_{\bL_1}\hat{t}^{n}_{\bL_2} W_{\mathbf{0},b\tau}(\br),\\
\chi_{k m r}(\br)&=&\frac{1}{\sqrt{\ell L_m }}\frac{1}{\sqrt{\mathcal{N}}}
\sum_{s\in \mathbb{Z}} e^{2\pi i s k_1}\hat{t}^s_{\bL_1}\left(\hat{t}^s_{\bL_1}e^{2\pi i(k_2+\frac{r}{q})\frac{y}{L_2}}\varphi_m\left(x-(k_2+\frac{r}{q})\frac{2\pi\ell^2}{L_2}\right)\right),
\end{eqnarray}
with normalization $\mathcal{N}=s_{tot}n_{tot}$. The operators $f_{b\tau k rs}$ and $c_{a\tau krms}$ denote the
annihilation operators for these single-particle $c$ and $f$ fermion states at finite $\bB$, respectively. The Landau gauge LLs are given by
\begin{eqnarray}
\varphi_m(x)&=&\frac{1}{\pi^{\frac{1}{4}}}\frac{1}{\sqrt{2^mm!}}e^{-x^2/2\ell^2}H_m(x/\ell),
\end{eqnarray}
where $m$ denotes the LL index.
The LL upper cutoffs $m_{a,\tau}$ for $\phi/\phi_0=1/q$ sequence are set by $m_\ast\leq q/2$ \cite{singh2024topological}   such that $m_{1,+1}=m_{2,-1}=m_{\ast}+1$, $m_{2,+1}=m_{1,-1}=m_{\ast}$, $m_{3,+1}=m_{4,-1}=m_{\ast}+2$ and $m_{4,+1}=m_{3,-1}=m_{\ast}-1$. The label $k=(k_1,k_2)\in[0,1)\otimes[0,1/q)$ represents magnetic momentum corresponding to $\hat{t}_{\bL_1}$ and $\hat{t}^q_{\bL_2}$ respectively: 
\begin{eqnarray}
\hat{t}_{\bL_1}\eta_{b\tau kr'}&=&e^{-2\pi i k_1} \eta_{b\tau kr'},\label{f in Mt 1}\\
\hat{t}^q_{\bL_2}\eta_{b\tau kr'}&=&e^{-2\pi i qk_2} \eta_{b\tau kr'}, \label{f in Mt 2}\\
\hat{t}_{\bL_1}\chi_{krm}&=&e^{-2\pi i k_1} \chi_{krm},\label{c in Mt 1}\\
\hat{t}^q_{\bL_2}\chi_{krm}&=&e^{-2\pi i qk_2} \chi_{krm}. \label{c in Mt 2}
\end{eqnarray}
 The labels $r' \in \{0,\ldots q-1\}$ and $r \in \{0,\ldots,p-1\}$, denote the magnetic strip $[\frac{r'}{q},\frac{r'+1}{q})$ and $[\frac{r}{q},\frac{r+1}{q})$ along $\tilde{\bg}_2$ for $\eta$ and $\chi$ respectively. Thus for each flavor (spin, valley and orbital) of $f$ and $c$, we have exactly $q$ and $p$ states respectively, 
 per magnetic moir\'e Brillouin  Zone (mBZ), i.e. $\tilde{\bg}_1\otimes \tilde{\bg}_2/q$ . In other words, we have exactly $1$ $f$-fermion state and $\phi/\phi_0 = p/q$ $c$-fermion LL states per moir\'e unit cell for each flavor, as expected.

%---------------------Section change----

\subsection{Non-interacting topological heavy fermions in magnetic field}
The single particle topological heavy fermion model in magnetic field for given spin $s$ reads  \cite{singh2024topological}
\begin{eqnarray}
    \hat{H}_0 &=& \sum_{\tau} \int d^2 \br \psi^{\dagger}_{\tau s}(\br)H^{\tau}_{BM}\left(\bp-\frac{e}{c}\bA\right)\psi_{\tau s}(\br) = \sum_{\tau} \hat{H}^{\tau}_{cc} + \sum_{\tau} \left(\hat{H}^{\tau}_{cf}+ \text{h.c.}\right),
\end{eqnarray}
 where $h.c.$ represents Hermitian conjugate and 
 \begin{eqnarray}
\hat{H}^{\tau}_{cc}&=&\sum_{k\in[0,1)\otimes [0,\frac{1}{q})} \sum_{a,a'=1}^4 \sum_{m=0}^{m_{a,\tau}}\sum_{m'=0}^{m_{a',\tau}}\sum_{r,\tilde{r}=0}^{p-1}\tilde{h}^{\tau}_{[amr],[a'm'\tilde{r}]}(k) c^\dagger_{a \tau k r m s}
c_{a'\tau k \tilde{r} m' s},\label{Eq:APPX:Unstrained_CC}\\
\hat{H}^{\tau}_{cf}&=&\sum_{k\in[0,1)\otimes [0,\frac{1}{q})}\sum_{a=1}^4 \sum_{b=1}^2 \sum_{m=0}^{m_{a,\tau}}
 \sum_{r=0}^{p-1}\sum_{r'=0}^{q-1}h^{\tau}_{[amr],[br']}(k)c^\dagger_{a\tau k r m s}
f_{b\tau k r' s}\label{Eq:APPX:Unstrained_CF},
\end{eqnarray}
The $f$-$f$ hybridization is trivial in absence of heterostrain.
The $c$-$c$ matrix elements read  \cite{singh2024topological}
\begin{eqnarray}
\tilde{h}^{\tau}_{[amr],[a'm'\tilde{r}]}(k) = \delta_{r\tilde{r}}\left(\begin{array}{cc}
0_{2\times 2} & h^{\tau,c}_{mm'} \\
\sigma_x h^{\tau,c}_{mm'} \sigma_x & M\delta_{mm'}\sigma_x
\end{array}\right)_{aa'},\label{Eq:appx:Unstrained_cc}
\end{eqnarray}
where the Pauli matrix $\sigma_x$ acts on the $c$ orbitals and
\begin{eqnarray}
h^{+1,c}_{mm'} = i\frac{\sqrt{2} v_{\ast}}{\ell}\left(\begin{array}{cc}
-\sqrt{m'}\delta_{m+1,m'} & 0 \\
0 & \sqrt{m}\delta_{m,m'+1}
\end{array}\right)
\end{eqnarray}
with $h^{-1,c}_{mm'}=-\sigma_xh^{+1,c}_{mm'}\sigma_x$. The matrix element between the $f$- and $c$- electrons reads  \cite{singh2024topological}
\begin{eqnarray}
   h^\tau_{[amr],[br']} =  \left(\begin{array}{c}
\gamma I^0_{[mr],[r']}(k) \sigma_0+v'_\ast(\tau I^x_{[mr],[r']}(k)\sigma_x+I^y_{[mr],[r']}(k)\sigma_y)\\
0_{2\times 2}
\end{array}\right)_{ab}\label{Eq:appx:Unstrained_cf}
\end{eqnarray}
where 
\begin{eqnarray}
I^0_{[mr],[r']}(k_1,k_2)&=&
 \frac{\sqrt{L_{1x}}}{\sqrt{\ell}}\sum_{j\in \mathbb{Z}}\sum_{s\in \mathbb{Z}}\delta_{r'-r,jq+sp}
e^{-2\pi i s k_1}e^{2\pi i s\left(k_2+\frac{r}{q}\right)\frac{L_{1y}}{L_2}}e^{\pi i s(s-1) \frac{p}{q}\frac{L_{1y}}{L_2}}
e^{-2\pi^2 \left(k_2+\frac{r}{q}+ s\frac{p}{q}\right)^2\frac{\lambda^2}{L^2_2}}\nonumber\\
&&\times\mathcal{F}_m\left(\lambda,\left(s+\frac{r}{p}+k_2\frac{q}{p}\right)L_{1x}\right)\label{I0}\\
I^x_{[mr],[r']}(k_1,k_2)&=&\frac{i}{\sqrt{2}\ell}\frac{\sqrt{L_{1x}}}{\sqrt{\ell}}\sum_{j\in \mathbb{Z}}\sum_{s\in \mathbb{Z}}\delta_{r'-r,jq+sp}
e^{-2\pi i s k_1}e^{2\pi i s\left(k_2+\frac{r}{q}\right)\frac{L_{1y}}{L_2}}e^{\pi i s(s-1) \frac{p}{q}\frac{L_{1y}}{L_2}}
e^{-2\pi^2 \left(k_2+\frac{r}{q}+ s\frac{p}{q}\right)^2\frac{\lambda^2}{L^2_2}}\nonumber\\
&&\times\left(
\sqrt{m}\mathcal{F}_{m-1}\left(\lambda,\left(s+\frac{r}{p}+k_2\frac{q}{p}\right)L_{1x}\right)
-\sqrt{m+1}\mathcal{F}_{m+1}\left(\lambda,\left(s+\frac{r}{p}+k_2\frac{q}{p}\right)L_{1x}\right)
\right)\label{Ix},\\
I^y_{[mr],[r']}(k_1,k_2)&&= \frac{\sqrt{L_{1x}}}{\sqrt{\ell}}\sum_{j\in \mathbb{Z}}\sum_{s\in \mathbb{Z}}\delta_{r'-r,jq+sp}
e^{-2\pi i s k_1}e^{2\pi i s\left(k_2+\frac{r}{q}\right)\frac{L_{1y}}{L_2}}e^{\pi i s(s-1) \frac{p}{q}\frac{L_{1y}}{L_2}}
\left(k_2+\frac{r}{q}+ s\frac{p}{q}\right)\frac{2\pi}{L_2}
\nonumber\\
&&\times e^{-2\pi^2 \left(k_2+\frac{r}{q}+ s\frac{p}{q}\right)^2\frac{\lambda^2}{L^2_2}}
\mathcal{F}_m\left(\lambda,\left(s+\frac{r}{p}+k_2\frac{q}{p}\right)L_{1x}\right) \label{Iy},
\end{eqnarray}
with 
\begin{eqnarray}
    \mathcal{F}_m(\lambda,x_0) &=&\frac{1}{\pi^{\frac{1}{4}}\sqrt{2^{m}m!}}\sqrt{\frac{\ell^2}{\ell^2+\lambda^2}}
e^{-\frac{1}{2}x_0^2/(\ell^2+\lambda^2)}
\mathcal{H}_m\left(-2x_0\frac{\ell}{\ell^2+\lambda^2},\frac{2\lambda^2}{\ell^2+\lambda^2}-1\right) 
\end{eqnarray}
and the two-variable Hermite polynomial is given as
\begin{eqnarray}\label{2varHermite}
\mathcal{H}_n(x,y)=n!\sum_{k=0}^{\left[\frac{n}{2}\right]}\frac{x^{n-2k}y^k}{(n-2k)!k!}, 
\end{eqnarray}
where $[n]$ denotes the floor function at $n$. 
%In the main text, we use the notation $\Upsilon_{mr'}\equiv \Upsilon_{m0,r'}$ where $\Upsilon_{mr,r'}$ is defined in Eq.(\ref{I0}).

%---------------------Section change----

\subsection{Heterostrain in THF in magnetic field}\label{Appx:Heterostrain_THF_B=0andB}
In this section we discuss the promotion of strained THF model derived in Ref\cite{herzog2025topological} to finite $\bB$. We start by noting that heterostrain deforms the lattice structure. Specifically, in the presence of uniaxial heterostrain, the moir\'e $\bq_j$ vectors (defined below Eq.(\ref{Eq:Appx:moire_q})) get deformed into
\begin{eqnarray}
  \tilde{\bq}_j =   \bq_j -\mathcal{E}^T \left(R_{2\pi/3}\right)^{j-1} \bK_G
\end{eqnarray}
where $\bK_G=\frac{4\pi}{3a_0}\hat{x}$ is Dirac momentum of monolayer graphene, $R_\vartheta$ is the two-dimensional anticlockwise rotation 
matrix and the strain tensor is given by
\begin{eqnarray}
    \mathcal{E} &=&R^T_{\tilde{\phi}}\left(\begin{array}{cc}
       -\epsilon  &0  \\
       0  & \nu_G\epsilon\end{array}\right)R_{\tilde{\phi}}
\end{eqnarray}
where $\nu_G= 0.16$ is the Poisson ratio of graphene and $\{\epsilon,\tilde{\phi}\}$ parametrize the magnitude and orientation (with respect to $\hat{x}$) of the uniaxial heterostrain. The deformed moir\'e reciprocal lattice vectors read 
\begin{eqnarray}
\tilde{\bg}_1&=&\tilde{\bq}_2-\tilde{\bq}_3\\\tilde{\bg}_2&=&\tilde{\bq}_3-\tilde{\bq}_1
\end{eqnarray}
The corresponding moir\'e lattice vectors $\tilde{\bL}_i$ can be calculated by noting that $\tilde{\bg}_i\cdot\tilde{\bL}_j=2\pi\delta_{ij}$. 
For a generic value of $(\epsilon,\tilde{\phi})$, the vector $\tilde{\bL}_2$ will not be $\parallel \hat{y}$. However to use the Landau gauge formulation developed in \cite{singh2024topological} and reviewed in previous subsection, we need $\tilde{\bL}_2$ to be parallel to $y$-axis. We start by defining the angle $\theta_0$ such that $\ R(\theta_0)\tilde{\bL}_2\parallel \hat{y}$. For example,  at the parameterization $(\epsilon,\tilde{\phi})=(0.15\%,0)$, we find that $e^{i\theta_0}\sim0.999+i0.013$. Rotating the coordinate frame by $\theta_0$ then makes $\tilde{\bL}_2\parallel\hat{y}$ and enables us to employ the Landau gauge formulation to solve the strained problem. If the $\bB=0$ strained THF model reads $H_0(k,\bar{k})$, where $k=k_x+ik_y,\bar{k}=k^\ast$ such that $(k_x,k_y)$ is defined with respect to unrotated coordinate frame then with respect to the $k'=k'_x+ik'_y,\bar{k}'=k'^\ast$
in the rotated coordinate frame, the strained THF model is given by $H_0(e^{i\theta_0}k',e^{-i\theta_0}\bar{k}')$. This follows from noting that $e^{i\theta_0}k'=k$. We now redefine the unprimed notation to be with respect to the rotated coordinate frame. The $\bB=0$ THF model in rotated frame thus reads \cite{herzog2025topological}
\begin{eqnarray}
  h^{\tau=+1}_{\bB=0}(k,\bar{k}) = \left(\begin{array}{cc} h_{cc}(k,\bar{k})&h_{cf}(k,\bar{k})\\h^\dagger_{cf}(k,\bar{k}) & h_{ff}\end{array}\right)\label{Eq:Appx:Landmark_notation}
\end{eqnarray}
at valley $\bK$  where 
\begin{eqnarray}
h_{cc}(k,\bar{k}) &=& \left(\begin{array}{cccc}
0&c(\epsilon_{xy}-i\epsilon_{-})&v_\ast e^{i\theta_0}k&c'(\epsilon_{xy}+i\epsilon_{-})  \\
c(\epsilon_{xy}+i\epsilon_{-})&0&c'(\epsilon_{xy}-i\epsilon_{-})&v_\ast e^{-i\theta_0}\bar{k} \\
v_\ast e^{-i\theta_0}\bar{k}&c'(\epsilon_{xy}+i\epsilon_{-})&0&M-iM'\epsilon_{+} \\c'(\epsilon_{xy}-i\epsilon_{-})&v_\ast e^{i\theta_0}k&M+iM'\epsilon_{+}&0
\end{array}\right)\\
    h_{cf}(k,\bar{k})&=& e^{-\frac{1}{2}\bk^2\lambda^2}\left(\begin{array}{cccc}
    \gamma+i\gamma'\epsilon_{+}& v_\ast'e^{-i\theta_
    0}\bar{k}\\v_\ast'e^{i\theta_
    0}k&\gamma- i\gamma'\epsilon_{+}\\
     c''(\epsilon_{xy}-i\epsilon_{-})&0\\
    0&c''(\epsilon_{xy}+i\epsilon_{-})
\end{array}\right)\\
h_{ff} &=& M_f\left(\begin{array}{cc}0&\epsilon_{xy}-i\epsilon_-\\\epsilon_{xy}+i\epsilon_-&0\end{array}\right)\label{Eq:appx:ff_strained_B=0}
\end{eqnarray}
and $h^{\tau=-1}_{\bB=0}(k,\bar{k})$ is related by time reversal. The parameters $\epsilon_{\pm},\epsilon_{xy}$ can be identified via the decomposition $\mathcal{E}=\epsilon_{+}\sigma_0 + \epsilon_-\sigma_z + \epsilon_{xy}\sigma_x$. In particular, $(\epsilon_{xy},\epsilon_-)=(\sin(2\tilde{\phi}), -\cos(2\tilde{\phi}))\epsilon(\nu_G+1)/2$ and $\epsilon_+= \epsilon(\nu_G-1)/2$. 
We can now systematically promote the above strained THF model to finite $\bB$. The model is given by 
\begin{eqnarray}
    \hat{H}^\tau_{0,\bB} = \hat{H}^\tau_{cc}  + \left(\hat{H}^\tau_{cf}+h.c.\right)+\hat{H}^\tau_{ff} \label{Eq:appx:Single_particle_B}
\end{eqnarray}
The couplings in the unstrained expressions of $\hat{H}^\tau_{cc}$ and $\hat{H}^\tau_{cf}$ in Eqs.(\ref{Eq:APPX:Unstrained_CC}) and (\ref{Eq:APPX:Unstrained_CF}) respectively now get modified due to strain as explained below. We start by noting that it was proved in \cite{singh2024topological} that the $c$-$c$ coupling can be promoted to finite $\bB$ by straightforward canonical substitution, $k\rightarrow -i\frac{\sqrt{2}}{\ell}\hat{a}$, as long as the upper cutoff of $c$-LL index $m_\ast\leq q/2$ for a $\phi/\phi_0=1/q$ sequence. Thus heterostrain modifies  Eq.(\ref{Eq:appx:Unstrained_cc}) to be
\begin{eqnarray}
\tilde{h}^{\tau,\epsilon}_{[amr],[a'm'\tilde{r}]}(k) = \delta_{r\tilde{r}}\left(\begin{array}{cc}
c(\epsilon_{xy}\sigma_x + \tau\epsilon_-\sigma_y)\delta_{mm'} & h^{\tau,c,\epsilon}_{mm'} + c'(\epsilon_{xy}\sigma_x-\tau\epsilon_-\sigma_y)\delta_{mm'}\\
\sigma_x h^{\tau,c,\epsilon}_{mm'} \sigma_x + c'(\epsilon_{xy}\sigma_x-\tau\epsilon_-\sigma_y)\delta_{mm'} & (M\sigma_x+\tau M'\epsilon_+\sigma_y)\delta_{mm'}
\end{array}\right)_{aa'},\label{Eq:appx:Strained_cc}
\end{eqnarray}
where
\begin{eqnarray}
  h^{+1,c,\epsilon}_{mm'} =   i\frac{\sqrt{2} v_{\ast}}{\ell}\left(\begin{array}{cc}
-e^{i\theta_0}\sqrt{m'}\delta_{m+1,m'} & 0 \\
0 & e^{-i\theta_0}\sqrt{m}\delta_{m,m'+1}
\end{array}\right)
\end{eqnarray}
and $h^{-1,c,\epsilon}_{mm'}=\sigma_xh^{+1,c,\epsilon}_{mm'}\sigma_x$. We next note that the heterostrain introduces only non-derivative coupling terms in $c$-$f$ hybridization (see Supplementary notes 4 B 1 in \cite{singh2024topological}), which can formally be promoted to finite $\bB$ by $e^{-\frac{1}{2}\bk^2\lambda^2}C\rightarrow CI^0_{[mr],[r']}(k)$, for any $c$-number $C$. Thus heterostrain modifies Eq.(\ref{Eq:appx:Unstrained_cf}) to be
\begin{eqnarray}
    h^{\tau,\epsilon}_{[amr],[br']} = \left(\begin{array}{cc}(\gamma\sigma_0+i\gamma'\epsilon_+\sigma_z)I^0_{[mr],[r']} +h^{\tau,cf,\epsilon}_{[mr],[r']}\\ c''(\epsilon_{xy}\sigma_0-i\tau\epsilon_-\sigma_z)I^0_{[mr],[r']}
    \end{array}\right)_{ab}
\end{eqnarray}
where
\begin{eqnarray}
    h^{\tau,cf,\epsilon}_{[mr],[r']} = v_\ast'\left(\begin{array}{cc} 0& e^{-i\tau\theta_0}(\tau I^x_{[mr],[r']}-iI^y_{[mr],[r']})\\e^{i\tau\theta_0}(\tau I^x_{[mr],[r']}+iI^y_{[mr],[r']})&0\end{array}\right)
\end{eqnarray}
\begin{figure}
    \centering
    \includegraphics[width=0.8\linewidth]{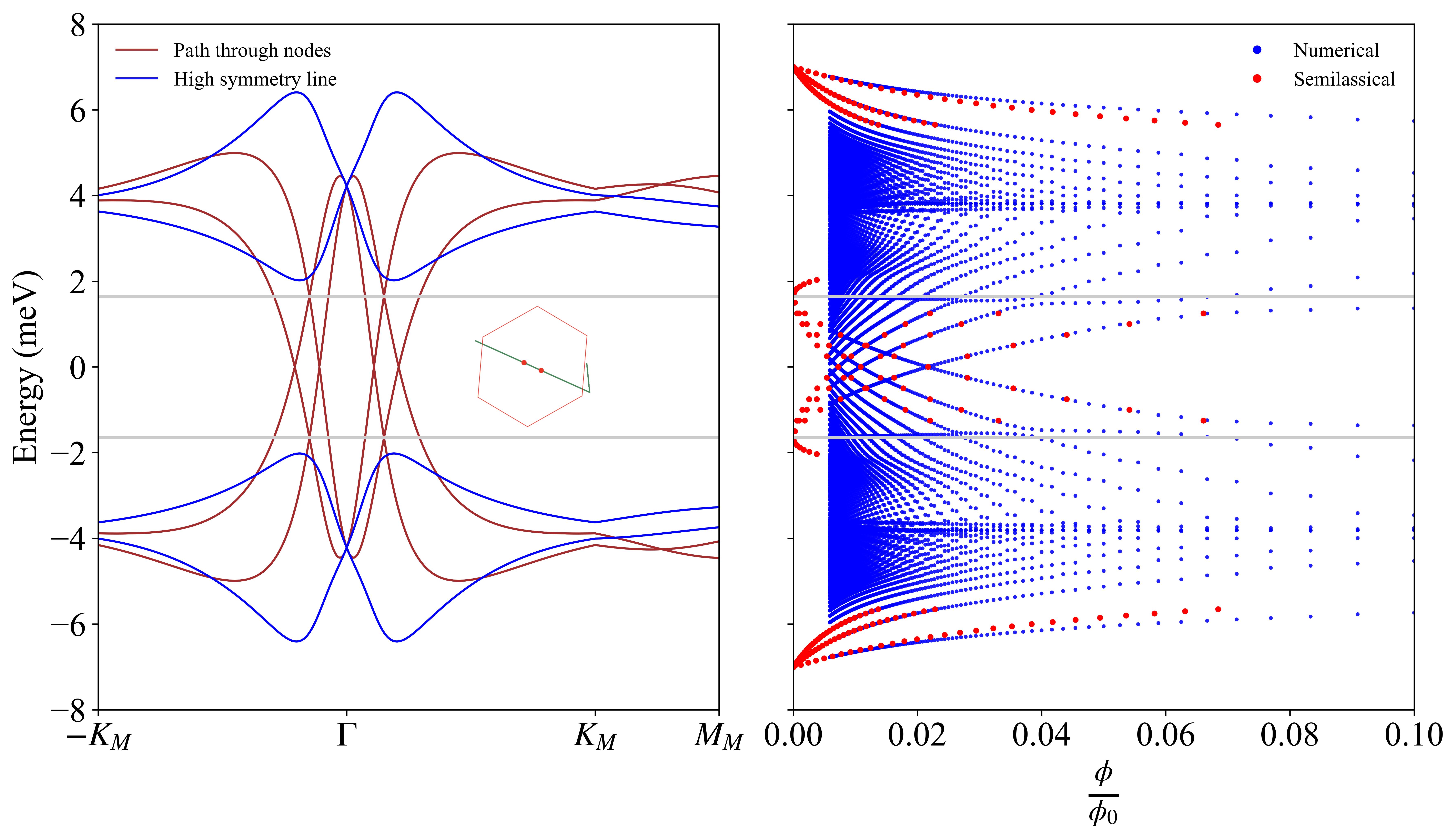}
    \caption{The strained THF model bandstructure in the left panel and its continuation in $\bB$ on the right panel. On the left panel, the bands along the high-symmetry line cut are in \textcolor{blue}{blue} while that along a rotated path through Dirac nodes is in \textcolor{brown}{brown}. The path through the nodes is shown in the inset in left panel, with location of nodes marked in \textcolor{red}{red}. The semiclassically (see Appx(\ref{Appx:Semiclassical:Prescription})) obtained LLs in \textcolor{red}{red} are plotted on top of the numerically obtained spectrum in \textcolor{blue}{blue} on the right panel. Solid \textcolor{gray}{grey} horizontal lines are plotted at the Dirac crossing points in both panels.}
    \label{fig:Non_itr_strain_B}
\end{figure}
The strained $f$-$f$ hybridization at $\bB$ for given spin $s$ can be calculated as
\begin{eqnarray}
   \hat{H}^\tau_{ff} &=&\sum_{k\in[0,1)\otimes [0,\frac{1}{q})} \sum_{b',b=1}^2 \sum_{r',r''=0}^{q-1}\bar{h}^{\tau}_{[br'],[b'r'']}(k)f^\dagger_{b\tau k r's}
f_{b'\tau k r'' s},
\end{eqnarray}
where
\begin{eqnarray}
  \bar{h}^{\tau}_{[br'],[b'r'']}(k) &=& \int d^2\br \eta^\dagger_{b\tau k r'}(\br)H^{\tau}_{BM}(\bp-\frac{e}{c}\bA)  \eta_{b'\tau k r''}(\br)\\
&=&\frac{1}{\sqrt{\mathcal{N}}}\sum_{s'n'}e^{2\pi i (s'k_1+n'(k_2+\frac{r''}{q}))} \int d^2\br \eta^\dagger_{b\tau k r'}(\br)H^{\tau}_{BM}(\bp-\frac{e}{c}\bA)\hat{t}^{s'}_{\tilde{\bL}_1 } \hat{t}^{n'}_{\tilde{\bL}_2 }  W_{\bf{0},b'\tau}(\br)\\
&=& \frac{1}{\sqrt{\mathcal{N}}}\sum_{s'n'}e^{2\pi i (s'k_1+n'(k_2+\frac{r''}{q}))} \int d^2\br \eta^\dagger_{b\tau k r'}(\br)\hat{t}^{s'}_{\tilde{\bL}_1 }H^{\tau}_{BM}(\bp-\frac{e}{c}\bA) \hat{t}^{n'}_{\tilde{\bL}_2 }  W_{\bf{0},b'\tau}(\br)\\
&=& \frac{s_{tot}}{\sqrt{\mathcal{N}}}\sum_{n'}e^{2\pi i n'(k_2+\frac{r''}{q})} \int d^2\br \eta^\dagger_{b\tau k r'}(\br)\hat{t}^{n'}_{\tilde{\bL}_2 } H^{\tau}_{BM}(\bp-\frac{e}{c}\bA)  W_{\bf{0},b'\tau}(\br)\\
&\approx& \frac{s_{tot}}{\sqrt{\mathcal{N}}}\sum_{n'}e^{2\pi i n'(k_2+\frac{r''}{q})} \int d^2\br \left(\hat{t}^{-n'}_{\tilde{\bL}_2 }\eta_{b\tau k r'}(\br)\right)^\dagger H^{\tau}_{BM}(\bp)  W_{\bf{0},b'\tau}(\br)
\end{eqnarray}
as the well localized Wannier state is exponentially suppressed in the region where $\bA = (0,Bx)$ is appreciable. 
Now note that $\hat{t}^{-1}_{\tilde{\bL}_2}\hat{t}_{\tilde{\bL}_1}=e^{2\pi i \frac{p}{q}}\hat{t}_{\tilde{\bL}_1}\hat{t}^{-1}_{\tilde{\bL}_2}$ and thus
\begin{eqnarray}
  \hat{t}^{-n'}_{\tilde{\bL}_2 }\eta_{b\tau k r'}(\br)  &=& \frac{1}{\sqrt{\mathcal{N}}}\sum_{sn}e^{2\pi i \left(sk_1+n(k_2+\frac{r'}{q})\right)} \hat{t}^{-n'}_{\tilde{\bL}_2 }\hat{t}^{s}_{\tilde{\bL}_1 } \hat{t}^{n}_{\tilde{\bL}_2 } W_{\bf{0},b\tau}(\br)\\
  &=& \frac{1}{\sqrt{\mathcal{N}}}\sum_{sn}e^{2\pi i \left(sk_1+n(k_2+\frac{r'}{q})\right)} e^{2\pi i sn'\frac{p}{q}}\hat{t}^{s}_{\tilde{\bL}_1 } \hat{t}^{n-n'}_{\tilde{\bL}_2 } W_{\bf{0},b\tau}(\br)\\
  &=& e^{2\pi i n'\left(k_2+\frac{r'}{q}\right)}\eta_{b\tau [k_1+n'\frac{p}{q}]_1k_2 r'}(\br)
\end{eqnarray}
Thus
\begin{eqnarray}
 \bar{h}^{\tau}_{[br'],[b'r'']}(k) &=& \frac{s_{tot}}{\sqrt{\mathcal{N}}}\sum_{n'}e^{2\pi i n'(\frac{r''-r'}{q})} \int d^2\br \eta_{b\tau [k_1+n'\frac{p}{q}]_1k_2 r'}^\dagger (\br)H^{\tau}_{BM}(\bp)  W_{\bf{0},b'\tau}(\br)\\
  &=& \frac{s_{tot}}{\mathcal{N}}\sum_{n'sn} e^{2\pi i n'(\frac{r''-r'}{q})}e^{-2\pi i s(k_1+\frac{n'p}{q})}e^{-2\pi i n(k_2+\frac{r'}{q})}\int d^2\br \left(\hat{t}^{s}_{\tilde{\bL}_1 } \hat{t}^{n}_{\tilde{\bL}_2 }W_{\bf{0},b\tau}(\br)\right)^\dagger  H^{\tau}_{BM}(\bp)  W_{\bf{0},b'\tau}(\br)\\
  &=& \frac{s_{tot}}{\mathcal{N}}\sum_{n'sn} e^{2\pi i n'(\frac{r''-r'}{q})}e^{-2\pi i s(k_1+\frac{n'p}{q})}e^{-2\pi i n(k_2+\frac{r'}{q})}\delta
  _{s,0}\delta_{n,0}\int d^2\br W_{\bf{0},b\tau}^\dagger(\br)  H^{\tau}_{BM}(\bp)  W_{\bf{0},b'\tau}(\br)
\end{eqnarray}
since the $W_{\bf{0},b\tau}$ are well localized at the origin. We thus have
\begin{eqnarray}
    \bar{h}^{\tau}_{[br'],[b'r'']}(k) &=& \frac{s_{tot}}{{\mathcal{N}}}\sum_{n'}e^{2\pi i n'(\frac{r''-r'}{q})} \int d^2\br W_{\bf{0},b\tau}^\dagger (\br) H^{\tau}_{BM}(\bp)  W_{\bf{0},b'\tau}(\br)\\
    &=& \delta_{r'',r'} \int d^2\br W_{\bf{0},b\tau}^\dagger  (\br)H^{\tau}_{BM}(\bp)  W_{\bf{0},b'\tau}(\br)\\
     &=& \delta_{r'',r'} M_f\left(\epsilon_{xy}\sigma_x + \tau\epsilon_-\sigma_y\right)_{bb'}
\end{eqnarray}
The strained THF Hofstadter spectrum is shown in Fig.(\ref{fig:Non_itr_strain_B})-right panel.

%---------------------Section change----

\section{Self-consistent Hartree Fock scheme for THF in magnetic field}\label{Appx:SCF_DETAILS}

In this appendix, we discuss the details of the Hartree-Fock self-consistency method employed to obtain the Chern gaps for the strain-induced anisotropic semimetal in magnetic field at CNP.

%---------------------Section change----

\subsection{Density operator }

We start by noting that the low energy field Eq.(\ref{Eq:appx:Low_energy_expansion}) projected to the heavy fermionic degree of freedom reads
    \begin{eqnarray}
 \hat{\psi}^f_{\tau s}(\br)&=& \sum_{k\in[0,1)\otimes [0,\frac{1}{q})}\sum_{b=1}^2\sum_{r'= 0}^{q-1} \eta_{b\tau k r'}(\br) f_{b\tau k r's} \\
&=& \sum_{k}\sum_{b=1}^2\sum_{r'= 0}^{q-1}  \sum_{s,n} e^{2\pi i\left(sk_1+n(k_2+\frac{r'}{q})\right)}\hat{t}^s_{\tilde{\bL}_1}\hat{t}^n_{\tilde{\bL}_2} W_{\bf{0},b\tau}(\br) f_{b\tau k r's}\\
&=&\sum_{b} \sum_{sn}\left(\hat{t}^s_{\tilde{\bL}_1}\hat{t}^n_{\tilde{\bL}_2} W_{b\tau}(\br)\right) \left(\sum_{kr'} \frac{1}{\sqrt{\mathcal{N}}} e^{2\pi i \left(sk_1+n(k_2+\frac{r'}{q})\right)} \hat{f}_{b\tau kr's}\right).\label{Eq:appx:lowfield_f_stop1}
\end{eqnarray}
Now, since any lattice vector $\bR$ can be decomposed as $\bR= s\tilde{\bL}_1+ n\tilde{\bL}_2$, with $s=\tilde{\bg}_1\cdot\bR/2\pi$ and $n=\tilde{\bg}_2\cdot\bR/2\pi$, and \cite{singh2024topological} $\hat{t}^s_{\tilde{\bL}_1} = e^{-i\pi s(s-1)\frac{p}{q}\frac{\tilde{L}_{1_y}}{\tilde{L}_m}} e^{2\pi is \frac{p}{q}\frac{y}{\tilde{L
}_m}}$
with $\tilde{L
}_m=\tilde{L}_2=|\tilde{\bL}_2|$, we can rewrite Eq.(\ref{Eq:appx:lowfield_f_stop1})  as 
\begin{eqnarray}
    \hat{\psi}^f_{\tau s}(\br)&=&\sum_{\bR}e^{i\mathcal{M}_R(\br)} W_{b\tau}(\br-\bR)d_{\bR b\tau s} \equiv \sum_{\bR} \tilde{W}_{\bR b\tau}(\br)d_{\bR b\tau s}\label{Eq:appx:precursor:Lattice_annihilation_at_B}.
\end{eqnarray}
Here the magnetic phase is given by 
\begin{eqnarray}
   \mathcal{M}_R(\br) &=& \frac{p}{q}\frac{\tilde{\bg}_1\cdot\bR}{\tilde{L}_m}\left(y + \frac{\tilde{L}_{1y}}{2}-\frac{\tilde{\bg}_1\cdot\bR}{4\pi}\tilde{L}_{1y}\right).\label{MTG phase to shifted Wannier} 
\end{eqnarray}
for $\phi/\phi_0=p/q$ and the lattice space annihilation operator reads
\begin{eqnarray}
  d^\dagger_{\bR b\tau s}  = \sum_{kr'} \frac{1}{\sqrt{\mathcal{N}}} e^{ i \left(\tilde{\bg}_1 k_1+\tilde{\bg}_2(k_2+\frac{r'}{q})\right)\cdot\bR} \hat{f}_{b\tau kr's}\label{Eq:appx:Lattice_annihilation_at_B}
\end{eqnarray}
Thus we can alternatively express Eq.(\ref{Eq:appx:Low_energy_expansion}) as
\begin{eqnarray}
    \hat{\psi}_{\tau s}(\br) &=& \sum_{\bR} \tilde{W}_{\bR b\tau}(\br)d^\dagger_{\bR b\tau s}
+\sum_{k\in[0,1)\otimes [0,\frac{1}{q})}\sum_{a=1}^4 \sum_{m=0}^{m_{a,\tau}}\sum_{r = 0}^{p-1} \Psi_{a\tau}(\br)\chi_{k r m}(\br)
c_{a\tau k r m s} \label{Eq:appx:Alternative_Low_energy_expansion}. 
\end{eqnarray}
and the real space density operator is given by 
\begin{eqnarray}
   \hat{\rho}(\br) = \sum_{\tau s} \hat{\psi}^\dagger_{\tau s} (\br)\hat{\psi}_{\tau s}(\br)
\end{eqnarray}
We note that for a $\phi/\phi_0=1/q$ sequence, the quantum number $r$ in $\chi_{krms}(\br)$ is always zero as $p=1$. We will thus hide this quantum label for brevity and represent the MTG-LL state by $\chi_{kms}(\br)$ and the corresponding annihilation operator by $c_{a\tau kms}$.

%---------------------Section change----

\subsection{Coulomb interaction}

Closely following the derivation in \cite{song2022magic}, we consider double-gate-screened Coulomb interaction 
\begin{eqnarray}
    V(\br) &=& U_\xi\sum^\infty_{n=-\infty}\frac{(-1)^n}{\sqrt{\left(|\br|/{\xi}\right)^2+n^2}} \\
    &=& \pi \xi^2 U_\xi\int \frac{d^2\bq}{(2\pi)^2} e^{-i\bq\cdot\br} \frac{\text{tanh}(\xi|\bq|/2)}{\xi|\bq|/2}
\end{eqnarray}
and the interaction Hamiltonian can be given as
\begin{eqnarray}
    \hat{H}_{I} = \frac{1}{2}\int d^2\br_1 d^2\br_2 V(\br_1-\br_2):\hat{\rho}(\br_1)::\hat{\rho}(\br_2):
\end{eqnarray}
where $:\hat{O}:=\hat{O}-\langle\hat{O}\rangle_{G_0}$ represents normal ordering with respect to a state $|G_0\rangle$ at charge neutrality point such that
\begin{eqnarray}
   \langle f^\dagger_{b_2k_2^fr_2\tau_2 s_2} f_{b_1k_1^fr_1\tau_1 s_1}\rangle_{G_0}&=& \frac{1}{2}\delta_{b_2b_1}\delta_{k_2^fk_1^f}\delta_{r_2r_1}\delta_{\tau_2\tau_1}\delta_{s_2 s_1},\label{Eq:appx:NormalOrder_f}\\
    \langle c^\dagger_{a_2k^c_2m_2\tau_2 s_2}c_{a_1k^c_1m_1\tau_1 s_1}\rangle_{G_0}&=&\frac{1}{2}\delta_{a_2a_1} \delta_{k_2^ck_1^c}\delta_{m_2m_1}\delta_{\tau_2\tau_1}\delta_{s_2 s_1},\\
  \langle c^\dagger_{a_2k^c_2m_2\tau_2 s_2}  f_{b_1k_1^fr_1\tau_1 s_1}\rangle_{G_0}&=&0.
\end{eqnarray}
It follows from  Eq.(\ref{Eq:appx:Lattice_annihilation_at_B}) and Eq.(\ref{Eq:appx:NormalOrder_f})
\begin{eqnarray}
    :d^\dagger_{\bR\alpha\tau s}d_{\bR'\alpha'\tau' s'}: &=& \frac{1}{\mathcal{N}}\sum_{\substack{kr\\k'r'}}e^{-i\left(k_1\bg_1 + (k_2+\frac{r}{q})\bg_2\right)\cdot\bR}e^{i\left(k_1'\bg_1 + (k_2'+\frac{r'}{q})\bg_2\right)\cdot\bR'}:f^\dagger_{\alpha kr\tau s} f_{\alpha'k'r'\tau' s'}:\\
    &=&\frac{1}{\mathcal{N}}\sum_{\substack{kr\\k'r'}}e^{-i\left(k_1\bg_1 + (k_2+\frac{r}{q})\bg_2\right)\cdot\bR}e^{i\left(k_1'\bg_1 + (k_2'+\frac{r'}{q})\bg_2\right)\cdot\bR'}\left(f^\dagger_{\alpha kr\tau s} f_{\alpha'k'r'\tau's'}-\frac{1}{2}\delta_{\alpha\alpha'}\delta_{kk'}\delta_{rr'}\delta_{\tau \tau'}\delta_{s s'}\right)\\
    &=& d^\dagger_{\bR\alpha\tau s}d_{\bR'\alpha'\tau' s'} - \frac{1}{2\mathcal{N}}\sum_{kr} e^{i\left(k_1\bg_1 + (k_2+\frac{r}{q})\bg_2\right)\cdot(\bR'-\bR)}\delta_{\alpha\alpha'}\delta_{\tau \tau'}\delta_{s s'}\\
    &=& d^\dagger_{\bR\alpha\tau s}d_{\bR'\alpha'\tau' s'} - \frac{1}{2}\delta_{\bR\bR'}\delta_{\alpha\alpha'}\delta_{\tau \tau'}\delta_{s s'}\label{Eq:Normal_order_d}
\end{eqnarray}
and thus
\begin{eqnarray}
    \langle d^\dagger_{\bR\alpha\tau s}d_{\bR'\alpha'\tau' s'}\rangle_{G_0} = \frac{1}{2}\delta_{\bR\bR'}\delta_{\alpha\alpha'}\delta_{\tau \tau'}\delta_{s s'}
\end{eqnarray}

%---------------------Section change----

\subsection{Interaction terms}\label{Appx:Interaction_terms_T=0}

Below we derive the density-density interaction terms at $\bB$. The lattice vectors $\tilde{\bL}_i$ and $\tilde{\bg}_i$ are defined with respect to the coordinate axis in which $\tilde{\bL}_2\parallel \hat{y}$.  
\subsubsection{\texorpdfstring{$ff$-$ff$}{ff-ff} density-density interaction}
The density-density interaction for the $f$s reads
\begin{eqnarray}
    \hat{H}_{U,\bB} &=& \frac{1}{2}\int d^2\br_1 \int d^2\br_2 V(\br_1-\br_2):\hat{\rho}_{ff,\bB}::\hat{\rho}_{ff,\bB}:
\end{eqnarray}
where the $f$-$f$ density is
\begin{eqnarray}
 \hat{\rho}_{ff,\bB}(\br) &=& \sum_{\substack{\tau s}} \sum_{\substack{\alpha_1 \alpha_2\\k^f_1 k^f_2\\r_1 r_2}}\eta^\dagger_{\alpha_1 k_1^f r_1 \tau_1 s_1}(\br) \eta_{\alpha_2 k_2^f r_2 \tau s} (\br)f^\dagger_{\alpha_1k^f_1r_1\tau_s} f_{\alpha_2 k^f_2r_2\tau s}\label{Eq:appx:ff_dens_atB}\\
 &=& \sum_{\substack{\tau s}} \sum_{\substack{\bR_1\bR_2\\\alpha_1\alpha_2}}\tilde{W}^\dagger_{\bR_1\alpha_1\tau}(\br)\tilde{W}_{\bR_2\alpha_2\tau}(\br)(\br)d^\dagger_{\bR_1\alpha_1 \tau s} d_{\bR_2\alpha_2 \tau s}\label{Eq:appx:dd_dens_atB}
\end{eqnarray}
Since $\tilde{W}^\dagger_{\bR\alpha\tau}(\br)$ is still a sharp gaussian, we can omit the contributions from different sites and thus
\begin{eqnarray}
     \hat{\rho}_{ff,\bB}(\br) &\approx& \sum_{\substack{\tau s \bR}} \sum_{\substack{\alpha_1\alpha_2}}\tilde{W}^\dagger_{\bR\alpha_1\tau}(\br)\tilde{W}_{\bR\alpha_2\tau}(\br)d^\dagger_{\bR\alpha_1 \tau s} d_{\bR\alpha_2 \tau s}\\
     &=&\sum_{\substack{\tau s \bR \alpha}} \tilde{W}^\dagger_{\bR\alpha\tau}(\br)\tilde{W}_{\bR\alpha\tau}(\br)(\br)d^\dagger_{\bR\alpha \tau s} d_{\bR\alpha \tau s}\\
     &=&\sum_{\bR}n_f(\br-\bR)\sum_{\substack{\tau s\alpha}} d^\dagger_{\bR\alpha \tau s} d_{\bR\alpha \tau s}\label{Eq:appx:dd_dens_atB_simplified}
\end{eqnarray}
as \cite{song2022magic}
\begin{eqnarray}
  \tilde{W}^\dagger_{\bR\alpha_1\tau}(\br)\tilde{W}_{\bR\alpha_2\tau}(\br)  = W^\dagger_{\alpha_1\tau}(\br-\bR)W_{\alpha_2\tau}(\br-\bR) = \delta_{\alpha_1\alpha_2}n_{f}(\br-\bR)
\end{eqnarray}
is independent of $\tau$. Thus 
\begin{eqnarray}
     \hat{H}_{U,\bB} &=& \frac{1}{2}\sum_{\substack{\bR_1\bR_2\\\alpha_1\alpha_2\\\tau_1\tau_2\\s_1s_2}}\int d^2\br_1 \int d^2\br_2 V(\br_1-\br_2)n_f(\br_1-\bR_1)n_f(\br_2-\bR_2)\nonumber\\&&:d^\dagger_{\bR_1\alpha_1\tau_1s_1}d_{\bR_1\alpha_1\tau_1s_1}::d^\dagger_{\bR_2\alpha_2\tau_2s_2}d_{\bR_2\alpha_2\tau_2s_2}:\\
&=&\frac{1}{2}\sum_{\substack{\bR_1\bR_2\\\alpha_1\alpha_2\\\tau_1\tau_2\\s_1s_2}}\int d^2\br_1 \int d^2\br_2 V(\br_1-\br_2+\bR_1-\bR_2)n_f(\br_1)n_f(\br_2)\nonumber\\&&:d^\dagger_{\bR_1\alpha_1\tau_1s_1}d_{\bR_1\alpha_1\tau_1s_1}::d^\dagger_{\bR_2\alpha_2\tau_2s_2}d_{\bR_2\alpha_2\tau_2s_2}:\\
&=&\frac{1}{2}\sum_{\substack{\bR_1\bR_2\\\alpha_1\alpha_2\\\tau_1\tau_2\\s_1s_2}}U(\bR_2-\bR_1):d^\dagger_{\bR_1\alpha_1\tau_1s_1}d_{\bR_1\alpha_1\tau_1s_1}::d^\dagger_{\bR_2\alpha_2\tau_2s_2}d_{\bR_2\alpha_2\tau_2s_2}:
\end{eqnarray}
where the integral 
\begin{eqnarray}
    U(\bR)=\int d^2\br_1 \int d^2\br_2 V(\br_1-\br_2-\bR)n_f(\br_1)n_f(\br_2)
\end{eqnarray}
was calculated in \cite{song2022magic}. Allowing only on-site and nearest neighbor contribution, we have 
\begin{eqnarray}
     \hat{H}_{U,\bB} &=& \frac{U_1}{2}\sum_{\bR}\sum_{\substack{\alpha_1\tau_1s_1\\\alpha_2\tau_2s_2}} :d^\dagger_{\bR\alpha_1\tau_1s_1}d_{\bR\alpha_1\tau_1s_1}::d^\dagger_{\bR\alpha_2\tau_2s_2}d_{\bR\alpha_2\tau_2s_2}: \nonumber\\&+& \frac{U_2}{2}\sum_{\langle\bR_1\bR_2\rangle}\sum_{\substack{\alpha_1\tau_1s_1\\\alpha_2\tau_2s_2}} :d^\dagger_{\bR_1\alpha_1\tau_1s_1}d_{\bR_1\alpha_1\tau_1s_1}::d^\dagger_{\bR_2\alpha_2\tau_2s_2}d_{\bR_2\alpha_2\tau_2s_2}: 
\end{eqnarray}
where $\langle\bR_1\bR_2\rangle$ represents sum over nearest neighbor sites. Using the fact that $\{d^\dagger_{\bR\ldots},d_{\bR'\ldots}\}=0$ for $\bR\neq\bR'$ and that triangular lattice has six nearest neighbours, we get
\begin{eqnarray}
 \hat{H}_{U,\bB}    &=& (8\mathcal{N}U_1+48\mathcal{N}U_2) - (4U_1+24U_2)\sum_{\bR\alpha\tau s}d^\dagger_{\bR\alpha\tau s} d_{\bR\alpha \tau s}+\frac{U_1}{2}\sum_\bR  \sum_{\substack{\tau,\tilde{\tau}\\ s,\tilde{ s}\nonumber\\\alpha,\tilde{\alpha}}} d^\dagger_{\bR\alpha\tau s} d_{\bR\alpha \tau s} d^\dagger_{\bR\tilde{\alpha} \tilde{\tau} \tilde{ s}} d_{\bR\tilde{\alpha} \tilde{\tau} \tilde{ s}}\\
&+&\frac{U_2}{2}\sum_{\langle\bR\bR'\rangle}  \sum_{\substack{\tau,\tilde{\tau}\\ s,\tilde{ s}\\\alpha,\tilde{\alpha}}} d^\dagger_{\bR\alpha\tau s} d^\dagger_{\bR'\tilde{\alpha} \tilde{\tau} \tilde{ s}} d_{\bR'\tilde{\alpha} \tilde{\tau} \tilde{ s}} d_{\bR\alpha \tau s} 
\end{eqnarray}
Since
\begin{eqnarray}
    d_{\bR\alpha \tau s} d^\dagger_{\bR\tilde{\alpha} \tilde{\tau} \tilde{ s}} d_{\bR\tilde{\alpha} \tilde{\tau} \tilde{ s}} &=& \left( -d^\dagger_{\bR\tilde{\alpha} \tilde{\tau} \tilde{ s}}d_{\bR\alpha \tau s} + \delta_{\alpha\tilde{\alpha}}\delta_{\tau\tilde{\tau}}\delta_{ s\tilde{ s}}\right)d_{\bR\tilde{\alpha} \tilde{\tau} \tilde{ s}}= d^\dagger_{\bR\tilde{\alpha} \tilde{\tau} \tilde{ s}}d_{\bR\tilde{\alpha} \tilde{\tau} \tilde{ s}}d_{\bR\alpha \tau s} + \delta_{\alpha\tilde{\alpha}}\delta_{\tau\tilde{\tau}}\delta_{ s\tilde{ s}}d_{\bR\tilde{\alpha} \tilde{\tau} \tilde{ s}}
\end{eqnarray}
we finally have 
\begin{eqnarray}
    \hat{H}_{U,\bf{B}}&=& (8\mathcal{N}U_1+48\mathcal{N}U_2) - (3.5U_1+24U_2)\sum_{\bR\alpha\tau s}d^\dagger_{\bR\alpha\tau s} d_{\bR\alpha \tau s}+\frac{U_1}{2}\sum_\bR  \sum_{\substack{\tau,\tilde{\tau}\\ s,\tilde{ s}\\\alpha,\tilde{\alpha}}} d^\dagger_{\bR\alpha\tau s}  d^\dagger_{\bR\tilde{\alpha} \tilde{\tau} \tilde{ s}} d_{\bR\tilde{\alpha} \tilde{\tau} \tilde{ s}}d_{\bR\alpha \tau s}\nonumber\\
&+&\frac{U_2}{2}\sum_{\langle\bR\bR'\rangle}  \sum_{\substack{\tau,\tilde{\tau}\\ s,\tilde{ s}\\\alpha,\tilde{\alpha}}} d^\dagger_{\bR\alpha\tau s} d^\dagger_{\bR'\tilde{\alpha} \tilde{\tau} \tilde{ s}} d_{\bR'\tilde{\alpha} \tilde{\tau} \tilde{ s}} d_{\bR\alpha \tau s}\label{Eq:ff_Interaction_full}
\end{eqnarray}
We note that $\hat{H}_{U,\bf{B}}$ takes the same form as $\hat{H}_{U}$ derived in Ref\cite{song2022magic}, except that the $\bB=0$ lattice space annihilation operator, $f_{\bR \alpha \tau s}$, corresponding to  the $\bB=0$ Wannier state $W_{\mathbf{0},\alpha \tau}(\br-\bR)$,  gets replaced by $d_{\bR \alpha \tau s}$  in Eq.(S157) of Ref\cite{song2022magic}, where $d_{\bR \alpha \tau s}$ is defined in Eq.(\ref{Eq:appx:Lattice_annihilation_at_B}) (also see Eq.(\ref{Eq:appx:precursor:Lattice_annihilation_at_B})).

%---------------------Section change----

\subsubsection{\texorpdfstring{$cc$-$cc$}{cc-cc} density-density interaction}
The density-density interaction for the $c$s reads
\begin{eqnarray}
    \hat{H}_{V,\bB} &=& \frac{1}{2}\int d^2\br_1 \int d^2\br_2 V(\br_1-\br_2):\hat{\rho}_{cc,\bB}(\br_1)::\hat{\rho}_{cc,\bB}(\br_2):
\end{eqnarray}
where the $c$-$c$ density is
\begin{eqnarray}
\hat{\rho}_{cc,\mathbf{B}}(\br_2) &=&  \sum_{\substack{\tau_2 s_2\\a_1 a_2}}\sum_{\substack{ k^c_1, m_1\\k^c_2, m_2} } \Psi^\dagger_{a_2\tau}(\br_2)\chi^\ast_{k^c_2 m_2}(\br_2) \Psi_{a_1\tau}(\br_1)\chi_{k^c_1 m_1}(\br_2):c_{ a_2 k^c_2 m_2 \tau_2 s_2}^\dagger c_{ a_1 k^c_1 m_1 \tau_2 s_2} :\label{Eq:appx:cc_dens_atB}.
\end{eqnarray}
Note that we skip the magnetic strip label $r$ in the MTG-LLs $\chi_{mk}(\br)$, i.e. focus on a $1/q$ sequence. We next have
\begin{eqnarray}
\hat{H}_{V,\bf{B}} &=&  \frac{1}{2}\int d^2\br_1\int d^2\br_2 V(\br_1-\br_2) \hat{\rho}_{cc,\mathbf{B}}(\br_1)\hat{\rho}_{cc,\mathbf{B}}(\br_2)  \\
&=&\sum_{\substack{\tau_1s_1\\\tau_2s_2}}\sum_{\substack{a_1 a_1'\\m_1 m_1'\\k^1_c k^{1'}_c}} \sum_{\substack{a_2 a_2'\\m_2 m_2'\\k^2_c k^{2'}_c}} \frac{1}{2}\int d^2\br_1\int d^2\br_2 V(\br_1-\br_2) \left(\Psi^\dagger_{a_1\tau_1}(\br_1)\chi^\ast_{k^c_1 m_1}(\br_1) \Psi_{a_1'\tau_1}(\br_1)\chi_{k^{c'}_1 m_1'}(\br_1)\right)\nonumber\\
&&\left(\Psi^\dagger_{a_2'\tau_2}(\br_2)\chi^\ast_{k^{c'}_2 m_2'}(\br_2) \Psi_{a_2\tau_2}(\br_2)\chi_{k^{c}_2 m_2}(\br_2)\right):c_{ a_1 k^c_1 m_1 \tau_1 s_1}^\dagger c_{ a_1' k^{c'}_1 m_1' \tau_1 s_1}::c_{ a_2' k^{c'}_2 m_2' \tau_2 s_2}^\dagger c_{ a_2 k^c_2 m_2 \tau_2 s_2}:\label{Eq:appx:cc-cc-1}
\end{eqnarray}
Let us rewrite
\begin{eqnarray}
 \chi^\ast_{k^{c'}_2 m_2}(\br_2)  \chi_{k^{c}_2 m_2'}(\br_2) &=& \int \frac{d^2\bq}{(2\pi)^2} e^{-i\bq\cdot\br_2} \int d^2\br e^{i\bq\cdot\br} \chi^\ast_{k^{c'}_2 m_2'}(\br)\chi_{k^{c}_2 m_2}(\br)\\
 &=& \int \frac{d^2\bq}{(2\pi)^2} e^{-i\bq\cdot\br_2} \mathcal{W}^{\bq}_{m_2' m_2}(k^{c'}_2,k^{c}_2)\label{Eq:appx:Fourier_W1}
 \end{eqnarray}
 and similarly
 \begin{eqnarray}
\chi^\ast_{k^c_1 m_1}(\br_1) \chi_{k^{c'}_1 m_1'}(\br_1)   = \int \frac{d^2\bq'}{(2\pi^2)} e^{-i\bq'\cdot\br_1} \mathcal{W}^{\bq'}_{m_1 m_1'}(k^{c}_1,k^{c'}_1)\label{Eq:appx:Fourier_W2}
 \end{eqnarray}
Now consider the Fourier decomposition of the $\bB=0$ $\bk\cdot\bp$ states
\begin{eqnarray}
\Psi^\dagger_{a_1\tau_1}(\br_1)\Psi_{a_1'\tau_1}(\br_1) = \sum_{\bG_1\bG_1'} e^{-i(\bG_1'-\bG_1)\cdot\br_1} u^{(\tau_1)\dagger}_{\bG_1a_1}(0) u^{(\tau_1)}_{\bG_1'a_1'}(0)\label{Eq:appx:Fourier_Psi_1}\\
\Psi^\dagger_{a_2'\tau_2}(\br_2)\Psi_{a_2\tau_2}(\br_2) = \sum_{\bG_2\bG_2'} e^{-i(\bG_2'-\bG_2)\cdot\br_2} u^{(\tau_2)\dagger}_{\bG_2'a_2'}(0) u^{(\tau_2)}_{\bG_2 a_2}(0)\label{Eq:appx:Fourier_Psi_2}
\end{eqnarray}
where $\bG,\bG'\in\mathbb{Z}\tilde{\bg}_1+\mathbb{Z}\tilde{\bg}_2$. We now substitute Eqs.(\ref{Eq:appx:Fourier_W1})-(\ref{Eq:appx:Fourier_Psi_2}) in Eq.(\ref{Eq:appx:cc-cc-1}), convert the $\bq$ integrals into summations $\frac{1}{\Omega_{tot}}\sum_\bq$ and get
\begin{eqnarray}
     \hat{H}_{V,\bf{B}} &=&  \sum_{\substack{\tau_1s_1\\\tau_2s_2}}\sum_{\substack{a_1 a_1'\\m_1 m_1'\\k^1_c k^{1'}_c}} \sum_{\substack{a_2 a_2'\\m_2 m_2'\\k^2_c k^{2'}_c}} \frac{1}{\Omega_{tot}}\sum_{\bq_2} \frac{1}{\Omega_{tot}}\sum_{\bq_1}
\mathcal{W}^{\bq_2}_{m_2' m_2}(k^{c'}_2,k^{c}_2) \mathcal{W}^{\bq_1}_{m_1 m_1'}(k^{c}_1,k^{c'}_1) \nonumber\\
&&\sum_{\substack{\bG_1,\bG_1'\\\bG_2\bG_2'}}\frac{1}{2}\int d^2\br_1\int d^2\br_2 V(\br_1-\br_2) \left(e^{-i(\bq_1+\bG_1'-\bG_1)\cdot\br_1}\right)\left(e^{-i(\bq_2+\bG_2'-\bG_2)\cdot\br_2}\right)\nonumber\\
&&\left(u^{(\tau_1)\dagger}_{\bG_1a_1}(0) u^{(\tau_1)}_{\bG_1'a_1'}(0)\right)\left(u^{(\tau_2)\dagger}_{\bG_2'a_2'}(0) u^{(\tau_2)}_{\bG_2 a_2}(0)\right):c_{ a_1 k^c_1 m_1 \tau_1 s_1}^\dagger c_{ a_1' k^{c'}_1 m_1' \tau_1 s_1}::c_{ a_2' k^{c'}_2 m_2' \tau_2 s_2}^\dagger c_{ a_2 k^c_2 m_2 \tau_2 s_2}:       
\end{eqnarray}
The real space integral factor above is
\begin{eqnarray}
&&\frac{1}{2}\int d^2\br_1\int d^2\br_2 V(\br_1-\br_2) \left(e^{-i(\bq_1+\bG_1'-\bG_1)\cdot\br_1}\right)\left(e^{-i(\bq_2+\bG_2'-\bG_2)\cdot\br_2}\right) \nonumber\\
=&&V(\bq_1+\bG_1'-\bG_1) \frac{1}{2}\int d^2\br_2  e^{-i(\bq_1+\bG_1'-\bG_1 + \bq_2+\bG_2'-\bG_2)\cdot\br_2}
\end{eqnarray}
Since the LL index $m\leq q/2$ (where the integer $q=\frac{\phi_0}{\phi}$),
$\bq_i$ cannot be huge-- 
$\mathcal{W}^{\bq_i}_{m_i m_i'}(k^{c}_i,k^{c'}_i)$ decays exponentially by $e^{-\bq_i^2\ell^2/2}$--so we separate the momentum conservations on $\bG$s and $\bq$s.  
\begin{eqnarray}
=&&V(\bq_1+\bG_1'-\bG_1) \frac{1}{2}\int d^2\br_2  e^{-i(\bq_1+\bG_1'-\bG_1 + \bq_2+\bG_2'-\bG_2)\cdot\br_2}    \\=&&\Omega_{tot}\frac{1}{2}V(\bq_1+\bG_1'-\bG_1)\delta_{\bq_2,-\bq_1} \delta_{\bG_1'-\bG_1,\bG_2'-\bG_2}
\end{eqnarray}
where $\Omega_{tot}$ is the area of the sample. Substituting above back in $ \hat{H}_{V,\bf{B}}$ gives 
\begin{eqnarray}
 \hat{H}_{V,\bf{B}} &=&  \frac{1}{2\Omega_{tot}}\sum_{\substack{\tau_1s_1\\\tau_2s_2}}\sum_{\substack{a_1 a_1'\\m_1 m_1'\\k^1_c k^{1'}_c}} \sum_{\substack{a_2 a_2'\\m_2 m_2'\\k^2_c k^{2'}_c}} \sum_{\bq} 
\mathcal{W}^{-\bq}_{m_2' m_2}(k^{c'}_2,k^{c}_2) \mathcal{W}^{\bq}_{m_1 m_1'}(k^{c}_1,k^{c'}_1) \nonumber \\
&&\sum_{\substack{\bG_1,\bG_1'\\\bG_2\bG_2'}}V(\bq+\bG_1'-\bG_1)\delta_{\bG_1'-\bG_1,\bG_2'-\bG_2}\left(u^{(\tau_1)\dagger}_{\bG_1a_1}(0) u^{(\tau_1)}_{\bG_1'a_1'}(0)\right)\left(u^{(\tau_2)\dagger}_{\bG_2'a_2'}(0) u^{(\tau_2)}_{\bG_2 a_2}(0)\right)\nonumber
\\&&:c_{ a_1 k^c_1 m_1 \tau_1 s_1}^\dagger c_{ a_1' k^{c'}_1 m_1' \tau_1 s_1}::c_{ a_2' k^{c'}_2 m_2' \tau_2 s_2}^\dagger c_{ a_2 k^c_2 m_2 \tau_2 s_2}:  \\
&=&  \frac{1}{2\Omega_{tot}}\sum_{\substack{\tau_1s_1\\\tau_2s_2}}\sum_{\substack{a_1 a_1'\\m_1 m_1'\\k^1_c k^{1'}_c}} \sum_{\substack{a_2 a_2'\\m_2 m_2'\\k^2_c k^{2'}_c}} \sum_{\bq} 
\mathcal{W}^{-\bq}_{m_2' m_2}(k^{c'}_2,k^{c}_2) \mathcal{W}^{\bq}_{m_1 m_1'}(k^{c}_1,k^{c'}_1)\nonumber\\
&&\sum_{\substack{\bG_1,\bG_2\\\bG}}V(\bq+\bG)\left(u^{(\tau_1)\dagger}_{\bG_1a_1}(0) u^{(\tau_1)}_{\bG_1-\bG a_1'}(0)\right)\left(u^{(\tau_2)\dagger}_{\bG_2-\bG a_2'}(0) u^{(\tau_2)}_{\bG_2 a_2}(0)\right)\nonumber
\\&&:c_{ a_1 k^c_1 m_1 \tau_1 s_1}^\dagger c_{ a_1' k^{c'}_1 m_1' \tau_1 s_1}::c_{ a_2' k^{c'}_2 m_2' \tau_2 s_2}^\dagger c_{ a_2 k^c_2 m_2 \tau_2 s_2}:
\end{eqnarray}
Now note that
\begin{eqnarray}
&&\sum_{\substack{\bG_1,\bG_2\\\bG}}V(\bq+\bG)\left(u^{(\tau_1)\dagger}_{\bG_1a_1}(0) u^{(\tau_1)}_{\bG_1-\bG a_1'}(0)\right)\left(u^{(\tau_2)\dagger}_{\bG_2-\bG a_2'}(0) u^{(\tau_2)}_{\bG_2 a_2}
(0)\right)\nonumber    \\
=&&\sum_{\bG} V(\bq+\bG)\langle \tilde{u}^{(\tau_1)}_{a_1}(0)|\tilde{u}^{(\tau_1)}_{a_1'}(\bG)\rangle  \langle \tilde{u}^{(\tau_2)}_{a_2'}(\bG)|\tilde{u}^{(\tau_2)}_{a_2'}(0)\rangle 
\end{eqnarray}
 is dominated by the $\bG=0$ contribution\cite{song2022magic}. We thus have
\begin{eqnarray}
 \hat{H}_{V,\bf{B}} &=&   \frac{1}{2\Omega_{tot}}\sum_{\substack{\tau_1s_1\\\tau_2s_2}}\sum_{\substack{a_1\\m_1 m_1'\\k^1_c k^{1'}_c}} \sum_{\substack{a_2 \\m_2 m_2'\\k^2_c k^{2'}_c}} \sum_{\bq} V(\bq)
\mathcal{W}^{\bq}_{m_1 m_1'}(k^{c}_1,k^{c'}_1)
\mathcal{W}^{-\bq}_{m_2' m_2}(k^{c'}_2,k^{c}_2) 
\nonumber\\&&:c_{ a_1 k^c_1 m_1 \tau_1 s_1}^\dagger c_{ a_1 k^{c'}_1 m_1' \tau_1 s_1}::c_{ a_2 k^{c'}_2 m_2' \tau_2 s_2}^\dagger c_{ a_2 k^c_2 m_2 \tau_2 s_2}:  \label{Eq:cc_Interaction_full}
\end{eqnarray}

The expression for $\mathcal{W}^{\bq}_{m m'}(k,k')$ can be found in Appendix(\ref{Appx:LLformfactor}).

%---------------------Section change----

\subsubsection{\texorpdfstring{$cc$-$ff$}{cc-ff} density-density interaction}
The density-density interaction between $c$ and $f$ reads
\begin{eqnarray}
    \hat{H}_{W,\bB} &=& \int d^2\br_1 \int d^2\br_2 V(\br_1-\br_2):\hat{\rho}_{ff,\bB}(\br_1)::\hat{\rho}_{cc,\bB}(\br_2):
\end{eqnarray}
where the density operators $\hat{\rho}_{ff,\bB}$ and $\hat{\rho}_{cc,\bB}$ are defined in Eqs.(\ref{Eq:appx:dd_dens_atB_simplified}) and (\ref{Eq:appx:cc_dens_atB}) respectively. So we have
\begin{eqnarray}
    \hat{H}_{W,\bB} &=&  \sum_{\bR \alpha}\sum_{\substack{a_1 k_1^c m_1\\a_1' k_1^{c'} m_1'}}\sum_{\substack{\tau_1 s_1\\\tau_2s_2}} \int \frac{d^2\bq}{(2\pi)^2}V(\bq) \left(\int d^2\br_1 e^{-i\bq\cdot\br_1}n_f(\br_1-\bR)\right)\nonumber \\
    &\times& \left(\int d^2\br_2 e^{i\bq\cdot\br_2} \Psi^\dagger_{a_1\tau_2}(\br_2)\chi^\ast_{k^c_1 m_1}(\br_2) \Psi_{a_1'\tau_2}(\br_2)\chi_{k^{c'}_1 m_1'}(\br_2)\right) :d^\dagger_{\bR \alpha \tau_1 s_1} d^\dagger_{\bR \alpha \tau_1 s_1}: :c_{ a_1 k^c_1 m_1 \tau_2 s_2}^\dagger c_{ a_1' k^{c'}_1 m_1' \tau_2 s_2}:
\end{eqnarray} 
We can also rewrite the above as 
\begin{eqnarray}
    \hat{H}_{W,\bB} &=&  \sum_{\bR \alpha}\sum_{\substack{a_1 k_1^c m_1\\a_1' k_1^{c'} m_1'}}\sum_{\substack{\tau_1 s_1\\\tau_2s_2}} \int \frac{d^2\bq'}{(2\pi)^2} \mathcal{W}^{\bq'}_{m_1 m_1'}(k_1^c,k^{c'}_1)\int d^2 \br_1\int d^2\br_2 V(\br_1-\br_2) n_f(\br_1-\bR)\nonumber\\
   &\times& e^{-i\bq'\cdot\br_2} \Psi^\dagger_{a_1\tau_2}(\br_2) \Psi_{a_1'\tau_2}(\br_2) :d^\dagger_{\bR \alpha \tau_1 s_1} d^\dagger_{\bR \alpha \tau_1 s_1}: :c_{ a_1 k^c_1 m_1 \tau_2 s_2}^\dagger c_{ a_1' k^{c'}_1 m_1' \tau_2 s_2}:
\end{eqnarray}
where $\mathcal{W}^\bq$ is defined in Eq.(\ref{Eq:appx:Fourier_W2}). Now using the Fourier expansion
\begin{eqnarray}
\Psi^\dagger_{a_1\tau_2}(\br_2) \Psi_{a_1'\tau_2}(\br_2) &=& \sum_{\bG,\bG'} e^{-i(\bG-\bG')\cdot\br_2} u^{(\tau_2)\dagger }_{\bG a_1}(0) u^{(\tau_2) }_{\bG' a_1'}(0)\\
&=& \sum_{\bG,\bG'} e^{-i\bG\cdot\br_2} u^{(\tau_2)\dagger }_{\bG a_1}(0) u^{(\tau_2) }_{\bG-\bG' a_1'}(0)
\end{eqnarray}
we have
\begin{eqnarray}
   \hat{H}_{W,\bB} &=&   \sum_{\bR \alpha}\sum_{\substack{a_1 k_1^c m_1\\a_1' k_1^{c'} m_1'}}\sum_{\substack{\tau_1 s_1\\\tau_2s_2}} \int \frac{d^2\bq'}{(2\pi)^2} \mathcal{W}^{\bq'}_{m_1 m_1'}(k_1^c,k^{c'}_1)\int d^2 \br_1\int d^2\br_2 V(\br_1-\br_2) n_f(\br_1-\bR)\nonumber\\
   &\times& \sum_{\bG,\bG'} e^{-i(\bG+\bq')\cdot\br_2} u^{(\tau)\dagger }_{\bG a_1}(0) u^{(\tau) }_{\bG-\bG' a_1'}(0) :d^\dagger_{\bR \alpha \tau_1 s_1} d^\dagger_{\bR \alpha \tau_1 s_1}: :c_{ a_1 k^c_1 m_1 \tau_2 s_2}^\dagger c_{ a_1' k^{c'}_1 m_1' \tau_2 s_2}:
\end{eqnarray}
Performing the $\br_2$ integral gives
\begin{eqnarray}
   \hat{H}_{W,\bB} &=&   \sum_{\bR \alpha}\sum_{\substack{a_1 k_1^c m_1\\a_1' k_1^{c'} m_1'}}\sum_{\substack{\tau_1 s_1\\\tau_2s_2}} \int \frac{d^2\bq'}{(2\pi)^2} \mathcal{W}^{\bq'}_{m_1 m_1'}(k_1^c,k^{c'}_1)\sum_{\bG\bG'}V(\bG+\bq') \int d^2 \br_1  e^{-i(\bG+\bq')\cdot\br_1} n_f(\br_1-\bR)\nonumber\\
   &\times& \left(u^{(\tau)\dagger }_{\bG a_1}(0) u^{(\tau) }_{\bG-\bG' a_1'}(0)\right) :d^\dagger_{\bR \alpha \tau_1 s_1} d^\dagger_{\bR \alpha \tau_1 s_1}: :c_{ a_1 k^c_1 m_1 \tau_2 s_2}^\dagger c_{ a_1' k^{c'}_1 m_1' \tau_2 s_2}: \\
   &=&   \sum_{\bR \alpha}\sum_{\substack{a_1 k_1^c m_1\\a_1' k_1^{c'} m_1'}}\sum_{\substack{\tau_1 s_1\\\tau_2s_2}} \int \frac{d^2\bq'}{(2\pi)^2} e^{-i\bq'\cdot\bR}\mathcal{W}^{\bq'}_{m_1 m_1'}(k_1^c,k^{c'}_1)\sum_{\bG\bG'}V(\bG+\bq')  n_f(\bG+\bq')\nonumber\\
   &\times& \left(u^{(\tau)\dagger }_{\bG a_1}(0) u^{(\tau) }_{\bG-\bG' a_1'}(0)\right) :d^\dagger_{\bR \alpha \tau_1 s_1} d^\dagger_{\bR \alpha \tau_1 s_1}: :c_{ a_1 k^c_1 m_1 \tau_2 s_2}^\dagger c_{ a_1' k^{c'}_1 m_1' \tau_2 s_2}:
\end{eqnarray}
where 
\begin{eqnarray}
    \int d^2 \br e^{-i\bq\cdot\br} n_f(r) \equiv n_f(\bq)
\end{eqnarray}
Recall that $\mathcal{W}^{\bq'}$ decays exponentially in $\bq'$, i.e. $e^{-\bq'^2\ell^2/2}$, which is much faster than $\sum_\bG V(\bq'+\bG)n_f(\bq'+\bG)$ as $\ell>>\tilde{L}_m$ at low $\bB$. We thus have
\begin{eqnarray}
     \hat{H}_{W,\bB} &\approx&  \sum_{\bR \alpha}\sum_{\substack{a_1 k_1^c m_1\\a_1' k_1^{c'} m_1'}}\sum_{\substack{\tau_1 s_1\\\tau_2s_2}} \left(\int \frac{d^2\bq'}{(2\pi)^2} e^{-i\bq'\cdot\bR}\mathcal{W}^{\bq'}_{m_1 m_1'}(k_1^c,k^{c'}_1)\right)\nonumber\\
   &\times& \left(\sum_{\bG\bG'}V(\bG)  n_f(\bG)u^{(\tau)\dagger }_{\bG a_1}(0) u^{(\tau) }_{\bG-\bG' a_1'}(0)\right) :d^\dagger_{\bR \alpha \tau_1 s_1} d^\dagger_{\bR \alpha \tau_1 s_1}: :c_{ a_1 k^c_1 m_1 \tau_2 s_2}^\dagger c_{ a_1' k^{c'}_1 m_1' \tau_2 s_2}:\\
   &=&\sum_{\bR \alpha}\sum_{\substack{a_1 k_1^c m_1\\a_1' k_1^{c'} m_1'}}\sum_{\substack{\tau_1 s_1\\\tau_2s_2}} \left(\chi^\ast_{k^c_1 m_1}(\bR) \chi_{k^{c'}_1 m_1'}(\bR)\right) \Omega_0 W_{a_1}\delta_{a_1a_1'}:d^\dagger_{\bR \alpha \tau_1 s_1} d^\dagger_{\bR \alpha \tau_1 s_1}: :c_{ a_1 k^c_1 m_1 \tau_2 s_2}^\dagger c_{ a_1' k^{c'}_1 m_1' \tau_2 s_2}:
\end{eqnarray}
where the last line follows from \cite{song2022magic}
\begin{eqnarray}
   \frac{1}{\Omega_0} \sum_{\bG,\bG'}  V(\bG)  n_f(\bG)u^{(\tau)\dagger }_{\bG a_1}(0) u^{(\tau) }_{\bG-\bG' a_1'}(0) = \delta_{a_1 a_1'} W_{a_1} 
\end{eqnarray}
where $\Omega_0$ is the strained unit cell area. We thus have
\begin{eqnarray}
 \hat{H}_{W,\bB}  &=& \Omega_0\sum_{\bR }\sum_{\substack{k_1^c m_1\\k_1^{c'} m_1'}}\left(\chi^\ast_{k^c_1 m_1}(\bR) \chi_{k^{c'}_1 m_1'}(\bR)\right) \sum_{\substack{a \alpha\\ \tau_1 s_1 \\ \tau_2 s_2}}W_a :d^\dagger_{\bR \alpha \tau_1 s_1} d^\dagger_{\bR \alpha \tau_1 s_1}: :c_{ a k^c_1 m_1 \tau_2 s_2}^\dagger c_{ a k^{c'}_1 m_1' \tau_2 s_2}: 
\end{eqnarray}
We can now rewrite the above in magnetic momentum space as
\begin{eqnarray}
     \hat{H}_{W,\bB}  &=&  \frac{\Omega_0}{\mathcal{N}}\sum_{\substack{\alpha \tau_1 s_1\\k_1^f k_2^f\\r_1r_2}}\sum_{\substack{a\tau_2 s_2\\k_1^c k_2^c\\m_1m_2}} \sum_\bR e^{i((k^f_{21}-k^f_{11})\tilde{\bg}_1+(k^f_{22}-k^f_{12} + \frac{r_2-r_1}{q})\tilde{\bg}_2)\cdot\bR}\left(\chi^\ast_{k^c_2 m_2}(\bR) \chi_{k^{c}_1 m_1}(\bR)\right)\nonumber\\
     &\times& W_a :f_{\alpha k^f_1 r_1 \tau_1 s_1}^\dagger f_{\alpha k^f_2 r_2 \tau_1 s_1}: :c_{ a k^c_2 m_2 \tau_2 s_2}^\dagger c_{ a k^c_1 m_1 \tau_2 s_2} :\label{Eq:Appx:MTG_density_sum_req}
\end{eqnarray}
where we use the notation $k^{f/c}_i = (k^{f/c}_{1i},k^{f/c}_{2i})$. Using the result in Appendix(\ref{appx:MTG density sum}), we can rewrite the above as
\begin{eqnarray}
     \hat{H}_{W,\bB}  &=&  \frac{1}{\mathcal{N}}\sum_{\substack{\alpha \tau_1 s_1\\k_1^f k_2^f\\r_1r_2}}\sum_{\substack{a\tau_2 s_2\\k_1^c k_2^c\\m_1m_2}} W_a\sum_{\substack{\bR\\l_1,l_2\in\mathcal{Z}}} \mathcal{C}^{\substack{k_1^f k_2^f;k_1^c k_2^c}}_{\substack{r_1r_2;s_1s_2}}(\bR) F_{m_1m_2}^{l_1l_2}(k_1^c k_2^c):f_{\alpha k^f_1 r_1 \tau_1 s_1}^\dagger f_{\alpha k^f_2 r_2 \tau_1 s_1}: :c_{ a k^c_2 m_2 \tau_2 s_2}^\dagger c_{ a k^c_1 m_1 \tau_2 s_2} : \label{Eq:appx:cc_ff_density-density}
\end{eqnarray}
where 
\begin{eqnarray}
     F^{l_1,l_2}_{m_1,m_2}(k_1^c,k_2^c) = e^{i(\phi^q_{l_1}(k^c_1)-\phi^q_{l_2}(k^c_2))} \frac{\tilde{L}_{1x}}{\ell}\varphi_{m_1}\left(-(l_1+qk^c_{21})\tilde{L}_{1x}\right) \varphi_{m_2}\left(-(l_2+qk^c_{22})L_{1x}\right) =(F^{l_2,l_1}_{m_2,m_1}(k_2^c,k_1^c))^\ast\label{Eq:Fdefn}
\end{eqnarray}
where the magnetic phase $\phi^q_{l_1}(k^c_1)$ is defined in Eq.(\ref{Eq:Defphase})
and the magnetic momentum conservation
\begin{eqnarray}
\mathcal{C}^{\substack{k_1^f k_2^f;k_1^c k_2^c}}_{\substack{r_1r_2;s_1s_2}}(\bR) &=& \frac{1}{\mathcal{N}}e^{i\left(\left(k^f_{12}-k^f_{11}+k^c_{11}-k^c_{12}\right)\tilde{\bg}_1 + \left(k_{21}^c-k^c_{22}+k_{22}^f-k^f_{21}+\frac{s_1-s_2+r_2-r_1}{q}\right)\tilde{\bg}_2\right)\cdot\bR} .\label{Eq:appx:Magnetic_momentum_conservation}
\end{eqnarray}

%---------------------Section change----

\subsubsection{Exchange interaction}
The exchange interaction between the $c$ and $f$ fermions is given by 
    \begin{eqnarray}
  \hat{H}_{{J},\bf{B}} &=& \frac{1}{2}\int d^2\br_1 d^2 \br_2 V(\br_1 -\br_2)  [\hat{\rho}_{fc,\bf{B}}(\br_1) \hat{\rho}_{cf,\bf{B}}(\br_2) + \hat{\rho}_{cf,\bf{B}}(\br_1) \hat{\rho}_{fc,\bf{B}}(\br_2)] 
\end{eqnarray}
where the $c$-$f$ density operator reads
\begin{eqnarray}
     \hat{\rho}_{cf,\bf{B}}(\br) &=& \sum_{\substack{\tau s} } \sum_{\substack{k^c m} } \sum_{\substack{k^f r} }   \sum_{\substack{b a} }\chi^\ast_{k^c,m} (\br)\Psi^\dagger_{a,\tau}(\br) \eta_{b,k^f,r,\tau}(\br)  c_{ a k^c m \tau s}^\dagger  f_{b_2 k^f r \tau s}\\
     &=&\sum_{\substack{\tau s} } \sum_{\substack{k^c m} } \sum_{\substack{\bR} }   \sum_{\substack{b_2 a_2} }\chi^\ast_{k^c_2,m} (\br)\Psi^\dagger_{a,\tau}(\br) \tilde{W}_{\bR b \tau}c_{ a k^c m \tau s}^\dagger d_{\bR b \tau s}\\
     &=& \hat{\rho}^\dagger_{fc,\bf{B}}(\br)  
\end{eqnarray}
We thus have 
\begin{eqnarray}
    \hat{H}_{{J},\bf{B}} &=& \frac{1}{2}\int d^2\br_1 d^2 \br_2 V(\br_1 -\br_2) \sum_{\substack{\bR_1 b_1 \bR_2 b_2\\k^c_1 m_1 a_1 k^c_2 m_2 a_2\\\tau_1 s_1 \tau_2 s_2}} \left(\tilde{W}^\dagger_{\bR_1 b_1 \tau_1}(\br_1) \chi_{k^c_1,m_1} (\br_1)\Psi_{a_1,\tau_1}(\br_1) \right)  \left(\chi^\ast_{k^c_2,m_2} (\br_2)\Psi^\dagger_{a_2,\tau_2}(\br_2)  \tilde{W}_{\bR_2 b_2 \tau_2}(\br_2)\right)\nonumber\\
    &\times&\left(d_{\bR_1 b_1 \tau_1 s_1}^\dagger c_{ a_1 k^c_1 m_1 \tau_1 s_1} c_{ a_2 k^c_2 m_2 \tau_2 s_2}^\dagger  d_{\bR_2 b_2 \tau_2 s_2} + c_{ a_2 k^c_2 m_2 \tau_2 s_2}^\dagger  d_{\bR_2 b_2 \tau_2 s_2} d_{\bR_1 b_1  \tau_1 s_1}^\dagger c_{ a_1 k^c_1 m_1 \tau_1 s_1}\right)\\
     &\approx& \frac{1}{2}\int d^2\br_1 d^2 \br_2 V(\br_1 -\br_2) \sum_{\substack{\bR b_1 b_2\\k^c_1 m_1 a_1 k^c_2 m_2 a_2\\\tau_1 s_1 \tau_2 s_2}} \left(\tilde{W}^\dagger_{\bR b_1 \tau_1}(\br_1) \chi_{k^c_1,m_1} (\br_1)\Psi_{a_1,\tau_1}(\br_1) \right)  \left(\chi^\ast_{k^c_2,m_2} (\br_2)\Psi^\dagger_{a_2,\tau_2}(\br_2)  \tilde{W}_{\bR b_2 \tau_2}(\br_2)\right)\nonumber\\
    &\times&\left(d_{\bR b_1 \tau_1 s_1}^\dagger c_{ a_1 k^c_1 m_1 \tau_1 s_1} c_{ a_2 k^c_2 m_2 \tau_2 s_2}^\dagger  d_{\bR b_2 \tau_2 s_2} + c_{ a_2 k^c_2 m_2 \tau_2 s_2}^\dagger  d_{\bR b_2 \tau_2 s_2} d_{\bR b_1  \tau_1 s_1}^\dagger c_{ a_1 k^c_1 m_1 \tau_1 s_1}\right)\\
    &=& \frac{1}{2} \int \frac{d^2\bq}{(2\pi)^2 } V(\bq) \sum_{\substack{\bR b_1 b_2\\k^c_1 m_1 a_1 k^c_2 m_2 a_2\\\tau_1 s_1 \tau_2 s_2}} \left(\int d^2\br_1 e^{-i\bq_1\cdot\br_1} \tilde{W}^\dagger_{\bR b_1 \tau_1}(\br_1) \chi_{k^c_1,m_1} (\br_1)\Psi_{a_1,\tau_1}(\br_1) \right) \nonumber \\
    &\times& \left(\int d^2\br_2 e^{i\bq\cdot\br_2} \chi^\ast_{k^c_2,m_2} (\br_2)\Psi^\dagger_{a_2,\tau_2}(\br_2)  \tilde{W}_{\bR b_2 \tau_2}(\br_2)\right)\nonumber\\
    &\times&\left(d_{\bR b_1 \tau_1 s_1}^\dagger c_{ a_1 k^c_1 m_1 \tau_1 s_1} c_{ a_2 k^c_2 m_2 \tau_2 s_2}^\dagger  d_{\bR b_2 \tau_2 s_2} + c_{ a_2 k^c_2 m_2 \tau_2 s_2}^\dagger  d_{\bR b_2 \tau_2 s_2} d_{\bR b_1  \tau_1 s_1}^\dagger c_{ a_1 k^c_1 m_1 \tau_1 s_1}\right)
\end{eqnarray}
as the interaction will be dominated by the on-site term because of the well-localized nature of the Wannier states \cite{song2022magic}. Now see that
\begin{eqnarray}
   \chi^\ast_{k^c_2,m_2} (\br_2)\Psi^\dagger_{a_2,\tau_2}(\br_2)  \tilde{W}_{\bR b_2 \tau_2}(\br_2) &=& \chi^\ast_{k^c_2,m_2} (\br_2)\Psi^\dagger_{a_2,\tau_2}(\br_2) e^{i{\mathcal{M}}_\bR(\br)} {W}_{\bR b_2 \tau_2}(\br_2)\\
    &=& \chi^\ast_{k^c_2,m_2} (\br_2)\Psi^\dagger_{a_2,\tau_2}(\br_2) \hat{t}^s_{\tilde{\bL}_1} \hat{t}^n_{\tilde{\bL}_2} W_{\mathbf{0}b_2\tau_2}(\br_2)
\end{eqnarray}
where integers $s,n$ are defined by $\bR=s\tilde{\bL}_1+n\tilde{\bL}_2$.
Thus 
\begin{eqnarray}
    \int d^2\br_2 e^{i\bq\cdot\br_2} \chi^\ast_{k^c_2,m_2} (\br_2)\Psi^\dagger_{a_2,\tau_2}(\br_2)  \tilde{W}_{\bR b_2 \tau_2}(\br_2) &=& \int d^2\br_2 e^{i\bq\cdot\br_2} \chi^\ast_{k^c_2,m_2} (\br_2)\Psi^\dagger_{a_2,\tau_2}(\br_2)e^{-i\pi s(s-1)\frac{\tilde{L}_{1y}}{q\tilde{L}_2}} e^{2\pi i s\frac{y}{q\tilde{L}_2}} \nonumber\\&\times& W_{\mathbf{0}b_2\tau_2}(\br_2-s\tilde{\bL}_1-n\tilde{\bL}_2)\\&=& e^{i\bq\cdot\bR} \int d^2\br_2 e^{i\bq\cdot\br_2} \chi^\ast_{k^c_2,m_2} (\br_2+\bR)\Psi^\dagger_{a_2,\tau_2}(\br_2)e^{-i\pi s(s-1)\frac{\tilde{L}_{1y}}{q\tilde{L}_2}} \nonumber\\&\times& e^{2\pi i s\frac{y+s\tilde{L}_{1y}+n\tilde{L}_2}{q\tilde{L}_2}}  W_{\mathbf{0}b_2\tau_2}(\br_2)
\end{eqnarray}
where we used $\Psi^\dagger_{a_2,\tau_2}(\br_2+\bR)=\Psi^\dagger_{a_2,\tau_2}(\br_2)$. Now note that
\begin{eqnarray}
   e^{-i\pi s(s-1)\frac{\tilde{L}_{1y}}{q\tilde{L}_2}} e^{2\pi i s\frac{y+s\tilde{L}_{1y}+n\tilde{L}_2}{q\tilde{L}_2}}   \chi^\ast_{k^c_2,m_2} (\br_2+\bR) &=& e^{2\pi i \frac{sn}{q}} e^{i\pi s(s+1)\frac{\tilde{L}_{1y}}{q\tilde{L}_2}}  e^{2\pi i s\frac{y}{q\tilde{L}_2}}  \hat{T}^{-s}_{\tilde{\bL}_1}\hat{T}^{-n}_{\tilde{\bL}_2}\chi^\ast_{k^c_2,m_2} (\br_2)\\
   &=& \left( e^{-2\pi i \frac{sn}{q}} e^{-i\pi s(s+1)\frac{\tilde{L}_{1y}}{q\tilde{L}_2}}  e^{-2\pi i s\frac{y}{q\tilde{L}_2}}  \hat{T}^{-s}_{\tilde{\bL}_1}\hat{T}^{-n}_{\tilde{\bL}_2}\chi_{k^c_2,m_2} (\br_2) \right)^\ast\\
   &=& \left( e^{-2\pi i \frac{sn}{q}}  \hat{t}^{-s}_{\tilde{\bL}_1}\hat{t}^{-n}_{\tilde{\bL}_2}\chi_{k^c_2,m_2} (\br_2) \right)^\ast \\&=& \left(\hat{t}^{-n}_{\tilde{\bL}_2}\hat{t}^{-s}_{\tilde{\bL}_1}\chi_{k^c_2,m_2} (\br_2) \right)^\ast\\
   &=& e^{-2\pi i s k^c_{12}} \left(\hat{t}^{-n}_{\tilde{\bL}_2}\chi_{k^c_2,m_2} (\br_2) \right)^\ast\\
   &=& e^{-2\pi i s k^c_{12}} e^{-2\pi i n k^c_{22}} \chi^\ast_{[k^c_{12}+\frac{n}{q}]_1,k^c_{22},m_2} (\br_2)
\end{eqnarray} 
where we used the algebra of magnetic translations $\hat{t}_{\tilde{\bL}_1}\hat{t}_{\tilde{\bL}_2}= e^{2\pi i/q}\hat{t}_{\tilde{\bL}_2}\hat{t}_{\tilde{\bL}_1} \Leftrightarrow \hat{t}^{-1}_{\tilde{\bL}_1}\hat{t}^{-1}_{\tilde{\bL}_2}= e^{2\pi i/q}\hat{t}^{-1}_{\tilde{\bL}_2}\hat{t}^{-1}_{\tilde{\bL}_1}\Leftrightarrow \hat{t}^{-1}_{\tilde{\bL}_2} \hat{t}_{\tilde{\bL}_1} = e^{2\pi i/q}\hat{t}_{\tilde{\bL}_1}\hat{t}^{-1}_{\tilde{\bL}_2}$ and $k^c_2 = (k^c_{12},k^c_{22})$. We thus have
\begin{eqnarray}
     \hat{H}_{{J},\bf{B}}
     &=& 
      \frac{1}{2} \int \frac{d^2\bq}{(2\pi)^2 } V(\bq) \sum_{\substack{\bR b_1 b_2\\k^c_1 m_1 a_1 k^c_2 m_2 a_2\\\tau_1 s_1 \tau_2 s_2}} e^{ i(k^c_{11}-k^c_{12})\tilde{\bg}_1\cdot\bR}e^{i (k^c_{21}-k^c_{22})\tilde{\bg}_2\cdot\bR}\nonumber\\
      &\times&\left(\int d^2\br_1 e^{-i\bq_1\cdot\br_1} {W}^\dagger_{\bf{0} b_1 \tau_1}(\br_1) \chi_{[k^c_{11}+\frac{\tilde{\bg}_2\cdot\bR}{2\pi q}]_1,k^c_{21},m_1} (\br_1)\Psi_{a_1,\tau_1}(\br_1) \right)\nonumber \\
      &\times& \left(\int d^2\br_2 e^{i\bq\cdot\br_2} \chi^\ast_{[k^c_{12}+\frac{\tilde{\bg}_2\cdot\bR}{2\pi q}]_1,k^c_{22},m_2} (\br_2)\Psi^\dagger_{a_2,\tau_2}(\br_2)  {W}_{\bf{0} b_2 \tau_2}(\br_2)\right)
     \nonumber \\
    &\times&\left(d_{\bR b_1 \tau_1 s_1}^\dagger c_{ a_1 k^c_1 m_1 \tau_1 s_1} c_{ a_2 k^c_2 m_2 \tau_2 s_2}^\dagger  d_{\bR b_2 \tau_2 s_2} + c_{ a_2 k^c_2 m_2 \tau_2 s_2}^\dagger  d_{\bR b_2 \tau_2 s_2} d_{\bR b_1  \tau_1 s_1}^\dagger c_{ a_1 k^c_1 m_1 \tau_1 s_1}\right)
\end{eqnarray}
Now note that at low $\bB$, $\ell>>\tilde{L}_m$ and thus the MTG-LL functions vary very slowly compared to the Wannier functions (which are localized at origin). We thus approximate
\begin{eqnarray}
     \hat{H}_{{J},\bf{B}}
     &\approx& 
      \sum_{\substack{\bR b_1 b_2\\k^c_1 m_1 a_1 k^c_2 m_2 a_2\\\tau_1 s_1 \tau_2 s_2}} e^{ i(k^c_{11}-k^c_{12})\tilde{\bg}_1\cdot\bR}e^{i (k^c_{21}-k^c_{22})\tilde{\bg}_2\cdot\bR} \chi^\ast_{[k^c_{12}+\frac{\tilde{\bg}_2\cdot\bR}{2\pi q}]_1,k^c_{22},m_2} (0) \chi_{[k^c_{11}+\frac{\tilde{\bg}_2\cdot\bR}{2\pi q}]_1,k^c_{21},m_1} (0)\nonumber\\
      &\times&\frac{1}{2} \int \frac{d^2\bq}{(2\pi)^2 } V(\bq) \left(\int d^2\br_1 e^{-i\bq_1\cdot\br_1} {W}^\dagger_{\bf{0} b_1 \tau_1}(\br_1) \Psi_{a_1,\tau_1}(\br_1) \right) \left(\int d^2\br_2 e^{i\bq\cdot\br_2} \Psi^\dagger_{a_2,\tau_2}(\br_2)  {W}_{\bf{0} b_2 \tau_2}(\br_2)\right)
     \nonumber \\
    &\times&\left(d_{\bR b_1 \tau_1 s_1}^\dagger c_{ a_1 k^c_1 m_1 \tau_1 s_1} c_{ a_2 k^c_2 m_2 \tau_2 s_2}^\dagger  d_{\bR b_2 \tau_2 s_2} + c_{ a_2 k^c_2 m_2 \tau_2 s_2}^\dagger  d_{\bR b_2 \tau_2 s_2} d_{\bR b_1  \tau_1 s_1}^\dagger c_{ a_1 k^c_1 m_1 \tau_1 s_1}\right)\\
    &=& \Omega_{0} \sum_{\substack{\bR b_1 b_2\\k^c_1 m_1 a_1 k^c_2 m_2 a_2\\\tau_1 s_1 \tau_2 s_2}} e^{ i(k^c_{11}-k^c_{12})\tilde{\bg}_1\cdot\bR}e^{i (k^c_{21}-k^c_{22})\tilde{\bg}_2\cdot\bR} \chi^\ast_{[k^c_{12}+\frac{\tilde{\bg}_2\cdot\bR}{2\pi q}]_1,k^c_{22},m_2} (0) \chi_{[k^c_{11}+\frac{\tilde{\bg}_2\cdot\bR}{2\pi q}]_1,k^c_{21},m_1} (0)\nonumber\\
    &\times& \mathcal{J}_{\tau_1 b_1 a_1, \tau_2 b_2 a_2} \left(d_{\bR b_1 \tau_1 s_1}^\dagger c_{ a_1 k^c_1 m_1 \tau_1 s_1} c_{ a_2 k^c_2 m_2 \tau_2 s_2}^\dagger  d_{\bR b_2 \tau_2 s_2} + c_{ a_2 k^c_2 m_2 \tau_2 s_2}^\dagger  d_{\bR b_2 \tau_2 s_2} d_{\bR b_1  \tau_1 s_1}^\dagger c_{ a_1 k^c_1 m_1 \tau_1 s_1}\right)
\end{eqnarray}
where $\mathcal{J}_{\tau_1 b_1 a_1, \tau_2 b_2 a_2}$ was calculated in Ref\cite{song2022magic}. Using the properties of  $\mathcal{J}_{\tau_1 b_1 a_1, \tau_2 b_2 a_2}$ Ref\cite{song2022magic} we can rewrite the above as 
\begin{eqnarray}
  \hat{H}_{{J},\bf{B}}    &=& -J\Omega_{0} \sum_{\substack{\bR \alpha \tau\\k^c_1 m_1 a_1 k^c_2 m_2 a_2\\s_1 s_2}} e^{ i(k^c_{11}-k^c_{12})\tilde{\bg}_1\cdot\bR}e^{i (k^c_{21}-k^c_{22})\tilde{\bg}_2\cdot\bR} \chi^\ast_{[k^c_{12}+\frac{\tilde{\bg}_2\cdot\bR}{2\pi q}]_1,k^c_{22},m_2} (0) \chi_{[k^c_{11}+\frac{\tilde{\bg}_2\cdot\bR}{2\pi q}]_1,k^c_{21},m_1} (0)\nonumber\\
 &\times& \left( :d^\dagger_{\bR\alpha\tau\bs_1} d_{\bR\alpha\tau\bs_2}::c^\dagger_{\alpha+2 k_2^cm_2\tau\bs_2} c_{\alpha+2 k_1^cm_1\tau\bs_1}:-d^\dagger_{\bR\alpha\tau\bs_1} d_{\bR\bar{\alpha}-\tau\bs_2} c^\dagger_{\bar{\alpha}+2 k_2^cm_2-\tau\bs_2} c_{\alpha+2 k_1^cm_1\tau\bs_1}\right)
\end{eqnarray}
where $J = \mathcal{J}_{\tau 1 3,\tau 1 3}= \mathcal{J}_{\tau 2 4,\tau 2 4}= -\mathcal{J}_{\tau 1 3,\bar{\tau} 2 4}=-\mathcal{J}_{\tau 2 4,\bar{\tau} 1 3}$ \cite{song2022magic}. We can simplify the prefactor as
\begin{eqnarray}
 \frac{1}{\mathcal{N}\Omega_0} \mathcal{F}^\bR_{m_1 m_2}(k^c_1,k^c_2) &=&   e^{ i(k^c_{11}-k^c_{12})\tilde{\bg}_1\cdot\bR}e^{i (k^c_{21}-k^c_{22})\tilde{\bg}_2\cdot\bR} \chi^\ast_{[k^c_{12}+\frac{\tilde{\bg}_2\cdot\bR}{2\pi q}]_1,k^c_{22},m_2} (0) \chi_{[k^c_{11}+\frac{\tilde{\bg}_2\cdot\bR}{2\pi q}]_1,k^c_{21},m_1} (0)\\
  &=& \frac{1}{\mathcal{N}\ell \tilde{L}_m}  e^{ i(k^c_{11}-k^c_{12})\tilde{\bg}_1\cdot\bR}e^{i (k^c_{21}-k^c_{22})\tilde{\bg}_2\cdot\bR} \sum_{l l'\in \mathbb{Z}} e^{i (\frac{l-l'}{q})\tilde{\bg}_2\cdot\bR } e^{i(\phi^q_{l}(k^c_1)-\phi^q_{l'}(k^c_2))}\nonumber\\
  &\times&\varphi_{m_1}\left(-(l_1+qk^c_{21})L_{1x}\right) \varphi_{m_2}\left(-(l_2+qk^c_{22})L_{1x}\right)
\end{eqnarray}
In a compact form
\begin{eqnarray}
 \frac{1}{\mathcal{N}\Omega_0} \mathcal{F}^\bR_{m_1 m_2}(k^c_1,k^c_2) &=& \sum_{l_1 l_2\in \mathbb{Z}}  \frac{1}{\Omega_0 \mathcal{N}}   e^{ i(k^c_{11}-k^c_{12})\tilde{\bg}_1\cdot\bR}e^{i (k^c_{21}-k^c_{22}+\frac{l_1-l_2}{q})\tilde{\bg}_2\cdot\bR}  F^{l_1 l_2}_{m_1 m_2}(k^c_1, k^c_2)
\end{eqnarray}
We thus have 
\begin{eqnarray}
     \hat{H}_{{J},\bf{B}}    &=& -\frac{J}{\mathcal{N}} \sum_{\substack{\bR \alpha \tau\\k^c_1 m_1 k^c_2 m_2 \\s_1 s_2}} \mathcal{F}^\bR_{m_1 m_2}(k^c_1,k^c_2) \left( :d^\dagger_{\bR\alpha\tau\bs_1} d_{\bR\alpha\tau s_2}::c^\dagger_{\alpha+2 k_2^cm_2\tau s_2} c_{\alpha+2 k_1^cm_1\tau s_1}:\right.\nonumber\\
 &-&\left. d^\dagger_{\bR\alpha\tau\bs_1} d_{\bR\bar{\alpha}-\tau s_2} c^\dagger_{\bar{\alpha}+2 k_2^cm_2-\tau s_2} c_{\alpha+2 k_1^cm_1\tau s_1}\right) \label{Eq:appx:cf_fc_Exchange}
\end{eqnarray}
In terms of the momentum space representation of $f$-modes, we have
\begin{eqnarray}
    \hat{H}_{{J},\bf{B}}    &=& -\frac{J}{\mathcal{N}^2} \sum_{\substack{\bR \alpha \tau\\k^c_1 m_1  k^c_2 m_2 \\k^f_1 k^f_2 r_2 r_1\\s_1 s_2}} \mathcal{F}^\bR_{m_1 m_2}(k^c_1,k^c_2) e^{i(k^f_{12}-k^f_{11})\tilde{\bg}_1\cdot\bR} e^{i(k^f_{22}-k^f_{21}+\frac{r_2-r_1}{q})\tilde{\bg}_1\cdot\bR}
    \nonumber\\&\times& \left(:f_{\alpha k^f_1 r_1 \tau s_1}^\dagger f_{\alpha k^f_2 r_2 \tau s_2}:c^\dagger_{\alpha+2 k_2^cm_2\tau\bs_2} c_{\alpha+2 k_1^cm_1\tau s_1}: - f_{\alpha k^f_1 r_1 \tau s_1}^\dagger f_{\bar{\alpha} k^f_2 r_2 -\tau s_2} c_{ \bar{\alpha}+2 k^c_2 m_2 \tau s_2}^\dagger c_{ \alpha+2 k^c_1 m_1 \tau s_1}\right).
\end{eqnarray}
Now see that
\begin{eqnarray}
 &&\frac{1}{\mathcal{N}}\mathcal{F}^\bR_{m_1 m_2}(k^c_1,k^c_2) e^{i(k^f_{12}-k^f_{11})\tilde{\bg}_1\cdot\bR} e^{i(k^f_{22}-k^f_{21}+\frac{r_2-r_1}{q})\tilde{\bg}_1\cdot\bR} \\&=&  \frac{1}{\mathcal{N}}  \sum_{l_1 l_2\in \mathbb{Z}}   e^{ i(k^c_{11}-k^c_{12}+ k^f_{12}-k^f_{11})\tilde{\bg}_1\cdot\bR}e^{i (k^c_{21}-k^c_{22}+k^f_{22}-k^f_{21}+\frac{l_1-l_2 +r_2-r_1}{q})\tilde{\bg}_2\cdot\bR}  F^{l_1 l_2}_{m_1 m_2}(k^c_1, k^c_2)\\&=& \sum_{l_1 l_2}\mathcal{C}^{\substack{k_1^f k_2^f;k_1^c k_2^c}}_{\substack{r_1r_2;l_1l_2}}(\bR) F_{m_1m_2}^{l_1l_2}(k_1^c, k_2^c)
\end{eqnarray}
where $F_{m_1m_2}^{l_1l_2}(k_1^c, k_2^c)$ is defined in Eq.(\ref{Eq:Fdefn}), and $\mathcal{C}^{\substack{k_1^f k_2^f;k_1^c k_2^c}}_{\substack{r_1r_2;l_1l_2}}(\bR)$ is the magnetic momentum conserving term defined in Eq.({\ref{Eq:appx:Magnetic_momentum_conservation}}). We thus have \begin{eqnarray}
    \hat{H}_{{J},\bf{B}}    &=& -\frac{J}{\mathcal{N}} 
    \sum_{\substack{\alpha k^c_1 m_1 k^c_2 m_2\\k^f_1 k^f_2 r_2 r_1\\\tau s_1 s_2}}
    \sum_{\substack{\bR\\l_1,l_2\in\mathcal{Z}}} \mathcal{C}^{\substack{k_1^f k_2^f;k_1^c k_2^c}}_{\substack{r_1r_2;l_1l_2}}(\bR) F_{m_1m_2}^{l_1l_2}(k_1^c, k_2^c)\left(:f_{\alpha k^f_1 r_1 \tau s_1}^\dagger f_{\alpha k^f_2 r_2 \tau s_2}:c^\dagger_{\alpha+2 k_2^cm_2\tau\bs_2} c_{\alpha+2 k_1^cm_1\tau s_1}: \right. \nonumber\\&-& \left.  f_{\alpha k^f_1 r_1 \tau s_1}^\dagger f_{\bar{\alpha} k^f_2 r_2 -\tau s_2} c_{ \bar{\alpha}+2 k^c_2 m_2 \tau s_2}^\dagger c_{ \alpha+2 k^c_1 m_1 \tau s_1}\right).
\end{eqnarray}

%---------------------Section change----

\subsection{Mean field decomposition}

In this section, we derive the mean field decomposition of the interaction terms discussed in Appx(\ref{Appx:Interaction_terms_T=0}). Let us start by defining the order parameters with respect to a magnetic momentum conserving variational Slater determinant state. Given the well localized nature of the Wannier states, we only consider on-site order parameters for $f$ fermions and define 
    \begin{eqnarray}
O^{d,\bR}_{\alpha\tau s,\alpha'\tau' s'}&=&\langle d^\dagger_{\bR\alpha\tau s} d_{\bR\alpha'\tau' s'} \rangle\\
&=& \frac{1}{\mathcal{N}}\sum_{\substack{kr\\k'r'}} e^{i\left((k'_1-k_1)\tilde{\bg}_1 + (k'_2-k_2+\frac{r'-r}{q})\tilde{\bg}_2\right)\cdot\bR}\langle f^\dagger_{\alpha k r\tau s} f_{\beta k' r'\tau' s'}\rangle\\
&\equiv& \sum_{\substack{r r'}} e^{i\left( \frac{r'-r}{q}\right)\tilde{\bg}_2\cdot\bR}[O^f(r,r')]_{\alpha \tau s, \beta \tau' s'}\label{Eq:appx:Real_space_f_order}
\end{eqnarray}
where $d_{\bR \alpha \tau s}$ is defined in Eq.(\ref{Eq:appx:Lattice_annihilation_at_B}) and
    \begin{eqnarray}
[O^f(r_1,r_2)]_{\beta\tau' s',\alpha\tau s}   = \frac{1}{\mathcal{N}}\sum_{\tilde{k}}\langle f^\dagger_{\beta\tilde{k}r_1\tau' s'}f_{\alpha\tilde{k}r_2\tau s}\rangle \label{Eq:appx:mom_space_f_order}
\end{eqnarray}
The $c$-$c$ order parameter is defined as
\begin{eqnarray}
    O^c_{\alpha m \tau  s,\beta m'\tau' s'}(k) &=& \langle :c^{\dagger}_{\alpha k m \tau  s} c_{\beta k m' \tau'  s'}:\rangle = 
    \langle c^{\dagger}_{\alpha k m \tau  s} c_{\beta k m' \tau'  s'}\rangle -\frac{1}{2}\delta_{\alpha\beta}\delta_{m_1m_2}\delta_{\tau\tau'}\delta_{ s s'}\label{Eq:appx:c_order}
\end{eqnarray}
and the $c$-$f$ order parameter is given by
\begin{eqnarray}
    O^{cd}_{\alpha k m \tau s,\bR\beta \tau' s'} &=& \langle c^\dagger_{\alpha k m \tau s} d_{\bR\beta \tau' s'}\rangle = \frac{1}{\sqrt{\mathcal{N}}}\sum_{r}e^{i(k_1\tilde{\bg}_1+(k_2+\frac{r}{q})\tilde{\bg}_2)\cdot\bR}\langle c^\dagger_{\alpha k m \tau s} f_{\beta k r \tau' s'}\rangle\\
    &\equiv&\sum_{r}e^{i(k_1\tilde{\bg}_1+(k_2+\frac{r}{q})\tilde{\bg}_2)\cdot\bR} O^{cf}_{\alpha m\tau s,\beta\tau' s'}(k,r)\label{Eq:appx:cf_order_real}
\end{eqnarray}
where
\begin{eqnarray}
  O^{cf}_{\alpha m\tau s,\beta\tau' s'}(k,r) = \frac{1}{\sqrt{\mathcal{N}}}   \langle c^\dagger_{\alpha k m \tau s} f_{\beta k r \tau' s'}\rangle\label{Eq:appx:cf_order_mom}
\end{eqnarray}

%---------------------Section change----

\subsubsection{\texorpdfstring{$ff$-$ff$}{ff-ff} density-density interaction}
The density-density interaction for $f$s obtained in Eq.(\ref{Eq:ff_Interaction_full}) is given by
\begin{eqnarray}
    \hat{H}_{U,\bf{B}}&=& (8\mathcal{N}U_1+48\mathcal{N}U_2) - (3.5U_1+24U_2)\sum_{\bR\alpha\tau s}d^\dagger_{\bR\alpha\tau s} d_{\bR\alpha \tau s}+\frac{U_1}{2}\sum_\bR  \sum_{\substack{\tau,\tilde{\tau}\\ s,\tilde{ s}\\\alpha,\tilde{\alpha}}} d^\dagger_{\bR\alpha\tau s}  d^\dagger_{\bR\tilde{\alpha} \tilde{\tau} \tilde{ s}} d_{\bR\tilde{\alpha} \tilde{\tau} \tilde{ s}}d_{\bR\alpha \tau s}\nonumber\\
&+&\frac{U_2}{2}\sum_{\langle\bR\bR'\rangle}  \sum_{\substack{\tau,\tilde{\tau}\\ s,\tilde{ s}\\\alpha,\tilde{\alpha}}} d^\dagger_{\bR\alpha\tau s} d^\dagger_{\bR'\tilde{\alpha} \tilde{\tau} \tilde{ s}} d_{\bR'\tilde{\alpha} \tilde{\tau} \tilde{ s}} d_{\bR\alpha \tau s}
\end{eqnarray}
Using Wick's theorem, we can decompose the above four fermion terms into bilinears and constants as follows\begin{eqnarray}
\frac{U_1}{2}\sum_\bR \sum_{\substack{\tau,\tilde{\tau}\\ s,\tilde{ s}\\\alpha,\tilde{\alpha}}} d^\dagger_{\bR\alpha\tau s}  d^\dagger_{\bR\tilde{\alpha} \tilde{\tau} \tilde{ s}} d_{\bR\tilde{\alpha} \tilde{\tau} \tilde{ s}}d_{\bR\alpha \tau s}  &\approx&  U_1\sum_\bR \sum_{\substack{\tau,\tilde{\tau}\\ s,\tilde{ s}\\\alpha,\tilde{\alpha}}} \left(\langle d^\dagger_{\bR\tilde{\alpha} \tilde{\tau} \tilde{ s}} d_{\bR\tilde{\alpha} \tilde{\tau}\tilde{ s}}\rangle d^\dagger_{\bR\alpha\tau s} d_{\bR\alpha\tau s}- d^\dagger_{\bR\tilde{\alpha} \tilde{\tau} \tilde{ s}} d_{\bR\alpha\tau s}\langle d^\dagger_{\bR\alpha\tau s}d_{\bR\tilde{\alpha} \tilde{\tau} \tilde{ s}} \rangle \right)\nonumber\\
&+&\frac{U_1}{2}\sum_\bR \sum_{\substack{\tau,\tilde{\tau}\\ s,\tilde{ s}\\\alpha,\tilde{\alpha}}} \left(\langle d^\dagger_{\bR\alpha\tau s}d_{\bR\tilde{\alpha} \tilde{\tau} \tilde{ s}}\rangle \langle d^\dagger_{\bR\tilde{\alpha} \tilde{\tau} \tilde{ s}}d_{\bR\alpha\tau s} 
\rangle- \langle d^\dagger_{\bR\tilde{\alpha} \tilde{\tau} \tilde{ s}} d_{\bR\tilde{\alpha} \tilde{\tau}\tilde{ s}}\rangle \langle d^\dagger_{\bR\alpha\tau s} d_{\bR\alpha\tau s}\rangle\right)
\end{eqnarray}
Similarly,
\begin{eqnarray}
\frac{U_2}{2}\sum_{\langle\bR\bR'\rangle}  \sum_{\substack{\tau,\tilde{\tau}\\ s,\tilde{ s}\\\alpha,\tilde{\alpha}}} d^\dagger_{\bR\alpha\tau s} d^\dagger_{\bR'\tilde{\alpha} \tilde{\tau} \tilde{ s}} d_{\bR'\tilde{\alpha} \tilde{\tau} \tilde{ s}} d_{\bR\alpha \tau s}&\approx&   \frac{U_2}{2}\sum_{\langle\bR\bR'\rangle}  \sum_{\substack{\tau,\tilde{\tau}\\ s,\tilde{ s}\\\alpha,\tilde{\alpha}}} \left(\langle d^\dagger_{\bR\alpha\tau s} d_{\bR\alpha\tau s} \rangle d^\dagger_{\bR'\tilde{\alpha} \tilde{\tau} \tilde{ s}} d_{\bR'\tilde{\alpha} \tilde{\tau} \tilde{ s}}  + \langle d^\dagger_{\bR'\tilde{\alpha} \tilde{\tau} \tilde{ s}} d_{\bR'\tilde{\alpha} \tilde{\tau} \tilde{ s}} 
 \rangle d^\dagger_{\bR\alpha\tau s} d_{\bR\alpha\tau s} \right) \nonumber\nonumber\\
&-&\frac{U_2}{2}\sum_{\langle\bR\bR'\rangle}  \sum_{\substack{\tau,\tilde{\tau}\\ s,\tilde{ s}\\\alpha,\tilde{\alpha}}}  \langle d^\dagger_{\bR\alpha\tau s} d_{\bR\alpha\tau s} \rangle \langle d^\dagger_{\bR'\tilde{\alpha} \tilde{\tau} \tilde{ s}} d_{\bR'\tilde{\alpha} \tilde{\tau} \tilde{ s}}\rangle
\end{eqnarray}
where we do not allow for order parameters of form $\langle d^\dagger_\bR d^\dagger_{\bR'}\rangle$ for $\bR\neq\bR'$. Note that the $\langle \hat{O}\rangle$ is with respect to the variational Slater determinant state.  In terms of order parameter  $O^{d,\bR}_{\alpha\tau s,\alpha'\tau' s'}$ defined in Eq(\ref{Eq:appx:Real_space_f_order}), we have
\begin{eqnarray}
\hat{H}^{MF}_{U,\bf{B}}&=&   - (3.5U_1+24U_2)\sum_{\bR\alpha\tau s}d^\dagger_{\bR\alpha\tau s} d_{\bR\alpha \tau s}   + U_1 \sum_{\bR}\left(Tr[O^{d,\bR}]\sum_{\alpha\tau s}d^\dagger_{\bR\alpha\tau s} d_{\bR\alpha\tau s}-\sum_{\substack{\alpha \tau s\\\beta \tau' s'}}d^\dagger_{\bR\beta\tau' s'}d_{\bR\alpha\tau s}O^{d,\bR}_{\alpha\tau s,\beta\tau' s'}\right)\nonumber\\
 &+& \frac{U_2}{2}\sum_{\langle\bR\bR'\rangle}  \left(Tr[O^{d,\bR}]\sum_{\beta\tau' s'}d^\dagger_{\bR' \beta\tau' s'}d_{\bR'\beta\tau' s'} + Tr[O^{d,\bR'}]\sum_{\alpha\tau s}d^\dagger_{\bR \alpha\tau s}d_{\bR\alpha\tau s}\right)\nonumber\\
 &+& (8\mathcal{N}U_1+48\mathcal{N}U_2) +\frac{U_1}{2}\sum_{\bR}\left(   Tr[O^{d,\bR}O^{d,\bR}]- Tr[O^{d,\bR}]Tr[O^{d,\bR}]\right) -\frac{U_2}{2}\sum_{\langle \bR\bR'\rangle} Tr[O^{d,\bR}]Tr[O^{d,\bR'}]
\end{eqnarray}
where $Tr[O^{d,\bR}]=\sum_{\alpha\tau s}\langle d^\dagger_{\bR \alpha\tau s}d_{\bR\alpha\tau s} \rangle$. We now rewrite the above mean-field term as a sum of bilinears and constants as
\begin{eqnarray}
    \hat{H}^{MF}_{U,\bf{B}} = \bar{H}_{U,\bf{B}}-E_{U,\bf{B}}
\end{eqnarray}
The bilinear part in terms of momentum space representation of $f$s reads
\begin{eqnarray}
 \bar{H}_{U,\bf{B}} &=&    - (3.5U_1+24U_2)\sum_{kr\alpha\tau s}f^\dagger_{\alpha kr\tau s} f_{\alpha kr\tau s} +U_1\sum_{r_1r_2}Tr[O^f(r_1, r_2)]\sum_{\substack{\alpha k\\rr'}}\sum_{\tau s}\delta_{r',[r+r_1-r_2]_q}f^\dagger_{\alpha kr \tau s} f_{\alpha kr' \tau s}\nonumber\\
 &-&U_1 \sum_{\substack{\alpha \tau s\\\beta \tau' s'}} \sum_{r_1r_2}[O^f(r_1,r_2)]_{\beta\tau' s',\alpha\tau s}\sum_{krr'}\delta_{r',[r+r_1-r_2]_q}f^\dagger_{\alpha kr \tau s} f_{\beta kr' \tau' s'} \nonumber\\
 &+& U_2\sum_{\bf{\delta}}\sum_{r_1r_2}e^{i(\frac{r_2-r_1}{q})\tilde{\bg}_2\cdot\bf{\delta}}Tr[O^f(r_1,r_2)]\sum_{\alpha\tau s}\sum_{krr'}\delta_{r',[r+r_1-r_2]_q}f^\dagger_{\alpha kr \tau s} f_{\alpha kr' \tau s}
\end{eqnarray}
with $\delta=\{\pm\tilde{\bL}_1,\pm\tilde{\bL}_2,\pm(\tilde{\bL}_1-\tilde{\bL}_2)\}$. The matrix $O^f(r_1,r_2)$ is defined in Eq.(\ref{Eq:appx:mom_space_f_order}) and $Tr[O^f(r_1,r_2)]$ is the trace of the matrix. 
The constant term reads 
\begin{eqnarray}
   E_{U,\bf{B}}&=& -(8\mathcal{N}U_1+ 48\mathcal{N}U_2) - \frac{\mathcal{N}U_1}{2}\left(Tr[O^f(r_1r_1')O^f(r_2,r_2')]-Tr[O^f(r_1,r_1')]Tr[O^f(r_2,r_2')]\right) \delta_{r_2',[r_2+r_1-r_1']_q}\nonumber\\
   &+&\frac{\mathcal{N}U_2}{2}\sum_{\substack{r_1 r_1'\\r_2r_2'}}\sum_{\bf{\delta}}e^{i(\frac{r_1'-r_1}{q})\tilde{\bg}_2\cdot\bf{\delta}}Tr[O^f(r_1,r_1')]Tr[O^f(r_2,r_2')]\delta_{r_2',[r_2+r_1-r_1']_q}
\end{eqnarray}

%---------------------Section change----

\subsubsection{\texorpdfstring{$cc$-$cc$}{cc-cc} density-density interaction}
The density-density interaction for $c$s obtained in Eq.(\ref{Eq:cc_Interaction_full}) is given by
\begin{eqnarray}
 \hat{H}_{V,\bf{B}} &=&   \frac{1}{2\Omega_{tot}}\sum_{\substack{\tau_1 s_1\\\tau_2 s_2}}\sum_{\substack{a_1\\m_1 m_1'\\k^1_c k^{1'}_c}} \sum_{\substack{a_2 \\m_2 m_2'\\k^2_c k^{2'}_c}} \sum_{\bq} V(\bq)
\mathcal{W}^{\bq}_{m_1 m_1'}(k^{c}_1,k^{c'}_1)
\mathcal{W}^{-\bq}_{m_2' m_2}(k^{c'}_2,k^{c}_2) 
\nonumber\\&&:c_{ a_1 k^c_1 m_1 \tau_1  s_1}^\dagger c_{ a_1 k^{c'}_1 m_1' \tau_1  s_1}::c_{ a_2 k^{c'}_2 m_2' \tau_2  s_2}^\dagger c_{ a_2 k^c_2 m_2 \tau_2  s_2}:  
\end{eqnarray}
Using the Wick's theorem, above can be decomposed as (only Hartree channel)
\begin{eqnarray}
\hat{H}^{MF}_{V,\bf{B}} &\approx&   \frac{1}{2\Omega_{tot}}\sum_{\substack{\tau_1 s_1\\\tau_2 s_2}}\sum_{\substack{a_1\\m_1 m_1'\\k^1_c k^{1'}_c}} \sum_{\substack{a_2 \\m_2 m_2'\\k^2_c k^{2'}_c}} \sum_{\bq} V(\bq)
\mathcal{W}^{\bq}_{m_1 m_1'}(k^{c}_1,k^{c'}_1)
\mathcal{W}^{-\bq}_{m_2' m_2}(k^{c'}_2,k^{c}_2) 
\nonumber\\&&\bigg[\langle :c_{ a_1 k^c_1 m_1 \tau_1  s_1}^\dagger c_{ a_1 k^{c'}_1 m_1' \tau_1  s_1}:\rangle :c_{ a_2 k^{c'}_2 m_2' \tau_2  s_2}^\dagger c_{ a_2 k^c_2 m_2 \tau_2  s_2}: + \langle :c_{ a_2 k^{c'}_2 m_2' \tau_2  s_2}^\dagger c_{ a_2 k^c_2 m_2 \tau_2  s_2}:\rangle :c_{ a_1 k^c_1 m_1 \tau_1  s_1}^\dagger c_{ a_1 k^{c'}_1 m_1' \tau_1  s_1}:\nonumber\\
    &&-\langle :c_{ a_1 k^c_1 m_1 \tau_1  s_1}^\dagger c_{ a_1 k^{c'}_1 m_1' \tau_1  s_1}:\rangle
\langle :c_{ a_2 k^{c'}_2 m_2' \tau_2  s_2}^\dagger c_{ a_2 k^c_2 m_2 \tau_2  s_2}:\rangle\bigg]
\end{eqnarray}
For a magnetic translation preserving state, we have
\begin{eqnarray}
\hat{H}^{MF}_{V,\bf{B}} =&&   \frac{1}{2\Omega_{tot}}\sum_{\substack{\tau_1 s_1\\\tau_2 s_2}}\sum_{\substack{a_1\\m_1 m_1'\\k^1_c k^{1'}_c}} \sum_{\substack{a_2 \\m_2 m_2'\\k^2_c k^{2'}_c}} \sum_{\bq} V(\bq)
\mathcal{W}^{\bq}_{m_1 m_1'}(k^{c}_1,k^{c'}_1)
\mathcal{W}^{-\bq}_{m_2' m_2}(k^{c'}_2,k^{c}_2) 
\nonumber\\&&\bigg[\delta_{k^c_1,k^{c'}_1}\langle :c_{ a_1 k^c_1 m_1 \tau_1  s_1}^\dagger c_{ a_1 k^{c'}_1 m_1' \tau_1  s_1}:\rangle :c_{ a_2 k^{c'}_2 m_2' \tau_2  s_2}^\dagger c_{ a_2 k^c_2 m_2 \tau_2  s_2}:\\ &&+ \delta_{k^{c}_2,k^{c'}_2}\langle :c_{ a_2 k^{c}_2 m_2' \tau_2  s_2}^\dagger c_{ a_2 k^c_2 m_2 \tau_2  s_2}:\rangle :c_{ a_1 k^c_1 m_1 \tau_1  s_1}^\dagger c_{ a_1 k^{c'}_1 m_1' \tau_1  s_1}:\nonumber\\
    &&-\delta_{k^{c}_1,k^{c'}_1}\delta_{k^{c}_2,k^{c'}_2}\langle :c_{ a_1 k^c_1 m_1 \tau_1  s_1}^\dagger c_{ a_1 k^{c}_1 m_1' \tau_1  s_1}:\rangle
\langle :c_{ a_2 k^{c}_2 m_2' \tau_2  s_2}^\dagger c_{ a_2 k^c_2 m_2 \tau_2  s_2}:\rangle\bigg]    
\end{eqnarray}
Well note that $\mathcal{W}^\bq_{m_1,m_2}(k,k)$ is non-zero only if $\bq$ is of form $n_1\tilde{\bg}_1+\frac{n_2}{q}\tilde{\bg}_2$ with $n_{i}\in\mathbb{Z}$ (see the discussion below Eq.(\ref{Appx:Eq:Condition_on_q})). Also since the basis of our $c$-modes orthogonality is that $\mathcal{W}^\bG_{m_1,m_2}(k_1,k_2)=\delta_{\bG,0}\delta_{m_1,m_2}\delta_{k_1k_2}$, we have
\begin{eqnarray}
    \hat{H}_{V,\bf{B}} =&&   \frac{1}{2\Omega_{tot}}\sum_{\substack{\tau_1 s_1\\\tau_2 s_2}}\sum_{\substack{a_1\\m_1 m_1'\\k^1_c }} \sum_{\substack{a_2 \\m_2 m_2'\\k^2_c}} \sum_{\bq=\frac{n_2}{q}\tilde{\bg}_2} V(\bq)
\mathcal{W}^{\bq}_{m_1 m_1'}(k^{c}_1,k^{c}_1)
\mathcal{W}^{-\bq}_{m_2' m_2}(k^{c}_2,k^{c}_2) 
\nonumber\\&&\bigg[\langle :c_{ a_1 k^c_1 m_1 \tau_1  s_1}^\dagger c_{ a_1 k^{c}_1 m_1' \tau_1  s_1}:\rangle :c_{ a_2 k^{c}_2 m_2' \tau_2  s_2}^\dagger c_{ a_2 k^c_2 m_2 \tau_2  s_2}:\nonumber\\ &&+\langle :c_{ a_2 k^{c}_2 m_2' \tau_2  s_2}^\dagger c_{ a_2 k^c_2 m_2 \tau_2  s_2}:\rangle :c_{ a_1 k^c_1 m_1 \tau_1  s_1}^\dagger c_{ a_1 k^{c}_1 m_1' \tau_1  s_1}:\nonumber\\
 &&-\langle :c_{ a_1 k^c_1 m_1 \tau_1  s_1}^\dagger c_{ a_1 k^{c}_1 m_1' \tau_1  s_1}:\rangle
\langle :c_{ a_2 k^{c}_2 m_2' \tau_2  s_2}^\dagger c_{ a_2 k^c_2 m_2 \tau_2  s_2}:\rangle\bigg]    
\end{eqnarray}
with $n_2=0,\pm1,\pm2,\ldots\pm(q-1)$.
Above can further be simplified to
\begin{eqnarray}
    \hat{H}^{MF}_{V,\bf{B}} =&&   \frac{1}{2\Omega_{tot}}\sum_{\substack{\tau_1 s_1\\\tau_2 s_2}}\sum_{\substack{a_1\\m_1 m_1'\\k^1_c }} \sum_{\substack{a_2 \\m_2 m_2'\\k^2_c}} \sum_{\bq=\frac{n_2}{q}\tilde{\bg}_2} V(\bq)
\mathcal{W}^{\bq}_{m_1 m_1'}(k^{c}_1,k^{c}_1)
\mathcal{W}^{-\bq}_{m_2' m_2}(k^{c}_2,k^{c}_2) 
\nonumber\\&&\bigg[2\langle :c_{ a_1 k^c_1 m_1 \tau_1  s_1}^\dagger c_{ a_1 k^{c}_1 m_1' \tau_1  s_1}:\rangle :c_{ a_2 k^{c}_2 m_2' \tau_2  s_2}^\dagger c_{ a_2 k^c_2 m_2 \tau_2  s_2}:\nonumber\\
 &&-\langle :c_{ a_1 k^c_1 m_1 \tau_1  s_1}^\dagger c_{ a_1 k^{c}_1 m_1' \tau_1  s_1}:\rangle
\langle :c_{ a_2 k^{c}_2 m_2' \tau_2  s_2}^\dagger c_{ a_2 k^c_2 m_2 \tau_2  s_2}:\rangle\bigg]    
\end{eqnarray}
We can similarly rewrite 
\begin{eqnarray}
\hat{H}^{MF}_{V,\bf{B}} \approx     \bar{H}_{V,\bf{B}} - E_{V,\bf{B}}
\end{eqnarray}
where
%\begin{eqnarray}
%    \bar{H}_{V,\bf{B}} =  \frac{1}{\Omega_{tot}}\sum_{\substack{\tau_1 s_1\\\tau_2 s_2}}\sum_{\substack{a_1\\m_1 m_1'\\k^1_c }} \sum_{\substack{a_2 \\m_2 m_2'\\k^2_c}}\sum_{n_2=0}^{q-1}\delta_{\bq,\pm \frac{n_2}{q} \bg_2}V(\bq)
%\mathcal{W}^{\bq}_{m_1 m_1'}(k^{c}_1,k^{c}_1)
%\mathcal{W}^{-\bq}_{m_2' m_2}(k^{c}_2,k^{c}_2) \langle :c_{ a_1 k^c_1 m_1 \tau_1  s_1}^\dagger c_{ a_1 k^{c}_1 m_1' \tau_1  s_1}:\rangle c_{ a_2 k^{c}_2 m_2' \tau_2  s_2}^\dagger c_{ a_2 k^c_2 m_2 \tau_2  s_2}\nonumber\\
%\end{eqnarray}
\begin{eqnarray}
    \bar{H}_{V,\bf{B}} =  \frac{1}{\Omega_{tot}}\sum_{\substack{\tau_1 s_1\\ \tau_2 s_2\\\bq}}\sum_{\substack{a_1\\m_1 m_1'\\k^1_c }} \sum_{\substack{a_2 \\m_2 m_2'\\k^2_c}}\sum_{n_2=0}^{q-1}\delta_{\bq,\pm \frac{n_2}{q} \tilde{\bg}_2}V(\bq)
\mathcal{W}^{\bq}_{m_1 m_1'}(k^{c}_1,k^{c}_1)
\mathcal{W}^{-\bq}_{m_2' m_2}(k^{c}_2,k^{c}_2) O^c_{a_1 m_1\tau_1 s_1, a_1 m_1' \tau_1 s_1}(k^c_1) c_{ a_2 k^{c}_2 m_2' \tau_2  s_2}^\dagger c_{ a_2 k^c_2 m_2 \tau_2  s_2}\nonumber\\
\end{eqnarray}
and 
\begin{eqnarray}
   E_{V,\bf{B}} &=&   \frac{1}{2\Omega_{tot}}\sum_{\substack{\tau_1 s_1\\\tau_2 s_2\\\bq}}\sum_{\substack{a_1\\m_1 m_1'\\k^1_c }} \sum_{\substack{a_2 \\m_2 m_2'\\k^2_c}} \sum_{n_2=0}^{q-1}\delta_{\bq,\pm \frac{n_2}{q} \tilde{\bg}_2} V(\bq)
\mathcal{W}^{\bq}_{m_1 m_1'}(k^{c}_1,k^{c}_1)
\mathcal{W}^{-\bq}_{m_2' m_2}(k^{c}_2,k^{c}_2) O^c_{a_1 m_1\tau_1 s_1, a_1 m_1' \tau_1 s_1}(k^c_1)
O^c_{a_2 m_2' \tau_2 s_2, a_2 m_2 \tau_2 s_2}(k^c_2)
\nonumber\\
&+&\frac{1}{2\Omega_{tot}}\sum_{\substack{\tau_1 s_1\\ \tau_2 s_2\\\bq}}\sum_{\substack{a_1\\m_1 m_1'\\k^1_c }} \sum_{\substack{a_2 m_2 k^c_2}}\sum_{n_2=0}^{q-1}\delta_{\bq,\pm \frac{n_2}{q} \tilde{\bg}_2}V(\bq)
\mathcal{W}^{\bq}_{m_1 m_1'}(k^{c}_1,k^{c}_1)
\mathcal{W}^{-\bq}_{m_2 m_2}(k^{c}_2,k^{c}_2) O^c_{a_1 m_1\tau_1 s_1, a_1 m_1' \tau_1 s_1}(k^c_1)
\end{eqnarray}
where $O^c$ has been defined in Eq.({\ref{Eq:appx:c_order}}). We can further simplify the problem by neglecting the $n_2\neq 0$ contributions in above, as both $V(\bq)$ and $\mathcal{W}^\bq$ fall fast with $\bq$.

%---------------------Section change----

\subsubsection{\texorpdfstring{$cc$-$ff$}{cc-ff} density-density interaction}\label{Appx:cc-ff-at B}
The density-density interaction between $c$ and $f$ derived in Eq.(\ref{Eq:appx:cc_ff_density-density}) reads
\begin{eqnarray}
     \hat{H}_{W,\bB}  &=&  \frac{1}{\mathcal{N}}\sum_{\substack{\alpha \tau_1 s_1\\k_1^f k_2^f\\r_1r_2}}\sum_{\substack{a\tau_2 s_2\\k_1^c k_2^c\\m_1m_2}} W_a\sum_{\substack{\bR\\l_1,l_2\in\mathcal{Z}}} \mathcal{C}^{\substack{k_1^f k_2^f;k_1^c k_2^c}}_{\substack{r_1r_2;s_1s_2}}(\bR) F_{m_1m_2}^{l_1l_2}(k_1^c k_2^c):f_{\alpha k^f_1 r_1 \tau_1 s_1}^\dagger f_{\alpha k^f_2 r_2 \tau_1 s_1}: :c_{ a k^c_2 m_2 \tau_2 s_2}^\dagger c_{ a k^c_1 m_1 \tau_2 s_2} :
\end{eqnarray}
Using the Wick's theorem we can decompose the above as
\begin{eqnarray}
\hat{H}^{MF}_{W,\mathbf{B}} 
&\approx&  \frac{1}{\mathcal{N}}\sum_{\substack{\alpha \tau_1 s_1\\k_1^f k_2^f\\r_1r_2}}\sum_{\substack{a\tau_2 s_2\\k_1^c k_2^c\\m_1m_2}} W_a\sum_{\substack{\bR\\l_1,l_2\in\mathcal{Z}}} \mathcal{C}^{\substack{k_1^f k_2^f;k_1^c k_2^c}}_{\substack{r_1r_2;l_1l_2}}(\bR) F_{m_1m_2}^{l_1l_2}(k_1^c k_2^c)\nonumber \\  &&\bigg[\delta_{k_2^fk_1^f}\left(\langle f_{\alpha k^f_1 r_1 \tau_1  s_1}^\dagger f_{\alpha k^f_1 r_2 \tau_1  s_1} \rangle-\frac{1}{2}\delta_{r_2r_1}\right)
    c_{ a k^c_2 m_2 \tau_2  s_2}^\dagger c_{ a k^c_1 m_1 \tau_2  s_2} \nonumber\\&&+ \delta_{k^c_1k^c_2}O^c_{am_2\tau_2 s_2,am_1\tau_2 s_2}(k^c_1)f_{\alpha k^f_1 r_1 \tau_1  s_1}^\dagger f_{\alpha k^f_2 r_2 \tau_1  s_1} +\frac{1}{4}\delta_{k_2^ck_1^c}\delta_{m_2m_1}\delta_{k_2^fk_1^f}\delta_{r_2r_1} \nonumber\\
    &&-\sqrt{\mathcal{N}}\delta_{k^c_1k^f_1}O^{cf\ast}_{a m_1 \tau_2  s_2,\alpha \tau_1  s_1}(k^c_1,r_1) c_{ a k^c_2 m_2 \tau_2  s_2}^\dagger f_{\alpha k^f_2 r_2 \tau_1  s_1} \nonumber\\&-&\sqrt{\mathcal{N}}\delta_{k^c_2k^f_2}O^{cf}_{a m_2 \tau_2  s_2,\alpha \tau_1  s_1}(k^c_2,r_2) f_{\alpha k^f_1 r_1 \tau_1  s_1}^\dagger c_{ a k^c_1 m_1 \tau_2  s_2}\nonumber\\
    &&+ \mathcal{N}\delta_{k^c_1k^f_1}\delta_{k^c_2k^f_2}O^{cf\ast}_{a m_1 \tau_2  s_2,\alpha \tau_1  s_1}(k_1^c,r_1)O^{cf}_{a m_2 \tau_2  s_2,\alpha \tau_1  s_1}(k_2^c,r_2) \nonumber\\&&-  \delta_{k^c_1k^c_2}\delta_{k^f_1k^f_2}\left(O^c_{am_2\tau_2 s_2,am_1\tau_2 s_2}(k^c_1)+\frac{1}{2}\delta_{m_1m_2}\right)\langle f_{\alpha k^f_1 r_1 \tau_1  s_1}^\dagger f_{\alpha k^f_1 r_2 \tau_1 s_1}\rangle\bigg]
\end{eqnarray}
where we assume that the magnetic momentum is preserved by the Slater determinant state. The order parameters $O^{c}$ and $O^{cf}$ are defined in Eqs.(\ref{Eq:appx:c_order}) and (\ref{Eq:appx:cf_order_mom})  We can similarly rewrite the above as
\begin{eqnarray}
\hat{H}^{MF}_{W,\bf{B}} \approx     \bar{H}_{W,\bf{B}} - E_{W,\bf{B}}
\end{eqnarray}
where
\begin{eqnarray}
    \bar{H}_{W,\bB}&=&\sum_{\substack{a\tau_2 s_2\\k^c m_1m_2}} W_a\sum_{\substack{l_1,l_2\in\mathcal{Z}}} F^{l_1,l_2}_{m_1m_2}(k^c,k^c)\left(\sum_{r_1r_2}Tr(O^f(r_1,r_2))\delta_{l_2,[l_1+r_2-r_1]_q}-4\delta_{l_2,[l_1]_q}\right)   c_{ a k^c m_2 \tau_2  s_2}^\dagger c_{ a k^c m_1 \tau_2  s_2}\nonumber\\
+&&\frac{1}{\mathcal{N}}\sum_{\substack{\alpha \tau_1 s_1 \\r_1r_2 k^f}}\sum_{\substack{a\tau_2 s_2\\ m_1m_2\\k^c}} W_a\sum_{\substack{l_1,l_2\in\mathcal{Z}}} \delta_{l_2,[l_1+r_2-r_1]_{q}} F^{l_1,l_2}_{m_1m_2}(k^c,k^c) O^c_{am_2\tau_2 s_2,am_1\tau_2 s_2}(k^c)f_{\alpha k^f r_1 \tau_1  s_1}^\dagger f_{\alpha k^f r_2 \tau_1  s_1}\nonumber\\
-&&\frac{1}{\mathcal{\sqrt{N}}} \sum_{\substack{\alpha \tau_1 s_1 \\r_1r_2\\ k^f}}\sum_{\substack{a\tau_2 s_2\\ m_1m_2\\k^c}} W_a\sum_{\substack{l_1,l_2\in\mathcal{Z}}} \delta_{l_2,[l_1+r_2-r_1]_{q}}\left( F^{l_1,l_2}_{m_1m_2}(k^c,k^f) O^{cf\ast}_{a m_1 \tau_2  s_2,\alpha \tau_1  s_1}(k^c,r_1) c_{ a k^f m_2 \tau_2  s_2}^\dagger f_{\alpha k^f r_2 \tau_1  s_1} + h.c.\right)
\end{eqnarray}
where $O^f(r_1,r_2)$ is defined in Eq.(\ref{Eq:appx:mom_space_f_order}) and 
\begin{eqnarray}
 E_{W,\bf{B}} =   -&&\sum_{\substack{\alpha \tau_1 s_1 \\r_1r_2}}\sum_{\substack{a\tau_2 s_2\\k^c_1 k^c_2 \\m_1m_2}} W_a\sum_{\substack{l_1,l_2\in\mathcal{Z}}} \delta_{l_2,[l_1+r_2-r_1]_{q}} F^{l_1,l_2}_{m_1m_2}(k^c_1,k^c_2)O^{cf\ast}_{a m_1 \tau_2  s_2,\alpha \tau_1  s_1}(k_1^c,r_1)O^{cf}_{a m_2 \tau_2  s_2,\alpha \tau_1  s_1}(k_2^c,r_2)\nonumber\\
+&&  \sum_{\substack{a\tau_2 s_2\\k^c m_1m_2}} W_a\sum_{\substack{l_1,l_2\in\mathcal{Z}}} F^{l_1,l_2}_{m_1m_2}(k^c,k^c)\left(\sum_{r_1r_2}Tr(O^f(r_1,r_2))\delta_{l_2,[l_1+r_2-r_1]_q}-4\delta_{l_2,[l_1]_q}\right)   O^c_{a k^c m_2 \tau_2  s_2,a k^c m_1 \tau_2  s_2}(k^c)\nonumber\\
+&& 2\sum_{\substack{a k^c m}} W_a \sum_{\substack{l_1,l_2\in\mathcal{Z}}} F^{l_1,l_2}_{mm}(k^c,k^c) \left(\sum_{r_1r_2}Tr(O^f(r_1,r_2))\delta_{l_2,[l_1+r_2-r_1]_q}-4\delta_{l_2,[l_1]q}\right) \nonumber\\
+&& 4  \sum_{\substack{a k^c \tau_2 s_2 \\m_1m_2}} W_a\sum_{\substack{l_1,l_2\in\mathcal{Z}}} \delta_{l_2,[l_1]_{q}} F^{l_1,l_2}_{m_1m_2}(k^c,k^c) O^c_{a k^c m_2 \tau_2  s_2,a k^c m_1 \tau_2  s_2}(k^c).\label{Eq:Approx_m_MF}
\end{eqnarray}
We can make further approximations to simplify the above terms before the numerical implementation. Note that the magnetic momentum index $k^c_i$ in $F^{l_1,l_2}_{m_1,m_2}(k_1^c,k_2^c)$ only fine tunes the position at which the harmonic oscillator $\varphi_{m_i}$ is evaluated, in vicinity of $l_i \tilde{L}_{1x}$, as $qk^c_{2i}\in[0,1)$. Moreover since $\varphi_m(x)$ is a function of $x/\ell$, at low $\bB$, we can ignore the $k^c_i$ dependence and approximate it by $F^{l_1,l_2}_{m_1,m_2}(k_1^c,k_2^c)\approx F^{l_1,l_2}_{m_1,m_2}(0,0)$. Moreover we drop the subleading terms in summations of the form $\sum_{l1,l_2\in \mathbb{Z}}F^{l_1,l_2}_{m_1,m_2}(0,0)\delta_{l_2,[l_1+u]_q}$, i.e. when $l_1\neq l_2$ and approximate it by $\approx \sum_{l\in \mathbb{Z}}F^{l,l}_{m_1,m_2}(0,0)\delta_{0,[u]_q}$ which further simplifies to $\approx \delta_{m_1,m_2}\delta_{0,[u]_q}$ at low $\bB$ as  $  \frac{\tilde{L}_{1x}}{\ell}\sum_{l\in\mathbb{Z}}\varphi_{m_1}\left(-l\tilde{L}_{1x}\right) \varphi_{m_2}\left(-l\tilde{L}_{1x}\right)\rightarrow \langle m_1|m_2\rangle = \delta_{m_1 m_2}$ .

%---------------------Section change----

\subsubsection{Exchange interaction}
The exchange interaction between $c$ and $f$ derived in Eq.(\ref{Eq:appx:cf_fc_Exchange}) reads
\begin{eqnarray}
     \hat{H}_{{J},\bf{B}}    &=& -\frac{J}{\mathcal{N}} \sum_{\substack{\bR \alpha \tau\\k^c_1 m_1 k^c_2 m_2 \\s_1 s_2}} \mathcal{F}^\bR_{m_1 m_2}(k^c_1,k^c_2) \left( :d^\dagger_{\bR\alpha\tau s_1} d_{\bR\alpha\tau s_2}::c^\dagger_{\alpha+2 k_2^cm_2\tau s_2} c_{\alpha+2 k_1^cm_1\tau s_1}:\right.\nonumber\\
 &-&\left. d^\dagger_{\bR\alpha\tau s_1} d_{\bR\bar{\alpha}-\tau s_2} c^\dagger_{\bar{\alpha}+2 k_2^cm_2-\tau s_2} c_{\alpha+2 k_1^cm_1\tau s_1}\right)
\end{eqnarray}
See that $:d^\dagger_{\bR\alpha\tau s_1} d_{\bR\alpha\tau s_2}::c^\dagger_{\alpha+2 k_2^cm_2\tau s_2} c_{\alpha+2 k_1^cm_1\tau s_1}:$ can be expanded as
\begin{eqnarray}
    =&& \left(d^\dagger_{\bR\alpha\tau s_1} d_{\bR\alpha\tau s_2} - \frac{1}{2}\delta_{ s_1 s_2}\right)   \left(c^\dagger_{\alpha+2 k_2^cm_2\tau s_2} c_{\alpha+2 k_1^cm_1\tau s_1} - \frac{1}{2}\delta_{k^c_1k^c_2}\delta_{m_1m_2}\delta_{ s_1 s_2}\right)\\
=&&d^\dagger_{\bR\alpha\tau s_1} c^\dagger_{\alpha+2 k_2^cm_2\tau s_2} c_{\alpha+2 k_1^cm_1\tau s_1}d_{\bR\alpha\tau s_2} -\frac{1}{2}\delta_{ s_1 s_2}\left(\delta_{k^c_1k^c_2}\delta_{m_1m_2} d^\dagger_{\bR\alpha\tau s_1} d_{\bR\alpha\tau s_2}+c^\dagger_{\alpha+2 k_2^cm_2\tau s_2} c_{\alpha+2 k_1^cm_1\tau s_1}\right) \nonumber\\
+&&\frac{1}{4}\delta_{k^c_1k^c_2}\delta_{m_1m_2} \delta_{ s_1 s_2}
\end{eqnarray}
Now using Wick's theorem, the above can be decomposed as 
\begin{eqnarray}
 &\approx&\langle d^\dagger_{\bR\alpha\tau s_1} d_{\bR\alpha\tau s_2}\rangle  c^\dagger_{\alpha+2 k_2^cm_2\tau s_2} c_{\alpha+2 k_1^cm_1\tau s_1} + \delta_{k^c_1k^c_2}\langle c^\dagger_{\alpha+2 k_1^cm_2\tau s_2} c_{\alpha+2 k_1^cm_1\tau s_1}\rangle  d^\dagger_{\bR\alpha\tau s_1} d_{\bR\alpha\tau s_2}\nonumber\\
 &-&\left(\langle d^\dagger_{\bR\alpha\tau s_1} c_{\alpha+2 k_1^cm_1\tau s_1}\rangle
c^\dagger_{\alpha+2 k_2^cm_2\tau s_2} d_{\bR\alpha\tau s_2} + \langle 
c^\dagger_{\alpha+2 k_2^cm_2\tau s_2} d_{\bR\alpha\tau s_2}\rangle
d^\dagger_{\bR\alpha\tau s_1} c_{\alpha+2 k_1^cm_1\tau s_1}\right)\nonumber\\
&+& \langle d^\dagger_{\bR\alpha\tau s_1} c_{\alpha+2 k_1^cm_1\tau s_1}\rangle
\langle c^\dagger_{\alpha+2 k_2^cm_2\tau s_2} d_{\bR\alpha\tau s_2}\rangle - \delta_{k^c_1k^c_2} \langle d^\dagger_{\bR\alpha\tau s_1} d_{\bR\alpha\tau s_2}\rangle  \langle c^\dagger_{\alpha+2 k_1^cm_2\tau s_2} c_{\alpha+2 k_1^cm_1\tau s_1}\rangle\nonumber\\
&-&\frac{1}{2}\delta_{ s_1 s_2}\left(\delta_{k^c_1k^c_2}\delta_{m_1m_2} d^\dagger_{\bR\alpha\tau s_1} d_{\bR\alpha\tau s_2}+c^\dagger_{\alpha+2 k_2^cm_2\tau s_2} c_{\alpha+2 k_1^cm_1\tau s_1}\right) 
+\frac{1}{4}\delta_{k^c_1k^c_2}\delta_{m_1m_2} \delta_{ s_1 s_2}\\
&=& \left(\langle d^\dagger_{\bR\alpha\tau s_1} d_{\bR\alpha\tau s_2}\rangle  -\frac{1}{2}\delta_{ s_1 s_2}\right)\left(c^\dagger_{\alpha+2 k_2^cm_2\tau s_2} c_{\alpha+2 k_1^cm_1\tau s_1}-\frac{1}{2}\delta_{m_1m_2}\delta_{k^c_1k^c_2}\delta_{ s_1 s_2}\right)\nonumber\\
&+& \delta_{k^c_1k^c_2}\left(\langle c^\dagger_{\alpha+2 k_1^cm_2\tau s_2} c_{\alpha+2 k_1^cm_1\tau s_1}\rangle  -\frac{1}{2}\delta_{ s_1 s_2}\delta_{m_1m_2}\right)\left(d^\dagger_{\bR\alpha\tau s_1} d_{\bR\alpha\tau s_2}-\frac{1}{2}\delta_{ s_1 s_2}\right)\nonumber\\
&-&\delta_{k^c_1k^c_2} \left(\langle d^\dagger_{\bR\alpha\tau s_1} d_{\bR\alpha\tau s_2}\rangle -\frac{1}{2}\delta_{ s_1 s_2}\right) \left(\langle c^\dagger_{\alpha+2 k_1^cm_2\tau s_2} c_{\alpha+2 k_1^cm_1\tau s_1}\rangle -\frac{1}{2}\delta_{m_1m_2}\delta_{ s_1 s_2}\right)\nonumber\\
&-&\left(\langle d^\dagger_{\bR\alpha\tau s_1} c_{\alpha+2 k_1^cm_1\tau s_1}\rangle
c^\dagger_{\alpha+2 k_2^cm_2\tau s_2} d_{\bR\alpha\tau s_2} + \langle 
c^\dagger_{\alpha+2 k_2^cm_2\tau s_2} d_{\bR\alpha\tau s_2}\rangle
d^\dagger_{\bR\alpha\tau s_1} c_{\alpha+2 k_1^cm_1\tau s_1}\right)\nonumber\\
&+& \langle d^\dagger_{\bR\alpha\tau s_1} c_{\alpha+2 k_1^cm_1\tau s_1}\rangle
\langle c^\dagger_{\alpha+2 k_2^cm_2\tau s_2} d_{\bR\alpha\tau s_2}\rangle
\end{eqnarray}
We finally have 
\begin{eqnarray}
&&:d^\dagger_{\bR\alpha\tau s_1} d_{\bR\alpha\tau s_2}::c^\dagger_{\alpha+2 k_2^cm_2\tau s_2} c_{\alpha+2 k_1^cm_1\tau s_1}:\nonumber\\
    \approx&& \left(O^{d\bR}_{\alpha\tau s_1,\alpha\tau s_2}-\frac{1}{2}\delta_{ s_1 s_2}\right):c^\dagger_{\alpha+2 k_2^cm_2\tau s_2} c_{\alpha+2 k_1^cm_1\tau s_1}: + \delta_{k^c_1k^c_2}O^c_{\alpha+2 m_2\tau s_2,\alpha+2 m_1\tau s_1}(k^c_1):d^\dagger_{\bR\alpha\tau s_1} d_{\bR\alpha\tau s_2}:\nonumber \\
-&&\delta_{k^c_1k^c_2}\left(O^{d\bR}_{\alpha\tau s_1,\alpha\tau s_2}-\frac{1}{2}\delta_{ s_1 s_2}\right)O^c_{\alpha+2 m_2\tau s_2,\alpha+2 m_1\tau s_1}(k^c_1)
-O^{cd\ast}_{\alpha+2 k_1^cm_1\tau s_1,\bR\alpha\tau s_1}c^\dagger_{\alpha+2 k_2^cm_2\tau s_2} d_{\bR\alpha\tau s_2}\nonumber\\-&&O^{cd}_{\alpha+2 k_2^cm_2\tau s_2,\bR\alpha\tau s_2} d^\dagger_{\bR\alpha\tau s_1} c_{\alpha+2 k_1^cm_1\tau s_1}
+O^{cd\ast}_{\alpha+2 k_1^cm_1\tau s_1,\bR\alpha\tau s_1}O^{cd}_{\alpha+2 k_2^cm_2\tau s_2,\bR\alpha\tau s_2}  
\end{eqnarray}
where
\begin{eqnarray}
    O^c_{\alpha m \tau  s,\beta m'\tau' s'}(k) &=& \langle c^{\dagger}_{\alpha k m \tau  s} c_{\beta k m' \tau'  s'}\rangle -\frac{1}{2}\delta_{\alpha\beta}\delta_{m_1m_2}\delta_{\tau\tau'}\delta_{ s s'}\\
    O^{cd}_{\alpha k m \tau s,\bR\beta \tau' s'} &=& \langle c^\dagger_{\alpha k m \tau s} d_{\bR\beta \tau' s'}\rangle = \frac{1}{\sqrt{\mathcal{N}}}\sum_{r}e^{i(k_1\tilde{\bg}_1+(k_2+\frac{r}{q})\tilde{\bg}_2)\cdot\bR}\langle c^\dagger_{\alpha k m \tau s} f_{\beta k r \tau' s'}\rangle\\
    &\equiv&\sum_{r}e^{i(k_1\tilde{\bg}_1+(k_2+\frac{r}{q})\tilde{\bg}_2)\cdot\bR} O^{cf}_{\alpha m\tau s,\beta\tau' s'}(k,r)
\end{eqnarray}
Now let us look at 
$-d^\dagger_{\bR\alpha\tau s_1} d_{\bR\bar{\alpha}-\tau s_2} c^\dagger_{\bar{\alpha}+2 k_2^cm_2-\tau s_2} c_{\alpha+2 k_1^cm_1\tau s_1}= -d^\dagger_{\bR\alpha\tau s_1} c^\dagger_{\bar{\alpha}+2 k_2^cm_2-\tau s_2} c_{\alpha+2 k_1^cm_1\tau s_1}d_{\bR\bar{\alpha}-\tau s_2}=$
\begin{eqnarray}
    -&& O^{d\bR} _{\alpha\tau s_1,\bar{\alpha}-\tau s_2} c^\dagger_{\bar{\alpha}+2 k_2^cm_2-\tau s_2} c_{\alpha+2 k_1^cm_1\tau s_1} - \delta_{k^c_1k^c_2}O^{c}_{\bar{\alpha}+2 m_2-\tau s_2,\alpha+2 m_1\tau s_1}(k^c_1)d^\dagger_{\bR\alpha\tau s_1} d_{\bR\bar{\alpha}-\tau s_2} \nonumber\\
    +&&\delta_{k^c_1k^c_2}O^{c}_{\bar{\alpha}+2 m_2-\tau s_2,\alpha+2 m_1\tau s_1}(k^c_1) O^{d\bR} _{\alpha\tau s_1,\bar{\alpha}-\tau s_2} - O^{cd\ast}_{\alpha+2 k_1^cm_1\tau s_1,\bR\alpha\tau s_1}O^{cd}_{\bar{\alpha}+2 k_2^cm_2-\tau s_2,\bR\bar{\alpha}-\tau s_2}\nonumber\\
    +&&O^{cd\ast}_{\alpha+2 k_1^cm_1\tau s_1,\bR\alpha\tau s_1}  c^\dagger_{\bar{\alpha}+2 k_2^cm_2-\tau s_2}  d_{\bR\bar{\alpha}-\tau s_2} + O^{cd}_{\bar{\alpha}+2 k_2^cm_2-\tau s_2,\bR\bar{\alpha}-\tau s_2}
d^\dagger_{\bR\alpha\tau s_1} c_{\alpha+2 k_1^cm_1\tau s_1}
\end{eqnarray}
So we have
\begin{eqnarray}
    \hat{H}^{MF}_{J,\bf{B}}&\approx&-\frac{J}{\mathcal{N}}\sum_{ s_1 s_2}\sum_{\tau\alpha\bR}\sum_{\substack{k_1^ck^c_2\\m_1m_2}} F^\bR_{m_1m_2}(k_1^c,k^c_2)\bigg[\left(O^{d\bR}_{\alpha\tau s_1,\alpha\tau s_2}-\frac{1}{2}\delta_{ s_1 s_2}\right):c^\dagger_{\alpha+2 k_2^cm_2\tau s_2} c_{\alpha+2 k_1^cm_1\tau s_1}: \nonumber\\&+& \delta_{k^c_1k^c_2}O^c_{\alpha+2 m_2\tau s_2,\alpha+2 m_1\tau s_1}(k^c_1):d^\dagger_{\bR\alpha\tau s_1} d_{\bR\alpha\tau s_2}: -O^{d\bR} _{\alpha\tau s_1,\bar{\alpha}-\tau s_2} c^\dagger_{\bar{\alpha}+2 k_2^cm_2-\tau s_2} c_{\alpha+2 k_1^cm_1\tau s_1}\nonumber\\
&-&\delta_{k^c_1k^c_2}O^{c}_{\bar{\alpha}+2 m_2-\tau s_2,\alpha+2 m_1\tau s_1}(k^c_1)d^\dagger_{\bR\alpha\tau s_1} d_{\bR\bar{\alpha}-\tau s_2} \nonumber\\ 
&-&O^{cd\ast}_{\alpha+2 k_1^cm_1\tau s_1,\bR\alpha\tau s_1}c^\dagger_{\alpha+2 k_2^cm_2\tau s_2} d_{\bR\alpha\tau s_2}-O^{cd}_{\alpha+2 k_2^cm_2\tau s_2,\bR\alpha\tau s_2} d^\dagger_{\bR\alpha\tau s_1} c_{\alpha+2 k_1^cm_1\tau s_1}\nonumber\\
    &+&O^{cd\ast}_{\alpha+2 k_1^cm_1\tau s_1,\bR\alpha\tau s_1}  c^\dagger_{\bar{\alpha}+2 k_2^cm_2-\tau s_2}  d_{\bR\bar{\alpha}-\tau s_2} + O^{cd}_{\bar{\alpha}+2 k_2^cm_2-\tau s_2,\bR\bar{\alpha}-\tau s_2} d^\dagger_{\bR\alpha\tau s_1} c_{\alpha+2 k_1^cm_1\tau s_1}\nonumber\\
&-&\delta_{k^c_1k^c_2}\left(O^{d\bR}_{\alpha\tau s_1,\alpha\tau s_2}-\frac{1}{2}\delta_{ s_1 s_2}\right)O^c_{\alpha+2 m_2\tau s_2,\alpha+2 m_1\tau s_1}(k^c_1) +O^{cd\ast}_{\alpha+2 k_1^cm_1\tau s_1,\bR\alpha\tau s_1}O^{cd}_{\alpha+2 k_2^cm_2\tau s_2,\bR\alpha\tau s_2}  \nonumber  
\\
&+&\delta_{k^c_1k^c_2}O^{c}_{\bar{\alpha}+2 m_2-\tau s_2,\alpha+2 m_1\tau s_1}(k^c_1) O^{d\bR} _{\alpha\tau s_1,\bar{\alpha}-\tau s_2} - O^{cd\ast}_{\alpha+2 k_1^cm_1\tau s_1,\bR\alpha\tau s_1}O^{cd}_{\bar{\alpha}+2 k_2^cm_2-\tau s_2,\bR\bar{\alpha}-\tau s_2}\bigg]
\end{eqnarray}
Rewriting the above in terms of the momentum space representation for $f$s gives us
\begin{eqnarray}
\hat{H}^{MF}_{J,\bf{B}}&\approx&-J\sum_{\substack{ s_1 s_2\\\tau\alpha\\k^c}}\sum_{\substack{ m_1m_2\\ l_1,l_2\in\mathbb{Z}}} F^{l_1,l_2}_{m_1,m_2}(k^c,k^c) \left(\sum_{r_1r_2}\delta_{l_2,[l_1+r_2-r_1]_q}O^f(r_1,r_2)_{\alpha\tau s_1,\alpha\tau s_2}-\frac{1}{2}\delta_{l_1,[l_2]_q}\delta_{ s_1 s_2}\right)\nonumber\\&\times&\left(:c^\dagger_{\alpha+2 k^cm_2\tau s_2} c_{\alpha+2 k^cm_1\tau s_1}:-O^c_{\alpha+2 m_2\tau s_2,\alpha+2 m_1\tau s_1}(k^c) \right)\nonumber
\\&& -\frac{J}{\mathcal{N}}\sum_{\substack{ s_1 s_2\\\tau\alpha\\k^c k^f}}\sum_{\substack{ m_1m_2\\ r_1r_2\\l_1,l_2\in\mathbb{Z}}} F^{l_1,l_2}_{m_1,m_2}(k^c,k^c) \delta_{l_2,[l_1+r_2-r_1]_q}O^c_{\alpha+2 m_2\tau s_2,\alpha+2 m_1\tau s_1}(k^c):f^\dagger_{\alpha k^f r_1 \tau s_1} f_{\alpha k^f r_2\tau s_2}:\nonumber\\
&&+J\sum_{\substack{ s_1 s_2\\\tau\alpha\\k^c}}\sum_{\substack{ m_1m_2\\ r_1r_2\\l_1,l_2\in\mathbb{Z}}} F^{l_1,l_2}_{m_1,m_2}(k^c,k^c) \delta_{l_2,[l_1+r_2-r_1]_q}O^f(r_1,r_2)_{\alpha\tau s_1,\bar{\alpha}-\tau s_2} c^\dagger_{\bar{\alpha}+2 k^c m_2-\tau s_2} c_{\alpha+2 k^c m_1\tau s_1}
\nonumber\\&& +\frac{J}{\mathcal{N}}\sum_{\substack{ s_1 s_2\\\tau\alpha\\k^c k^f}}\sum_{\substack{ m_1m_2\\ r_1r_2\\l_1,l_2\in\mathbb{Z}}} F^{l_1,l_2}_{m_1,m_2}(k^c,k^c) \delta_{l_2,[l_1+r_2-r_1]_q} O^c_{\bar{\alpha}+2 m_2-\tau s_2,\alpha+2 m_1\tau s_1}(k^c)f^\dagger_{\alpha k^f r_1 \tau s_1} f_{\bar{\alpha} k^f r_2-\tau s_2}\nonumber\\
&&+ \frac{J}{\mathcal{\sqrt{N}}}\sum_{\substack{ s_1 s_2\\\tau\alpha\\k^c_1 k^c_2}}\sum_{\substack{ m_1m_2\\ r_1r_2\\l_1,l_2\in\mathbb{Z}}} \left(F^{l_1,l_2}_{m_1,m_2}(k^c_1,k^c_2)\delta_{l_2,[l_1+r_2-r_1]_q} O^{cf\ast}_{\alpha+2m_1\tau s_1,\alpha\tau s_1}(k^c_1,r_1)c^\dagger_{\alpha+2k^c_2m_2\tau s_2} f_{\alpha k^c_2r_2\tau s_2} + h.c. \right)\nonumber\\
&&- \frac{J}{\mathcal{\sqrt{N}}}\sum_{\substack{ s_1 s_2\\\tau\alpha\\k^c_1 k^c_2}}\sum_{\substack{ m_1m_2\\ r_1r_2\\l_1,l_2\in\mathbb{Z}}} \left(F^{l_1,l_2}_{m_1,m_2}(k^c_1,k^c_2)\delta_{l_2,[l_1+r_2-r_1]_q} O^{cf\ast}_{\alpha+2m_1\tau s_1,\alpha\tau s_1}(k^c_1,r_1)c^\dagger_{\bar{\alpha}+2 k_2^cm_2-\tau s_2}  d_{\bR\bar{\alpha}-\tau s_2} + h.c. \right)\nonumber\\
%&&+ J\sum_{\substack{ s_1 s_2\\\tau\alpha\\k^c}}\sum_{\substack{ m_1m_2\\ s_1,s_2}} F^{s_1,s_2}_{m_1,m_2}(k^c,k^c) \left(\sum_{r_1r_2}\delta_{s_2,[s_1+r_2-r_1]_q}O^f(r_1,r_2)_{\alpha\tau s_1,\alpha\tau s_2}-\frac{1}{2}\delta_{s_1,[s_2]_q}\delta_{ s_1 s_2}\right)O^c_{\alpha+2 m_2\tau s_2,\alpha+2 m_1\tau s_1}(k^c) \nonumber\\
&&- J\sum_{\substack{ s_1 s_2\\\tau\alpha\\k^c}}\sum_{\substack{ m_1m_2\\ r_1r_2\\l_1,l_2\in\mathbb{Z}}} F^{l_1,l_2}_{m_1,m_2}(k^c,k^c) \delta_{l_2,[l_1+r_2-r_1]_q}O^f(r_1,r_2)_{\alpha\tau s_1,\bar{\alpha}-\tau s_2}O^{c}_{\bar{\alpha}+2 m_2-\tau s_2,\alpha+2 m_1\tau s_1}(k^c) \nonumber\\
&&-J\sum_{\substack{ s_1 s_2\\\tau\alpha\\k^c_1 k^c_2}}\sum_{\substack{ m_1m_2\\ r_1r_2\\l_1,l_2\in\mathbb{Z}}} F^{l_1,l_2}_{m_1,m_2}(k^c_1,k^c_2)\delta_{l_2,[l_1+r_2-r_1]_q} O^{cf\ast}_{\alpha+2m_1\tau s_1,\alpha\tau s_1}(k^c_1,r_1) O^{cf}_{\alpha+2m_2\tau s_2,\alpha\tau s_2}(k^c_2,r_2)\nonumber\\
&&+J\sum_{\substack{ s_1 s_2\\\tau\alpha\\k^c_1 k^c_2}}\sum_{\substack{ m_1m_2\\ r_1r_2\\l_1,l_2\in\mathbb{Z}}} F^{l_1,l_2}_{m_1,m_2}(k^c_1,k^c_2)\delta_{l_2,[l_1+r_2-r_1]_q} O^{cf\ast}_{\alpha+2m_1\tau s_1,\alpha\tau s_1}(k^c_1,r_1) O^{cf}_{\bar{\alpha}+2m_2-\tau s_2,\bar{\alpha}-\tau s_2}(k^c_2,r_2)
\end{eqnarray}
We can rewrite the above as a sum of bilinears and constants as
\begin{eqnarray}
\hat{H}_{J,\bf{B}}&\approx& \bar{H}_{J,\bf{B}}-E_{J,\bf{B}}
\end{eqnarray}
where 
\begin{eqnarray}
 \bar{H}_{J,\bf{B}} &=& -J\sum_{\substack{s_1 s_2\\\tau\alpha\\k^c}}\sum_{\substack{ m_1m_2\\ r_1r_2\\l_1,l_2\in \mathbb{Z}}} F^{l_1,l_2}_{m_1,m_2}(k^c,k^c) \left(\sum_{r_1r_2}\delta_{l_2,[l_1+r_2-r_1]_q}O^f(r_1,r_2)_{\alpha\tau s_1,\alpha\tau s_2}-\frac{1}{2}\delta_{l_1,[l_2]_q}\delta_{s_1 s_2}\right)c^\dagger_{\alpha+2 k^cm_2\tau s_2} c_{\alpha+2 k^cm_1\tau s_1}  \nonumber\\
&&+J\sum_{\substack{ s_1 s_2\\\tau\alpha\\k^c}}\sum_{\substack{ m_1m_2\\ r_1r_2\\l_1,l_2 \in \mathbb{Z}}} F^{l_1,l_2}_{m_1,m_2}(k^c,k^c) \delta_{l_2,[l_1+r_2-r_1]_q}O^f(r_1,r_2)_{\alpha\tau s_1,\bar{\alpha}-\tau s_2} c^\dagger_{\bar{\alpha}+2 k^c m_2-\tau s_2} c_{\alpha+2 k^c m_1\tau s_1}\nonumber\\ 
 && -\frac{J}{\mathcal{N}}\sum_{\substack{ s_1 s_2\\\tau\alpha\\k^c k^f}}\sum_{\substack{ m_1m_2\\ r_1r_2\\l_1,l_2 \in \mathbb{Z}}} F^{l_1,l_2}_{m_1,m_2}(k^c,k^c) \delta_{l_2,[l_1+r_2-r_1]_q}O^c_{\alpha+2 m_2\tau s_2,\alpha+2 m_1\tau s_1}(k^c)f^\dagger_{\alpha k^f r_1 \tau s_1} f_{\alpha k^f r_2\tau s_2}\nonumber
\\&& +\frac{J}{\mathcal{N}}\sum_{\substack{ s_1 s_2\\\tau\alpha\\k^c k^f}}\sum_{\substack{ m_1m_2\\ r_1r_2\\l_1,l_2 \in \mathbb{Z}}} F^{l_1,l_2}_{m_1,m_2}(k^c,k^c) \delta_{l_2,[l_1+r_2-r_1]_q} O^c_{\bar{\alpha}+2 m_2-\tau s_2,\alpha+2 m_1\tau s_1}(k^c)f^\dagger_{\alpha k^f r_1 \tau s_1} f_{\bar{\alpha} k^f r_2-\tau s_2}\nonumber\\
&&- \frac{J}{\mathcal{\sqrt{N}}}\sum_{\substack{ s_1 s_2\\\tau\alpha\\k^c_1 k^c_2}}\sum_{\substack{ m_1m_2\\ r_1r_2\\l_1,l_2 \in \mathbb{Z}}} \left(F^{l_1,l_2}_{m_1,m_2}(k^c_1,k^c_2)\delta_{l_2,[l_1+r_2-r_1]_q} O^{cf\ast}_{\alpha+2m_1\tau s_1,\alpha\tau s_1}(k^c_1,r_1)c^\dagger_{\bar{\alpha}+2 k_2^cm_2-\tau s_2}  f_{\bar{\alpha} k^c_2r_2-\tau s_2} + h.c. \right)\nonumber
\\
&&+ \frac{J}{\mathcal{\sqrt{N}}}\sum_{\substack{ s_1 s_2\\\tau\alpha\\k^c_1 k^c_2}}\sum_{\substack{ m_1m_2\\ r_1r_2\\l_1,l_2 \in \mathbb{Z}}} \left(F^{l_1,l_2}_{m_1,m_2}(k^c_1,k^c_2)\delta_{l_2,[l_1+r_2-r_1]_q} O^{cf\ast}_{\alpha+2m_1\tau s_1,\alpha\tau s_1}(k^c_1,r_1)c^\dagger_{\alpha+2k^c_2m_2\tau s_2} f_{\alpha k^c_2r_2\tau s_2} + h.c. \right)
\end{eqnarray}
and 
\begin{eqnarray}
    -E_{J,\bf{B}} &=& -\frac{J}{2}\sum_{\substack{ s\\\tau\alpha\\k^c}}\sum_{\substack{ m\\ r_1r_2\\l_1,l_2\in\mathbb{Z}}} F^{l_1,l_2}_{m,m}(k^c,k^c) \left(\sum_{r_1r_2}\delta_{l_2,[l_1+r_2-r_1]_q}O^f(r_1,r_2)_{\alpha\tau s,\alpha\tau s}-\frac{1}{2}\delta_{l_1,[l_2]_q}\right)\nonumber\\
&&- J\sum_{\substack{ s_1 s_2\\\tau\alpha\\k^c}}\sum_{\substack{ m_1m_2\\ r_1r_2\\l_1,l_2\in\mathbb{Z}}} F^{l_1,l_2}_{m_1,m_2}(k^c,k^c) \delta_{l_2,[l_1+r_2-r_1]_q}O^f(r_1,r_2)_{\alpha\tau s_1,\alpha\tau s_2}O^c_{\alpha+2 m_2\tau s_2,\alpha+2 m_1\tau s_1}(k^c)\nonumber\\
&&+ J\sum_{\substack{ s_1 s_2\\\tau\alpha\\k^c}}\sum_{\substack{ m_1m_2\\ r_1r_2\\l_1,l_2\in\mathbb{Z}}} F^{l_1,l_2}_{m_1,m_2}(k^c,k^c) \delta_{l_2,[l_1+r_2-r_1]_q}O^f(r_1,r_2)_{\alpha\tau s_1,\bar{\alpha}-\tau s_2}O^{c}_{\bar{\alpha}+2 m_2-\tau s_2,\alpha+2 m_1\tau s_1}(k^c) \nonumber\\
&&+J\sum_{\substack{ s_1 s_2\\\tau\alpha\\k^c_1 k^c_2}}\sum_{\substack{ m_1m_2\\ r_1r_2\\l_1,l_2\in\mathbb{Z}}} F^{l_1,l_2}_{m_1,m_2}(k^c_1,k^c_2)\delta_{l_2,[l_1+r_2-r_1]_q} O^{cf\ast}_{\alpha+2m_1\tau s_1,\alpha\tau s_1}(k^c_1,r_1) O^{cf}_{\alpha+2m_2\tau s_2,\alpha\tau s_2}(k^c_2,r_2)\nonumber\\
&&-J\sum_{\substack{ s_1 s_2\\\tau\alpha\\k^c_1 k^c_2}}\sum_{\substack{ m_1m_2\\ r_1r_2\\l_1,l_2\in\mathbb{Z}}} F^{l_1,l_2}_{m_1,m_2}(k^c_1,k^c_2)\delta_{l_2,[l_1+r_2-r_1]_q} O^{cf\ast}_{\alpha+2m_1\tau s_1,\alpha\tau s_1}(k^c_1,r_1) O^{cf}_{\bar{\alpha}+2m_2-\tau s_2,\bar{\alpha}-\tau s_2}(k^c_2,r_2)
\end{eqnarray}
We make the same approximations to the above terms discussed below Eq.(\ref{Eq:Approx_m_MF}) before numerical implementation.

%---------------------Section change----

\subsection{The one-shot spectrum and self-consistent calculation of gaps at CNP}\label{appx:One_shot_in_B_strain}

The mean-field interaction Hamiltonian is of the form 
\begin{eqnarray}
    \hat{H}_{I,\bB} - E_{I,\bB} = \bar{H}_{U,\bf{B}} +  \bar{H}_{V,\bf{B}} + \bar{H}_{W,\bf{B}} +  \bar{H}_{J,\bf{B}} - \left(E_{U,\bf{B}}+E_{V,\bf{B}}+E_{W,\bf{B}}+E_{J,\bf{B}}\right)\label{Eq:appx:Full_mean_field_interac}
\end{eqnarray}
as derived in the previous subsections, where $\hat{H}_{I,\bB}$ and $E_{I,\bB}$ are bilinears and constants respectively. Motivated by the parent state for the semimetal at $\bB=0$ \cite{herzog2025kekule}, we consider the parent state 
%    \begin{eqnarray}
%    |\Phi\rangle = \prod_{s=\uparrow\downarrow}\prod_{r=0}^{q-1}\prod_{k^f}f^\dagger_{\alpha k^f r\bf{K} s}|FS_B\rangle
%\end{eqnarray}
\begin{eqnarray}
    |\Phi\rangle = \prod_{s=\uparrow\downarrow}\prod_{\tau=\pm 1}\prod_{r=0}^{q-1}\prod_{k^f}\frac{f^\dagger_{1 k^f r \tau s}+i\tau f^\dagger_{2 k^f r \tau s}}{\sqrt{2}}|FS\rangle_B\label{Eq:appx::CNP_strain_candidate_B}
\end{eqnarray}
where $|FS\rangle_B$ is formed by half filling the $c$-LLs in absence of strain and $M$. In other words, 
\begin{eqnarray}
    |FS\rangle_B = \prod_{n}^{'} \prod_{k^c \tau s}u^\dagger_{k^c n \tau s}|0\rangle.
\end{eqnarray}
where $\prod_{n}^{'}$ represents the product over the filled c-Lls and $u^\dagger$ represents the corresponding creation operator. Moreover we can expand them as
\begin{eqnarray}
    u^\dagger_{k^c n \tau s} = \sum_{a}\sum_{m=0}^{m_{a\tau}} U^{\tau}_{am,n}(k^c) c^\dagger_{a m k^c \tau s}
\end{eqnarray}
where $U^{\tau}_{am,n}(k^c)$ are the wavefunctions. The wavefunctions (independent of $k^c$ as these are LLS) and the energies at valley $\bK$ (spin $\uparrow\downarrow$ degenerate ) read
\begin{eqnarray}
  &&\frac{1}{\sqrt{2}}[|m\rangle, 0,-i|m+1\rangle,0]^T ~~~ ;m\in\{0,\ldots, m_\ast+1\} ~~;E_m =-\frac{\sqrt{2(m+1)}v_\ast}{\ell}\\
  &&\frac{1}{\sqrt{2}}[0,|m+1\rangle, 0,i|m\rangle]^T ~~~~~~ ;m\in\{0,\ldots, m_\ast-1\} ~~~;E_m =-\frac{\sqrt{2(m+1)}v_\ast}{\ell}\\
  &&[0,0,|0\rangle,0]^T~~~~~~~~~~~~~~~~~~~~~~~~~~~~~~~~~~~~~~~~~~~~~~~~~~~~;E=0
\end{eqnarray}
and at valley $\bK'$ (spin $\uparrow\downarrow$ degenerate ) are
\begin{eqnarray}
&&\frac{1}{\sqrt{2}}[0,|m\rangle,0,i|m+1\rangle]^T ~~~~~~~~ ;m\in\{0,\ldots, m_\ast+1\} ~~;E_m =-\frac{\sqrt{2(m+1)}v_\ast}{\ell}\\
  &&\frac{1}{\sqrt{2}}[|m+1\rangle, 0,-i|m\rangle,0]^T ~~~~~~ ;m\in\{0,\ldots, m_\ast-1\} ~~~;E_m =-\frac{\sqrt{2(m+1)}v_\ast}{\ell}\\
  &&[0,0,0,|0\rangle]^T~~~~~~~~~~~~~~~~~~~~~~~~~~~~~~~~~~~~~~~~~~~~~~~~~~~~~~~;E=0    
\end{eqnarray}
It trivially follows that $\nu_c=Tr[O^c]=0$, $O^{cf}=0$
\begin{eqnarray}
   [O^f(r_1,r_2)]_{\beta \tau' s', \alpha \tau s}= \frac{1}{2}\delta_{ss'}\delta_{\tau \tau'} \delta_{rr'}\left(\sigma_0 - \tau\sigma_y\right)_{\beta \alpha}
\end{eqnarray} $\nu_f= Tr[O^f]-4 =0$ and thus the filling $\nu = \nu_c+\nu_f=0$ as expected. The bilinear part of the one-shot mean-field interaction evaluates to
\begin{eqnarray}
 \hat{H}_{I,\bB} =   - \sum_{\tau s r k^f\alpha \beta} f^\dagger_{\alpha k^f r \tau s} \tau U_1/2   \left(\sigma_y\right)_{\alpha,\beta} f^\dagger_{\beta k^f r \tau s}\label{Appx:Eq:Oneshot_MF_B}
\end{eqnarray}
The one-shot Hofstadter spectrum evaluated by diagonalizing $\hat{H}^\tau_{0,\bB}+\hat{H}_{I,\bB}$ is shown in right panel of Fig.(\ref{fig:main:fig2}b),
where $\hat{H}^\tau_{0,\bB}$ is given by Eq.(\ref{Eq:appx:Single_particle_B}). As can be seen in the figure, we find gaps with Chern number of form $\pm 4n$, where $n\in \mathbb{Z}_{\geq 0}$. The magnitude of the gap at filling $\nu = \pm 4n\frac{\phi}{\phi_0}$, with Chern number $\pm 4n$ can now be self-consistently calculated. We fill the one-shot LLs upto the filling $\nu$ and feed the  new order parameters into Eq.(\ref{Eq:appx:Full_mean_field_interac}) to generate a new mean field Hamiltonian. This process is repeated till the norm of order parameter (OP) converge, i.e. $||\Delta OP||_\infty\leq 10^{-7}$. We work on a $10\times 10$ mesh in the magnetic BZ. The magnitude of the self-consistently calculated gaps is shown in Fig.(\ref{fig:main:fig2}d) and the Streda plot is shown in Fig.(\ref{fig:main:fig2}f).

%---------------------Section change----

\section{MTG LL form factors}\label{Appx:LLformfactor}

In this section, we calculate the form factors for the MTG LLs
\begin{eqnarray}
    &&\mathcal{W}^{\bq}_{m_2m_1}(k_2^c,k_1^c) = \int d^2 \br e^{i\bq\cdot\br}\chi^{\ast}_{k^c_2m_2}(\br) \chi_{k^c_1m_1}(\br)\\
    =&& \frac{1}{\ell L_m \mathcal{N}}\sum_{s_1,s_2\in\mathcal{Z}}\int d^2 \br e^{i\bq\cdot\br} \left(e^{2\pi i s_2k_{12}^c}\hat{t}^{s_2}_{\tilde{\bL}_1}e^{2\pi i k^c_{22}\frac{y}{\tilde{L}_2}}\varphi_{m_2}(x-qk_{22}^c\tilde{L}_{1x})\right)^\ast \left(e^{2\pi i s_1k_{11}^c}\hat{t}^{s_1}_{\tilde{\bL}_1}e^{2\pi i k^c_{21}\frac{y}{\tilde{L}_2}}\varphi_{m_1}(x-qk_{21}^c\tilde{L}_{1x})\right)\\
 =&& \frac{1}{\ell L_m \mathcal{N}}\sum_{s_1,s_2\in\mathcal{Z}}\int d^2 \br e^{i\bq\cdot\br} \left(e^{2\pi i s_2k_{12}^c}\hat{t}^{s_2}_{\tilde{\bL}_1}e^{2\pi i k^c_{22}\frac{y}{\tilde{L}_2}}\varphi_{m_2}(x-qk_{22}^c\tilde{L}_{1x})\right)^\ast\nonumber\\&& \left(e^{2\pi i s_1k_{11}^c}e^{-i\pi s_1(s_1-1)\frac{L_{1y}}{q\tilde{L}_2}}e^{2\pi is_1\frac{y}{q\tilde{L}_2}}\hat{T}^{s_1}_{\tilde{\bL}_1}e^{2\pi i k^c_{21}\frac{y}{\tilde{L}_2}}\varphi_{m_1}(x-qk_{21}^c\tilde{L}_{1x})\right)\\
 =&& \frac{1}{\ell L_m \mathcal{N}}\sum_{s_1,s_2\in\mathcal{Z}}\int d^2 \br e^{i\bq\cdot(\br+ s_1\tilde{\bL}_1)} \hat{T}^{-s_1}_{\tilde{\bL}_1}\left(e^{2\pi i s_2k_{12}^c}\hat{t}^{s_2}_{\tilde{\bL}_1} e^{2\pi i k^c_{22}\frac{y}{\tilde{L}_2}}\varphi_{m_2}(x-qk_{22}^c\tilde{L}_{1x})\right)^\ast \nonumber\\&& \left(e^{2\pi i s_1k_{11}^c}e^{-i\pi s_1(s_1-1)\frac{L_{1y}}{q\tilde{L}_2}}e^{2\pi is_1\frac{y+s_1L_{1y}}{q\tilde{L}_2}}e^{2\pi i k^c_{21}\frac{y}{\tilde{L}_2}}\varphi_{m_1}(x-qk_{21}^c\tilde{L}_{1x})\right)\\
 =&& \frac{1}{\ell L_m \mathcal{N}} \sum_{s_1,s_2\in\mathcal{Z}}\int d^2 \br e^{i\bq\cdot(\br+ s_1\tilde{\bL}_1)} e^{2\pi is_1\frac{y}{q\tilde{L}_2}} e^{+i\pi s_1(s_1+1)\frac{L_{1y}}{q\tilde{L}_2}}\hat{T}^{-s_1}_{\tilde{\bL}_1} \left(e^{2\pi i s_2k_{12}^c}\hat{t}^{s_2}_{\tilde{\bL}_1} e^{2\pi i k^c_{22}\frac{y}{\tilde{L}_2}}\varphi_{m_2}(x-qk_{22}^c\tilde{L}_{1x})\right)^\ast \nonumber\\&& \left(e^{2\pi i s_1k_{11}^c}e^{2\pi i k^c_{21}\frac{y}{\tilde{L}_2}}\varphi_{m_1}(x-qk_{21}^c\tilde{L}_{1x})\right)\\
 =&& \frac{1}{\ell L_m \mathcal{N}} \sum_{s_1,s_2\in\mathcal{Z}}\int d^2 \br e^{i\bq\cdot(\br+ s_1\tilde{\bL}_1)}\left(e^{2\pi i s_2k_{12}^c} \hat{t}^{s_2-s_1}_{\tilde{\bL}_1} e^{2\pi i k^c_{22}\frac{y}{\tilde{L}_2}}\varphi_{m_2}(x-qk_{22}^c\tilde{L}_{1x})\right)^\ast \nonumber\\&& \left(e^{2\pi i s_1k_{11}^c}e^{2\pi i k^c_{21}\frac{y}{\tilde{L}_2}}\varphi_{m_1}(x-qk_{21}^c\tilde{L}_{1x})\right) 
 \end{eqnarray}
 We thus have
 \begin{eqnarray}
\mathcal{W}^{\bq}_{m_2m_1}(k_2^c,k_1^c) = && \delta_{[k^c_{11}+q_1]_1,k^c_{12}}\frac{s_{tot}}{\ell L_m \mathcal{N}} \sum_{s\in\mathcal{Z}}\int d^2 \br e^{i\bq\cdot\br}\left(e^{2\pi i sk_{12}^c}\hat{t}^{s}_{\tilde{\bL}_1} e^{2\pi i k^c_{22}\frac{y}{\tilde{L}_2}}\varphi_{m_2}(x-qk_{22}^c\tilde{L}_{1x})\right)^\ast\nonumber\\&& \left(e^{2\pi i k^c_{21}\frac{y}{\tilde{L}_2}}\varphi_{m_1}(x-qk_{21}^c\tilde{L}_{1x})\right)
\end{eqnarray}
where $q_1$ is defined through $\bq=q_1\tilde{\bg}_1+q_2\tilde{\bg}_2$. We then have
\begin{eqnarray}
\mathcal{W}^{\bq}_{m_2m_1}(k_2^c,k_1^c)=&& \delta_{[k^c_{11}+q_1]_1,k^c_{12}}\frac{s_{tot}}{\ell L_m \mathcal{N}} \sum_{s\in\mathcal{Z}}\int d^2 \br e^{i\bq\cdot\br}e^{-2\pi i sk_{12}^c}\nonumber\\&&\left(e^{-i\pi s(s-1)\frac{L_{1y}}{q\tilde{L}_2}} e^{2\pi i s\frac{y}{q\tilde{L}_2}}e^{2\pi i k^c_{22}\frac{y-sL_{1y}}{\tilde{L}_2}}\varphi_{m_2}(x-(s+qk_{22}^c)\tilde{L}_{1x})\right)^\ast\nonumber\\&& \left(e^{2\pi i k^c_{21}\frac{y}{\tilde{L}_2}}\varphi_{m_1}(x-qk_{21}^c\tilde{L}_{1x})\right)    
\end{eqnarray}
This leads to the $y$ integral 
\begin{eqnarray}
    \int dy e^{i(q_y+\frac{2\pi}{L_m}(k_{21}^c-k_{22}^c) - \frac{2\pi s}{qL_m})y} &=& e^{2\pi i \frac{y}{L_m}\left(\frac{q_yL_m}{2\pi} + k_{21}^c-k_{22}^c -\frac{s}{q}\right)}\\
    &=& n_{tot}L_m\delta_{k_{21}^c-k_{22}^c,\frac{s}{q}-\frac{q_yL_m}{2\pi} } 
\end{eqnarray}
Thus
\begin{eqnarray}
 &&\mathcal{W}^{\bq}_{m_2m_1}(k_2^c,k_1^c) =  \delta_{[k^c_{11}+q_1]_1,k^c_{12}}  \sum_{s\in\mathcal{Z}} \delta_{k_{21}^c-k_{22}^c,\frac{s}{q}-\frac{q_yL_m}{2\pi} } e^{i\pi s(s-1)\frac{L_{1y}}{q\tilde{L}_2}} e^{2\pi i sk^c_{22}\frac{L_{1y}}{\tilde{L}_2}} e^{-2\pi i sk_{12}^c}\nonumber\\
 &&\frac{1}{\ell}\int dx e^{iq_xx}\varphi_{m_2}(x-(s+qk_{22}^c)\tilde{L}_{1x}) \varphi_{m_1}(x-qk_{21}^c\tilde{L}_{1x})\\
 &&=  \delta_{[k^c_{11}+q_1]_1,k^c_{12}}  \sum_{s\in\mathcal{Z}} \delta_{k_{21}^c-k_{22}^c,\frac{s}{q}-\frac{q_yL_m}{2\pi} } e^{i\pi s(s-1)\frac{L_{1y}}{q\tilde{L}_2}} e^{2\pi i sk^c_{22}\frac{L_{1y}}{\tilde{L}_2}} e^{-2\pi i sk_{12}^c} e^{iq_x(s+qk^c_{22})\tilde{L}_{1x}}\nonumber\\
 &&\frac{1}{\ell}\int dx e^{iq_xx}\varphi_{m_2}(x) \varphi_{m_1}(x+(s+q(k^c_{22}-k_{21}^c)\tilde{L}_{1x})\\
 &&=  \delta_{[k^c_{11}+q_1]_1,k^c_{12}}  \sum_{s\in\mathcal{Z}} \delta_{k_{21}^c-k_{22}^c,\frac{s}{q}-\frac{q_yL_m}{2\pi} } e^{i\pi s(s-1)\frac{L_{1y}}{q\tilde{L}_2}} e^{2\pi i sk^c_{22}\frac{L_{1y}}{\tilde{L}_2}} e^{-2\pi i sk_{12}^c} e^{iq_xq_y\ell^2}
 e^{ik^c_{21}q_x\frac{2\pi\ell^2}{L_m}}\nonumber\\
 &&\frac{1}{\ell}\int dx e^{iq_xx}\varphi_{m_2}(x) \varphi_{m_1}(x+q_y\ell^2)\\
\end{eqnarray}
where we used the delta condition and the fact that $q\tilde{L}_{1x}L_m=2\pi\ell^2$ in last step. Now see that
\begin{eqnarray}
&&\frac{1}{\ell}\int dx e^{iq_xx}\varphi_{m_2}(x) \varphi_{m_1}(x+q_y\ell^2) = \frac{1}{\ell}    \int dx \varphi_{m_2}(x) e^{iq_xx} e^{q_y\ell^2\partial_x}\varphi_{m_1}(x) \\
&&= \frac{1}{\ell}    \int dx \varphi_{m_2}(x) e^{iq_xx + q_y\ell^2\partial_x} e^{\frac{1}{2}[iq_x x,q_y\ell^2\partial_x]}\varphi_{m_1}(x) \\
&&=\frac{1}{\ell} e^{-\frac{i}{2}q_xq_y\ell^2}   \int dx \varphi_{m_2}(x) e^{iq_xx + q_y\ell^2\partial_x} \varphi_{m_1}(x)
\end{eqnarray}
Since in h.o. basis, $x=\frac{\ell}{\sqrt{2}}(\hat{a}+\hat{a}^\dagger)$ , $\partial_x=\frac{1}{\sqrt{2}\ell}(\hat{a}-\hat{a}^\dagger)$, we have
\begin{eqnarray}
e^{iq_xx + q_y\ell^2\partial_x} &=& e^{iq_x\frac{\ell}{\sqrt{2}}(\hat{a}+\hat{a}^\dagger)+ q_y\ell^2\frac{1}{\sqrt{2}\ell}(\hat{a}-\hat{a}^\dagger)}\\
&=& e^{\frac{i\ell}{\sqrt{2}}\left(\bar{q_c}\hat{a}+q_c\hat{a}^\dagger\right)}
\end{eqnarray}
where $q_c=q_x+iq_y$ and $\bar{q}_c = q^\ast_c$. Thus
\begin{eqnarray}
\mathcal{W}^{\bq}_{m_2m_1}(k_2^c,k_1^c) &=&  \delta_{[k^c_{11}+q_1]_1,k^c_{12}}  \sum_{s\in\mathcal{Z}} \delta_{k_{21}^c-k_{22}^c,\frac{s}{q}-\frac{q_yL_m}{2\pi} } e^{i\pi s(s-1)\frac{L_{1y}}{q\tilde{L}_2}} e^{2\pi i s\left(k^c_{22}\frac{L_{1y}}{\tilde{L}_2}-k_{12}^c\right)}e^{\frac{i}{2}q_xq_y\ell^2}
e^{ik^c_{21}q_x\frac{2\pi\ell^2}{L_m}}\nonumber \\   
&&\times \langle m_2|e^{\frac{i\ell}{\sqrt{2}}\left(\bar{q_c}\hat{a}+q_c\hat{a}^\dagger\right)}|m_1\rangle
\end{eqnarray}
where
\begin{eqnarray}\label{small}
\bra{n} e^{c_+a^\dagger + c_-a}\ket{m} = 
\begin{cases}
      e^{\frac{1}{2}c_+c_-}\sqrt{\frac{m!}{n!}}(c_+)^{n-m}L^{n-m}_{m}(-c_+c_-) & \text{for}~~ n \geq m,\\
      e^{\frac{1}{2}c_+c_-}\sqrt{\frac{n!}{m!}}(c_-)^{m-n}L^{m-n}_{n}(-c_+c_-) & \text{for}~~ n < m,
    \end{cases}   
\end{eqnarray}
and $L^{m-n}_{n}(x)$ is the associated Laguerre polynomial, 
\begin{eqnarray}
L^{m}_{N}(x) = \sum_{k=0}^{N} \frac{(N+m)!}{(N-k)!(m+k)!k!}(-x)^k.
\end{eqnarray}
Since $g_{1y}=0$, $\frac{q_yL_m}{2\pi}=q_2 g_{2_y}\frac{2\pi}{L_m}=q_2$. Also see that if $k_2^c=k_1^c$, 
\begin{eqnarray}
\delta_{[k^c_{11}+q_1]_1,k^c_{12}}  \sum_{s\in\mathcal{Z}} \delta_{k_{21}^c-k_{22}^c,\frac{s}{q}-q_2 }    \rightarrow  \delta_{[q_1]_1,0}  \sum_{s\in\mathcal{Z}} \delta_{s,qq_2 } \label{Appx:Eq:Condition_on_q}
\end{eqnarray}
Thus $q_{1},qq_2\in\mathcal{Z}$ or $\bq = n_1\tilde{\bg}_1+ n_2\tilde{\bg}_2/q$ with $n_{1(2)}\in\mathbb{Z}$, if $k_2^c=k_1^c$.

%---------------------Section change----

\section{MTG LL density summed over lattice sites}\label{appx:MTG density sum}
In this section, we calculate the sum of the MTG LL density over the moir\'e lattice sites. We use the result obtained in this section to simplify the Eq.(\ref{Eq:Appx:MTG_density_sum_req}). The summation of interest reads 
\begin{eqnarray}
   &&\Omega_0\sum_{\bR} e^{-i\left(k^f_{11}-k^f_{12}\right)\tilde{\bg}_1\cdot\bR}  e^{-i\left(k^f_{21}-k^f_{22} + \frac{r_1-r_2}{q}\right)\tilde{\bg}_2   \cdot\bR}  \chi^\ast_{k^c_2 m_2}(\bR)\chi_{k^c_1 m_1}(\bR)\nonumber\\
   =&& \Omega_0\sum_{s_0,n_0} e^{-2\pi i s_0(k^f_{11}-k^f_{12})} e^{-2\pi i n_0 \left(k^f_{21}-k^f_{22} + \frac{r_1-r_2}{q}\right)} \chi^\ast_{k^c_2 m_2}(s_0\tilde{\bL}_1+n_0\tilde{\bL}_2)\chi_{k^c_1 m_1}(s_0\tilde{\bL}_1+n_0\tilde{\bL}_2)
\end{eqnarray}

To simplify the above summation, we start by first noting that
\begin{eqnarray}
&&\chi^\ast_{k^c_2 m_2}(s_0\tilde{\bL}_1+n_0\tilde{\bL}_2)\chi_{k^c_1 m_1}(s_0\tilde{\bL}_1+n_0\tilde{\bL}_2) \nonumber \\
&=& \frac{1}{\ell\mathcal{N}\tilde{L}_2} \sum_{s_2,s_1}e^{-2\pi i s_2k^c_{12}} e^{2\pi i s_2k^c_{22}\frac{L_{1y}}{L_m}}e^{+i\pi s_2(s_2-1)\frac{L_{1y}}{qL_m}}e^{-2\pi i (s_2+qk^c_{22})(\frac{n_0}{q}+s_0\frac{L_{1y}}{qL_m})}\varphi_{m_2}\left(-\{(s_2-s_0)+qk^c_{22}\}\tilde{L}_{1x}\right)
\nonumber \\
&\times&e^{2\pi i s_1k^c_{11}} e^{-2\pi i s_1k^c_{21}\frac{L_{1y}}{L_m}}e^{-i\pi s_1(s_1-1)\frac{L_{1y}}{qL_m}}e^{2\pi i (s_1+qk^c_{21})(\frac{n_0}{q}+s_0\frac{L_{1y}}{qL_m})}\varphi_{m_1}\left(-\{(s_1-s_0)+qk^c_{21}\}\tilde{L}_{1x}\right)
\\
&=&\frac{1}{\ell\mathcal{N}\tilde{L}_2} \sum_{s_2,s_1} e^{-2\pi i s_0k^c_{12}}e^{2\pi i s_0k^c_{22}\frac{L_{1y}}{L_m}}e^{2\pi i s_0k^c_{11}} e^{-2\pi i s_0k^c_{21}\frac{L_{1y}}{L_m}}\nonumber \\
&\times&e^{-2\pi i s_2k^c_{12}} e^{2\pi i s_2k^c_{22}\frac{L_{1y}}{L_m}}e^{+i\pi (s_2+s_0)(s_2+s_0-1)\frac{L_{1y}}{qL_m}}e^{-2\pi i (s_2+qk^c_{22})(\frac{n_0}{q}+s_0\frac{L_{1y}}{qL_m})}\varphi_{m_2}\left(-\{s_2+qk^c_{22}\}\tilde{L}_{1x}\right)
\nonumber \\
&\times&e^{2\pi i s_1k^c_{11}} e^{-2\pi i s_1k^c_{21}\frac{L_{1y}}{L_m}}e^{-i\pi (s_1+s_0)(s_1+s_0-1)\frac{L_{1y}}{qL_m}}e^{2\pi i (s_1+qk^c_{21})(\frac{n_0}{q}+s_0\frac{L_{1y}}{qL_m})}\varphi_{m_1}\left(-\{s_1+qk^c_{21}\}\tilde{L}_{1x}\right)
\\
&=&\frac{1}{\ell\mathcal{N}\tilde{L}_2}  e^{-2\pi i s_0k^c_{12}}e^{2\pi i s_0k^c_{11}} \sum_{s_2,s_1} e^{2\pi i (s_1-s_2+q(k^c_{21}-k^c_{22}))\frac{n_0}{q}}\nonumber \\
&\times&e^{-2\pi i s_2k^c_{12}} e^{2\pi i s_2k^c_{22}\frac{L_{1y}}{L_m}}e^{+i\pi s_2(s_2-1)\frac{L_{1y}}{qL_m}}\varphi_{m_2}\left(-\{s_2+qk^c_{22}\}\tilde{L}_{1x}\right)
\nonumber \\
&\times&e^{2\pi i s_1k^c_{11}} e^{-2\pi i s_1k^c_{21}\frac{L_{1y}}{L_m}}e^{-i\pi s_1(s_1-1)\frac{L_{1y}}{qL_m}}\varphi_{m_1}\left(-\{s_1+qk^c_{21}\}\tilde{L}_{1x}\right)
\end{eqnarray}
Now let us do the $n_0$ sum. We have
\begin{eqnarray}
    \sum_{n_0} e^{2\pi i (s_1-s_2+q(k^c_{21}-k^c_{22}))\frac{n_0}{q}} e^{-2\pi i n_0 \left(k^f_{21}-k^f_{22} + \frac{r_1-r_2}{q}\right)}
    &=&\sum_{n_0} e^{2\pi i n_0\left(\frac{s_1-s_2+r_2-r_1}{q} + k^c_{21}-k^c_{22} - k^f_{21}+k^f_{22}\right)}\\
=&&n_{tot}\sum_{l,j\in\mathcal{Z}} \delta_{k^c_{21}-k^c_{22} - k^f_{21}+k^f_{22},\frac{l}{q}}\delta_{s_1-s_2,jq + r_1-r_2-l}
\end{eqnarray}
For the $s_0$ sum we have
\begin{eqnarray}
    &&\sum_{s_0} e^{2\pi i s_0\left((k^c_{11}-k^c_{12}) - (k^f_{11}-k^f_{12})\right)}= s_{tot}\delta_{k^c_{11},[k^c_{12} + k^f_{11}-k^f_{12}]_1}
\end{eqnarray} 
Thus 
\begin{eqnarray}
   &&\Omega_0\sum_{\bR} e^{-i\left(k^f_{11}-k^f_{12}\right)\tilde{\bg}_1\cdot\bR}  e^{-i\left(k^f_{21}-k^f_{22} + \frac{r_1-r_2}{q}\right)\tilde{\bg}_2   \cdot\bR}  \chi^\ast_{k^c_2 m_2}(\bR)\chi_{k^c_1 m_1}(\bR)\nonumber\\
   &&=\frac{\Omega_0}{\ell L_m} \sum_{s_1,s_2}\sum_{j,l} \delta_{k^c_{21}-k^c_{22} - k^f_{21}+k^f_{22},\frac{l}{q}}\delta_{s_1-s_2,jq + r_1-r_2-l}\delta_{k^c_{11},[k^c_{12} - k^f_{12}+k^f_{11}]_1}\nonumber\\
   &\times&\left[e^{i(\phi^q_{s_1}(k^c_1)-\phi^q_{s_2}(k^c_2))} \varphi_{m_2}(-(s_2+qk^c_{22})\tilde{L}_{1x}) \varphi_{m_1}(-(s_1+qk^c_{21})\tilde{L}_{1x})\right]
   \end{eqnarray}
where
\begin{eqnarray}
   e^{i\phi^q_{s}(k^c)}  &\equiv& e^{2\pi i sk^c_1} e^{-2\pi i sk^c_2\frac{L_{1y}}{L_m}}e^{-i\pi s(s-1)\frac{L_{1y}}{qL_m}} \label{Eq:Defphase}
\end{eqnarray}

%---------------------Section change----

\section{Effective model for the strained semimetal state at CNP}

In this section, we discuss an effective low-energy description of the strained semimetal state in vicinity of $\Gamma$. We closely follow the derivation employing Schrieffer-Wolff transformation (SWT) presented in Ref\cite{singh2024topological}. We present the derivation for the valley $\bK$ as $\bK'$ is related by time reversal. Let us start by reviewing the one-shot interacting model.
%---------------------Section change----
\subsection{One-shot bands}\label{Appx:Parent_state_strain_CNP:at_angle}

The full non-interacting problem for strain axis at $\tilde{\phi}$ with respect to $\hat{x}$ reads

\begin{eqnarray}
    H_0^\tau(\bk,\epsilon,\tilde{\phi}) = h^{0,\tau}(\bk)+h^0_{M}+ h^\tau_\epsilon(\tilde{\phi})
\end{eqnarray}
where
\begin{eqnarray}
    h^{0,\bK}(\bk) = \left(\begin{array}{ccc}
     0_{2\times2} &v_\ast (k_x\sigma_0+ik_y\sigma_z)&\gamma\sigma_0+v'_\ast \left(k_x\sigma_x+k_y\sigma_y\right)\\
     v_\ast (k_x\sigma_0-ik_y\sigma_3)& 0_{2\times2}& 0_{2\times2}\\
   \gamma\sigma_0+v'_\ast \left(k_x\sigma_x+k_y\sigma_y\right)&0_{2\times2} &0_{2\times2}\\
    \end{array}\right)
\end{eqnarray}
where we consider only up to $\mathcal{O}(k)$, the mass term
\begin{eqnarray}
 h^0_M =    \left(\begin{array}{ccc}
     0_{2\times2} &0_{2\times2}&0_{2\times2}\\
     0_{2\times2}& M\sigma_x& 0_{2\times2}\\
   0_{2\times2}&0_{2\times2} &0_{2\times2}\\
    \end{array}\right),
\end{eqnarray}
the strain perturbation 
\begin{eqnarray}
    h^\bK_\epsilon(\tilde{\phi}) = \left(\begin{array}{ccc} 
    c \frac{\epsilon}{2}\left(\nu_G+1\right)\left(\begin{array}{cc}0&ie^{-2i\tilde{\phi}}\\-ie^{+2i\tilde{\phi}}&0\end{array}\right)
    &c' \frac{\epsilon}{2}\left(\nu_G+1\right)\left(\begin{array}{cc}0&-ie^{2i\tilde{\phi}}\\ie^{-2i\tilde{\phi}}&0\end{array}\right)
    &i\gamma'\frac{\epsilon}{2}\left(\nu_G-1\right)\sigma_3 \\c' \frac{\epsilon}{2}\left(\nu_G+1\right)\left(\begin{array}{cc}0&-ie^{2i\tilde{\phi}}\\ie^{-2i\tilde{\phi}}&0\end{array}\right)& M'\frac{\epsilon}{2}\left(\nu_G-1\right)\sigma_y&c''\frac{\epsilon}{2}(\nu_G+1)\left(\begin{array}{cc}
        +ie^{-2i\tilde{\phi}} &0  \\
        0 & -ie^{2i\tilde{\phi}}
    \end{array}\right) \\-i\gamma'\frac{\epsilon}{2}\left(\nu_G-1\right)\sigma_3&c''\frac{\epsilon}{2}(\nu_G+1)\left(\begin{array}{cc}
        -ie^{2i\tilde{\phi}} &0  \\
        0 & +ie^{-2i\tilde{\phi}}
    \end{array}\right)
    & M_f \frac{\epsilon}{2}\left(\nu_G+1\right)\left(\begin{array}{cc}0&ie^{-2i\tilde{\phi}}\\-ie^{+2i\tilde{\phi}}&0\end{array}\right)
      \end{array}\right)
\end{eqnarray} 
which clearly is uniaxial as $h_\epsilon(\tilde{\phi}+\pi)=h_\epsilon(\tilde{\phi})$. The model at $\bK'$ is related by time reversal. 
Let us abbreviate $M_f\frac{\epsilon}{2}(\nu_G+1)$ by $\varepsilon$ ($>0$ at magic angle for BM model). Then the decoupled $f$ sector at valley $\tau$ reads
\begin{eqnarray}
    h^\tau_{ff} = |\varepsilon|\left(\begin{array}{cc}
        0 & \tau i e^{-i\tau 2\tilde{\phi}} \\
     -\tau i e^{i\tau 2\tilde{\phi}}    & 
    \end{array}\right)
\end{eqnarray}
with eigenvalues $-|\varepsilon|$ being \begin{eqnarray}
    |\phi_{\bR \tau s}\rangle_{-} = \frac{1}{\sqrt{2}}\left(\begin{array}{cc}
         e^{-i\tau\tilde{\phi}}\\
         i\tau e^{i\tau\tilde{\phi}} 
    \end{array}\right) ~~\text{or} ~~\frac{e^{-i\tau\tilde{\phi}}f^\dagger_{\bR 1 \tau s}+i\tau e^{i\tau\tilde{\phi}}f^\dagger_{\bR 2 \tau s}}{\sqrt{2}}|0\rangle.
\end{eqnarray}
Now motivated by Ref\cite{herzog2025kekule}, the one shot state can thus be constructed by filling the above $f$ modes on top of half filled $c$s (without strain $\equiv|FS\rangle$)
\begin{eqnarray}
    |GS,\nu=0\rangle = \prod_{\bR}\prod_{s=\uparrow\downarrow}\prod_{\tau=\pm 1}\frac{e^{-i\tau\tilde{\phi}}f^\dagger_{\bR 1 \tau s}+i\tau e^{i\tau\tilde{\phi}}f^\dagger_{\bR 2 \tau s}}{\sqrt{2}}|FS\rangle
\end{eqnarray}
and the order parameter evaluates to 
\begin{eqnarray}
    O^{f,\bR}_{[b \tau s],[b' \tau' s']}(\tilde{\phi}) &=& \frac{1}{2}\left(\sigma_0-\sin(2\tilde{\phi})\sigma_x\right)\tau_0\zeta_0 -\frac{1}{2}\cos(2\tilde{\phi})\sigma_y \tau_z \zeta_0
\end{eqnarray}
where $\sigma,\tau,\zeta$ are Pauli matrices for orbital, valley and spin respectively. The one-shot interactions can now be evaluated and reads 
\begin{eqnarray}
      H_{int} = \left(\begin{array}{ccc}
       0_{8\times 8}   & 0_{8\times 8}   &0_{8\times 8}  \\0_{8\times 8}  &0_{8\times 8}  
        & 0_{8\times 8}  \\0_{8\times 8}  & 0_{8\times 8}  & -U_1(P-\frac{1}{2})  \end{array}\right) = \left(\begin{array}{ccc}
       0_{4\times 4}   & 0_{4\times 4}   &0_{4\times 4}  \\0_{4\times 4}  &0_{4\times 4}  
        & 0_{4\times 4}  \\0_{4\times 4}  & 0_{4\times 4}  & \left(\begin{array}{cc}
           h^{\tau=+1}_U(\tilde{\phi})  &  0\\
           0  & h^{\tau=-1}_U(\tilde{\phi})
        \end{array} \right)\end{array}\right)\otimes \zeta_0
\end{eqnarray}
where $P = [O^{f,\bR}]^T$  and the on-site $U_1$ interaction for $f$s take the form 
\begin{eqnarray}
    h^\tau_U(\tilde{\phi}) = \tau U_1/2 \left(\begin{array}{cc}
           0  & ie^{-2 \tau i\tilde{\phi}}\\
           -ie^{2 \tau i\tilde{\phi}}  & 0
        \end{array}\right)
\end{eqnarray}
Thus the full interacting (one-shot) model takes the form
\begin{eqnarray}
    H^\tau(\bk,\epsilon,\tilde{\phi}) =  H_0^\tau(\bk,\epsilon,\tilde{\phi}) + H^\tau_U(\tilde{\phi}) \label{Appx:Eq:Full_one_shot_model}
\end{eqnarray}
where
\begin{eqnarray}
   H^\tau_U(\tilde{\phi}) &=& \left(\begin{array}{ccc}
     0_{2\times2} &0_{2\times2}&0_{2\times2}\\
     0_{2\times2}& 0_{2\times2}& 0_{2\times2}\\
   0_{2\times2}&0_{2\times2} &h^\tau_U(\tilde{\phi})\\
    \end{array}\right),
\end{eqnarray} 
We note in passing that
\begin{eqnarray}
    H^\bK(\bk,\epsilon,\tilde{\phi}) = D^\dagger_{C_{3z}} H^\bK(C_{3z}\bk,\epsilon,\tilde{\phi}+\pi/3) D_{C_{3z}}\label{Appx:Eq:Strain_Symmetry}
\end{eqnarray}
where 
\begin{eqnarray}
    D(C_{3z}) = \left(\begin{array}{ccc}
        e^{i2\pi/3\sigma_z} & 0& 0\\
        0 &\sigma_0 &0\\
        0 & 0&e^{i2\pi/3\sigma_z} 
    \end{array}\right)
\end{eqnarray}.
We thus see that changing the orientation of the strain axis by $\pi/3$ is equivalent to rotating the coordinate frame by $-2\pi/3$. Fig(\ref{fig:Different_mag_orientation}) shows the band structure at CNP for different orientations and magnitude of heterostrain. 
\begin{figure}
    \centering
\includegraphics[width=0.95\linewidth]{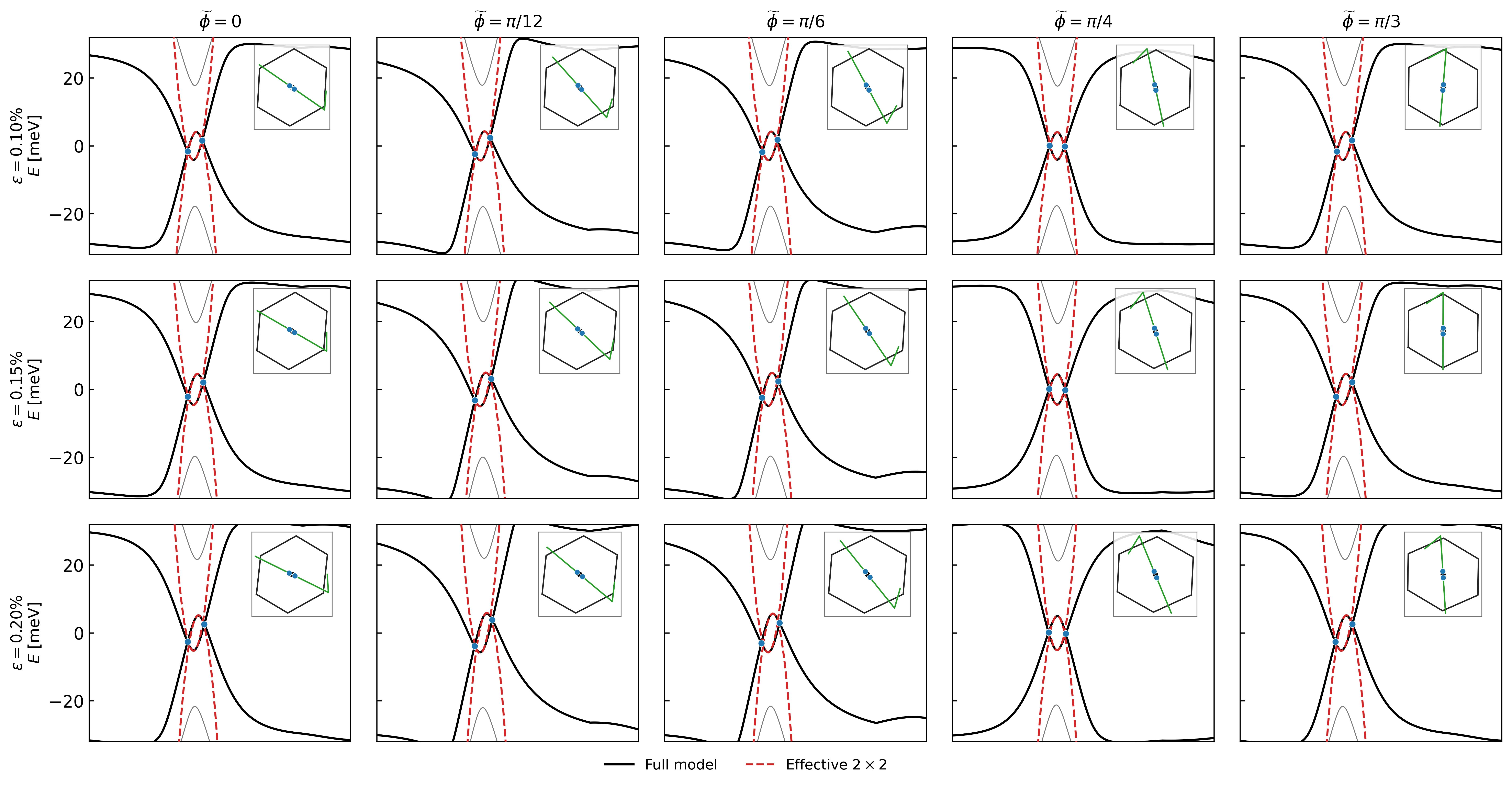}
    \caption{\textbf{Strain-induced anisotropic semimetal:} The comparison of the band structure obtained by diagonalizing the full one shot model and the effective model, given by Eqs.(\ref{Appx:Eq:Full_one_shot_model}) and (\ref{Appx:Eq:Effec_model}) respectively for different orientations and magnitude of heterostrain at $\tau=+1$. We  promote $\gamma,\gamma',c''\rightarrow e^{-\bk^2\lambda^2/2}\gamma,e^{-\bk^2\lambda^2/2}\gamma',e^{-\bk^2\lambda^2/2}c''$ before plotting.}
    \label{Supp:fig:EffectiveModel}
\end{figure}

%---------------------Section change----

\subsection{Effective model calculation}\label{Appx:Effective_model_CNP_oneshot}

In order to derive an effective model for the one-shot bands , we employ the SWT technique discussed in Ref\cite{singh2024topological}. Since we are interested in the low energy bands in vicinity of $\Gamma$, we treat $h^{0,\tau}(\bk)-h^{0,\tau}(0)$, $C_{3z}$ (and generically $C_{2x}$) breaking strain couplings $c,c',c''$ and the mass terms $M,M'$ as perturbations to the model ($\tau=+1$)
\begin{eqnarray}
 H_0(\tilde{\phi}) &=&  \left(\begin{array}{ccc}
     0_{2\times2} &0_{2\times2}&\gamma\sigma_0+i\gamma'\epsilon_b\sigma_3\\
     0_{2\times2}& 0_{2\times2}& 0_{2\times2}\\
   \gamma\sigma_0-i\gamma'\epsilon_b\sigma_3&0_{2\times2} &\frac{\tilde{U}_f}{2}\left(\begin{array}{cc}
           0  & ie^{-2 i\tilde{\phi}}\\
           -ie^{2  i\tilde{\phi}}  & 0
        \end{array}\right)\\
    \end{array}\right),
\end{eqnarray}
where $\tilde{U}_f=+M_f\epsilon(\nu_G+1)+U_1$ and $\epsilon_b=\frac{\epsilon}{2}(\nu_G-1)$.
The eigenvalues and eigenvectors of the above model are
\begin{align}
\lambda_1 &= 0, &
\mathbf v_1 &= (0,\,0,\,0,\,1,\,0,\,0)^T, \\[4pt]
\lambda_2 &= 0, &
\mathbf v_2 &= (0,\,0,\,1,\,0,\,0,\,0)^T, \\[6pt]
\lambda_3 &= \tfrac14\!\left(\tilde{U}_f - \Delta\right), &
\mathbf v_3 &= \frac{1}{\sqrt{2(1+\frac{|\gamma+i\gamma'\epsilon_b|^2}{\lambda_3^2})}}\biggl(
\frac{\ii e^{-2\ii\tilde{\phi}}(\gamma+i\gamma'\epsilon_b)}{\lambda_3},\,
\frac{\gamma-i\gamma'\epsilon_b}{\lambda_3},\,
0,\,0,\,
\ii e^{-2\ii\tilde{\phi}},\,
1
\biggr)^T, \\[6pt]
\lambda_4 &= \tfrac14\!\left(-\tilde{U}_f + \Delta\right), &
\mathbf v_4 &= \frac{1}{\sqrt{2(1+\frac{|\gamma+i\gamma'\epsilon_b|^2}{\lambda_4^2})}}\biggl(
-\frac{\ii e^{-2\ii\tilde{\phi}}(\gamma+i\gamma'\epsilon_b)}{\lambda_4},\,
\frac{\gamma-i\gamma'\epsilon_b}{\lambda_4},\,
0,\,0,\,
-\ii e^{-2\ii\tilde{\phi}},\,
1
\biggr)^T, \\[6pt]
\lambda_5 &= \tfrac14\!\left(-\tilde{U}_f - \Delta\right), &
\mathbf v_5 &= \frac{1}{\sqrt{2(1+\frac{|\gamma+i\gamma'\epsilon_b|^2}{\lambda_5^2})}}\biggl(
-\frac{\ii e^{-2\ii\tilde{\phi}}(\gamma+i\gamma'\epsilon_b)}{\lambda_5},\,
\frac{\gamma-i\gamma'\epsilon_b}{\lambda_5},\,
0,\,0,\,
-\ii e^{-2\ii\tilde{\phi}},\,
1
\biggr)^T, \\[6pt]
\lambda_6 &= \tfrac14\!\left(\tilde{U}_f + \Delta\right), &
\mathbf v_6 &= \frac{1}{\sqrt{2(1+\frac{|\gamma+i\gamma'\epsilon_b|^2}{\lambda_6^2})}}\biggl(
\frac{\ii e^{-2\ii\tilde{\phi}}(\gamma+i\gamma'\epsilon_b)}{\lambda_6},\,
\frac{\gamma-i\gamma'\epsilon_b}{\lambda_6},\,
0,\,0,\,
\ii e^{-2\ii\tilde{\phi}},\,
1
\biggr)^T.
\end{align}
where 
\begin{equation}
    \Delta = \sqrt{\tilde{U}_f^2 + 16(\gamma^2+\gamma'^2\epsilon_b^2)}\,.
\end{equation}
and the perturbations read
    \begin{eqnarray}
V(\bk,\tilde{\phi},\epsilon) &=& h^0_M  + \left(\begin{array}{ccc}
     0_{2\times2} &v_\ast (k_x\sigma_0+ik_y\sigma_z)&v'_\ast \left(k_x\sigma_x+k_y\sigma_y\right)\\
     v_\ast (k_x\sigma_0-ik_y\sigma_3)& 0_{2\times2}& 0_{2\times2}\\
   v'_\ast \left(k_x\sigma_x+k_y\sigma_y\right)&0_{2\times2} &0_{2\times2}\\
    \end{array}\right),\\
    &+& \left(\begin{array}{ccc} 
    c \epsilon_a\left(\begin{array}{cc}0&ie^{-2i\tilde{\phi}}\\-ie^{+2i\tilde{\phi}}&0\end{array}\right)
    &c' \epsilon_a\left(\begin{array}{cc}0&-ie^{2i\tilde{\phi}}\\ie^{-2i\tilde{\phi}}&0\end{array}\right)
    &0_{2\times 2}\\c' \epsilon_a\left(\begin{array}{cc}0&-ie^{2i\tilde{\phi}}\\ie^{-2i\tilde{\phi}}&0\end{array}\right)& M'\epsilon_b\sigma_y&c''\epsilon_a\left(\begin{array}{cc}
        +ie^{-2i\tilde{\phi}} &0  \\
        0 & -ie^{2i\tilde{\phi}}
    \end{array}\right) \\0_{2\times 2}&c''\epsilon_a\left(\begin{array}{cc}
        -ie^{2i\tilde{\phi}} &0  \\
        0 & +ie^{-2i\tilde{\phi}}
    \end{array}\right)
    & 0_{2\times 2}
      \end{array}\right)
\end{eqnarray}
where $\epsilon_a=\epsilon(\nu_G+1)/2$.
We want to project the model onto the Hilbert space spanned by vectors $\bf{v}_1$ and $\bf{v}_2$. Upto $\mathcal{O}(\bk^2)$, it is given by
\begin{eqnarray}
    H^{eff}_{ij} = \langle \textbf{v}_i|H_0(\tilde{\phi})|\textbf{v}_i \rangle \delta_{ij} + \langle v_i|V(\bk,\tilde{\phi},\epsilon)|\textbf{v}_j \rangle - \sum_{l=3}^6 \frac{\langle \textbf{v}_i|V(\bk,\tilde{\phi},\epsilon)|\textbf{v}_l\rangle \langle \textbf{v}_l|V(\bk,\tilde{\phi},\epsilon) | \textbf{v}_j\rangle}{\lambda_l }
\end{eqnarray}
with $i,j\in\{1,2\}$. In matrix form
%\begin{eqnarray}
%    H^{eff}_{ij}(\bk,\tilde{\phi}) = \left(\begin{array}{cc}
%       A(\tilde{\phi}) k +A^\ast(\tilde{\phi})\bar{k} &  F_0(\tilde{\phi}) + F_2(\tilde{\phi})k^2\\
%       F^\ast_0(\tilde{\phi}) + F^\ast_2(\tilde{\phi})\bar{k}^2  & A(\tilde{\phi}) k +A^\ast(\tilde{\phi})\bar{k}\end{array}\right)\label{Appx:Eq:Effec_model}
%\end{eqnarray}
%where $A(\tilde{\phi})= iA_1e^{-2i\tilde{\phi}} + A_2e^{4i\tilde{\phi}}$, $F_0(\tilde{\phi})= M_2+iM_1-i Ce^{-6i\tilde{\phi}}$, $F_2(\tilde{\phi})= -iDe^{2i\tilde{\phi}}$, and the constants $A_1,A_2,M_1,M_2,C,D>0$. The model at $\tau=-1$ is related by time reversal. The symmetry of the full model with respect to $\tilde{\phi}$ given in Eq.(\ref{Appx:Eq:Strain_Symmetry}) is obeyed by the effective model as well because
%$H^{eff}_{ij}(\bk,\tilde{\phi})= H^{eff}_{ij}(C_{3z}\bk,\tilde{\phi}+\pi/3)$. The values of the constants are $A_1=507.147$ $\mathrm{meV.\mathring{A}}$, $ A_2=361.295$ $\mathrm{meV.\mathring{A}}$, $M_1=2.885$ $meV$, $M_2= 3.279$ $meV$, $ C=0.150$ $meV$ and $D=871077.897$ $\mathrm{meV.\mathring{A}^2}$ at magic angle for the BM model. The comparison of the band structure calculated using the effective model with the full model is shown in Fig.(\ref{Supp:fig:EffectiveModel}) at $\tilde{\phi}=0$ and $\tau=+1$.
\begin{eqnarray}
    H^{\bK,\tilde{\phi}}_{eff}(\bk)=\left(\alpha_{\tilde{\phi} }k+\alpha^\ast_{\tilde{\phi}}\bar{k}\right)\sigma_0 + M\sigma_x-M'\epsilon_+\sigma_y+\left(\begin{array}{cc}
 0    &  g_{\tilde{\phi}}(\bk)\\
  g_{\tilde{\phi}}^\ast(\bk)   & 0
\end{array}\right)
\end{eqnarray}
where 
\begin{eqnarray}
   \alpha_{\tilde{\phi}} &=& -\frac{\epsilon_a v_\ast\tilde{\gamma}}{2|\tilde{\gamma}|^4}\left(\tilde{U}_fc'\tilde{\gamma}e^{+4i\tilde{\phi}}+2ic''|\tilde{\gamma}|^2 e^{-2i\tilde{\phi}}\right)\\
    g_{\tilde{\phi}}(\bk) &=& -\frac{\tilde{U}_f}{2}\left(\frac{v_\ast^2e^{2i\tilde{\phi}}k^2}{\tilde{\gamma}^{\ast 2}}+\frac{c'^2\epsilon_a^2 e^{-6i\tilde{\phi}}}{\tilde{\gamma}^2}\right) - \frac{2c'c''\epsilon_a^2}{\tilde{\gamma}}\label{Appx:Eq:Effec_model}
\end{eqnarray}
where $\tilde{\gamma}=\gamma-i\epsilon_b\gamma'$. The symmetry of the full model with respect to $\tilde{\phi}$ given in Eq.(\ref{Appx:Eq:Strain_Symmetry}) is obeyed by the effective model as well because
$H^{eff}_{ij}(\bk,\tilde{\phi})= H^{eff}_{ij}(C_{3z}\bk,\tilde{\phi}+\pi/3)$. The comparison of the band structure calculated using the effective model with the full one-shot model is shown in Fig.(\ref{Supp:fig:EffectiveModel})for $\tau=+1$.

%---------------------Section change----

\section{Prescription for semiclassical quantization}\label{Appx:Semiclassical:Prescription}
In this section, we discuss the prescription for obtaining the low $\bB$ LLs semiclassically. According to the Onsager quantization rule\cite{onsager1952london,lifschitz1954theory,lifshitz1956theory}
\begin{eqnarray}
    A(E_{n+1})-A(E_n) = \frac{2\pi e}{\hbar c} B
\end{eqnarray}
where $A(E)$ is the $\bk$-space area enclosed by the constant energy contour of energy $E$ at $\bB=0$.  In other words, exactly $\phi/\phi_0$ number of states at $\bB=0$ then form one LL after Landau quantization  
\begin{eqnarray}
    &&\frac{\Delta A(E)}{A_{BZ}} = \frac{\phi}{\phi_0}\\
    \implies && A(E_n)=A_{BZ}(n+\gamma_M) \frac{\phi}{\phi_0}
\end{eqnarray}
In our Figures, we use $\gamma_M=0$ and $1/2$ for Dirac-like and Band maxima/minima like Fermi surfaces respectively.

%---------------------Section change----

\section{Robustness of the features for the strain-induced anisotropic semimetal in heterostrain parameters}\label{Appx:Robust_Strain_Semimetal}
\begin{figure}
    \centering
    \includegraphics[width=0.990\linewidth]{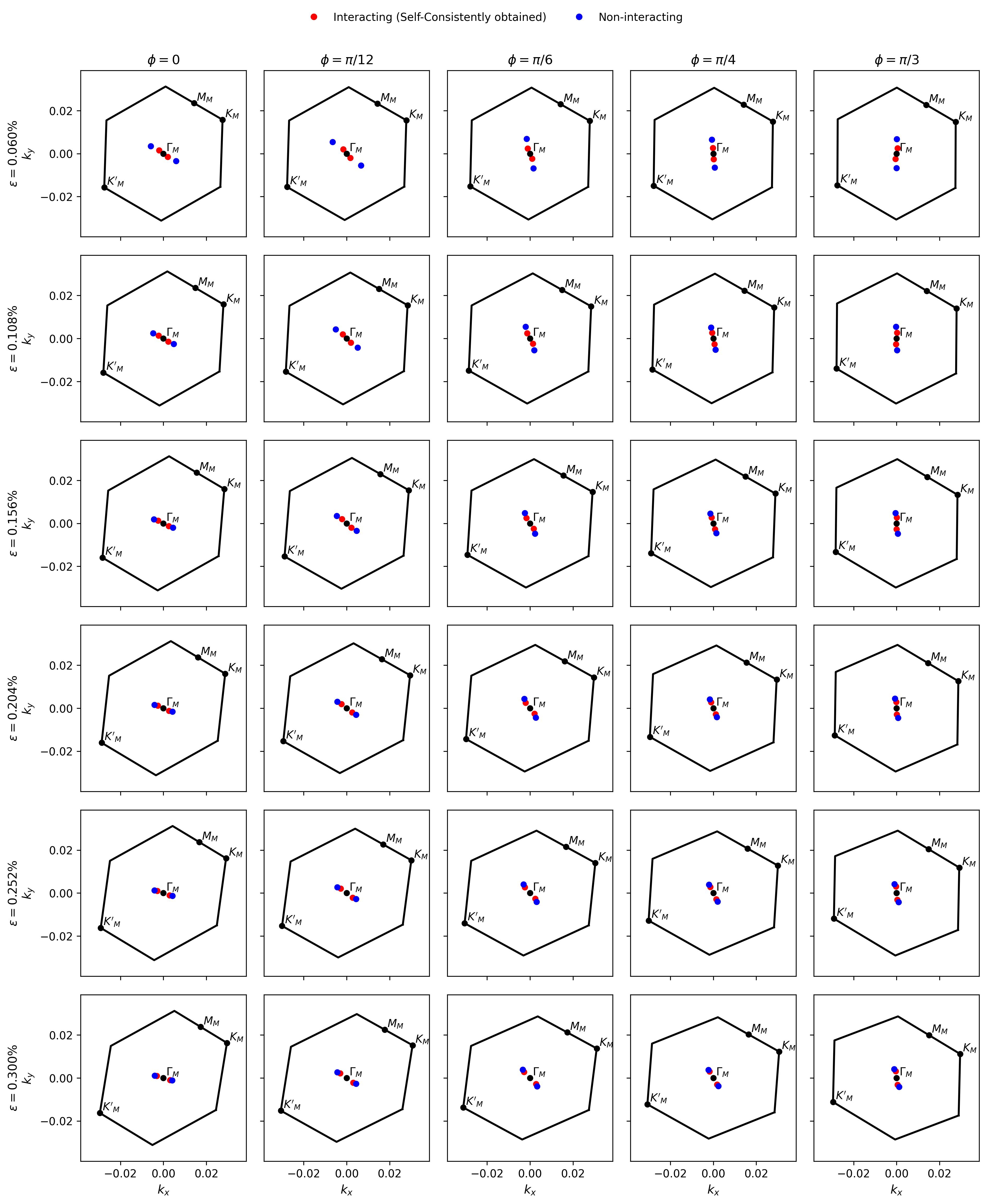}
    \caption{\textbf{Strain-induced anisotropic semimetal:} The location of Dirac nodes for the \textcolor{red}{interacting} (using self consistent Hartree Fock) and the \textcolor{blue}{non-interacting} model for different magnitudes $\epsilon$ and orientation $\phi$ of heterostrain at magic angle and $w_0/w_1=0.8$ at valley $\bK$.}
\label{fig:Different_mag_orientation_nodes}
\end{figure}

\begin{figure}
    \centering
    \includegraphics[width=0.990\linewidth]{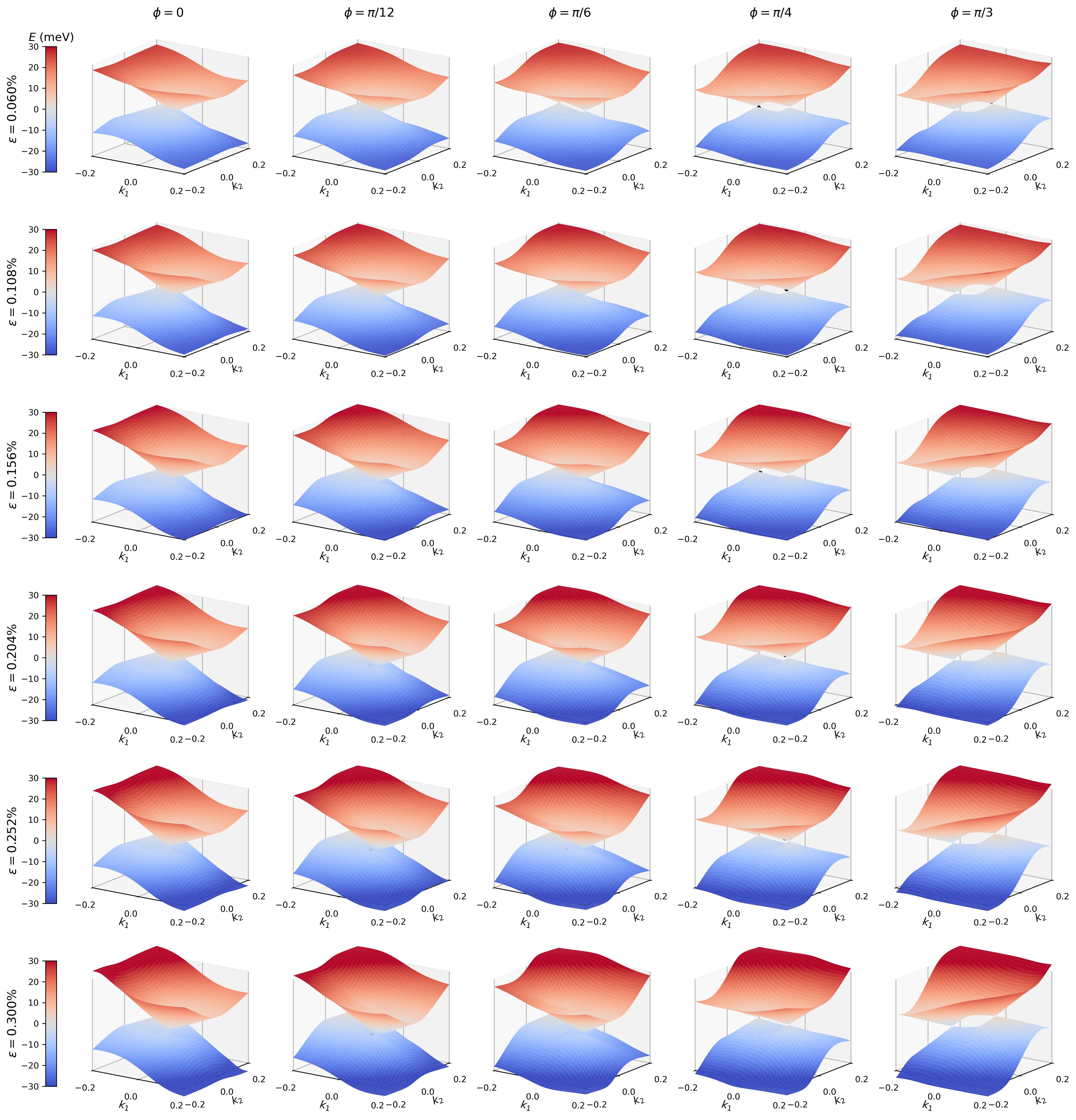}
    \caption{\textbf{Strain-induced anisotropic semimetal:} The surface plots for the bands carrying Dirac nodes for different magnitudes $\epsilon$ and orientation $\phi$ of heterostrain at magic angle and $w_0/w_1=0.8$ at valley $\bK$ (along $k_1\tilde{\bg}_1+k_2\tilde{\bg}_2$), obtained through self-consistent Hartree Fock.}
\label{fig:Different_mag_orientation}\end{figure}

\begin{figure}
    \centering
\includegraphics[width=0.990\linewidth]{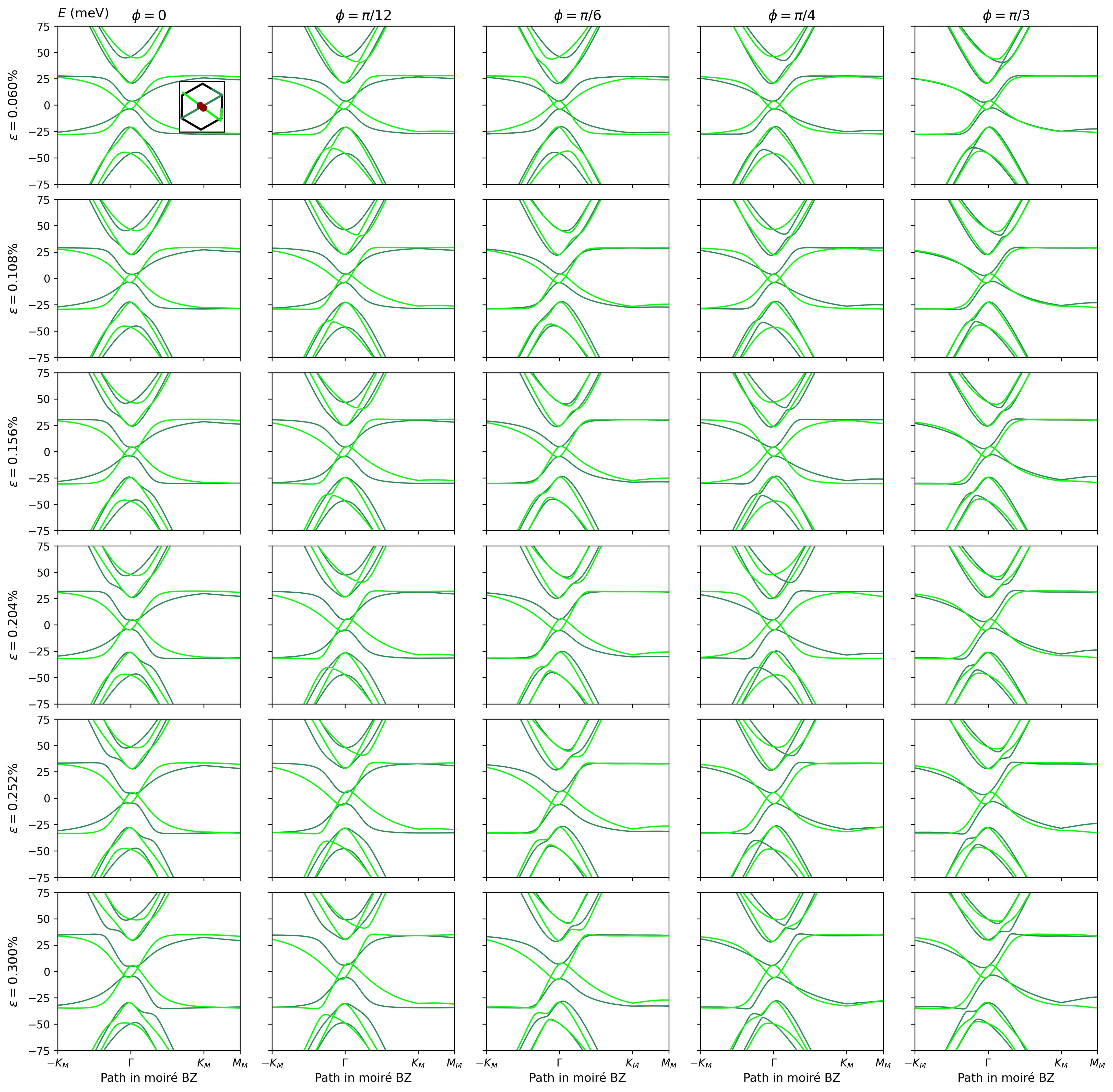}
    \caption{\textbf{Strain-induced anisotropic semimetal:} The line cuts along high symmetry points and along Dirac nodes (see inset of top left panel) for the bands carrying Dirac nodes for different magnitudes $\epsilon$ and orientation $\phi$ of heterostrain at magic angle and $w_0/w_1=0.8$ at valley $\bK$, obtained using self-consistent Hartree Fock. }
\label{fig:Different_mag_orientation_linecuts}\end{figure}

\begin{figure}
    \centering
    \includegraphics[width=0.95\linewidth]{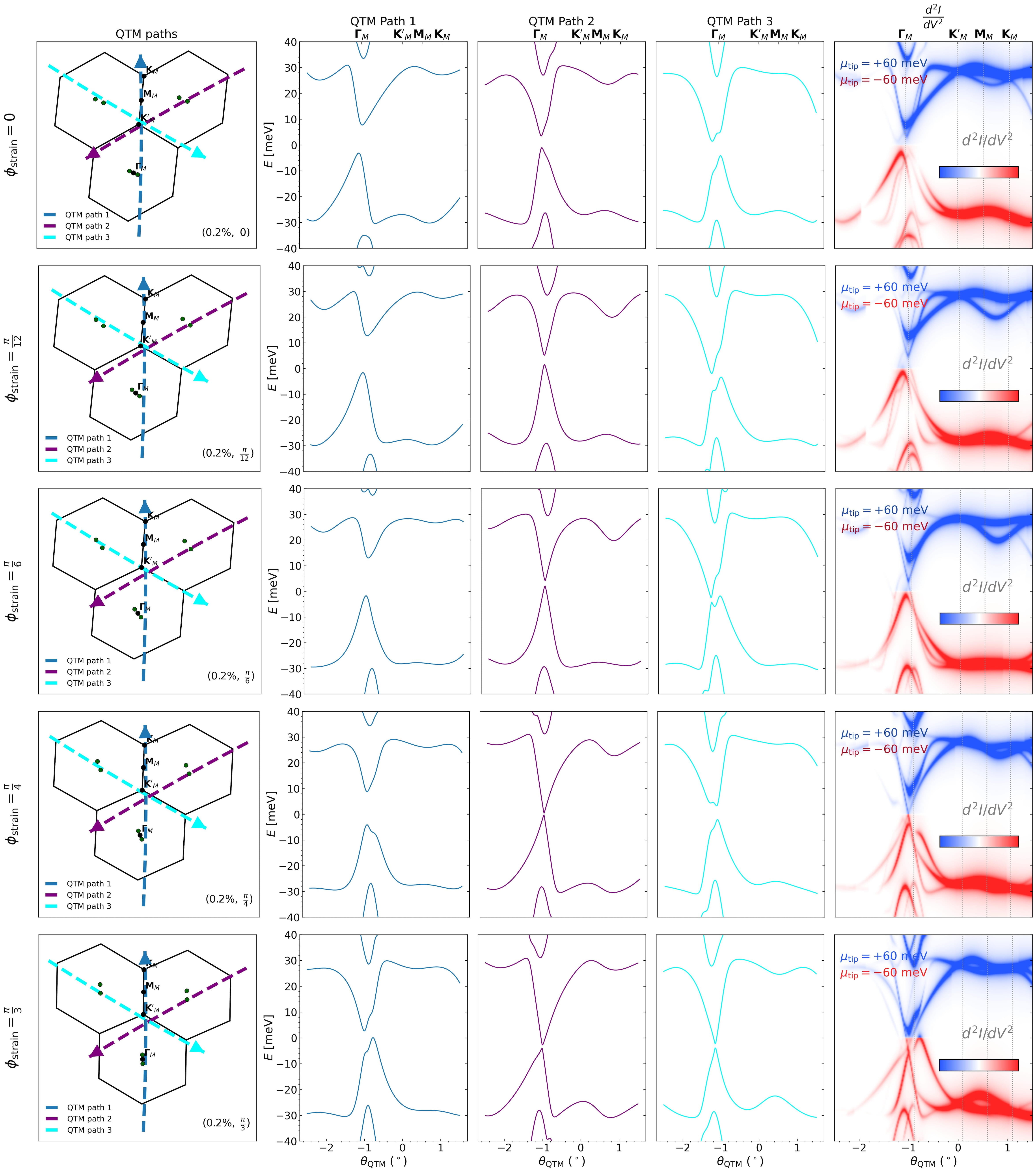}
    \caption{\textbf{Strain-induced anisotropic semimetal:} The leftmost panel of each row shows the paths in reciprocal space scanned by the quantum twisting microscope (QTM) for different strain orientations (mentioned in the bottom right of these panels) for a magnitude of  0.2\% , as a function of the relative angle between the top graphene layer of the MATBG sample and the monolayer graphene (MLG) tip, labeled by $\theta_{QTM}(^\circ)$\cite{wei2025dirac}. We mark the high symmetry points that comes closest to the QTM path 1 as $\theta_{QTM}$ varies (these points are also marked in panel 1 for each row). Different rows correspond to different strain orientations ($\phi_{strain}$). The panels 2,3 and 4 in each row shows the band structure along the three QTM paths, while the rightmost panel shows the calculated $d^2I/dV^2$ map\cite{wei2025dirac} at $T=4$K. We choose $\mu_{\rm sample}=0$ and thus  probing sample bands with negative(positive) energy requires hole(particle) doped MLG tip ($\mu_{\rm tip}=\mp 60 ~\text{meV}$). The $d^2I/dV^2>0$  (in red) for $\mu_{\rm tip}=-60$ meV while $d^2I/dV^2<0$  (in blue) for $\mu_{\rm tip}=+60$ meV, respectively. The corresponding bias between the layer and sample is given by $-eV = E+\mu_{\rm tip}$\cite{wei2025dirac}, for band energy $E$ (the $y$-axis in panels 2,3 and 4). The Dirac points are labeled in dark green and marked in panel 1 of each row. }
    \label{fig:QTM_0.2percent}
\end{figure}

\begin{figure}
    \centering
\includegraphics[width=0.95\linewidth]{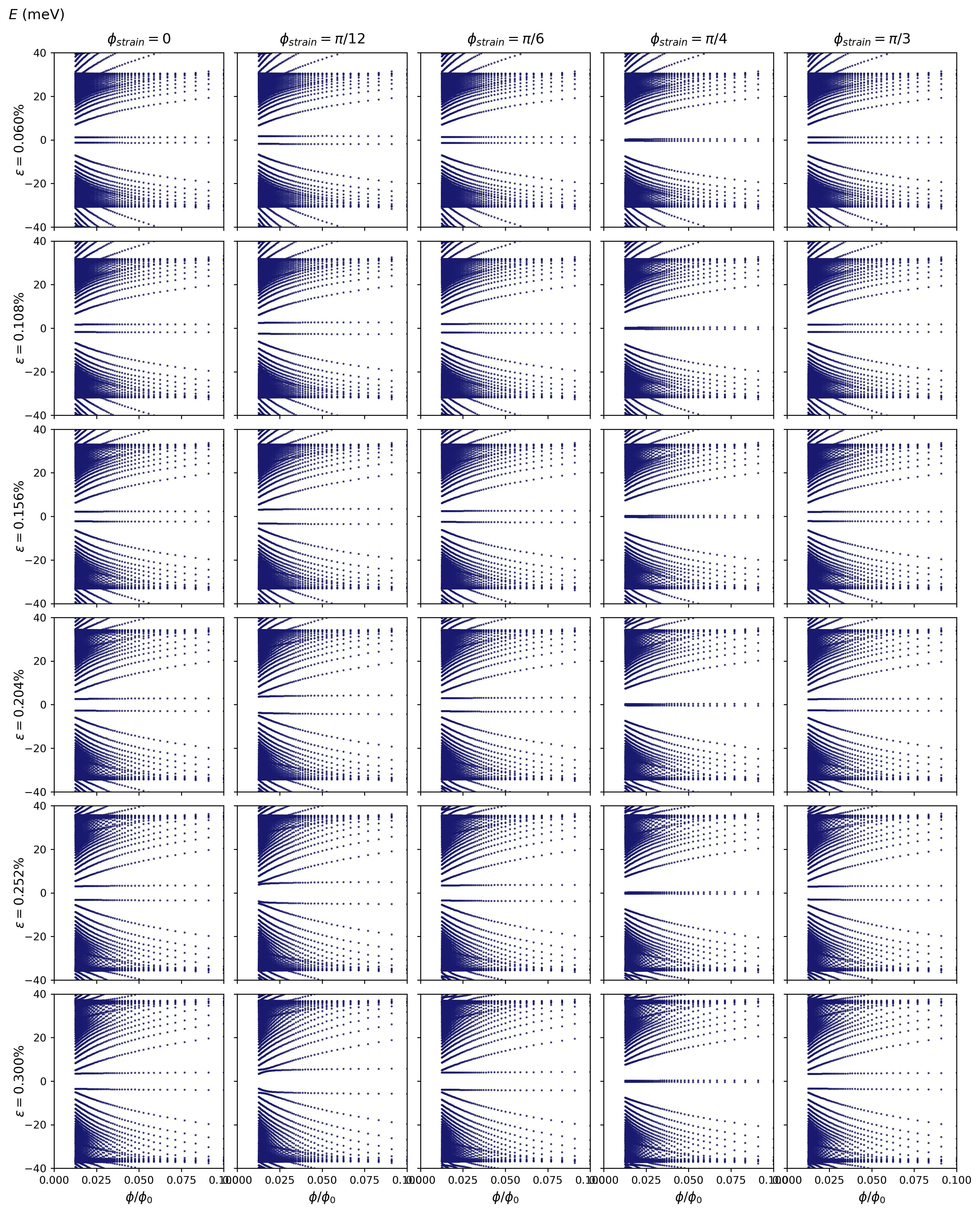}
    \caption{\textbf{Strain-induced anisotropic semimetal:}The Hofstadter spectrum for the strain-induced anisotropic semimetal for different magnitudes $\epsilon$ and orientation $\phi_{strain}$ of heterostrain at magic angle and $w_0/w_1=0.8$. As discussed in the main text, the Chern $C=0$ gap, set by difference between the Dirac node energies,  is sensitive to the orientation, and almost vanishes near $\phi_{strain}\sim \pi/4$.}
    \label{fig:Robust LL spectrum}
\end{figure}

\begin{figure}
    \centering
    \includegraphics[width=0.95\linewidth]{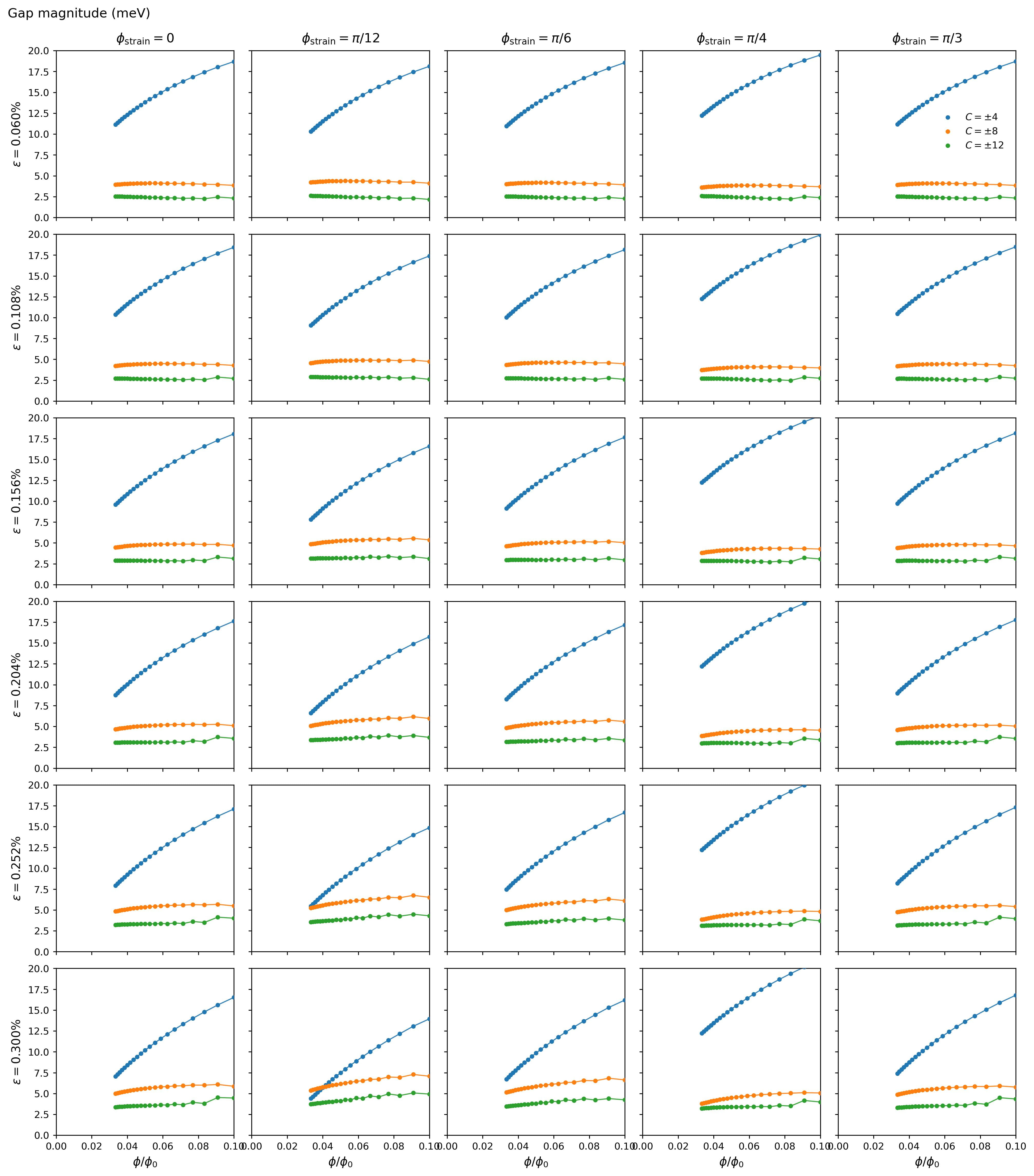}
    \caption{\textbf{Strain-induced anisotropic semimetal:} The magnitude of Chern $\pm 4,\pm 8,\pm 12$ gaps as a function of $\frac{\phi}{\phi_0}$ for different magnitudes $\epsilon$ and orientation $\phi_{strain}$ of heterostrain at magic angle and $w_0/w_1=0.8$. The hierarchy of the magnitudes of gaps is robust in $\bB$: Gap with  $C=\pm 4>C=\pm 8>C=\pm 12$. The only exception to this is at extremely low $\bB$ for large $\epsilon =0.3\%$ and $\phi_{strain}=\pi/12$, where the $C=\pm 4$ gets suppressed (also see analogous spectrum in Fig.(\ref{fig:Robust LL spectrum})). Since the gap between the Dirac nodes at $\bB=0\sim 2E_D(\pi/12)$ is large (which sets the $C=0$ gap), we have a crossing between the first $+\sqrt{B}$ LL emanating out of the lower Dirac node (at -$E_D(\pi/12)$) and the zero mode emanating out of the Dirac node at $+E_D(\pi/12)$. Such crossing occurs for all other considered $(\epsilon,\phi_{strain})$ combinations as well but at more smaller $\phi/\phi_0$.  However note that $C=\pm 8$ is still larger than $C=\pm 12$ gap within the considered $\phi/\phi_0$ range. }
    \label{fig:Robust_heirarchy}
\end{figure}

\begin{figure}
    \centering
    \includegraphics[width=0.95\linewidth]{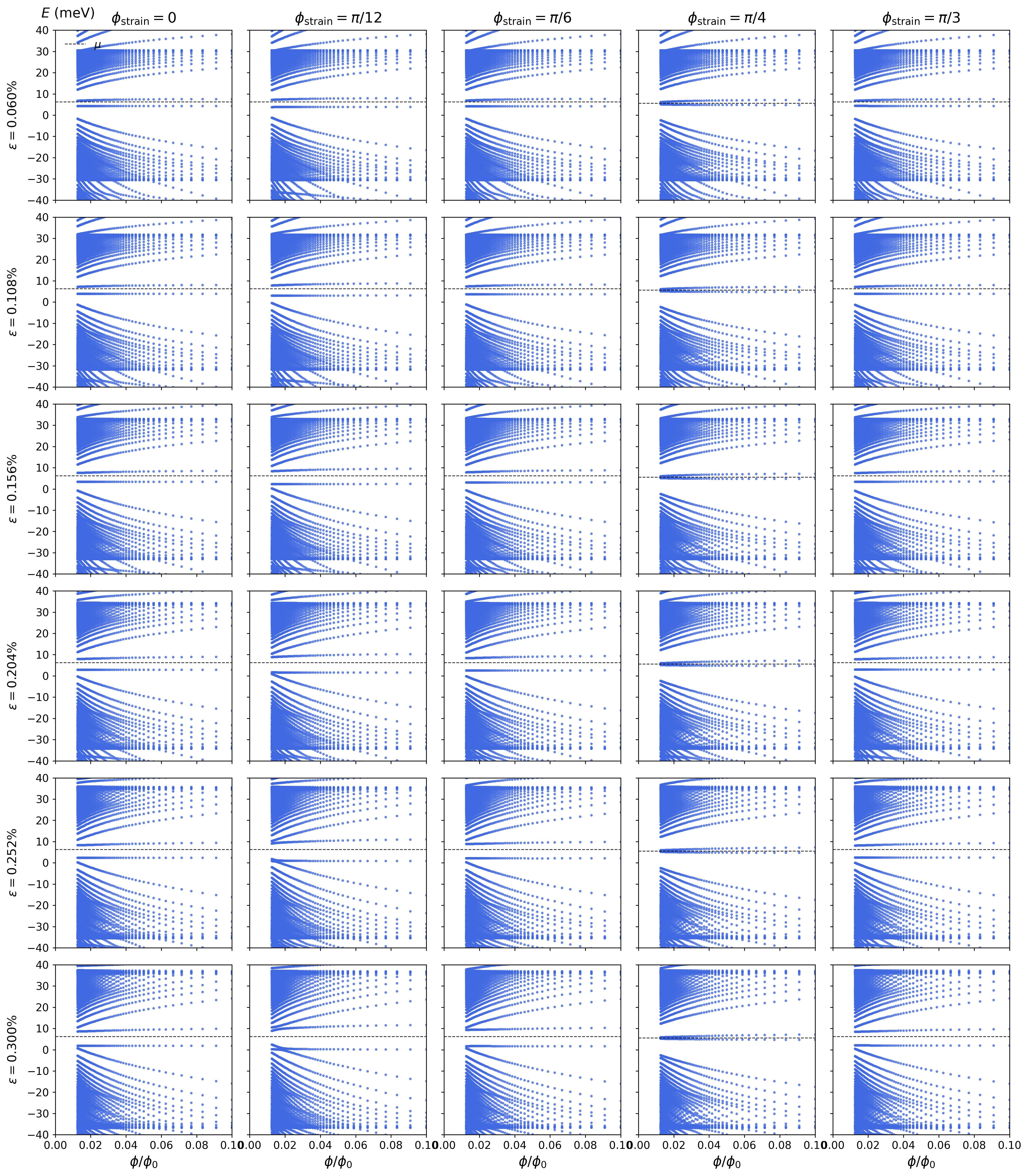}
    \caption{\textbf{Strain-induced anisotropic semimetal:}The Hofstadter spectrum with relaxation effects at different $(\epsilon,\phi_{strain})$ at magic angle and $w_0/w_1=0.8$. It exhibits qualitatively similar spectrum as that in absence of relaxation. The chemical potential $\mu$ in dashed line is chosen to lie in the Chern $0$ gap.}
    \label{fig:Strain_and_relaxation}
\end{figure}

\begin{figure}
    \centering
    \includegraphics[width=0.95\linewidth]{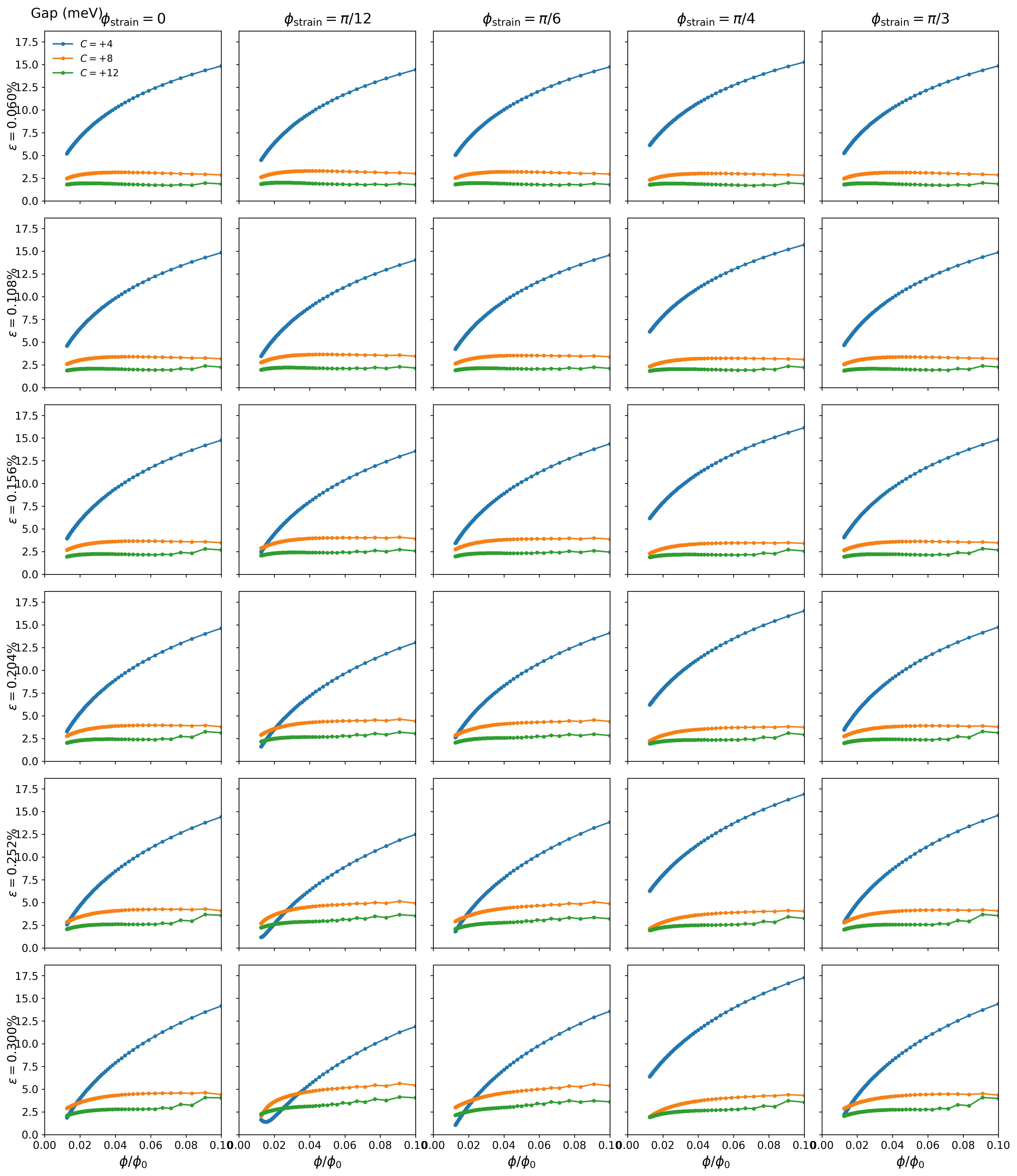}
    \caption{\textbf{Strain-induced anisotropic semimetal:}The magnitude of Chern $C=4,8$ and $12$ (see Fig.(\ref{fig:Strain_and_relaxation})) as a function of $\phi/\phi_0$ at different $(\epsilon,\phi_{strain})$ at magic angle and $w_0/w_1=0.8$, with effects of lattice relaxation included. We find the Chern gap hierarchy $\Delta_{C=8}>\Delta_{C=12}$ robust even in presence of relaxation.}
\label{fig:Strain_and_relaxation_Chern_plus}
\end{figure}

\begin{figure}
    \centering
    \includegraphics[width=0.95\linewidth]{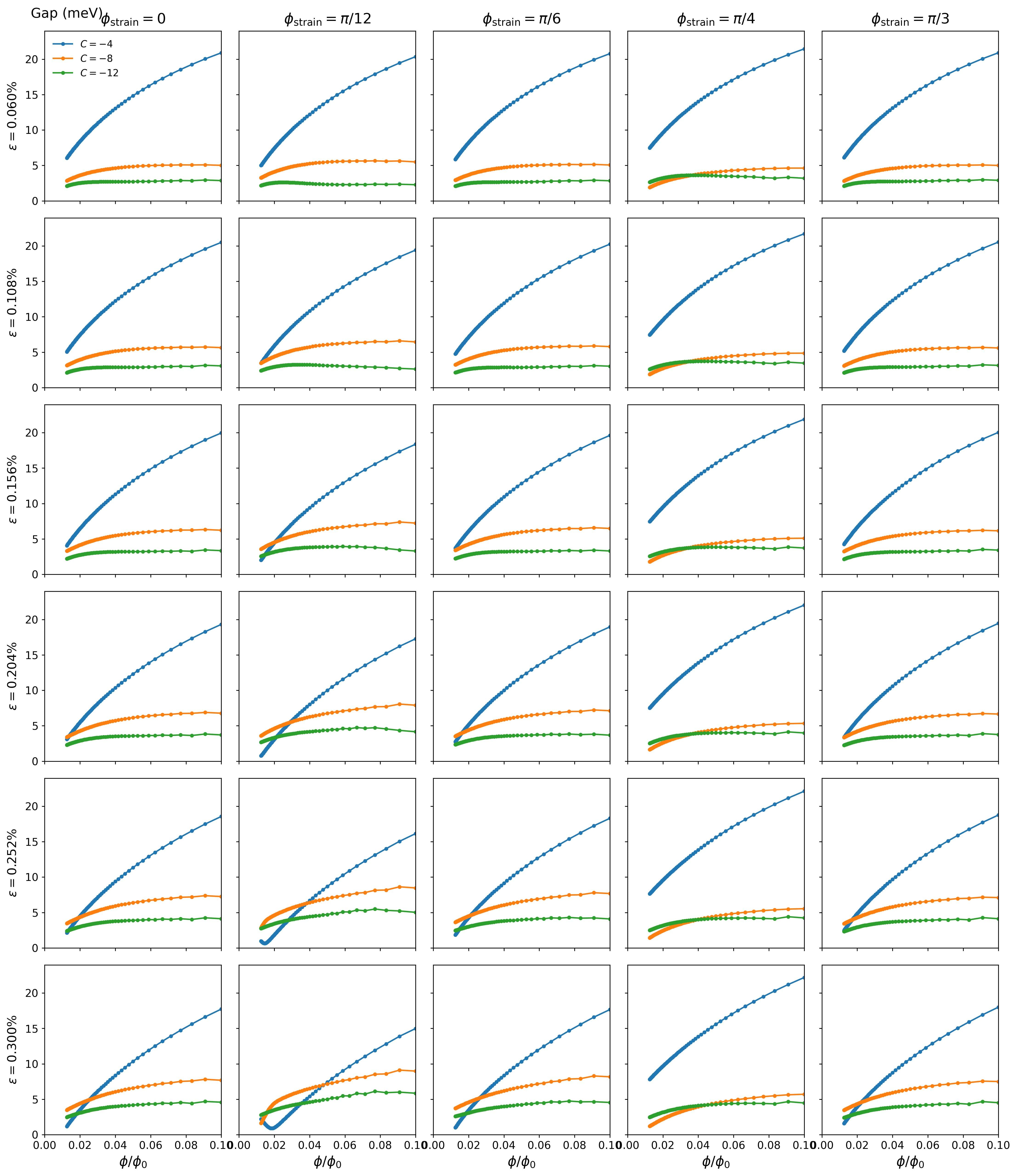}
    \caption{\textbf{Strain-induced anisotropic semimetal:} The magnitude of Chern $C=-4,-8$ and $-12$ (see Fig.(\ref{fig:Strain_and_relaxation})) as a function of $\phi/\phi_0$ at different $(\epsilon,\phi_{strain})$ at magic angle and $w_0/w_1=0.8$, with effects of relaxation effects included. We find the Chern gap hierarchy $\Delta_{C=-8}>\Delta_{C=-12}$ robust even in presence of relaxation, except for $\phi_{strain}\sim \pi/4$ when $\phi/\phi_0<0.4$.}
\label{fig:Strain_and_relaxation_Chern_minus}
\end{figure}

In this section, we highlight the robustness of the features for the strain-induced anisotropic semimetal across broad range of magnitude and orientation of the uniaxial heterostrain. The investigated range is $\epsilon\in\{0.060\%,0.108\%,0.156\%, 0.204\%, 0.252\%, 0.300\% \}$ and $\phi_{strain}\in\{0, \pi/12,\pi/6,\pi/4,\pi/3\}$.
\begin{itemize}
\item \textbf{Location of Dirac nodes} The Fig.(\ref{fig:Different_mag_orientation_nodes}) shows the location of Dirac nodes for the strain-induced anisotropic semimetal in red dots. This highlights that the dip of the bands near $\Gamma$ is always expected, because of the vicinity of these nodes to $\Gamma$.  
\item \textbf{Surface Plots for the bands carrying Dirac nodes}
The surface plot for the bands carrying Dirac nodes at valley $\bK$ along $k_1\tilde{\bg}_1+k_2\tilde{\bg}_2$ is shown in Fig.(\ref{fig:Different_mag_orientation}). We note that the nodes are always within $k_{1,2}\in(-0.2,+0.2)$ in the moir\'e BZ, or in other words, always close to $\Gamma$.
\item \textbf{Line cuts for the bands} The line cuts for the bands along a line through the high symmetry points (of the unstrained problem) is shown in dark green in Fig.(\ref{fig:Different_mag_orientation_linecuts}). The bands along the rotated path, which goes through the Dirac nodes of the semimetal is shown in bright green simultaneously. Clearly, the bands are flat everywhere in the moir\'e BZ except in the vicinity of $\Gamma$. 
\item \textbf{QTM paths and $d^2I/dV^2$ simulations} The paths (or momentum arcs) in reciprocal space scanned by the QTM experiment, the band structures along this path and the resulting singularities in $d^2I/dV^2$ for different orientations of strain at $0.2\%$ magnitude is presented in Fig.(\ref{fig:QTM_0.2percent}) using the results derived in \cite{wei2025dirac}. More specifically, we use 
\begin{eqnarray}
   & \frac{d^2 I}{dV^2}(\theta_{QTM})\Big|_{\rm sing}
\;\propto\;
\Big[f(-eV-\mu_T)-f(-\mu_T)\Big]\;\sum_{n=1}^{3}\sum_{\tilde{\lambda}}
\Theta\!\big(v_D-v_{ n\tilde{\lambda}}(\theta)\big)\,\Big(1-\frac{v_{ n\tilde{\lambda}}^2(\theta_{QTM})}{v_D^2}\Big)^{-3/2}
\nonumber\\
&\times
\sum_{\chi=0,1,2}
\Big[1+\frac{v_{ n\tilde{\lambda}}(\theta_{QTM})}{v_D}\cos\!\Big(\theta_{ n\tilde{\lambda}}(\theta_{QTM})+\frac{2\pi\chi}{3}\Big)\Big]\;
\mathcal{W}_{ n\tilde{\lambda}\chi}(\theta)\;
\delta\!\Big(eV+\mu_T+E_{n\tilde{\lambda}}(\theta_{QTM})\Big).
\end{eqnarray}
where 
$f(\varepsilon)=\big(e^{\varepsilon/(k_BT)}+1\big)^{-1}$ is the Fermi function (we use $T=4K$ in the plots), $\theta_{QTM}(\circ)$ is the angle between the probing monolayer graphene tip and top layer of the sample, $\Theta$ is the Heaviside step
function, $\tilde{\lambda}$ runs over the band index, n=$\{1,2,3\}$ labels the three path scanned by QTM in reciprocal space (shown in the left most panel of each row in Fig.(\ref{fig:QTM_0.2percent})), $v_D$ is Dirac velocity, $\textbf{v}_{n\tilde{\lambda}}(\theta_{QTM})$ (and $v_{n\tilde{\lambda}}(\theta_{QTM}) = |\textbf{v}_{n\tilde{\lambda}}(\theta_{QTM})|$ )is the local group velocity of the sample band $\lambda$ at the probed momentum, $\theta_{n\tilde{\lambda}}$ is the angle between $\textbf{v}_{ n\lambda}(\theta_{QTM})$
and the corresponding tip Dirac wave-vector direction, $E_{n\tilde{\lambda}}(\theta_{QTM})$ is the energy of sample band $\tilde{\lambda}$ at the probed momentum,  and the interference 
\begin{eqnarray}
  \mathcal{W}_{ n\tilde{\lambda}\chi}(\theta_{QTM})   \equiv \left|
e^{i\frac{2\pi\tau(\chi+n-1)}{3}}\,\psi^{\tilde{\lambda},n}_{tA}\!\big(\theta_{QTM}\big)
+
\psi^{\tilde{\lambda},n}_{tB}(\theta_{QTM})
\right|^2.
\end{eqnarray}
where $\psi^{\tilde{\lambda}n}_{tA}$ and $\psi^{\tilde{\lambda}n}_{tB}$ is the top layer (i.e. the one rotated anti-clockwise \cite{wei2025dirac}) sublattice A/B projected wavefunctions corresponding to the sample band $\tilde{\lambda}$ at the probed momentum. The plots features high energy flat bands everywhere along the QTM path except near the QTM's $\theta_{QTM}=-1^\circ$ (which is closest $\Gamma$) where the bands dip, consistent with Ref\cite{inbar2023quantum}. 
\item \textbf{Hofstadter spectrum }
The Hofstadter spectra for the strain-induced anisotropic semimetal are shown in Fig.(\ref{fig:Robust LL spectrum}). This figure shows the robustness of the spectrum in heterostrain parameters, except for the $\phi_{strain}$ sensitive energies of anomalous modes discussed in main text. 
\item \textbf{Chern Gaps}
The finite $\bB$ induced Chern gaps, $\Delta_C,$ are presented in Fig(\ref{fig:Robust_heirarchy}). We highlight that Chern gap magnitude follow the robust hierarchy of $\Delta_{C=\pm 4}>\Delta_{C=\pm 8}>\Delta_{C=\pm 12}$. The only exception within the  flux range considered is for $(\epsilon,\phi_{strain}) = (0.3\%, \pi/12)$, which occurs at extremely small $\phi/\phi_0<0.025$, where the $C=\pm 4$ gap gets suppressed. At such a small $\bB$, the spectrum can be interpreted as Landau quantization of the two Dirac nodes of the strain-induced anisotropic semimetal (see main-text). The suppression occurs from the fact that we have a crossing between the first $+\sqrt{B}$ LL emanating out of the lower Dirac node (at -$E_D(\phi_{strain})$) and the zero mode emanating out of the Dirac node at $+E_D(\phi_{strain})$ (see Fig.(\ref{fig:Robust LL spectrum})).    
%(\ref{fig:Robust_heirarchy}))(\ref{fig:Robust LL spectrum})
\item \textbf{Relaxation} The Hofstadter spectra including relaxation effects for different orientation  $\phi_{strain}$ and magnitude $\epsilon$ of heterostrain is shown in Fig.(\ref{fig:Strain_and_relaxation}). It highlights that including relaxation in our analysis offers qualitatively the same spectrum as that without relaxation. The Figs.(\ref{fig:Strain_and_relaxation_Chern_plus}) and (\ref{fig:Strain_and_relaxation_Chern_minus}) show the magnitudes of the Chern $\pm 4,\pm 8$ and $\pm 12$ gaps. We find the hierarchy of gaps with $C=4>8>12$ maintained (see Fig.(\ref{fig:Strain_and_relaxation_Chern_plus})), except at very low $\phi/\phi_0\lesssim 0.03 $ for $\epsilon= 0.252\%$ and $0.3\%$, where the magnitude of $C=4$ gap becomes smaller than $8,12$. Similarly, the hierarchy of gaps with $C=-4>-8>-12$ maintained (see Fig.(\ref{fig:Strain_and_relaxation_Chern_minus})), except at  low $\phi/\phi_0\lesssim 0.04$, where $C=-4$ gap can get smaller than $C=-8,-12$, and $C=-12$ can get larger than $C=-8$ for $\phi_{strain}\sim\pi/4$. The suppression of $C=\pm4$ gap occurs from the fact that we have a crossing between the first $\pm\sqrt{B}$ LL emanating out of the lower Dirac node (at $\mp E_D(\phi_{strain})$) and the zero mode emanating out of the Dirac node at $\pm E_D(\phi_{strain})$

\end{itemize}

%\begin{figure}
%    \centering
%    \includegraphics[width=0.85\linewidth]{Figures/Strain_relaxation_5panel.jpg}
%    \caption{The Hofstadter spectrum with relaxation effects at $\epsilon=0.2\%$ at different $\phi_{strain}$. It exhibits qualitatively similar spectrum as that in absence of relaxation.}
%    \label{fig:Strain_and_relaxation}
%\end{figure}

%------------Section change--------------

\begin{figure}
    \centering
    \includegraphics[width=0.95\linewidth]{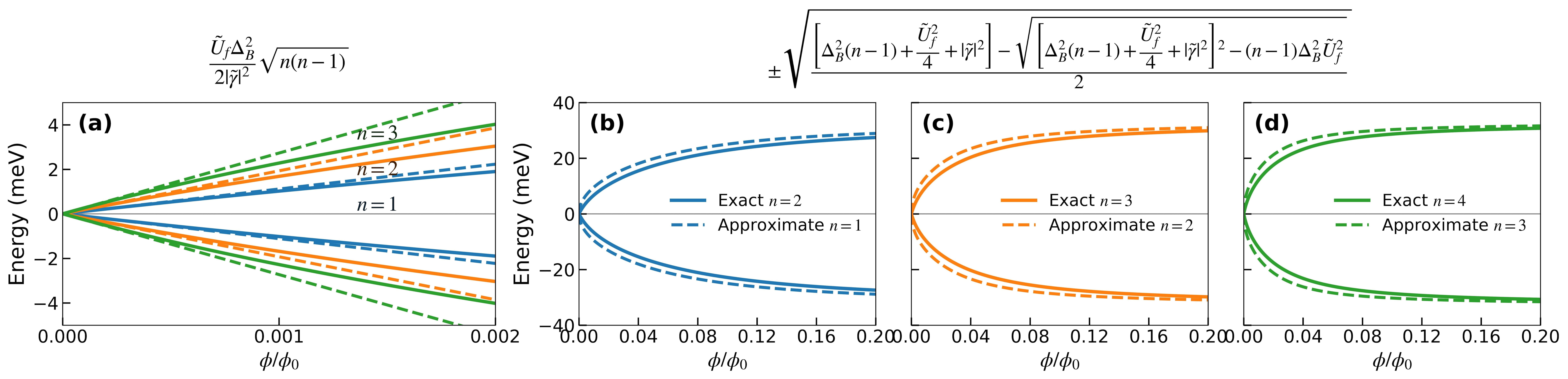}
    \caption{\textbf{Strain-induced anisotropic semimetal:} The match between the approximate expressions for the lowest lying excitations closest to CNP for the strain-induced anisotropic semimetal for the simple model in Eq.(\ref{Eq:appx:h6:SImple_strain}) obtained assuming (a)$\Delta_B\ll|\tilde{\gamma}|$ and (b) $\Delta_B\gg|\tilde{\gamma}|$. We set $\Sigma_{n}=1$.}
    \label{fig:appx:Leading:Strain}
\end{figure}
\section{Flat-chiral limit for the strain-induced anisotropic semimetal}\label{Appx:Simple:model:Strain}
In this section we discuss the flat-chiral limit $M=M'=v_\ast'=0$ of the strain-induced anisotropic semimetal discussed in the main text. We additionally drop the $C_{3z}$ (and generically $C_{2x}$) breaking strain couplings $c,c',c''$. The couplings $c',c''$ are mainly responsible for splitting the quadratic band touching at $\Gamma$ into two Dirac nodes (low energy effect $\sim$ ($-4$meV,$+4$meV) at $\epsilon=0.15\%$ ) in the flat limit. The $\bB=0$ model in this limit is given by ($\tau=+1$)
\begin{eqnarray}
    H^{\tilde{\phi}}_{\bK}(\bk)&=&  \left(\begin{array}{ccc}
     0_{2\times2} &v_\ast (k_x\sigma_0+ik_y\sigma_z)&\left(\begin{array}{cc}
       \tilde{\gamma}^\ast   & 0 \\
        0  & \tilde{\gamma}
     \end{array}\right)e^{-\bk^2\lambda^2/2}\\
     v_\ast (k_x\sigma_0-ik_y\sigma_z)& 0_{2\times2}& 0_{2\times2}\\
   e^{-\bk^2\lambda^2/2}\left(\begin{array}{cc}
       \tilde{\gamma}   & 0 \\
        0  & \tilde{\gamma}^\ast
     \end{array}\right)&0_{2\times2} &\frac{\tilde{U}_f}{2}\left(\begin{array}{cc}
           0  & ie^{-2 i\tilde{\phi}}\\
           -ie^{2  i\tilde{\phi}}  & 0
        \end{array}\right)\\
    \end{array}\right)\label{Appx:Eq:simple_mode_strain}
\end{eqnarray}
where $\tilde{\gamma}=\gamma-i\gamma'\frac{\epsilon}{2}(\nu_G-1)$, $\tilde{U}_f=+M_f\epsilon(\nu_G+1)+U_1$ and $\tilde{\phi}$ represents the orientation of uniaxial heterostrain with respect to the $x$-axis. We first note that $H^{\tilde{\phi}}_{\bK}(\bk)$ is independent of $\tilde{\phi}$ as $D^\dagger_{\tilde{\phi}}H^{\tilde{\phi}}_{\bK}(\bk)D_{\tilde{\phi}}=H^{\tilde{\phi}=0}_{\bK}(\bk)$ where
\begin{eqnarray}
    D_{\tilde{\phi}}=\left(\begin{array}{ccc}
       \left(\begin{array}{ccc}
       e^{-i\tilde{\phi}}  &0  \\
       0  & e^{+i\tilde{\phi}}
    \end{array}\right)  &  &\\
         & \mathbb{I}_2&\\
         & & \left(\begin{array}{ccc}
       e^{-i\tilde{\phi}}  &0  \\
       0  & e^{+i\tilde{\phi}}
    \end{array}\right)
    \end{array}\right)
\end{eqnarray}
We thus drop the $\tilde{\phi}$ dependence in the analysis below. Using the methods reviewed in Appendix(\ref{Appx:Heterostrain_THF_B=0andB}) we promote the $c$-$f$ coupling in Eq.(\ref{Appx:Eq:simple_mode_strain}) at finite $\bB$ to   
\begin{eqnarray}
   h^{\bK}_{[am],[br]}(k) = \left(\begin{array}{cc}
       \tilde{\gamma}^\ast   & 0 \\
        0  & \tilde{\gamma}\\
    0&0\\
    0&0
     \end{array}\right)_{ab}\Upsilon_{mr}(k)
\end{eqnarray}
where $k=(k_1,k_2)$ represents magnetic momentum, $a=\{1,\ldots,4\}$, $b=\{1,2\}$, $m$ denote the MTG LL index, $r=\{0\ldots,q-1\}$  and $\Upsilon_{mr}(k)=I^{0}_{[m],[r]}(k)$ where $I^{0}_{[m],[r]}(k)$ is defined in Eq.(\ref{I0}). We also neglect the phase from coordinate rotation $e^{i\theta_0}\sim 1$ for simplicity. Let us identify $h_{ab}=\gamma\sigma_0+i\gamma^\prime \frac{\epsilon}{2}(\nu_G-1)\sigma_z$ then for given spin projection $s$
\begin{eqnarray}
    \sum_{ambr}h^{\bK}_{[am],[br]}(k) c^\dagger_{a km \bK s} f_{b k r \bK s}&=& \sum_{ambr}\left(c^\dagger_{a km \bK s} h_{ab}\Upsilon_{mr}(k) f_{b k r \bK s}\right) \\
    &=& \sum_{ambr}\left(c^\dagger_{a km \bK s} h_{ab}  \sum_{r'} \Sigma_{m r'} V_{r'r} f_{b k r \bK s}\right)\\
    &=& \sum_{ambr}\left(c^\dagger_{a km \bK s}  h_{ab} \sum_{r'} \Sigma_{m} \delta_{mr'}V_{r'r} f_{b k r \bK s}\right)\\
    &=&\sum_{am b} h_{ab} \Sigma_{m}  c^\dagger_{a km \bK s}  \tilde{f}_{b k m \bK s}
\end{eqnarray}
where we used the singular value decomposition derived in \cite{singh2024topological}, $\Upsilon_{mr}=\sum_{r'}\Sigma_{mr'}V_{r'r}=\sum_{r'}\Sigma_{m}\delta_{mr'}V_{r'r}$, with singular value $\Sigma_m$ 
and $\tilde{f}_{b k m \bK s}=\sum_{r}V_{mr}f_{b k r \bK s}$. Since  $V$ is unitary matrix, the coupling between $\tilde{f}$ modes is still given by $h^{\tilde{f},\bK}_{[br],[br']}= -\frac{\tilde{U}_f}{2}\sigma_y\delta_{rr'}$. From the SVD decomposition we also see that $b=1$ $\tilde{f}$ modes with $r>m_{11}=m_\ast+1$ and $b=2$ $\tilde{f}$ modes with $ r>m_{21}=m_\ast$
 decouple from $a=1$ and $a=2$ $c$ fermions at valley $\bK$ respectively, where $m_{a\tau}$ represents upper cutoff on $m$ for $c$-MTG LL annihilation operator $c_{akm\tau s}$ with $m_{1,+1}=m_{2,-1}=m_{\ast}+1$, $m_{2,+1}=m_{1,-1}=m_{\ast}$,$m_{3,+1}=m_{4,-1}=m_{\ast}+2$ and $m_{4,+1}=m_{3,-1}=m_{\ast}-1$\cite{singh2024topological}. In other words, $\tilde{f}_{b k r \bK s}$ modes with $r>m_\ast+1$ are fully decoupled from the $c$ modes and will give rise to modes at $\pm \tilde{U}_f/2$. The coupled $c$-$f$ modes are described by computing the eigenvalues of 
 \begin{eqnarray}
&&   \hat{h}_{\bK} = \left(\begin{array}{cccccc}
0&0&-i\sqrt{2}\frac{v_{\ast}}{\ell}\hat{a}&0&\tilde{\gamma}^\ast \Sigma(\hat{a}^{\dagger}\hat{a})&0 \\
0&0&0&i\sqrt{2}\frac{v_{\ast}}{\ell}\hat{a}^{\dagger}&0 &\tilde{\gamma}\Sigma(\hat{a}^{\dagger}\hat{a}) \\
i\sqrt{2}\frac{v_{\ast}}{\ell}\hat{a}^{\dagger}&0&0&0&0&0\\0&-i\sqrt{2}\frac{v_{\ast}}{\ell}\hat{a}&0&0&0&0\\
\tilde{\gamma} \Sigma(\hat{a}^{\dagger}\hat{a})    & 0&0&0&0 &-i\tilde{U}_f/2\\
0 &  \tilde{\gamma}^\ast\Sigma(\hat{a}^{\dagger}\hat{a})&0&0&i\tilde{U}_f/2&0
\end{array}
\right)_{\alpha,\alpha'}
\end{eqnarray} 
in the ansatz 
\begin{eqnarray}
    \Theta_n= \left(c^n_1|n-1\rangle,c^n_2|n-1\rangle,c^n_3|n\rangle,c^n_4|n-2\rangle,c^n_5|n-1\rangle,c^n_6|n-1\rangle\right)^T
\end{eqnarray}
where $n=0,\ldots, m_\ast+2$ and  $c^{n<1}_1,c^{n<1}_2,c^{n<0}_3,c^{n<2}_4,c^{n<1}_5,c^{n<1}_6=0$ and $c^{n>m_\ast+1}_2,c^{n>m_\ast+1}_4=0$ and $\Sigma(m)=\Sigma_m\approx e^{-(m+\frac{1}{2})\lambda^2/\ell^2}$. We have two zero modes 
\begin{eqnarray}
    \Theta_{0} &=& \left(0,0,|0\rangle,0,0,0\right)^T\\
    \Theta_{1} &=& \left(0,i\tilde{U}_f\Delta_B|0\rangle,-2i\tilde{\gamma}^{\ast 2}\Sigma_0^2|1\rangle,0,2\tilde{\gamma}^\ast\Delta_B\Sigma_0|0\rangle,0\right)^T
\end{eqnarray}
where $\Delta_B=\sqrt{2}v_\ast\ell^{-1}$. The non-anomalous modes can be obtained by diagonalizing
\begin{eqnarray}
 h^{6,\bK}_{n} =   \left(\begin{array}{cccccc}
0&0&-i\Delta_{B}\sqrt{n} &0&\tilde{\gamma}^\ast \Sigma_{n-1}&0 \\
0&0&0&i\Delta_{B}\sqrt{n-1} &0 &\tilde{\gamma}\Sigma_{n-1} \\
i\Delta_{B}\sqrt{n}&0&0&0&0&0\\0&-i\Delta_{B}\sqrt{n-1}&0&0&0&0\\
\tilde{\gamma} \Sigma_{n-1}    & 0&0&0&0&-i\tilde{U}_f/2 \\
0 &  \tilde{\gamma}^\ast\Sigma_{n-1}&0&0&i\tilde{U}_f/2&0
\end{array}
\right)\label{Eq:appx:h6:SImple_strain}
\end{eqnarray}
Approximate analytical expressions for the lowest-energy single particle excitation states can be obtained as follows. We first examine the behavior of the non-anomalous modes at low fluxes $\phi/\phi_0\ll 0.04$ where $|\tilde{\gamma}|\gg\Delta_B$. We first set $\Sigma_m$ to 1 since we are interested in small values of $m$ at small magnetic field. The two zero modes of Eq.(\ref{Eq:appx:h6:SImple_strain}) for $\Delta_B=0$ read $(0,0,1,0,0,0)^T$, $(0,0,0,1,0,0)^T$. Projecting the model in Eq.(\ref{Eq:appx:h6:SImple_strain})
onto the subspace spanned by these zero modes up to $\mathcal{O}(\Delta^2_B)$ gives
\begin{eqnarray}
    h^{n\geq 2, low-\bB}_{eff,\bK}&=& \frac{\tilde{U}_fv_\ast^2}{\ell^2}\sqrt{n(n-1)} \left(\begin{array}{cc}
       0  & i/\tilde{\gamma}^2 \\
      -i/\tilde{\gamma}^{\ast 2}   & 0
    \end{array}\right)
\end{eqnarray}
In other words the non-anomalous modes approach $\frac{\tilde{U}_fv_\ast^2}{\ell^2|\tilde{\gamma}|^2}\sqrt{n(n-1)} $ as $\bB\rightarrow 0$. Note that in the main text we instead use the convention $\sqrt{n(n+1)}$ with $n\geq1$ for simplicity. Let us now focus on the flux regime $\phi/\phi_0\gg0.04$ where $|\tilde{\gamma}|\ll\Delta_B$. For $\tilde{\gamma}=0$, the modes closest to CNP are $\pm \Delta_B\sqrt{n-1}$ and their eigenvectors are  $|E^a_\pm\rangle=\left(0,1,0,\pm i,0,0\right)/\sqrt{2}$. The coupling $\tilde{\gamma}$ is responsible for the hybridization of these low lying modes to the high energy $\tilde{f}$ modes at $\pm \tilde{U}_f/2$ with eigenvectors $|E^b_\pm\rangle=\left(0,0,0,0,1,\pm i\right)/\sqrt{2}$. To obtain approximate expression for the modes closest to CNP for $\phi/\phi_0\gg0.04$ we project Eq.(\ref{Eq:appx:h6:SImple_strain}) onto the subspace spanned by $\{|E^a_\pm\rangle,|E^b_\pm\rangle\}$ and diagonalize the resulting matrix. The modified lowest energy excitation read
\begin{eqnarray}
    E^\pm_{n\geq 2} = \pm\sqrt{\frac{\left(\Delta_B^2(n-1)+\frac{\tilde{U}^2_f}{4}+|\tilde{\gamma}|^2\Sigma^2_{n-1}\right)-\sqrt{\left(\Delta_B^2(n-1)+\frac{\tilde{U}^2_f}{4}+|\tilde{\gamma}|^2\Sigma^2_{n-1}\right)^2-(n-1)\Delta_B^2\tilde{U}^2_f}}{2}}
\end{eqnarray}
The match between the approximate energy and numerically obtained lowest energy excitations near CNP is show in Fig.(\ref{fig:appx:Leading:Strain}).

%--------------Section Change--------------

\section{Auxiliary fermion mapping}\label{subsec:Auxiliary-fermion-mapping}

In this section, we review the auxiliary fermion mapping which we use
extensively to determine the location of the poles of the full Green's
function and to unravel their topological properties. To leave the
discussion as general as possible, we consider a generic multi-orbital
system with electron operators $\hat{\gamma}_{i}^{\dagger}$, where
the index $i$ may represent spin, momentum, orbital, or sub-lattice
labels. The system is described by the interacting Hamiltonian

\begin{equation}
\hat{H}=\sum_{i,j}t_{ij}\hat{\gamma}_{i}^{\dagger}\hat{\gamma}_{j}+\sum_{i,j,l,m}U_{ijlm}\hat{\gamma}_{i}^{\dagger}\hat{\gamma}_{j}\hat{\gamma}_{l}^{\dagger}\hat{\gamma}_{m}\label{eq:full_ham_aux}
\end{equation}

where $t_{ij}$ denotes the hopping-matrix elements, and $U_{ijlm}$
encodes the interaction terms. We can define the self-energy of the
system as usual by the Dyson relation:

\begin{equation}
\mathcal{G}_{ij}\left(i\omega_{n}\right)=\left[\left(i\omega_{n}\mathbb{I}-t-h^{\text{MF}}-\Sigma\left(i\omega_{n}\right)\right)^{-1}\right]_{ij},\label{eq:manyBGreen's}
\end{equation}
where $h_{ij}^{\text{MF}}$ is the mean field contribution from Eq.(\ref{eq:full_ham_aux}).
The dynamical self-energy admits a spectral representation 

\begin{equation}
\Sigma_{ij}\left(i\omega_{n}\right)=\int_{-\infty}^{\infty}d\omega\frac{\rho_{ij}^{\Sigma}(\omega)}{i\omega_{n}-\omega},
\end{equation}

with $\rho_{ij}^{\Sigma}(\omega)$ denoting the spectral function
of the self-energy, which is given explicitly by 

\begin{equation}
\rho^{\Sigma}(\omega)=-\frac{1}{2\pi i}\left(\Sigma\left(\omega+i0^{+}\right)-\Sigma^{\dagger}\left(\omega+i0^{+}\right)\right)
\end{equation}

Note that $\rho^{\Sigma}(\omega)$ is a Hermitian, positive semi-definite
matrix which allows it to be expressed as a sum of $\delta$-functions
multiplied by positive semi-definite matrices. These matrices can
be written as outer products of vectors. In general such a representation
can include an infinite sum of $\delta$-functions, but we can obtain
an arbitrarily good approximation by including sufficiently many terms.
Keeping a finite number of $\delta$-functions we approximate $\rho_{ij}^{\Sigma}(\omega)$
by:

\begin{equation}
\rho_{ij}^{\Sigma}(\omega)\approx\sum_{\alpha}v_{i}^{\alpha}\left(v_{j}^{\alpha}\right)^{*}\delta(\omega-\epsilon^{\alpha})
\end{equation}

where $v_{i}^{\alpha}$ are complex vectors and $\epsilon^{\alpha}$
are real frequencies. The parameters $v_{i}^{\alpha}$ and $\epsilon^{\alpha}$
are introduced to accurately describe the spectral function $\rho^{\Sigma}$
to any desired accuracy. In the following, we assume that a sufficient
number of terms have been included. Using the spectral representation
of the self-energy, we then obtain

\begin{equation}
\Sigma_{ij}\left(i\omega_{n}\right)=\sum_{\alpha}\frac{v_{i}^{\alpha}\left(v_{j}^{\alpha}\right)^{*}}{i\omega_{n}-\epsilon^{\alpha}}\label{eq:approxSelf}
\end{equation}

Within this approximation, the interacting Green's function in Eq.(\ref{eq:manyBGreen's})
is equivalent to the Green's function of a suitably defined auxiliary
non-interacting system. The proof follows by considering the non-interacting
Hamiltonian

\begin{equation}
\hat{H}^{\text{Aux}}=\sum_{ij}\left(t_{ij}+h_{ij}^{\text{MF}}\right)\hat{\gamma}_{i}^{\dagger}\hat{\gamma}_{j}+\sum_{\alpha}\epsilon^{\alpha}\hat{a}_{\alpha}^{\dagger}\hat{a}_{\alpha}+\sum_{\alpha i}\left(v_{i}^{\alpha}\hat{a}_{\alpha}^{\dagger}\hat{\gamma}_{i}+\text{h.c.}\right).\label{eq:auxmap}
\end{equation}

We then compute the Green's function of the non-interacting Hamiltonian
$\hat{H}^{\text{Aux}}$ in the basis of the combined $\hat{\gamma}_{i}$
and $\hat{a}_{\alpha}$ fermions defined by the spinor $\ensuremath{\hat{\Psi}^{\dagger}=\begin{pmatrix}\hat{\gamma}_{1}^{\dagger},\hat{\gamma}_{2}^{\dagger},\dots,\hat{a}_{1}^{\dagger},\hat{a}_{2}^{\dagger},\dots\end{pmatrix}}$:

\begin{equation}
\mathcal{G}^{\text{Aux}}\left(i\omega_{n}\right)=\begin{pmatrix}i\omega_{n}\mathbb{I}-t-h^{\text{MF}} & -V\\
-V^{\dagger} & i\omega_{n}\mathbb{I}-\mathcal{E}
\end{pmatrix}^{-1},\label{eq:auxgreen}
\end{equation}

where the hybridization matrix $V$ has elements $v_{i}^{\alpha}$
in the $i$-th row and $\alpha$-th column, while $\left[\mathcal{E}\right]_{\alpha\alpha'}=\delta_{\alpha\alpha'}\epsilon^{\alpha}$
is a diagonal matrix with entries determined by the poles of the self-energy.
To we can obtain the Green's function for the $\hat{\gamma}_{i}^{\dagger}$
fermions by inverting Eq.(\ref{eq:auxgreen}) using the block matrix
inversion formula (valid when $A$ and $D$ are invertible):

\begin{equation}
    \begin{pmatrix}A & B\\
C & D
\end{pmatrix}^{-1}={\displaystyle \begin{bmatrix}\left(A-BD^{-1}C\right)^{-1}\\
 & \left({D}-{C}{A}^{-1}{B}\right)^{-1}
\end{bmatrix}\begin{bmatrix}\mathbb{I} & -{B}{D}^{-1}\\
-{C}{A}^{-1} & \mathbb{I}
\end{bmatrix}.}
\end{equation}

Applying this identity to $\mathcal{G}^{\text{Aux}}\left(i\omega_{n}\right)$,
we obtain the Green's function of the physical $\hat{\gamma}_{i}^{\dagger}$
electrons under the Hamiltonian $\hat{H}^{\text{Aux}}$:
\begin{align}
\mathcal{G}^{\text{Aux},\gamma}\left(i\omega_{n}\right) & =-\int_{0}^{\beta}d\tau e^{i\omega_{n}\tau}\left\langle \mathcal{T}_{\tau}\hat{\gamma}_{i}^{\dagger}(\tau)\hat{\gamma}_{j}(0)\right\rangle _{\text{Aux}}=\left[i\omega_{n}\mathbb{I}-t-h^{\text{MF}}-V\left(i\omega_{n}\mathbb{I}-\mathcal{E}\right)^{-1}V^{\dagger}\right]^{-1}\label{eq:almost_auxiliary}
\end{align}

Equivalently, we can write Eq.(\ref{eq:almost_auxiliary}) as

\begin{align}
\left[\left(\mathcal{G}^{\text{Aux},\gamma}\left(i\omega_{n}\right)\right)^{-1}\right]_{ij} & =i\omega_{n}\delta_{ij}-t_{ij}-h_{ij}^{\text{MF}}-\sum_{\alpha}\frac{v_{i}^{\alpha}\left(v_{j}^{\alpha}\right)^{*}}{i\omega_{n}-\epsilon^{\alpha}}\\
 & =\left[\left(\mathcal{G}\left(i\omega_{n}\right)\right)^{-1}\right]_{ij}.
\end{align}

We then conclude that the full many-body Green's function can be expressed
in terms of a non-interacting Green's function of auxiliary Fermions
by isolating the $\gamma$ block of $\mathcal{G}^{\text{Aux}}\left(i\omega_{n}\right)$.
In general, the complexity of the self-energy dictates the utility
of the above mapping. Within the Hubbard-I approximation the auxiliary
fermion is especially simple as the self-energy can be easily recast
in the form of Eq.(\ref{eq:approxSelf}) with a few simple poles
which provides easy access to the topological properties of the THF
model under a magnetic field. We now calculate the auxiliary fermion
Hamiltonian in the THF model both with and without strain.

Throughout the remainder of the appendices, we use $\eta$ to represent the valley quantum number.

\section{The Mott semimetal state at charge neutrality point within Hubbard-I approximation}\label{sec:Landau-Level-spectrum} 
In this section we discuss the spectrum of the unstrained Mott semimetal at CNP within the Hubbard-I. The atomic self-energy $\Sigma^{\text{At}}$ \cite{aux_map_B2026} in Matsubara frequencies reads
\begin{equation}
\Sigma^{\text{At}}\left(i\omega_{n}\right)=\begin{cases}
\tilde{\Sigma}^{\text{At}}\left(i\omega_{n}\right)\delta_{ii'}\delta_{\eta\eta'}\delta_{ss;} & \text{if }i=5,6\\
0 & \text{otherwise}
\end{cases}
\end{equation}
in matrix form. Here $i=\{1,\ldots,4\}$ and $i=\{5,6\}$ 
represent the 4 $c$ fermion and 2 $f$ fermion orbitals at given spin and valley.
The scalar self energy for integer filling of $f$ electrons
in the ground state is given by
\begin{equation}
\tilde{\Sigma}^{\text{At}}\left(i\omega_{n}\right)=U_{1}^{2}\frac{\left(N_f/2+\nu_{f}\right)\left(N_f/2-\nu_{f}\right)}{N_f^{2}}\frac{1}{i\omega-\left[\frac{U_{1}}{N_f}\left(N_f+1\right)\nu_{f}+\epsilon_{f}\right]},\ \ \text{for }\nu_{f}=r-N_f/2,
\end{equation}
and where $U_{1}$ is the strength of the Hubbard interaction, $N_{f}=8$
is the number of $f$-electron flavors, $\nu_{f}$ is the filling
fraction of the $f$-electrons. Additionally, the mean-field Hamiltonian
is given by 
\begin{equation}
h_{i\eta s;i'\eta's'}^{\text{MF}}(\mathbf{k})=\delta_{\eta,\eta'}\delta_{s,s'}\tilde{h}_{i,i',\eta}^{\text{MF}}(\mathbf{k})
\end{equation}
with the orbital blocks of the Hamiltonian given by
\begin{align}
&\tilde{h}_{i,i',\eta}^{\text{MF}}=\nonumber\\&\begin{pmatrix}\epsilon_{c,1} & 0 & v_{\ast}(\eta k_{x}+ik_{y}) & 0 & e^{-\frac{\lambda^2\bk^2}{2}}\gamma & e^{-\frac{\lambda^2\bk^2}{2}}v'_{\star}(\eta k_{x}-ik_{y})\\
0 & \epsilon_{c,1} & 0 & v_{\ast}(\eta k_{x}-ik_{y}) & e^{-\frac{\lambda^2\bk^2}{2}}v'_{\star}(\eta k_{x}+ik_{y}) & e^{-\frac{\lambda^2\bk^2}{2}}\gamma\\
v_{\ast}(\eta k_{x}-ik_{y}) & 0 & \epsilon_{c,2} & M & 0 & 0\\
0 & v_{\ast}(\eta k_{x}+ik_{y}) & M & \epsilon_{c,2} & 0 & 0\\
e^{-\frac{\lambda^2\bk^2}{2}}\gamma & e^{-\frac{\lambda^2\bk^2}{2}}v'_{\star}(\eta k_{x}-ik_{y}) & 0 & 0 & \epsilon_{f}+\frac{U_{1}}{N_f}\left(N_f-1\right)\nu_{f} & 0\\
e^{-\frac{\lambda^2\bk^2}{2}}v'_{\star}(\eta k_{x}+ik_{y}) & e^{-\frac{\lambda^2\bk^2}{2}}\gamma & 0 & 0 & 0 & \epsilon_{f}+\frac{U_{1}}{N_f}\left(N_f-1\right)\nu_{f}
\end{pmatrix}\label{Eq:MF for Hubb-I}.
\end{align}
In the above, we only consider the Hartree potentials $\epsilon_{c,1}=\frac{V^{(0)}}{\Omega_{0}}\nu_{c}+W_{1}\nu_{f},$
$\epsilon_{c,2}=\frac{V^{(0)}}{\Omega_{0}}\nu_{c}+W_{3}\nu_{f}$
and $\epsilon_{f}=W_{1}(\nu_{c,1}+\nu_{c,2})+W_{3}(\nu_{c,2}+\nu_{c,3})$,
with $\nu_{c,a}$ denote the filling of the $c$ electrons with orbital
index $a$ and $\nu_{c}$ is the total filling fraction of the $c$
electrons. In addition to the mean-field terms given by Eq.(\ref{Eq:MF for Hubb-I}), our analysis also includes the effects of lattice relaxation is given by\cite{herzog2025topological}
\begin{eqnarray}
    h^{rel}_{i,i',\eta}(\bk) =\left(\begin{array}{ccc}
       \mu_1\sigma_0 + v_2(\eta k_x,k_y)\cdot\sigma & v_1(\eta k_x,k_y)\cdot\sigma^\ast  & 0\\
       & \mu_2\sigma_0 &0\\
     h.c.    &0 &0
    \end{array}\right)\label{Eq:Relaxation_for_Hubb1}
\end{eqnarray}
One can reproduce the pole structure of the Green's function
by introducing an additional coupling to auxiliary fermions $\hat{a}_{\mathbf{k}\alpha\eta s}$.
The coupled $f-a$ block of the Hamiltonian in momentum space is given
by 
\begin{align}
\hat{H}^{\text{Aux},fa} & =\sum_{\mathbf{k}\alpha\eta s}\left[\frac{U_{1}}{N_f}\left(N_f+1\right)\nu_{f}+\epsilon_{f}\right]\hat{f}_{\mathbf{k}\alpha\eta s}^{\dagger}\hat{f}_{\mathbf{k}\alpha\eta s}+\sum_{\mathbf{k}\alpha\eta s}\left[\frac{U_{1}}{N_f}\left(N_f+1\right)\nu_{f}+\epsilon_{f}\right]\hat{a}_{\mathbf{k}\alpha\eta s}^{\dagger}\hat{a}_{\mathbf{k}\alpha\eta s}\\
 & +\sum_{\mathbf{k}\alpha\eta s}\frac{U_{1}\sqrt{\left(N_{f}/2-\nu_{f}\right)\left(N_{f}/2+\nu_{f}\right)}}{N_f}\left(\hat{a}_{\mathbf{k}\alpha\eta s}^{\dagger}\hat{f}_{\mathbf{k}\alpha\eta s}+\text{h.c}\right),\label{Eq:Aux_mapping}
\end{align}
where $\hat{f}_{\mathbf{k}\alpha\eta s}^{\dagger}$ and $\hat{a}_{\mathbf{k}\alpha\eta s}^{\dagger}$
are the creation operators for the $f$ and auxiliary fermions at
momentum $\mathbf{k}$, valley $\eta$ and spin $s$, respectively. At the charge neutrality point, $\nu_c=\nu_f=0$, and all the Hartree corrections vanish. Moreover, the auxiliary fermion hybridization for the f-modes simplifies to (in matrix form)
\begin{eqnarray}
    h^{aux}_{i,i',\eta}= \left(\begin{array}{cccc}
        0 & 0&0 &0 \\
         0& 0& 0& 0\\
        0&0 & 0&U_1/2\sigma_0 \\
        0 & 0&U_1/2\sigma_0  &0 
    \end{array}\right)\label{Eq:aux_for_hubb1_CNP}
\end{eqnarray}
In the above matrix notation, the first $6\times 6$ block corresponds to the $c$ and $f$ fermions, followed by the $2\times 2$ block for the auxiliary fermions. Clearly, the auxiliary fermions only couple to the $f$- fermions (as expected from Eq.(\ref{Eq:Aux_mapping})). The excitation spectrum within the Hubbard-I approximation at the charge neutrality point can thus be obtained by diagonalizing
\begin{eqnarray}
    H^{aux}_{{Hubbard-I}_{[\eta,i,i']}}(\bk) = \tilde{h}^{MF}_{i,i',\eta}(\bk) + h^{rel}_{i,i',\eta}(\bk) + h^{aux}_{i,i',\eta}\label{Eq:Relaxed_noninteracting_Hubbard-I model}
\end{eqnarray}
where $\tilde{h}^{MF}_{i,i',\eta}(\bk),\tilde{h}^{rel}_{i,i',\eta}(\bk),h^{aux}_{i,i',\eta}$ are defined in 
Eq.(\ref{Eq:MF for Hubb-I}),(\ref{Eq:Relaxation_for_Hubb1}) and (\ref{Eq:aux_for_hubb1_CNP}) respectively with the convention that $\tilde{h}^{MF}_{i,i',\eta}(\bk),\tilde{h}^{rel}_{i,i',\eta}(\bk)$ are padded with zeros on the last two rows and columns matching the dimensions of $ h^{aux}_{i,i',\eta}$ once the auxiliary fermions are introduced. The corresponding spectral function can be obtained from the Green's function of the physical fermions (i.e. the $c$ and $f$ fermions) and is presented in Fig.(\ref{fig:main2_withrelax}a) (see Appendix(\ref{Appx:Spectral_map_color}) for details).
\begin{figure}
    \centering
    \includegraphics[width=0.95\linewidth]{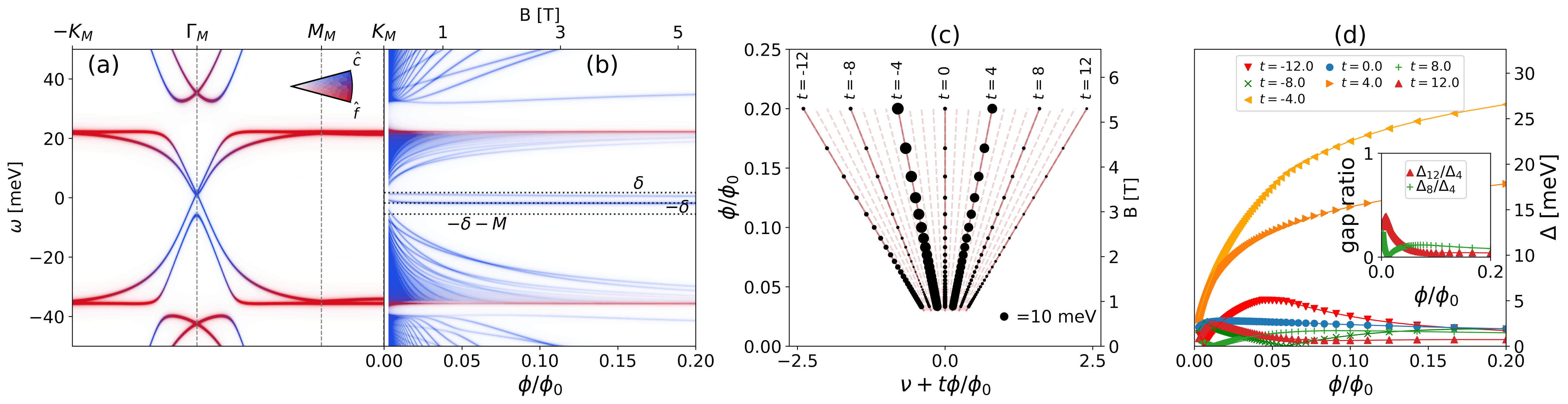}
    \caption{\textbf{Mott Semimetal:} Spectral function of the Mott semimetal within Hubbard-I at (a)$\bB=0$ and at (b)$\bB\neq0$ with effects of lattice relaxation included. The color scheme for the spectral function is explained in Appendix(\ref{Appx:Spectral_map_color}). (c) Single-electron excitation gap as a function of electron filling $\nu =s + t\frac{\phi}{\phi_0}$, along $t=0,\pm 4, \pm8,\pm12$ for $s=0$ (CNP). The area of the solid red dots scale with the size of the gaps. (d) The magnitudes of obtained Chern gaps as a function of $\frac{\phi}{\phi_0}$. The ratio of the magnitudes of $ C=12$ and $C=8$ with respect to $C=4$ gap shown as inset. Here $\delta=\mu_{1}/(1+\frac{4\gamma^2}{U_1^2})-\mu_2$. The $\omega$ axis shared by panels (a,b) is shifted by an
    overall chemical potential of $(\mu_{1}/(1+\frac{4\gamma^2}{U_1^2})+\mu_2$)/2.}
    \label{fig:main2_withrelax}
\end{figure}
\subsection{Promoting the auxiliary fermion construction to magnetic field in absence of Zeeman}
In the presence of an out-of-plane magnetic field $\bB$, the bands further quantize giving rise to Landau levels. The corresponding spectrum can be obtained by diagonalizing the matrix $\Xi_{m\alpha,m'\alpha'} = \langle m |\hat{h}_{\alpha,\alpha'}|m'\rangle$, where  \cite{singh2024topological}
\begin{eqnarray}
   &\tilde{h}_{i,i',\eta=+}^{\text{MF}}=\nonumber\\ &\left(\begin{array}{cccccccc}
\mu_{1} & i\sqrt{2}\frac{v_{2}}{\ell}\hat{a}^{\dagger} & -i\sqrt{2}\frac{v_{\ast}}{\ell}\hat{a} & -i\sqrt{2}\frac{v_{1}}{\ell}\hat{a} & \gamma\Sigma(\hat{a}^{\dagger}\hat{a}) & i\sqrt{2}\frac{v'_{\ast}}{\ell}\hat{a}^{\dagger}\Sigma(\hat{a}^{\dagger}\hat{a}) & 0 & 0\\
-i\sqrt{2}\frac{v_{2}}{\ell}\hat{a} & \mu_{1} & i\sqrt{2}\frac{v_{1}}{\ell}\hat{a}^{\dagger} & i\sqrt{2}\frac{v_{\ast}}{\ell}\hat{a}^{\dagger} & -i\sqrt{2}\frac{v'_{\ast}}{\ell}\hat{a}\Sigma(\hat{a}^{\dagger}\hat{a}) & \gamma\Sigma(\hat{a}^{\dagger}\hat{a}) & 0 & 0\\
i\sqrt{2}\frac{v_{\ast}}{\ell}\hat{a}^{\dagger} & -i\sqrt{2}\frac{v_{1}}{\ell}\hat{a} & \mu_{2} & M & 0 & 0 & 0 & 0\\
i\sqrt{2}\frac{v_{1}}{\ell}\hat{a}^{\dagger} & -i\sqrt{2}\frac{v_{\ast}}{\ell}\hat{a} & M & \mu_{2} & 0 & 0 & 0 & 0\\
\gamma\Sigma(\hat{a}^{\dagger}\hat{a}) & i\sqrt{2}\frac{v'_{\ast}}{\ell}\Sigma(\hat{a}^{\dagger}\hat{a})\hat{a}^{\dagger} & 0 & 0 & 0 & 0 & \frac{U_{1}}{2} & 0\\
-i\sqrt{2}\frac{v'_{\ast}}{\ell}\Sigma(\hat{a}^{\dagger}\hat{a})\hat{a} & \gamma\Sigma(\hat{a}^{\dagger}\hat{a}) & 0 & 0 & 0 & 0 & 0 & \frac{U_{1}}{2}\\
0 & 0 & 0 & 0 & \frac{U_{1}}{2} & 0 & 0 & 0\\
0 & 0 & 0 & 0 & 0 & \frac{U_{1}}{2} & 0 & 0
\end{array}\right),\label{Appx:Eq:Hubbar_I_magnetic_after_aux}
\end{eqnarray}
at valley $\bK$ (though the spectrum is spin-valley degenerate), $\hat{a}$ is a simple harmonic oscillator lowering operator and $|m\rangle$ is the simple harmonic oscillator eigenstate with $\Sigma(m)=\Sigma_m$. The LL index $m$ take the values $m\in\{0,\cdots,m_\alpha\}$ for given spinor index $\alpha\in\{1,2,\cdots,8\}$, where $m_{\alpha=1,\ldots, 4}=m_{\alpha,+1}$, $m_{5,7}=m_{\ast}+1$ and $m_{6,8}=m_{\ast}$. The numerically obtained spectrum is shown in right panel of Fig.(\ref{fig:main:fig2}a) in absence of lattice relaxation effects, i.e. $\mu_{1,2}$ and $v_{1,2}$.  The numerically obtained spectrum with lattice relaxation effects is shown in Fig.(\ref{fig:main2_withrelax}). We approximate $\Sigma_m\approx e^{-(m+\frac{1}{2})\frac{\lambda^2}{\ell^2}}$\cite{singh2024topological} and use LL upper cutoff of $m_\ast = m_x$ if $m_x = int\left(\lfloor(\phi/\phi_0)^{-1}+3\rfloor\times \frac{4}{3}\right)>30$ and $30$ otherwise.

\subsection{Analytical results at CNP for the Landau level spectrum}

Consider the case where $v_{1}=v_{2}=\mu_1=\mu_2=0$, at charge neutrality, where the
Hartree corrections vanish, setting $\nu_{c}=\nu_{f}=0$. 
To determine the spectrum at finite $B-$field, we start by performing
naive minimal substitution. We then make the replacement $(k_{x}+ik_{y})\rightarrow-i\frac{\sqrt{2}}{\ell}\hat{a}$
, where $\hat{a}$ is the Landau level annihilation operator, $\ell=\sqrt{\hbar/eB}$
is the magnetic length, and we define the cyclotron energy $\Delta_{B}=\sqrt{2}v_{\ast}/\ell$.
With these, definitions the auxiliary Hubbard-I Hamiltonian takes
the following form:
\begin{equation}
\tilde{h}_{i,i',\eta=+}^{\text{MF}}\left(B\right)=\begin{pmatrix}0 & 0 & -i\Delta_{B}\hat{a} & 0 & \gamma\Sigma\left(\hat{a}^{\dagger}\hat{a}\right) & i\Delta_{B}^{\prime}\hat{a}^{\dagger}\Sigma\left(\hat{a}^{\dagger}\hat{a}\right) & 0 & 0\\
0 & 0 & 0 & i\Delta_{B}\hat{a}^{\dagger} & -i\Delta_{B}^{\prime}\hat{a}\Sigma\left(\hat{a}^{\dagger}\hat{a}\right) & \gamma\Sigma\left(\hat{a}^{\dagger}\hat{a}\right) & 0 & 0\\
i\Delta_{B}\hat{a}^{\dagger} & 0 & 0 & M & 0 & 0 & 0 & 0\\
0 & -i\Delta_{B}\hat{a} & M & 0 & 0 & 0 & 0 & 0\\
\gamma\Sigma\left(\hat{a}^{\dagger}\hat{a}\right) & i\Delta_{B}^{\prime}\Sigma\left(\hat{a}^{\dagger}\hat{a}\right)\hat{a}^{\dagger} & 0 & 0 & 0 & 0 & \frac{U_{1}}{2} & 0\\
-i\Delta_{B}^{\prime}\Sigma\left(\hat{a}^{\dagger}\hat{a}\right)\hat{a} & \gamma\Sigma\left(\hat{a}^{\dagger}\hat{a}\right) & 0 & 0 & 0 & 0 & 0 & \frac{U_{1}}{2}\\
0 & 0 & 0 & 0 & \frac{U_{1}}{2} & 0 & 0 & 0\\
0 & 0 & 0 & 0 & 0 & \frac{U_{1}}{2} & 0 & 0
\end{pmatrix}_{ii'}
\end{equation}
we can work in units of the cyclotron energy, $\Delta_{B}$, by defining
\begin{equation}
\tilde \gamma \equiv \frac{\gamma}{\Delta_B},\qquad
\tilde U \equiv \frac{U}{\Delta_B},\qquad
\tilde M \equiv \frac{M}{\Delta_B},\qquad
\tilde\Delta' \equiv \frac{\Delta_B'}{\Delta_B}.
\end{equation} \label{eqn:dimensionless_params}

so that the Hamiltonian becomes
\begin{equation}
\tilde{h}_{i,i',\eta=+}^{\text{MF}}\left(B\right)=\Delta_{B}\begin{pmatrix}0 & 0 & -i\hat{a} & 0 & \tilde{\gamma}\Sigma\left(\hat{a}^{\dagger}\hat{a}\right) & i\tilde{\Delta}^{\prime}\hat{a}^{\dagger}\Sigma\left(\hat{a}^{\dagger}\hat{a}\right) & 0 & 0\\
0 & 0 & 0 & i\hat{a}^{\dagger} & -i\tilde{\Delta}^{\prime}\hat{a}\Sigma\left(\hat{a}^{\dagger}\hat{a}\right) & \tilde{\gamma}\Sigma\left(\hat{a}^{\dagger}\hat{a}\right) & 0 & 0\\
i\hat{a}^{\dagger} & 0 & 0 & \tilde{M} & 0 & 0 & 0 & 0\\
0 & -i\hat{a} & \tilde{M} & 0 & 0 & 0 & 0 & 0\\
\tilde{\gamma}\Sigma\left(\hat{a}^{\dagger}\hat{a}\right) & i\tilde{\Delta}^{\prime}\Sigma\left(\hat{a}^{\dagger}\hat{a}\right)\hat{a}^{\dagger} & 0 & 0 & 0 & 0 & \frac{\tilde{U}}{2} & 0\\
-i\tilde{\Delta}^{\prime}\Sigma\left(\hat{a}^{\dagger}\hat{a}\right)\hat{a} & \tilde{\gamma}\Sigma\left(\hat{a}^{\dagger}\hat{a}\right) & 0 & 0 & 0 & 0 & 0 & \frac{\tilde{U}}{2}\\
0 & 0 & 0 & 0 & \frac{\tilde{U}}{2} & 0 & 0 & 0\\
0 & 0 & 0 & 0 & 0 & \frac{\tilde{U}}{2} & 0 & 0
\end{pmatrix}_{ii'}.\label{eq:chiralAuxB}
\end{equation}
We can solve for the zero-modes of this Hamiltonian exactly we will
further use these exact solutions to include the effects of relaxation
and chiral symmetry breaking perturbatively at neutrality. This perturbation is carried out in Sec.(\ref{sec:anomalous}) 

\subsubsection{Zero modes of the chiral Hubbard-I Hamiltonian}\label{subsec:Zero-modes-exact}
We can obtain analytically the zero modes by simply considering the
action of the Hamiltonian in Eq.(\ref{eq:chiralAuxB}) We have a
single zero mode that is obvious 

\begin{equation}
|\psi_{0,1}\rangle=\frac{1}{\sqrt{1+\left(\frac{U_{1}}{2\gamma\Sigma\left(0\right)}\right)^{2}+\left(\frac{\Delta_{B}^{\prime}\Sigma\left(1\right)}{\gamma\Sigma\left(0\right)}\right)^{2}}}\left(\begin{array}{cccccccc}
0, & -\frac{U_{1}}{2\gamma\Sigma\left(0\right)}|0\rangle, & 0, & 0, & 0, & 0, & i\frac{\Delta_{B}^{\prime}\Sigma\left(1\right)}{\gamma\Sigma\left(0\right)}|1\rangle, & |0\rangle\end{array}\right)\label{eq:zeromodenonchiral}
\end{equation}

\begin{equation}
|\psi_{0,2}\rangle=\frac{1}{\sqrt{1+\left(\frac{U_{1}}{2\gamma\Sigma\left(1\right)}\right)^{2}+\left(\frac{U_{1}\Delta_{B}}{2M\gamma\Sigma\left(1\right)}\right)^{2}+2\left(\frac{\Delta_{B}^{\prime}\Sigma\left(2\right)}{\gamma\Sigma\left(1\right)}\right)^{2}}}\left(\begin{array}{cccccccc}
0, & -\frac{U_{1}}{2\gamma\Sigma\left(1\right)}|1\rangle, & -\frac{iU_{1}\Delta_{B}}{2\gamma\Sigma\left(1\right)M}|0\rangle, & 0, & 0, & 0, & \frac{i\sqrt{2}\Delta_{B}^{\prime}\Sigma\left(2\right)}{\gamma\Sigma\left(1\right)}|2\rangle, & |1\rangle\end{array}\right)\label{eq:zeromodenonchiral2}
\end{equation}
When $M=0$ Eq.(\ref{eq:zeromodenonchiral}) still yields a valid zero mode
while Eq.(\ref{eq:zeromodenonchiral2}) will have a finite limit:
\begin{equation}
|\psi_{0,2}\rangle=\left(\begin{array}{cccccccc}
0, & 0, & |0\rangle, & 0, & 0, & 0, & 0, & 0\end{array}\right).\label{eq:zeromodeM0}
\end{equation}
We note that in the strong couplig limit the auxiliary fermions and
the $f$-electrons are pushed to high energy such that $|\psi_{0,1}\rangle$
has weight mostly on the $\Gamma_{3}$ $c-$fermions. We also find
that when $\tilde{M}\neq0$ the weight on different orbitals of $|\psi_{0,2}\rangle$
depends on $B$, with the weight on the $\Gamma_{1}\oplus\Gamma_{2}$
$c-$fermions being zero at $B=0$ and saturating to $1$ as $B\rightarrow\infty$.
In contrast both $|\psi_{0,1}\rangle$ and the $M=0$ zero mode in
Eq.(\ref{eq:zeromodeM0}) are $B$-independent. 

\subsubsection{Exact solution in the chiral-flat limit}
We can solve for the full spectrum of Eq.(\ref{eq:chiralAuxB}) when
$M=0$. We take the definition 
\begin{equation}
   \tilde{h}_{i,i',\eta=+,0}^{\text{MF}}= \lim_{M\rightarrow0,v^\prime_\ast\rightarrow0}\tilde{h}_{i,i',\eta=+}^{\text{MF}}
   \label{Appx:Eq:Hubbar_I_magnetic_after_aux0}
\end{equation}
First, we note that the transformation 
\begin{equation}
R=\left(\begin{array}{cccccccc}
1 & 0 & 0 & 0 & 0 & 0 & 0 & 0\\
0 & 0 & 1 & 0 & 0 & 0 & 0 & 0\\
0 & 0 & 0 & 0 & 1 & 0 & 0 & 0\\
0 & 0 & 0 & 0 & 0 & 0 & 1 & 0\\
0 & 1 & 0 & 0 & 0 & 0 & 0 & 0\\
0 & 0 & 0 & 1 & 0 & 0 & 0 & 0\\
0 & 0 & 0 & 0 & 0 & 1 & 0 & 0\\
0 & 0 & 0 & 0 & 0 & 0 & 0 & 1
\end{array}\right)
\label{eqn:rot_block}
\end{equation}
Block-diagonalizes the Hamiltonian:
\begin{equation}
R\tilde{h}_{i,i',\eta}^{\text{MF}}\left(B\right)R^{T}=\Delta_{B}\begin{pmatrix}0 & -i\hat{a} & \tilde{\gamma}\Sigma\left(\hat{a}^{\dagger}\hat{a}\right) & 0 & 0 & 0 & 0 & 0\\
i\hat{a}^{\dagger} & 0 & 0 & 0 & 0 & 0 & 0 & 0\\
\tilde{\gamma}\Sigma\left(\hat{a}^{\dagger}\hat{a}\right) & 0 & 0 & \frac{\tilde{U}}{2} & 0 & 0 & 0 & 0\\
0 & 0 & \frac{\tilde{U}}{2} & 0 & 0 & 0 & 0 & 0\\
0 & 0 & 0 & 0 & 0 & i\hat{a}^{\dagger} & \tilde{\gamma}\Sigma\left(\hat{a}^{\dagger}\hat{a}\right) & 0\\
0 & 0 & 0 & 0 & -i\hat{a} & 0 & 0 & 0\\
0 & 0 & 0 & 0 & \tilde{\gamma}\Sigma\left(\hat{a}^{\dagger}\hat{a}\right) & 0 & 0 & \frac{\tilde{U}}{2}\\
0 & 0 & 0 & 0 & 0 & 0 & \frac{\tilde{U}}{2} & 0
\end{pmatrix}.\label{eq:rotChiral}
\end{equation}
Such that the spectrum decouples into two $4\times4$ blocks which
can be solved for independently.
\subsubsection{Full rank \texorpdfstring{$m>0$}{m>0} eigenstates}
We now focus on the top block in
Eq.(\ref{eq:rotChiral}), which we denote block $A$:
\begin{equation}
\left(\begin{array}{cccc}
0 & -i\hat{a} & \tilde{\gamma}\Sigma\left(\hat{a}^{\dagger}\hat{a}\right) & 0\\
i\hat{a}^{\dagger} & 0 & 0 & 0\\
\tilde{\gamma} \Sigma\left(\hat{a}^{\dagger}\hat{a}\right)& 0 & 0 & \frac{\tilde{U}}{2}\\
0 & 0 & \frac{\tilde{U}}{2} & 0
\end{array}\right)\left(\begin{array}{c}
\sum_{m}\psi_{c3,m}^{1}|m\rangle\\
\sum_{m}\psi_{c12,m}^{1}|m\rangle\\
\sum_{m}\psi_{f,m}^{1}|m\rangle\\
\sum_{m}\psi_{a,m}^{1}|m\rangle
\end{array}\right)=\left(\begin{array}{c}
\sum_{m}\psi_{c12,m}^{1}\left(-i\right)\sqrt{m}|m-1\rangle+\tilde{\gamma}\sum_{m}\Sigma\left(m\right)\psi_{f,m}^{1}|m\rangle\\
\sum_{m}\psi_{c3,m}^{1}i\sqrt{m+1}|m+1\rangle\\
\tilde{\gamma}\sum_{m}\Sigma\left(m\right)\psi_{c3,m}^{1}|m\rangle+\frac{\tilde{U}}{2}\sum_{m}\psi_{a,m}^{1}|m\rangle\\
\frac{\tilde{U}}{2}\sum_{m}\psi_{f,m}^{1}|m\rangle
\end{array}\right)
\end{equation}
We identify the fact that the Hamiltonian becomes solvable by choosing
an ansatz spinor where each component has has a definite LL index
$m\geq1$ given by:
\begin{align}
\left(\begin{array}{cccc}
0 & -i\hat{a} & \tilde{\gamma} \Sigma\left(m-1\right)& 0\\
i\hat{a}^{\dagger} & 0 & 0 & 0\\
\tilde{\gamma} \Sigma\left(m-1\right)& 0 & 0 & \frac{\tilde{U}}{2}\\
0 & 0 & \frac{\tilde{U}}{2} & 0
\end{array}\right)\left(\begin{array}{c}
a_{1}|m-1\rangle\\
a_{2}|m\rangle\\
a_{3}|m-1\rangle\\
a_{4}|m-1\rangle
\end{array}\right) & =\left(\begin{array}{c}
\left(a_{2}\left(-i\right)\sqrt{m}+\tilde{\gamma}\Sigma\left(m-1\right)a_{3}\right)|m-1\rangle\\
a_{1}i\sqrt{m}|m\rangle\\
\left(\tilde{\gamma}\Sigma\left(m-1\right)a_{1}+\frac{\tilde{U}}{2}a_{4}\right)|m-1\rangle\\
\frac{\tilde{U}}{2}a_{3,}|m-1\rangle
\end{array}\right)\\
 & =E\left(\begin{array}{c}
a_{1}|m-1\rangle\\
a_{2}|m\rangle\\
a_{3}|m-1\rangle\\
a_{4}|m-1\rangle
\end{array}\right)
\end{align}
We then find it convenient to define the following notation for the hybridization:
\begin{equation}
\Gamma_m \equiv \tilde\gamma\,\Sigma(m).
\label{eq:Gamma-def}
\end{equation}
Thus, we have the following set of equations 
\begin{align}
a_{2}\left(-i\right)\sqrt{m}+\Gamma_{m-1} a_{3} & =Ea_{1}\label{eq:a1sol}\\
a_{1}i\sqrt{m} & =Ea_{2}\label{eq:a2sol}\\
\Gamma_{m-1} a_{1}+\frac{\tilde{U}}{2}a_{4} & =Ea_{3}\label{eq:a3sol}\\
\frac{\tilde{U}}{2}a_{3,} & =Ea_{4}\label{eq:a4sol}\\
\left|a_{1}\right|^{2}+\left|a_{2}\right|^{2}+\left|a_{3}\right|^{2}+\left|a_{4}\right|^{2} & =1\label{eq:normasol}
\end{align}
Now we solve for the different coefficients
\begin{align}
a_{2} & =\frac{i\sqrt{m}}{E}a_{1}\label{eq:a2_solvedE}\\
a_{3} & =\frac{1}{\Gamma_{m-1}}\left(E-\frac{m}{E}\right)a_{1}\label{eq:a3_solvedE}\\
a_{4} & =\frac{\tilde{U}}{2\Gamma_{m-1} E}\left(E-\frac{m}{E}\right)a_{1}\label{eq:a4_solvedE}
\end{align}
where we used the Eq.(\ref{eq:a2sol}), Eq.(\ref{eq:a1sol}) and
Eq.(\ref{eq:a4sol}) respectively. We can get an equation for the
eigenvalues, $E$, using Eq.(\ref{eq:a2_solvedE}), Eq.(\ref{eq:a3_solvedE})
and Eq.(\ref{eq:a4_solvedE}) together with Eq.(\ref{eq:a3sol})
which yields
\begin{equation}
\tilde{\gamma}a_{1}+\frac{\tilde{U}^{2}}{4\Gamma_{m-1}E}\left(E-\frac{m}{E}\right)a_{1}=\frac{E}{\Gamma_{m-1}}\left(E-\frac{m}{E}\right)a_{1}
\end{equation}
Assuming $a_{1}$, is non-zero we get the equation 
\begin{equation}
\Gamma_{m-1}+\left(\frac{\tilde{U}^{2}}{4}-\frac{\tilde{U}^{2}}{4E^{2}}m\right)=E^{2}-m.
\end{equation}
This equation is a quadratic equation for $E^{2}$ which we can solve
readily, so the energies are given by 
\begin{equation}
E_{A,msr}=s\sqrt{x_{Amr}}
\label{eqn:anaspecchiralflatA}
\end{equation}
where 
\begin{equation}
x_{A,mr}=\frac{1}{2}\left[
m+\Gamma_{m-1}^2+\frac{\tilde U^2}{4}
+r\sqrt{\left(m+\Gamma_{m-1}^2+\frac{\tilde U^2}{4}\right)^2
-\tilde U^2m}\right].
\label{eq:xA}
\end{equation}
and $r,s=\pm1$. We can further determine the form of the coefficients
setting $a_{1}=1$ and then obtaining the normalization, overall we
find the eigenstate in the unrotated basis is 
\begin{equation}
\psi_{A,msr}=N_{A,msr}\left(\begin{array}{cccccccc}
a_{1}^{msr}|m-1\rangle & 0 & a_{2}^{msr}|m\rangle & 0 & a_{3}^{msr}|m-1\rangle & 0 & a_{4}^{msr}|m-1\rangle & 0\end{array}\right),\ 
\label{eqn:Aspinorsol}
\end{equation}
where the coefficients are given by:
\begin{align}
a_{1}^{msr} &=1,\\
a_{2}^{msr} &=\frac{i\sqrt{m}}{E_{A,msr}},\\
a_{3}^{msr} &=\frac{1}{\Gamma_{m-1}}
\left(E_{A,msr}-\frac{m}{E_{A,msr}}\right),\\
a_{4}^{msr} &=\frac{\tilde U}{2E_{A,msr}}a_3^{msr}
=\frac{\tilde U}{2\Gamma_{m-1} E_{A,msr}}
\left(E_{A,msr}-\frac{m}{E_{A,msr}}\right).\\
N_{A,msr}&=\left[
1+\frac{m}{E_{A,msr}^2}
+\frac{1}{\Gamma_{m-1}^2}\left(E_{A,msr}-\frac{m}{E_{A,msr}}\right)^2
+\left(\frac{\tilde U}{2\Gamma_{m-1} E_{A,msr}}\right)^2
\left(E_{A,msr}-\frac{m}{E_{A,msr}}\right)^2
\right]^{-1/2}.
\end{align}
 For the second block in Eq.(\ref{eq:rotChiral}), which we denote
block $B$, we can proceed analogously as in the previous case. Solving for the different coefficients of the eigenstates and using the same substitutions as for block A, we find an equation for
the energies given by 
\begin{equation}
E_{B,msr}=s\sqrt{x_{Bmr}}
\label{eqn:anaspecchiralflatB}
\end{equation}
where 
\begin{equation}
x_{B,mr}=\frac{1}{2}\left[
 m+\Gamma_m^2+\frac{\tilde U^2}{4}
+r\sqrt{\left(m+\Gamma_m^2+\frac{\tilde U^2}{4}\right)^2
-\tilde U^2m}\right].
\label{eq:xB}
\end{equation}
with $s,r=\pm$. Note, in the limit $\Gamma_m\rightarrow1$, neglecting the momentum-dependent hybridization,  we find $x_{Bmr}=x_{Amr}$, meaning the two chiral sectors become degenerate for $m\geq1$. The eigenstates for this block in the unrotated basis
are then given by 
\begin{equation}
\psi_{B,msr}=N_{B,msr}\left(\begin{array}{cccccccc}
0 & b_{1}^{msr}|m\rangle & 0 & b_{2}^{msr}|m-1\rangle & 0 & b_{3}^{msr}|m\rangle & 0 & b_{4}^{msr}|m\rangle\end{array}\right)
\end{equation}
where the coefficients and the normalization are obtained readily
as follows:
\begin{align}
b_{1}^{msr} &=1,\\
b_{2}^{msr} &=\frac{-i\sqrt{m}}{E_{B,msr}},\\
b_{3}^{msr} &=\frac{1}{\Gamma_m}
\left(E_{B,msr}-\frac{m}{E_{B,msr}}\right),\\
b_{4}^{msr} &=\frac{\tilde U}{2E_{B,msr}}b_3^{msr}
=\frac{\tilde U}{2\Gamma_m E_{B,msr}}
\left(E_{B,msr}-\frac{m}{E_{B,msr}}\right).\\
N_{B,msr}&=\left[
1+\frac{m}{E_{B,msr}^2}
+\frac{1}{\Gamma_m^2}\left(E_{B,msr}-\frac{m}{E_{B,msr}}\right)^2
+\left(\frac{\tilde U}{2\Gamma_m E_{B,msr}}\right)^2
\left(E_{B,msr}-\frac{m}{E_{B,msr}}\right)^2
\right]^{-1/2}.
\end{align}
We can verify that as $\tilde{U}/\gamma\rightarrow\infty$ the spectrum
decouples into a double Dirac node and a set of flat LLs at energy
$\pm U/2$. Overall, we can then use these expressions to find the
Green's function and the spectral function in the chiral flat limit.
We can also use these expressions as a starting point for perturbation
theory away from the chiral-flat limit (see Appendix(\ref{sec:pertawaychiral})).
We additionally consider the set of $m=0$ states in both chiral sectors.

\subsubsection{Reduced-rank states and energies at \texorpdfstring{$m=0$}{m=0}}

For $m=0$ we can start directly from the Hamiltonian in Eqn.(\ref{eq:chiralAuxB}) setting $v^\prime_\ast=M=0$. The first zero mode is obtained from the rank-1 action of the Hamiltonian on: 
\begin{equation}
\psi_{A,0,0}
=
\begin{pmatrix}
0&
0&
|0\rangle &
0 &
0 &
0 &
0 &
0
\end{pmatrix},
\qquad
E_{A,0,0}=0.
\label{eq:Azero-anomalous-state}
\end{equation}
The last label $0$ is not a band index.  It only keeps track that this state has
zero unperturbed energy and has no $s,r$ labels.

The reduced-rank $B$-sector states at $m=0$ are written as
\begin{equation}
\psi_{B,0,\lambda}
=
N_{B,0,\lambda}
\begin{pmatrix}
0 &
\beta_1^\lambda |0\rangle &
0 &
0 &
0 &
\beta_3^\lambda |0\rangle &
0 &
\beta_4^\lambda |0\rangle
\end{pmatrix}.
\label{eq:Bzero-reduced-rank-state}
\end{equation}
Here, $\lambda$ labels the three eigenstates of the reduced-rank block.  The
coefficients obey
\begin{align}
E_{B,0,\lambda}\beta_1^\lambda
&=
\Gamma_0\beta_3^\lambda,
\\
E_{B,0,\lambda}\beta_3^\lambda
&=
\Gamma_0\beta_1^\lambda+\frac{\tilde U}{2}\beta_4^\lambda,
\\
E_{B,0,\lambda}\beta_4^\lambda
&=
\frac{\tilde U}{2}\beta_3^\lambda,
\end{align}
with
\begin{equation}
N_{B,0,\lambda}
=
\left(
|\beta_1^\lambda|^2
+
|\beta_3^\lambda|^2
+
|\beta_4^\lambda|^2
\right)^{-1/2}.
\end{equation}
For the two finite-energy states one may choose
\begin{equation}
\lambda=(s,+),
\qquad
E_{B,0,s,+}
=
s\sqrt{\Gamma_0^2+\frac{\tilde U^2}{4}},
\qquad
\beta_1^{s,+}=1,
\qquad
\beta_3^{s,+}
=
\frac{E_{B,0,s,+}}{\Gamma_0},
\qquad
\beta_4^{s,+}
=
\frac{\tilde U}{2\Gamma_0}.
\end{equation}
There this corresponds to the limit $M=v^\prime_\ast=0$ of the second zero mode found in Eqn.(\ref{eq:zeromodenonchiral2}).  For $\tilde U\ne0$ we obtain
\begin{equation}
\lambda=z,
\qquad
E_{B,0,z}=0,
\qquad
\beta_1^z=1,
\qquad
\beta_3^z=0,
\qquad
\beta_4^z=-\frac{2\Gamma_0}{\tilde U}.
\end{equation}
The set of reduced-rank $B$ states will be denoted by
\begin{equation}
\mathcal I_0=\{(+,+),(-,+),z\}.
\end{equation}

\begin{figure}
    \centering
    \includegraphics[width=0.8\linewidth]{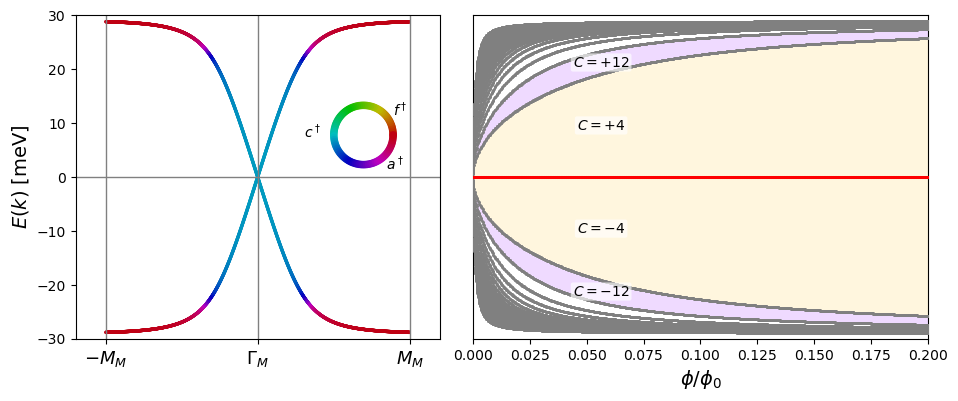}
    \caption{\textbf{Mott semimetal:} Location of the poles of the Mott semimetal Green's function in the chiral-flat limit with parameters taken from Tab.(\ref{tab:model_params}) at $B=0$ (a) and $B>0$ (b) at total electron filling $\nu=0$. The finite field spectra is obtained analytically from Eq.(\ref{eqn:anaspecchiralflatA}) and Eq.(\ref{eqn:anaspecchiralflatB}).The color wheel in (a) denotes the fermion character at each k-point, obtained from the expectation value of $Z_{\rm char} = e^{i\pi}\sigma_0 \oplus e^{i\pi}\sigma_0 \oplus e^{i\pi/3}\sigma_0 \oplus e^{-i\pi/3}\sigma_0$ such that $c-$electrons are colored cyan, $f$-electrons are colored yellow, and $a-$electrons are colored magenta. The red line on (b) denotes the two zero modes in the Landau-level spectrum stemming from the doubly degenerate Dirac points at $\Gamma_M$ in the $B=0$ spectra.}
    \label{fig:exact_chiral}
\end{figure}

\subsubsection{Green's function in the chiral limit without strain for \texorpdfstring{$m\geq 1$}{m>=1} states}

Now that we have calculated the dispersion and eigenstates in the
Landau basis exactly for the auxiliary fermion system, we can then
calculate the Green's function of the physical electrons from the
Green's function of the auxiliary system. Since the system can be block
diagonalized as we did above, we can simply calculate the Green's function
for each $4\times4$ block and take the $3\times3$ block that corresponds
to the physical electrons. This can be easily done by calculating
the Green's function in the Landau basis starting with the top block,
which we denote block $A$, we have 
\begin{align}
\left(\begin{array}{cccc}
0 & -ia & \tilde{\gamma}\Sigma\left(\hat{a}^{\dagger}\hat{a}\right) & 0\\
ia^{\dagger} & 0 & 0 & 0\\
\tilde{\gamma}\Sigma\left(\hat{a}^{\dagger}\hat{a}\right) & 0 & 0 & \frac{\tilde{U}}{2}\\
0 & 0 & \frac{\tilde{U}}{2} & 0
\end{array}\right)\left(\begin{array}{c}
a_{1}|m-1\rangle\\
a_{2}|m\rangle\\
a_{3}|m-1\rangle\\
a_{4}|m-1\rangle
\end{array}\right) & =\left(\begin{array}{cccc}
0 & -i\sqrt{m} & \Gamma_{m-1} & 0\\
i\sqrt{m} & 0 & 0 & 0\\
\Gamma_{m-1} & 0 & 0 & \frac{\tilde{U}}{2}\\
0 & 0 & \frac{\tilde{U}}{2} & 0
\end{array}\right)\left(\begin{array}{c}
a_{1}|m-1\rangle\\
a_{2}|m\rangle\\
a_{3}|m-1\rangle\\
a_{4}|m-1\rangle
\end{array}\right)\\
 & =H_{m}^{A}\left(\begin{array}{c}
a_{1}|m-1\rangle\\
a_{2}|m\rangle\\
a_{3}|m-1\rangle\\
a_{4}|m-1\rangle
\end{array}\right)
\end{align}
The Hamiltonian is thus block diagonal in the Landau level index $m$
for $m\geq0$ and we can calculate the Green's function for the auxiliary
system as
\begin{align}
\mathcal{G}_{m}^{A,\text{aux}}\left(E\right) & =\left(E\mathbb{I}-H_{m}^{A}\right)^{-1}\\
 & =\frac{1}{E^{4}-\left( m+\Gamma_{m-1}^{2}+\frac{\tilde{U}^{2}}{4}\right)E^{2}+\frac{\tilde{U}^{2}}{4} m}\times\\
 & \times\left(\begin{array}{cccc}
E\left(E^{2}-\frac{\tilde{U}^{2}}{4}\right) & \frac{1}{4}i\sqrt{m}\left(4E^{2}-\tilde{U}^{2}\right) & \Gamma_{m-1}E^{2} & \frac{\Gamma_{m-1}E\tilde{U}}{2}\\
-\frac{1}{4}i\sqrt{m}\left(4E^{2}-\tilde{U}^{2}\right) & E\left[E^{2}-\frac{1}{4}\left(4\Gamma_{m-1}^{2}+\tilde{U}^{2}\right)\right] & -i\Gamma_{m-1}E\sqrt{m} & -\frac{1}{2}i\Gamma_{m-1}\sqrt{m}\tilde{U}\\
\Gamma_{m-1}E^{2} & i\Gamma_{m-1}E\sqrt{m} & E\left(E^{2}- m\right) & \frac{1}{2}\tilde{U}\left(E^{2}-m-1\right)\\
\frac{\Gamma_{m-1}E\tilde{U}}{2} & \frac{1}{2}i\Gamma_{m-1}\sqrt{m}\tilde{U} & \frac{1}{2}U\left(E^{2}-m-1\right) & E\left(-\Gamma_{m-1}^{2}+E^{2}-m-1\right)
\end{array}\right)
\end{align}
The Green's function of the physical fermions corresponds to the top
$3\times3$ block above which becomes 
\begin{align}
\mathcal{G}_{m}^{A}\left(E\right)\\
 & =\frac{1}{E^{4}-\left( m+\Gamma_{m-1}^{2}+\frac{\tilde{U}^{2}}{4}\right)E^{2}+\frac{\tilde{U}^{2}}{4} m}\times\\
 & \times\left(\begin{array}{ccc}
E\left(E^{2}-\frac{\tilde{U}^{2}}{4}\right) & \frac{1}{4}i\sqrt{m}\left(4E^{2}-\tilde{U}^{2}\right) & \Gamma_{m-1}E^{2}\\
-\frac{1}{4}i\sqrt{m}\left(4E^{2}-\tilde{U}^{2}\right) & E\left[E^{2}-\frac{1}{4}\left(4\Gamma_{m-1}^{2}+\tilde{U}^{2}\right)\right] & -i\Gamma_{m-1}E\sqrt{m}\\
\Gamma_{m-1}E^{2} & i\Gamma_{m-1}E\sqrt{m} & E\left(E^{2}- m\right)
\end{array}\right)
\end{align}
We can proceed in an analogous way with the bottom block, which we
denote block $B$, we find 
\begin{align}
\left(\begin{array}{cccc}
0 & ia^{\dagger} & \tilde{\gamma}\Sigma\left(a^\dagger a\right) & 0\\
-ia & 0 & 0 & 0\\
\tilde{\gamma}\Sigma\left(a^\dagger a\right) & 0 & 0 & \frac{\tilde{U}}{2}\\
0 & 0 & \frac{\tilde{U}}{2} & 0
\end{array}\right)\left(\begin{array}{c}
b_{1}|m\rangle\\
b_{2}|m-1\rangle\\
b_{3}|m\rangle\\
b_{4}|m\rangle
\end{array}\right) & =\left(\begin{array}{cccc}
0 & i\sqrt{m} & \Gamma_{m} & 0\\
-i\sqrt{m} & 0 & 0 & 0\\
\Gamma_{m} & 0 & 0 & \frac{\tilde{U}}{2}\\
0 & 0 & \frac{\tilde{U}}{2} & 0
\end{array}\right)\left(\begin{array}{c}
b_{1}|m\rangle\\
b_{2}|m-1\rangle\\
b_{3}|m\rangle\\
b_{4}|m\rangle
\end{array}\right)\\
 & =H_{m}^{B}\left(\begin{array}{c}
b_{1}|m\rangle\\
b_{2}|m-1\rangle\\
b_{3}|m\rangle\\
b_{4}|m\rangle
\end{array}\right)
\end{align}
In this case, the Hamiltonian is block diagonal in the Landau level
index $m$ for $m\geq1$. Calculating the auxiliary Green's function
for the $B$ block and isolating the physical electron sector of the
Green's function, we obtain:
\begin{align}
\mathcal{G}_{m}^{B}\left(E\right) & =\left(E\mathbb{I}-H_{m}^{B}\right)^{-1}\\
 & =\frac{1}{E^{4}-\left(m+\Gamma_{m}^{2}+\frac{\tilde{U}^{2}}{4}\right)E^{2}+\frac{\tilde{U}^{2}}{4}m}\times\\
 & \times\left(\begin{array}{ccc}
E\left(E^{2}-\frac{\tilde{U}^{2}}{4}\right) & -\frac{1}{4}i\sqrt{m}\left(4E^{2}-\tilde{U}^{2}\right) & \Gamma_{m}E^{2}\\
\frac{1}{4}i\sqrt{m}\left(4E^{2}-\tilde{U}^{2}\right) & E\left[E^{2}-\frac{1}{4}\left(4\Gamma_{m}^{2}+\tilde{U}^{2}\right)\right] & i\Gamma_{m}E\sqrt{m}\\
\Gamma_{m}E^{2} & -i\Gamma_{m}E\sqrt{m} & E\left(E^{2}-m\right)
\end{array}\right).
\end{align}
\subsubsection{Green's function in the chiral limit without strain for \texorpdfstring{$m= 0$}{m=0} states}
We can additionally consider the Green's functions for the reduced-rank states. For the $A$ sector we only have a zero energy state such that
\begin{equation}
    \mathcal{G}_{0}^{A} (E)= 1/E
\end{equation}
For the reduced-rank states of the $B$ sector, we instead have the action of the Hamiltonian in the $3\times 3$
\begin{equation}
    \mathcal{G}_{0}^{B,\mathrm{aux}}(E)=\frac{1}{E^3-\frac{1}{4} E \left(4 \Gamma_{m} ^2+\tilde{U}^2\right)}\left(
\begin{array}{ccc}
 E^2-\frac{\tilde{U}^2}{4} & \Gamma_{m}  E & \frac{\Gamma_{m}  \tilde{U}}{2} \\
 \gamma  E & E^2 & \frac{E \tilde{U}}{2} \\
 \frac{\Gamma_{m}  \tilde{U}}{2} & \frac{E \tilde{U}}{2} & E^2-\Gamma_{m} ^2 \\
\end{array}
\right),
\end{equation}
such that projecting back to the $c-$ and $f-$ electron degrees of freedom we obtain
\begin{equation}
    \mathcal{G}_{0}^{B}(E)=\frac{1}{E^3-\frac{1}{4} E \left(4 \Gamma_{m} ^2+\tilde{U}^2\right)}\left(
\begin{array}{ccc}
 E^2-\frac{\tilde{U}^2}{4} & \Gamma_{m}  E \\
 \Gamma_{m}  E & E^2   \\
\end{array}
\right).
\end{equation}
In $\mathcal{G}_{0}^{B}(E)$ the Green's function has support only on the $\Gamma_3$ $c-$ fermions and the $f$ electrons. Meanwhile, $\mathcal{G}_{0}^{A} (E)$ is exclusively supported on the $\Gamma_1\oplus \Gamma_2$ $c-$fermions. 
\subsubsection{Spectral functions}
Note that we can obtain the spectral function directly from the auxiliary
fermion Hamiltonian, 
\begin{align}
\mathcal{A}_{\alpha mA}\left(E\right) & =-\text{Im}\left[\mathcal{G}_{m}^{A,\text{aux}}\left(E\right)\right]_{\alpha\alpha}=-\text{Im}\left[\mathcal{G}_{m}^{A}\left(E\right)\right]_{\alpha\alpha}\\
 & =\sum_{sr}\delta\left(E-E_{A,msr}\right)\left|a_{\alpha}^{msr}\right|^{2}
\end{align}
Similarly, we can calculate the orbitally resolved spectral function
for the $B$ sector. Note that the projection onto the physical $c$ and $f$ electrons is enforced by the fact that the index $\alpha$ only runs over the physical degrees of freedom, i.e, omitting the auxiliary fermion indices (also see Appendix(\ref{Appx:Spectral_map_color})).

\section{Perturbation theory away from the flat-chiral limit}\label{sec:pertawaychiral}
In this section we calculate the second order change in the energy away from the chiral-flat limit shown in Sec.{\ref{sec:Landau-Level-spectrum}}. To simplify the analysis, we do not consider relaxation, strain or Zeeman effects in this section and focus on the simplest terms deviating from the chiral-flat limit omited in Sec.{\ref{sec:Landau-Level-spectrum}}, $v_*^\prime$ and $M$. We start from the Hubbard-I Hamiltonian in a magnetic field shown in Eqn.(\ref{Appx:Eq:Hubbar_I_magnetic_after_aux}). With these conventions, we split this Hamiltonian as
\begin{equation}
\tilde{h}_{i,i',\eta=+}^{\text{MF}}\left(B\right)= \tilde{h}_{i,i',\eta=+,0}^{\text{MF}}\left(B\right)+\tilde{W}_{i,i',\eta=+}\left(B\right),
\qquad
\tilde{W}_{\eta=+}\left(B\right)=\tilde M_{\rm op}+\tilde{V}_{\rm op,\eta=+}\left(B\right), 
\end{equation}
where we define 
\begin{eqnarray}
    \tilde{W}_{i,i',\eta=+}\left(B\right)=\left(\begin{array}{cccccccc}
0 & 0 & 0 & 0 & 0 & i\tilde{\Delta}^{\prime}\hat{a}^{\dagger}\Sigma(\hat{a}^{\dagger}\hat{a}) & 0 & 0\\
0 & 0 & 0 & 0 & -i\tilde{\Delta}^{\prime}\hat{a}\Sigma(\hat{a}^{\dagger}\hat{a}) & 0 & 0 & 0\\
0 & 0 & 0 & \tilde{M} & 0 & 0 & 0 & 0\\
0 & 0 & \tilde{M} & 0 & 0 & 0 & 0 & 0\\
0 & i\tilde{\Delta}^{\prime}\Sigma(\hat{a}^{\dagger}\hat{a})\hat{a}^{\dagger} & 0 & 0 & 0 & 0 & 0 & 0\\
-i\tilde{\Delta}^{\prime}\Sigma(\hat{a}^{\dagger}\hat{a})\hat{a} & 0 & 0 & 0 & 0 & 0 & 0 & 0\\
0 & 0 & 0 & 0 & 0 & 0 & 0 & 0\\
0 & 0 & 0 & 0 & 0 & 0 & 0 & 0
\end{array}\right)_{i,i'}.\label{eq:full-dimensionless-H}
\end{eqnarray}
In the above, we define $\tilde M_{\rm op}$ as the matrix that includes entries proportional to $M$ in $\tilde{W}_{i,i',\eta=+}\left(B\right)$. Similarly we define $\tilde{V}_{\rm op,\eta=+}\left(B\right)$ corresponds to the matrix with entries proportional to $\tilde{\Delta}^\prime$.

\subsection{Chiral eigenstates with the \texorpdfstring{$\Sigma(m)$}{Sigma(m)} dependence included}

For $\tilde M=\tilde\Delta'=0$, the Hamiltonian preserves the two chiral
sectors.  We first write the ordinary rank-four states, valid for $m>0$, and
then treat the reduced-rank states with $m=0$. First we calculate the effect of a finite $M$. Then we investigate the effect of including a chiral-symmetry breaking $v^\prime_\ast$.

\subsection{Matrix elements of the \texorpdfstring{$M$}{M} perturbation}
Since the mass term connects matrix elements that reside in different blocks of the Hamiltonian in Eqn.(\ref{eq:chiralAuxB}) the
diagonal matrix elements vanish identically when taking the expectation value on the states in Eqn(\ref{eqn:Aspinorsol}):
\begin{equation}
M_{AA,msr,m's'r'}=0,
\qquad
M_{BB,msr,m's'r'}=0.
\end{equation}
\subsubsection{Rank-four mass matrix elements}

The off-diagonal elements are
\begin{align}
M_{AB,msr,m's'r'}
&\equiv \langle \psi_{A,msr}|\tilde M_{\rm op}|\psi_{B,m's'r'}\rangle\\
&=\tilde M\,N_{A,msr}N_{B,m's'r'}
\overline{a_2^{msr}}\,b_2^{m's'r'}\,\delta_{m,m'-1},
\label{eq:MAB}
\end{align}
and
\begin{align}
M_{BA,msr,m's'r'}
&\equiv \langle \psi_{B,msr}|\tilde M_{\rm op}|\psi_{A,m's'r'}\rangle\\
&=\tilde M\,N_{B,msr}N_{A,m's'r'}
\overline{b_2^{msr}}\,a_2^{m's'r'}\,\delta_{m-1,m'}.
\label{eq:MBA}
\end{align}
Therefore the selection rules are
\begin{equation}
M_{AB}\neq0 \Longleftrightarrow m'=m+1,
\qquad
M_{BA}\neq0 \Longleftrightarrow m'=m-1.
\end{equation}

\subsubsection{Reduced-rank mass matrix elements}

The mass perturbation connects the anomalous $A$ state only to the ordinary
$B$ block with $m=1$:
\begin{align}
M_{A,0,0;B,msr}
&\equiv
\langle \psi_{A,0,0}|\tilde M_{\rm op}|\psi_{B,msr}\rangle
=
\tilde M\,N_{B,msr}\,b_2^{msr}\,\delta_{m,1},
\\
M_{B,msr;A,0,0}
&\equiv
\langle \psi_{B,msr}|\tilde M_{\rm op}|\psi_{A,0,0}\rangle
=
\tilde M\,N_{B,msr}\,\overline{b_2^{msr}}\,\delta_{m,1}.
\label{eq:M-anomalous-A-to-B}
\end{align}
The reduced-rank $B$ states at $m=0$ have no component $4$.  Therefore the
mass perturbation does not connect them to the ordinary $A$-sector states:
\begin{equation}
M_{A,msr;B,0,\lambda}=0,
\qquad
M_{B,0,\lambda;A,msr}=0 .
\label{eq:M-reduced-B-zero}
\end{equation}
The corresponding absolute squares are
\begin{align}
|M_{A,0,0;B,1sr}|^2
&=
\tilde M^2 N_{B,1sr}^2 |b_2^{1sr}|^2,
\\
|M_{B,1sr;A,0,0}|^2
&=
\tilde M^2 N_{B,1sr}^2 |b_2^{1sr}|^2 .
\end{align}

\subsection{Matrix elements of the \texorpdfstring{$\tilde\Delta'$}{tildeDelta'} perturbation}

The chiral-breaking perturbation is the part of Eq.~\eqref{eq:full-dimensionless-H}
proportional to $\tilde\Delta'$.  It is also off diagonal between the two
chiral sectors, so
\begin{equation}
V_{AA,msr,m's'r'}=0,
\qquad
V_{BB,msr,m's'r'}=0.
\end{equation}

\subsubsection{Rank-four \texorpdfstring{$\tilde\Delta'$}{tildeDelta'} matrix elements}

For the $A\leftarrow B$ matrix element, applying $\tilde{V}_{\rm op,\eta=+}\left(B\right)$ to
$\psi_{B,m's'r'}$ gives nonzero components only in the $A$ sector:
\begin{align}
V_{AB,msr,m's'r'}
&\equiv \langle \psi_{A,msr}|\tilde{V}_{\rm op,\eta=+}\left(B\right)|\psi_{B,m's'r'}\rangle\\
&=i\tilde\Delta'\sqrt{m'+1}\,\delta_{m-1,m'+1}
N_{A,msr}N_{B,m's'r'}
\left[
\Sigma(m')b_3^{m's'r'}\overline{a_1^{msr}}
+\Sigma(m'+1)b_1^{m's'r'}\overline{a_3^{msr}}
\right].
\label{eq:VAB}
\end{align}
For the reverse matrix element,
\begin{align}
V_{BA,msr,m's'r'}
&\equiv \langle \psi_{B,msr}|\tilde{V}_{\rm op,\eta=+}\left(B\right)|\psi_{A,m's'r'}\rangle\\
&=-i\tilde\Delta'\sqrt{m'-1}\,\delta_{m,m'-2}
N_{B,msr}N_{A,m's'r'}
\left[
\Sigma(m'-1)a_3^{m's'r'}\overline{b_1^{msr}}
+\Sigma(m'-2)a_1^{m's'r'}\overline{b_3^{msr}}
\right].
\label{eq:VBA}
\end{align}
Equivalently, using the on-shell selection rules,
\begin{equation}
V_{AB}\neq0 \Longleftrightarrow m'=m-2,
\qquad
V_{BA}\neq0 \Longleftrightarrow m'=m+2.
\end{equation}

\subsubsection{Reduced-rank \texorpdfstring{$\tilde\Delta'$}{tildeDelta'} matrix elements}

The $\tilde\Delta'$ perturbation does not connect to
$|\psi_{A,0,0}\rangle$, because this state has support only on component
$3$:
\begin{equation}
V_{A,0,0;B,msr}=0,
\qquad
V_{B,msr;A,0,0}=0 .
\label{eq:V-anomalous-A-zero}
\end{equation}
The $\tilde\Delta'$ perturbation does connect the reduced-rank $B_0$ states to
the ordinary $A$-sector block, but only to $m=2$.  The two directions are
\begin{align}
V_{A,msr;B,0,\lambda}
&\equiv
\langle \psi_{A,msr}|\tilde{V}_{\rm op,\eta=+}\left(B\right)|\psi_{B,0,\lambda}\rangle
\nonumber\\
&=
i\tilde\Delta'\,N_{A,msr}N_{B,0,\lambda}\,\delta_{m,2}
\left[
\Sigma(0)\beta_3^\lambda\overline{a_1^{msr}}
+
\Sigma(1)\beta_1^\lambda\overline{a_3^{msr}}
\right],
\\
V_{B,0,\lambda;A,msr}
&\equiv
\langle \psi_{B,0,\lambda}|\tilde{V}_{\rm op,\eta=+}\left(B\right)|\psi_{A,msr}\rangle
\nonumber\\
&=
-i\tilde\Delta'\,N_{B,0,\lambda}N_{A,msr}\,\delta_{m,2}
\left[
\Sigma(1)a_3^{msr}\overline{\beta_1^\lambda}
+
\Sigma(0)a_1^{msr}\overline{\beta_3^\lambda}
\right].
\label{eq:V-reduced-B-to-A}
\end{align}
These expressions reduce to each other by Hermitian conjugation when
$\tilde\Delta'$, $\Sigma(m)$, and the unperturbed coefficients are real.  The
corresponding absolute squares are
\begin{align}
|V_{A,2sr;B,0,\lambda}|^2
&=
(\tilde\Delta')^2
N_{A,2sr}^2N_{B,0,\lambda}^2
\left|
\Sigma(0)\beta_3^\lambda\overline{a_1^{2sr}}
+
\Sigma(1)\beta_1^\lambda\overline{a_3^{2sr}}
\right|^2,
\\
|V_{B,0,\lambda;A,2sr}|^2
&=
(\tilde\Delta')^2
N_{B,0,\lambda}^2N_{A,2sr}^2
\left|
\Sigma(1)a_3^{2sr}\overline{\beta_1^\lambda}
+
\Sigma(0)a_1^{2sr}\overline{\beta_3^\lambda}
\right|^2 .
\end{align}

\subsection{Second-order perturbation theory}

Because both perturbations are off diagonal in the chiral-sector label,
all first-order energy corrections vanish:
\begin{equation}
\delta E_{A,msr}^{(1)}=0,
\qquad
\delta E_{B,msr}^{(1)}=0.
\end{equation}
For nonresonant levels, the dimensionless second-order shift is
\begin{equation}
\delta E_n^{(2)}=\sum_{\ell\neq n}
\frac{|\langle \ell|\tilde W|n\rangle|^2}{E_n^{(0)}-E_\ell^{(0)}}.
\end{equation}
The physical shift is obtained only at the end:
\begin{equation}
\delta\varepsilon_n^{(2)}=\Delta_B\,\delta E_n^{(2)}.
\end{equation}

It is useful to define the on-shell matrix elements after imposing the
selection rules:
\begin{align}
\mathcal M^A_{msr;s'r'}
&\equiv M_{AB,msr,m+1,s'r'}
=\tilde M\,N_{A,msr}N_{B,m+1,s'r'}
\overline{a_2^{msr}}\,b_2^{m+1,s'r'},\\
\mathcal V^A_{msr;s'r'}
&\equiv V_{AB,msr,m-2,s'r'}\\
&=i\tilde\Delta'\sqrt{m-1}\,
N_{A,msr}N_{B,m-2,s'r'}
\left[
\Sigma(m-2)b_3^{m-2,s'r'}\overline{a_1^{msr}}
+\Sigma(m-1)b_1^{m-2,s'r'}\overline{a_3^{msr}}
\right],\\
\mathcal M^B_{msr;s'r'}
&\equiv M_{BA,msr,m-1,s'r'}
=\tilde M\,N_{B,msr}N_{A,m-1,s'r'}
\overline{b_2^{msr}}\,a_2^{m-1,s'r'},\\
\mathcal V^B_{msr;s'r'}
&\equiv V_{BA,msr,m+2,s'r'}\\
&=-i\tilde\Delta'\sqrt{m+1}\,
N_{B,msr}N_{A,m+2,s'r'}
\left[
\Sigma(m+1)a_3^{m+2,s'r'}\overline{b_1^{msr}}
+\Sigma(m)a_1^{m+2,s'r'}\overline{b_3^{msr}}
\right].
\end{align}
The terms with negative oscillator labels are omitted.  We encode this by
\begin{equation}
\chi_{m\ge m_0}=\begin{cases}
1,&m\ge m_0,\\
0,&m<m_0.
\end{cases}
\end{equation}

For an ordinary $A$-sector state, the second-order energy correction is
\begin{align}
\delta E_{A,msr}^{(2)}
&=\sum_{s',r'}
\frac{|\mathcal M^A_{msr;s'r'}|^2}
{E_{A,msr}-E_{B,m+1,s'r'}}
+\chi_{m\ge3}\sum_{s',r'}
\frac{|\mathcal V^A_{msr;s'r'}|^2}
{E_{A,msr}-E_{B,m-2,s'r'}}.
\label{eq:second-A-compact}
\end{align}
For a $B$-sector state,
\begin{align}
\delta E_{B,msr}^{(2)}
&=\chi_{m\ge2}\sum_{s',r'}
\frac{|\mathcal M^B_{msr;s'r'}|^2}
{E_{B,msr}-E_{A,m-1,s'r'}}
+\sum_{s',r'}
\frac{|\mathcal V^B_{msr;s'r'}|^2}
{E_{B,msr}-E_{A,m+2,s'r'}}.
\label{eq:second-B-compact}
\end{align}
There is no mixed $\tilde M\tilde\Delta'$ term in the diagonal
second-order energy correction.  For an $A_m$ state, $\tilde M$ connects
to $B_{m+1}$ whereas $\tilde V$ connects to $B_{m-2}$.  For a $B_m$
state, $\tilde M$ connects to $A_{m-1}$ whereas $\tilde V$ connects to
$A_{m+2}$.  Thus the two perturbations never reach the same intermediate
state in the second-order diagonal sum.

Using $a_1=b_1=1$ and the explicit forms of $a_2$ and $b_2$, the absolute
squares are
\begin{align}
|\mathcal M^A_{msr;s'r'}|^2
&=\tilde M^2 N_{A,msr}^2N_{B,m+1,s'r'}^2
\frac{m(m+1)}{E_{A,msr}^2E_{B,m+1,s'r'}^2},\\
|\mathcal V^A_{msr;s'r'}|^2
&=(\tilde\Delta')^2 (m-1)\,N_{A,msr}^2N_{B,m-2,s'r'}^2
\left|
\Sigma(m-2)b_3^{m-2,s'r'}+\Sigma(m-1)\overline{a_3^{msr}}
\right|^2,\\
|\mathcal M^B_{msr;s'r'}|^2
&=\tilde M^2 N_{B,msr}^2N_{A,m-1,s'r'}^2
\frac{m(m-1)}{E_{B,msr}^2E_{A,m-1,s'r'}^2},\\
|\mathcal V^B_{msr;s'r'}|^2
&=(\tilde\Delta')^2 (m+1)N_{B,msr}^2N_{A,m+2,s'r'}^2
\left|
\Sigma(m+1)a_3^{m+2,s'r'}+\Sigma(m)\overline{b_3^{msr}}
\right|^2.
\end{align}
Since $a_3$ and $b_3$ are real for real $\Sigma$, $\tilde\gamma$, $\tilde U$,
and $E$, the bars in the last two equations may be dropped in that case.

The fully explicit dimensionless shifts are therefore
\begin{align}
\delta E_{A,msr}^{(2)}
&=\sum_{s',r'}
\frac{\tilde M^2 N_{A,msr}^2N_{B,m+1,s'r'}^2}
{E_{A,msr}-E_{B,m+1,s'r'}}
\frac{m(m+1)}{E_{A,msr}^2E_{B,m+1,s'r'}^2}
\nonumber\\
&\quad+\chi_{m\ge3}\sum_{s',r'}
\frac{(\tilde\Delta')^2 (m-1)\,N_{A,msr}^2N_{B,m-2,s'r'}^2}
{E_{A,msr}-E_{B,m-2,s'r'}}
\left|
\Sigma(m-2)b_3^{m-2,s'r'}+\Sigma(m-1)\overline{a_3^{msr}}
\right|^2,
\label{eq:second-A-explicit}
\end{align}
and
\begin{align}
\delta E_{B,msr}^{(2)}
&=\chi_{m\ge2}\sum_{s',r'}
\frac{\tilde M^2 N_{B,msr}^2N_{A,m-1,s'r'}^2}
{E_{B,msr}-E_{A,m-1,s'r'}}
\frac{m(m-1)}{E_{B,msr}^2E_{A,m-1,s'r'}^2}
\nonumber\\
&\quad+\sum_{s',r'}
\frac{(\tilde\Delta')^2 (m+1)N_{B,msr}^2N_{A,m+2,s'r'}^2}
{E_{B,msr}-E_{A,m+2,s'r'}}
\left|
\Sigma(m+1)a_3^{m+2,s'r'}+\Sigma(m)\overline{b_3^{msr}}
\right|^2.
\label{eq:second-B-explicit}
\end{align}

At the bottom of the oscillator ladder, the second-order formulas must be
supplemented by the reduced-rank intermediate states displayed above.  The
ordinary sums over $B_{m-2}$ and $A_{m-1}$ should be understood as sums over
rank-four blocks only; the missing boundary blocks are added separately.  For an
ordinary $A$-sector state with $m\ge1$,
\begin{align}
\delta E_{A,msr}^{(2)}
&=
\sum_{s',r'}
\frac{|\mathcal M^A_{msr;s'r'}|^2}
{E_{A,msr}-E_{B,m+1,s'r'}}
\nonumber\\
&\quad+
\chi_{m\ge3}
\sum_{s',r'}
\frac{|\mathcal V^A_{msr;s'r'}|^2}
{E_{A,msr}-E_{B,m-2,s'r'}}
\nonumber\\
&\quad+
\delta_{m,2}
\sum_{\lambda\in\mathcal I_0}
\frac{|V_{A,2sr;B,0,\lambda}|^2}
{E_{A,2sr}-E_{B,0,\lambda}} .
\label{eq:second-A-with-reduced-B}
\end{align}
For an ordinary $B$-sector state with $m\ge1$,
\begin{align}
\delta E_{B,msr}^{(2)}
&=
\chi_{m\ge2}
\sum_{s',r'}
\frac{|\mathcal M^B_{msr;s'r'}|^2}
{E_{B,msr}-E_{A,m-1,s'r'}}
\nonumber\\
&\quad+
\delta_{m,1}
\frac{|M_{B,1sr;A,0,0}|^2}
{E_{B,1sr}-E_{A,0,0}}
\nonumber\\
&\quad+
\sum_{s',r'}
\frac{|\mathcal V^B_{msr;s'r'}|^2}
{E_{B,msr}-E_{A,m+2,s'r'}} .
\label{eq:second-B-with-anomalous-A}
\end{align}
Finally, the reduced-rank states themselves have the shifts
\begin{align}
\delta E_{A,0,0}^{(2)}
&=
\sum_{s',r'}
\frac{|M_{A,0,0;B,1s'r'}|^2}
{E_{A,0,0}-E_{B,1s'r'}},
\\
\delta E_{B,0,\lambda}^{(2)}
&=
\sum_{s',r'}
\frac{|V_{B,0,\lambda;A,2s'r'}|^2}
{E_{B,0,\lambda}-E_{A,2s'r'}} .
\label{eq:second-reduced-rank-external}
\end{align}
Equations~\eqref{eq:second-A-with-reduced-B} and
\eqref{eq:second-B-with-anomalous-A} are the supplemented versions of the
ordinary second-order formulas.  The $m=2$ term in
Eq.~\eqref{eq:second-A-with-reduced-B} replaces the formal $B_{0,s'r'}$ sum,
and the $m=1$ term in Eq.~\eqref{eq:second-B-with-anomalous-A} is the missing
coupling to the anomalous $A_{0}$ zero mode.

The physical corrections are
\begin{equation}
\delta\varepsilon_{A,msr}^{(2)}=\Delta_B\delta E_{A,msr}^{(2)},
\qquad
\delta\varepsilon_{B,msr}^{(2)}=\Delta_B\delta E_{B,msr}^{(2)}.
\end{equation}
where we reintroduced the scaling with $\Delta_B$.
 
\subsection{Vanishing of the second-order energy corrections to the zero modes}
\label{sec:zero-mode-vanishing}

We now show that the two zero-energy reduced-rank states have vanishing
second-order energy corrections.  This statement concerns the energy
correction.  The first-order correction to the wavefunction need not vanish.

First consider the anomalous $A$ zero mode $|\psi_{A,0,0}\rangle$.  Its
first-order energy correction is zero because both $\tilde M_{\rm op}$ and
$\tilde{V}_{\rm op,\eta=+}\left(B\right)$ are off diagonal between the chiral sectors, and because
$\tilde{V}_{\rm op,\eta=+}\left(B\right)$ has no matrix element to this state.  The only possible
second-order contribution is
\begin{equation}
\delta E_{A,0,0}^{(2)}
=
\sum_{s',r'}
\frac{\tilde M^2 N_{B,1s'r'}^2|b_2^{1s'r'}|^2}
{-E_{B,1s'r'}}.
\end{equation}
For each fixed branch $r'$, the two particle-hole partners have
\begin{equation}
E_{B,1,+,r'}=e_{B,1r'},
\qquad
E_{B,1,-,r'}=-e_{B,1r'},
\qquad
N_{B,1,+,r'}^2|b_2^{1,+,r'}|^2
=
N_{B,1,-,r'}^2|b_2^{1,-,r'}|^2 .
\end{equation}
The equality of the absolute squares follows because $N_{B,1sr'}$ depends
only on $E_{B,1sr'}^2$, while $b_2^{1sr'}=-i/E_{B,1sr'}$ changes sign only by
an overall phase under $E_{B,1sr'}\to -E_{B,1sr'}$.  Therefore the $s'=+$
and $s'=-$ contributions cancel pairwise:
\begin{equation}
\frac{\tilde M^2 N_{B,1,+,r'}^2|b_2^{1,+,r'}|^2}{-e_{B,1r'}}
+
\frac{\tilde M^2 N_{B,1,-,r'}^2|b_2^{1,-,r'}|^2}{+e_{B,1r'}}
=0 .
\end{equation}
Thus
\begin{equation}
\delta E_{A,0,0}^{(2)}=0.
\end{equation}

Now consider the reduced-rank $B$ zero mode $|\psi_{B,0,z}\rangle$.  Its
first-order energy correction also vanishes.  The mass perturbation gives no
matrix element because the $B_0$ reduced-rank states have no component $4$,
and the only possible second-order contribution comes from the
$\tilde\Delta'$ coupling to the ordinary $A_2$ sector:
\begin{equation}
\delta E_{B,0,z}^{(2)}
=
\sum_{s',r'}
\frac{|V_{B,0,z;A,2s'r'}|^2}{-E_{A,2s'r'}} .
\end{equation}
Using $\beta_3^z=0$ and $\beta_1^z=1$, the relevant matrix element is
\begin{equation}
V_{B,0,z;A,2s'r'}
=
-i\tilde\Delta'\,N_{B,0,z}N_{A,2s'r'}
\Sigma(1)a_3^{2s'r'} .
\end{equation}
For each fixed branch $r'$, the two particle-hole partners satisfy
\begin{equation}
E_{A,2,+,r'}=e_{A,2r'},
\qquad
E_{A,2,-,r'}=-e_{A,2r'},
\qquad
N_{A,2,+,r'}^2|a_3^{2,+,r'}|^2
=
N_{A,2,-,r'}^2|a_3^{2,-,r'}|^2 .
\end{equation}
The last equality follows because $N_{A,2sr'}$ depends only on
$E_{A,2sr'}^2$, while
\begin{equation}
a_3^{2sr'}=
\frac{1}{\Gamma_1}
\left(E_{A,2sr'}-\frac{2}{E_{A,2sr'}}\right)
\end{equation}
is odd under $E_{A,2sr'}\to -E_{A,2sr'}$, so that $|a_3^{2sr'}|^2$ is even.
The $s'=+$ and $s'=-$ terms therefore cancel pairwise:
\begin{equation}
\frac{|V_{B,0,z;A,2,+,r'}|^2}{-e_{A,2r'}}
+
\frac{|V_{B,0,z;A,2,-,r'}|^2}{+e_{A,2r'}}
=0 .
\end{equation}
Hence
\begin{equation}
\delta E_{B,0,z}^{(2)}=0.
\end{equation}
The zero modes therefore remain pinned at zero energy through second order in
$\tilde M$ and $\tilde\Delta'$.

\subsection{Reduction to the \texorpdfstring{$m=1$}{m=1} Landau levels closest to charge neutrality}
\label{sec:m-one-low-LL-reduction}

We now specialize the general formulas to the pair of ordinary $m=1$ states
closest to charge neutrality.  In the shifted notation used in this note, the
ordinary $A$ state with label $m$ has oscillator content
$(m-1,m,m-1,m-1)$, while the ordinary $B$ state with the same label $m$ has
oscillator content $(m,m-1,m,m)$.  Thus, when $\Sigma(m)$ is set to a constant,
$A_m$ and $B_m$ have the same unperturbed energy.  This is the reason the
$m=1$ pair is the pair denoted by the lowest Landau level in the main text.

For the comparison with the main text, set
\begin{equation}
\Sigma(m)=1,
\qquad
\Gamma_m=\tilde\gamma,
\end{equation}
and choose the branch $r=\rho$ closest to charge neutrality.  We denote its
positive dimensionless energy by $e_n>0$, so that
\begin{equation}
E_{A,ns\rho}=E_{B,ns\rho}=s e_n,
\qquad
s=\pm .
\end{equation}
Equivalently,
\begin{equation}
e_n^2
=\frac{1}{2}\left[
 n+\tilde\gamma^2+\frac{\tilde U^2}{4}
+\rho\sqrt{\left(n+\tilde\gamma^2+\frac{\tilde U^2}{4}\right)^2
-\tilde U^2 n}
\right].
\end{equation}
Introduce the dimensionless quantities
\begin{align}
\mathfrak D_n
&\equiv e_n^2-\frac{\tilde U^2}{4},
\\
\widetilde{\mathfrak N}_n
&\equiv
\left[
\mathfrak D_n^2(e_n^2+n)
+\tilde\gamma^2 e_n^2
\left(e_n^2+\frac{\tilde U^2}{4}\right)
\right]^{1/2},
\\
\mathfrak B_n
&\equiv
\frac{\sqrt n\,\mathfrak D_n}{\widetilde{\mathfrak N}_n}.
\end{align}
The useful normalized amplitudes are then
\begin{align}
N_{A,ns\rho}a_1^{ns\rho}
&=N_{B,ns\rho}b_1^{ns\rho}
=\frac{e_n\mathfrak D_n}{\widetilde{\mathfrak N}_n},
\\
N_{A,ns\rho}a_2^{ns\rho}
&=is\,\mathfrak B_n,
\qquad
N_{B,ns\rho}b_2^{ns\rho}
=-is\,\mathfrak B_n,
\\
N_{A,ns\rho}a_3^{ns\rho}
&=N_{B,ns\rho}b_3^{ns\rho}
=s\frac{\tilde\gamma e_n^2}{\widetilde{\mathfrak N}_n}.
\end{align}
These relations are obtained directly from the coefficient formulas above and
the secular equation
\begin{equation}
\left(e_n^2-n\right)
\left(e_n^2-\frac{\tilde U^2}{4}\right)
=\tilde\gamma^2e_n^2 .
\end{equation}

First consider the positive-energy state $A_{1,+,\rho}$.  The
$\tilde\Delta'$ selection rule would send $A_m$ to $B_{m-2}$, so there is no
$\tilde\Delta'$ contribution for $m=1$.  The only contribution is the mass
coupling to $B_{2,\zeta,\rho}$ with $\zeta=\pm$:
\begin{align}
\delta E_{A,1,+,\rho}^{(2)}
&=
\sum_{\zeta=\pm}
\frac{
\left|
\tilde M
\left(N_{A,1,+,\rho}\overline{a_2^{1,+,\rho}}\right)
\left(N_{B,2,\zeta,\rho}b_2^{2,\zeta,\rho}\right)
\right|^2
}{e_1-\zeta e_2}
\\
&=
\sum_{\zeta=\pm}
\frac{\tilde M^2\mathfrak B_1^2\mathfrak B_2^2}{e_1-\zeta e_2}
\\
&=
\frac{2\tilde M^2\mathfrak B_1^2\mathfrak B_2^2e_1}
{e_1^2-e_2^2}.
\label{eq:A1-lowLL-shift-dimless}
\end{align}
The negative-energy partner has the opposite correction,
\begin{equation}
\delta E_{A,1,-,\rho}^{(2)}=-\delta E_{A,1,+,\rho}^{(2)}.
\end{equation}

Now consider $B_{1,+,\rho}$.  There are two terms.  The mass perturbation
connects $B_{1,+,\rho}$ to the anomalous zero mode $A_{0,0}$.  Since
$E_{A,0,0}=0$,
\begin{equation}
\delta E_{B,1,+,\rho;M}^{(2)}
=
\frac{\left|\tilde M N_{B,1,+,\rho}\overline{b_2^{1,+,\rho}}\right|^2}{e_1}
=
\frac{\tilde M^2\mathfrak B_1^2}{e_1}.
\label{eq:B1-lowLL-M-shift-dimless}
\end{equation}
The $\tilde\Delta'$ perturbation sends $B_m$ to $A_{m+2}$, so for $m=1$ the
intermediate ordinary states are $A_{3,\zeta,\rho}$.  The corresponding
matrix element is
\begin{align}
\mathcal V_{\zeta}
&\equiv
\mathcal V^{B}_{1,+,\rho;\zeta,\rho}
\\
&=
-i\sqrt2\,\tilde\Delta'
\left[
\left(N_{A,3,\zeta,\rho}a_3^{3,\zeta,\rho}\right)
\left(N_{B,1,+,\rho}\overline{b_1^{1,+,\rho}}\right)
+
\left(N_{A,3,\zeta,\rho}a_1^{3,\zeta,\rho}\right)
\left(N_{B,1,+,\rho}\overline{b_3^{1,+,\rho}}\right)
\right]
\\
&=
-i\sqrt2\,\tilde\Delta'
\frac{\tilde\gamma e_1e_3}{
\widetilde{\mathfrak N}_1\widetilde{\mathfrak N}_3}
\left(\mathfrak D_3e_1+\zeta\mathfrak D_1e_3\right).
\label{eq:B1-lowLL-Delta-matrix-element-dimless}
\end{align}
Therefore the $\tilde\Delta'$ contribution obtained from ordinary
second-order perturbation theory is
\begin{equation}
\delta E_{B,1,+,\rho;\Delta'}^{(2)}
=
2(\tilde\Delta')^2
\frac{\tilde\gamma^2e_1^2e_3^2}
{\widetilde{\mathfrak N}_1^2\widetilde{\mathfrak N}_3^2}
\sum_{\zeta=\pm}
\frac{
\left(\mathfrak D_3e_1+\zeta\mathfrak D_1e_3\right)^2
}{e_1-\zeta e_3}.
\label{eq:B1-lowLL-Delta-shift-dimless}
\end{equation}
Combining the two pieces,
\begin{equation}
\delta E_{B,1,+,\rho}^{(2)}
=
\frac{\tilde M^2\mathfrak B_1^2}{e_1}
+
2(\tilde\Delta')^2
\frac{\tilde\gamma^2e_1^2e_3^2}
{\widetilde{\mathfrak N}_1^2\widetilde{\mathfrak N}_3^2}
\sum_{\zeta=\pm}
\frac{
\left(\mathfrak D_3e_1+\zeta\mathfrak D_1e_3\right)^2
}{e_1-\zeta e_3}.
\label{eq:B1-lowLL-shift-dimless}
\end{equation}
Again,
\begin{equation}
\delta E_{B,1,-,\rho}^{(2)}=-\delta E_{B,1,+,\rho}^{(2)}.
\end{equation}

Finally we translate to the notation used in the main text.  Define physical
positive energies
\begin{equation}
\mathcal E_n\equiv \Delta_B e_n,
\end{equation}
and define
\begin{align}
\mathcal D_n
&\equiv
\mathcal E_n^2-\frac{U_1^2}{4},
\\
\widetilde N_n
&\equiv
\left[
\mathcal D_n^2\left(\mathcal E_n^2+n\Delta_B^2\right)
+\gamma^2\mathcal E_n^2
\left(\mathcal E_n^2+\frac{U_1^2}{4}\right)
\right]^{1/2},
\\
\mathcal B_n
&\equiv
\frac{\sqrt n\,\Delta_B\mathcal D_n}{\widetilde N_n}.
\end{align}
With these definitions, $\mathcal B_n=\mathfrak B_n$.  The physical energy
corrections are obtained from
\begin{equation}
E_{\rm corr}=\Delta_B\delta E^{(2)}.
\end{equation}
Thus the standard second-order perturbation-theory result following from the
Hamiltonian in this note is
\begin{equation}
E_{{\rm corr},LL}^{A}
=
\frac{2M^2\mathcal B_1^2\mathcal B_2^2\mathcal E_1}
{\mathcal E_1^2-\mathcal E_2^2},
\label{eq:EcorrA-lowLL-physical}
\end{equation}
and
\begin{equation}
E_{{\rm corr},LL}^{B}
=
\frac{M^2\mathcal B_1^2}{\mathcal E_1}
+
2\Delta_B'^2
\frac{\gamma^2\mathcal E_1^2\mathcal E_3^2}
{\widetilde N_1^2\widetilde N_3^2}
\sum_{\zeta=\pm}
\frac{
\left(\mathcal D_3\mathcal E_1+\zeta\mathcal D_1\mathcal E_3\right)^2
}{\mathcal E_1-\zeta\mathcal E_3}.
\label{eq:EcorrB-lowLL-physical}
\end{equation}
The modified low-energy levels are therefore
\begin{equation}
E_{A,\pm}=
\pm\left(\mathcal E_1+E_{{\rm corr},LL}^{A}\right),
\qquad
E_{B,\pm}=
\pm\left(\mathcal E_1+E_{{\rm corr},LL}^{B}\right).
\end{equation}

Equations~\eqref{eq:EcorrA-lowLL-physical} and
\eqref{eq:EcorrB-lowLL-physical} then correspond to the second order shifts quoted in the main text.

%------------------Section Change--------------------------------

\section{Projection onto the flat bands within Hubbard-I at CNP}\label{sec:projGreen's}

To understand analytically the behavior of the Landau level spectrum
we can appeal to the exactly solvable limit of the topological-heavy
fermion model when $\gamma>U$. In this limit we can effectively project
the Hamiltonian onto the flat bands to understand the effect of the
chiral perturbation and relaxation together. For that, consider the
non-interacting Hamiltonian for the $K$ valley
\begin{equation}
h_{0}\left(\mathbf{k}\right)=\left(\begin{array}{cccccc}
0 & 0 & e^{i\theta}\left|\mathbf{k}\right|v_{\ast} & 0 & \gamma & e^{-i\theta}\left|\mathbf{k}\right|v_{\ast}^{\prime}\\
0 & 0 & 0 & e^{-i\theta}\left|\mathbf{k}\right|v_{\ast} & e^{i\theta}\left|\mathbf{k}\right|v_{\ast}^{\prime} & \gamma\\
e^{-i\theta}\left|\mathbf{k}\right|v_{\ast} & 0 & 0 & 0 & 0 & 0\\
0 & e^{i\theta}\left|\mathbf{k}\right|v_{\ast} & 0 & 0 & 0 & 0\\
\gamma & e^{-i\theta}\left|\mathbf{k}\right|v_{\ast}^{\prime} & 0 & 0 & 0 & 0\\
e^{i\theta}\left|\mathbf{k}\right|v_{\ast}^{\prime} & \gamma & 0 & 0 & 0 & 0
\end{array}\right),
\end{equation}
we can reinstate the Gaussian damping by the replacements $\gamma\rightarrow\gamma e^{-\lambda^{2}\left|\mathbf{k}\right|^{2}/2}$,
$v_{\ast}^{\prime}\rightarrow v_{\ast}^{\prime}e^{-\lambda^{2}\left|\mathbf{k}\right|^{2}/2}$
but we set $\lambda=0$ in this appendix for simplicity. We also consider
the simplified model for relaxation where $v_{1}=v_{2}=0$. This Hamiltonian
has $SO\left(2\right)$ symmetry as such we can eliminate the angular
dependence by the unitary transformation 
\begin{equation}
U\left(\theta\right)=\left(\begin{array}{cccccc}
e^{i\theta} & 0 & 0 & 0 & 0 & 0\\
0 & e^{2i\theta} & 0 & 0 & 0 & 0\\
0 & 0 & 1 & 0 & 0 & 0\\
0 & 0 & 0 & e^{3i\theta} & 0 & 0\\
0 & 0 & 0 & 0 & e^{i\theta} & 0\\
0 & 0 & 0 & 0 & 0 & e^{2i\theta}
\end{array}\right)
\end{equation}
which yields the following rotated Hamiltonian where the angular dependence
was effectively removed:
\begin{equation}
U^{\dagger}\left(\theta\right)h_{0}\left(\mathbf{k}\right)U\left(\theta\right)=\left(\begin{array}{cccccc}
0 & 0 & v_{\ast}\left|\mathbf{k}\right| & 0 & \gamma & v_{\ast}^{\prime}\left|\mathbf{k}\right|\\
0 & 0 & 0 & v_{\ast}\left|\mathbf{k}\right| & v_{\ast}^{\prime}\left|\mathbf{k}\right| & \gamma\\
v_{\ast}\left|\mathbf{k}\right| & 0 & 0 & 0 & 0 & 0\\
0 & v_{\ast}\left|\mathbf{k}\right| & 0 & 0 & 0 & 0\\
\gamma & v_{\ast}^{\prime}\left|\mathbf{k}\right| & 0 & 0 & 0 & 0\\
v_{\ast}^{\prime}\left|\mathbf{k}\right| & \gamma & 0 & 0 & 0 & 0
\end{array}\right)
\end{equation}
The eigenvectors of the flat bands can then be obtained directly as
\begin{equation}
\psi_{f1}=\frac{1}{\sqrt{2}\sqrt{\left|\mathbf{k}\right|^{2}v_{\ast}^{2}+\left(\gamma+\left|\mathbf{k}\right|v_{\ast}^{\prime}\right)^{2}}}\left(0,0,-\gamma-v_{\ast}^{\prime}\left|\mathbf{k}\right|,-\gamma-v_{\ast}^{\prime}\left|\mathbf{k}\right|,v_{\ast}\left|\mathbf{k}\right|,v_{\ast}\left|\mathbf{k}\right|\right)
\end{equation}
\begin{equation}
\psi_{f2}=\frac{1}{\sqrt{2}\sqrt{\left|\mathbf{k}\right|^{2}v_{\ast}^{2}+\left(\gamma-\left|\mathbf{k}\right|v_{\ast}^{\prime}\right)^{2}}}\left(0,0,\gamma-v_{\ast}^{\prime}\left|\mathbf{k}\right|,v_{\ast}^{\prime}\left|\mathbf{k}\right|-\gamma,-v_{\ast}\left|\mathbf{k}\right|,v_{\ast}\left|\mathbf{k}\right|\right)
\end{equation}
\begin{figure}
    \centering
    \includegraphics[width=0.8\linewidth]{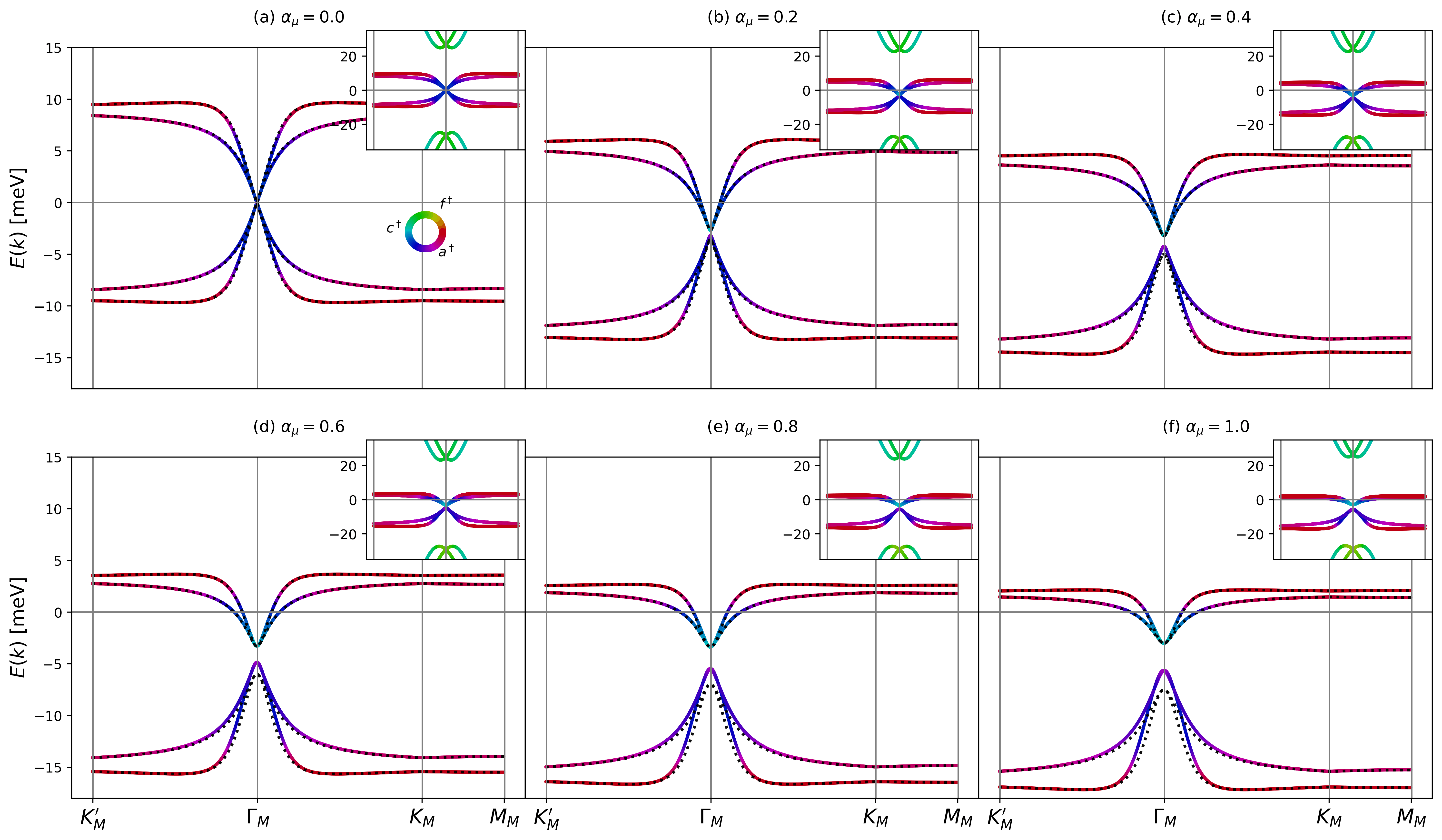}
    \caption{\textbf{Mott semimetal:}  Comparison of the dispersion of the full and the projected THF model with increasing relaxation where we multiply $\mu_1,\mu_2,v_1,v_2$ by an artificial prefactor $\alpha_\mu\in [0,1]$ for increasing values of $\alpha_\mu$  away from the chiral limit, $v_*^\prime\neq 0$ and with $\lambda = 0 $. The black dashed line denotes the projected Hubbard-I spectra obtained from Eq.\ref{eq:projected_spec}.  The color wheel in (a) denotes the fermion character at each k-point, obtained from the expectation value of $Z_{\rm char} = e^{i\pi}\sigma_0 \oplus e^{i\pi}\sigma_0 \oplus e^{i\pi/3}\sigma_0 \oplus e^{-i\pi/3}\sigma_0$ such that $c-$electrons are colored cyan, $f$-electrons are colored yellow, and $a-$electrons are colored magenta. The inset shows the same spectra over a wider energy range.} 
    \label{fig:projdispcomp}
\end{figure}
Using these wavefunctions we can project the full Green's function
in the Hubbard-I approximation to the flat-band subspace. We use $\psi_{f,\zeta=+}$ to label $\psi_{f1}$ and $\psi_{f,\zeta=-}$ to label $\psi_{f2}$:
\begin{equation}
\mathcal{G}_{f,\zeta\zeta'}\left(i\omega_{n},\mathbf{k}\right)^{-1}=\psi_{f,\zeta}^{\dagger}U\left(\theta\right)\left(i\omega_{n}-\tilde{h}_{\eta=+}^{\text{MF}}(\mathbf{k})-\Sigma^{\text{At}}\left(i\omega_{n}\right)\right)U^{\dagger}\left(\theta\right)\psi_{f,\zeta'}
\end{equation}
The components of the inverse Green's function can be readily obtained
individually as:
\begin{align}
\mathcal{G}_{f,++}\left(i\omega_{n},\mathbf{k}\right)^{-1} & =i\omega_{n}-\frac{1}{\left(\gamma+v'_{\star}\left|\mathbf{k}\right|\right)^{2}+v_{\ast}^{2}\left|\mathbf{k}\right|^{2}}\left(\epsilon_{c,2}\left(\gamma+v'_{\star}\left|\mathbf{k}\right|\right)^{2}+\epsilon_{f}\left|\mathbf{k}\right|^{2}v_{\ast}^{2}-M\cos(3\theta_{\mathbf{k}})\left(\gamma+v'_{\star}\left|\mathbf{k}\right|\right)^{2}\right)\nonumber \\
 & -\frac{\left|\mathbf{k}\right|^{2}v_{\ast}^{2}}{\left(\gamma+v'_{\star}\left|\mathbf{k}\right|\right)^{2}+\left|\mathbf{k}\right|^{2}v_{\ast}^{2}}U_{1}^{2}\frac{\left(N_f/2+\nu_{f}\right)\left(N_f/2-\nu_{f}\right)}{N_f^{2}}\frac{1}{i\omega_{n}-\left[\frac{U_{1}}{N_f}\left(N_f+1\right)\nu_{f}+\epsilon_{f}\right]}
\end{align}
\begin{align}
\mathcal{G}_{f,--}\left(i\omega_{n},\mathbf{k}\right)^{-1} & =i\omega_{n}-\frac{1}{\left(\gamma-v'_{\star}\left|\mathbf{k}\right|\right)^{2}+v_{\ast}^{2}\left|\mathbf{k}\right|^{2}}\left(\epsilon_{c,2}\left(\gamma-v'_{\star}\left|\mathbf{k}\right|\right)^{2}+\epsilon_{f}v_{\ast}^{2}\left|\mathbf{k}\right|^{2}-M\cos(3\theta_{\mathbf{k}})\left(\gamma-v'_{\star}\left|\mathbf{k}\right|\right)^{2}\right)\nonumber \\
 & -\frac{\left|\mathbf{k}\right|^{2}v_{\ast}^{2}}{\left(\gamma-v'_{\star}\left|\mathbf{k}\right|\right)^{2}+\left|\mathbf{k}\right|^{2}v_{\ast}^{2}} U_{1}^{2}\frac{\left(N_f/2+\nu_{f}\right)\left(N_f/2-\nu_{f}\right)}{N_f^{2}}\frac{1}{i\omega_{n}-\left[\frac{U_{1}}{N_f}\left(N_f+1\right)\nu_{f}+\epsilon_{f}\right]}
\end{align}
\begin{equation}
\mathcal{G}_{f,+-}\left(i\omega_{n},\mathbf{k}\right)^{-1}=-i\frac{M\sin(3\theta_{\mathbf{k}})\left(\gamma+v'_{\star}\left|\mathbf{k}\right|\right)\left(\gamma-v'_{\star}\left|\mathbf{k}\right|\right)}{\sqrt{\left(\left(\gamma-v'_{\star}\left|\mathbf{k}\right|\right)^{2}+\left|\mathbf{k}\right|^{2}v_{\ast}^{2}\right)\left(\left(\gamma+v'_{\star}\left|\mathbf{k}\right|\right)^{2}+\left|\mathbf{k}\right|^{2}v_{\ast}^{2}\right)}}
\end{equation}
\begin{equation}
\mathcal{G}_{f,-+}\left(i\omega_{n},\mathbf{k}\right)^{-1}=i\frac{M\sin(3\theta_{\mathbf{k}})\left(\gamma+v'_{\star}\left|\mathbf{k}\right|\right)\left(\gamma-v'_{\star}\left|\mathbf{k}\right|\right)}{\sqrt{\left(\left(\gamma-v'_{\star}\left|\mathbf{k}\right|\right)^{2}+\left|\mathbf{k}\right|^{2}v_{\ast}^{2}\right)\left(\left(\gamma+v'_{\star}\left|\mathbf{k}\right|\right)^{2}+\left|\mathbf{k}\right|^{2}v_{\ast}^{2}\right)}}
\end{equation}
From these components we can use the auxiliary fermion mapping to
determine the location of the poles of the Green's function effectively
which we can then use to get analytical understanding of the Landau
Level spectrum. We consider the case in which $M=0$ such that the
projected Green's function is diagonal, $\mathcal{G}_{f,\zeta\zeta'}\left(i\omega_{n},\mathbf{k}\right)^{-1}=\delta_{\zeta\zeta'}\mathcal{G}_{f,\zeta}\left(i\omega_{n},\mathbf{k}\right)^{-1}$.
We then identify the projected auxiliary fermion Hamiltonian 
\begin{equation}
h_{\text{aux}}^{\zeta}\left(\mathbf{k}\right)=\left(\begin{array}{cc}
\frac{\epsilon_{c,2}\left(\gamma+\zeta v'_{\star}\left|\mathbf{k}\right|\right)^{2}+v_{\ast}^{2}\left|\mathbf{k}\right|^{2}\epsilon_{f}}{\left(\gamma+\zeta v'_{\star}\left|\mathbf{k}\right|\right)^{2}+v_{\ast}^{2}\left|\mathbf{k}\right|^{2}} & \sqrt{\frac{\left(N_f/2+\nu_{f}\right)\left(N_f/2-\nu_{f}\right)}{\left(\gamma+\zeta v'_{\star}\left|\mathbf{k}\right|\right)^{2}+\left|\mathbf{k}\right|^{2}v_{\ast}^{2}}}\frac{U_{1}v_{\ast}\left|\mathbf{k}\right|}{N_f}\\
\sqrt{\frac{\left(N_f/2+\nu_{f}\right)\left(N_f/2-\nu_{f}\right)}{\left(\gamma+\zeta v'_{\star}\left|\mathbf{k}\right|\right)^{2}+v_{\ast}^{2}\left|\mathbf{k}\right|^{2}}}\frac{U_{1}v_{\ast}\left|\mathbf{k}\right|}{N_f} & \frac{U_{1}}{N_f}\left(N_f+1\right)\nu_{f}+\epsilon_{f}
\end{array}\right)\label{eq:auxiliaryHam_projected}
\end{equation}
The eigenvalues of the flat bands in the projected limit within the
Hubbard I approximation at neutrality $\nu_{f}=0$ are then given
by:
\begin{equation}
E_{\pm}^{\zeta}\left(\mathbf{k}\right)=\frac{1}{2}\left(\frac{\epsilon_{c,2}\left(\gamma+\zeta v'_{\star}\left|\mathbf{k}\right|\right)^{2}+v_{\ast}^{2}\left|\mathbf{k}\right|^{2}\epsilon_{f}}{\left(\gamma+\zeta v'_{\star}\left|\mathbf{k}\right|\right)^{2}+v_{\ast}^{2}\left|\mathbf{k}\right|^{2}}+\epsilon_{f}\pm\sqrt{\frac{(\epsilon_{c,2}-\epsilon_{f})^{2}\left(\gamma+\zeta v'_{\star}\left|\mathbf{k}\right|\right)^{4}}{\left(\left(\gamma+\zeta v'_{\star}\left|\mathbf{k}\right|\right)^{2}+v_{\ast}^{2}\left|\mathbf{k}\right|^{2}\right)^{2}}+\frac{v_{\ast}^{2}\left|\mathbf{k}\right|^{2}U_{1}^{2}}{\left(\gamma+\zeta v'_{\star}\left|\mathbf{k}\right|\right)^{2}+v_{\ast}^{2}\left|\mathbf{k}\right|^{2}}}\right)\label{eq:projected_spec}
\end{equation}
A comparison between the bands obtained by projection with the bands
obtained from the full Hamiltonian in the Hubbard-I approximation
as a function of increasing relaxation is shown in Fig.(\ref{fig:projdispcomp}).
To map the dependence of the spectra as a function of increasing relaxation,
we introduce an artificial prefactor $\alpha_{\mu}$ that multiplies
$\mu_{1}$, and $\mu_{2}$, where $\alpha_{\mu}\in[0,1]$. In the
absence of relaxation the projected model reproduces accurately the
Hubbard I spectra for the full model. As relaxation increases, the
gap between the remote and flat valence bands is reduced such that
the projection is less accurate for the valence band while the conduction
band remains described by the projected model. We further identify
the condition for having a gap at neutrality at the $\Gamma_{M}$
point, which occurs when we have $\epsilon_{c,2}\neq\epsilon_{f}$
when $M=0$, in this case the gap is set by $\Delta_{\Gamma_{M}}=\left|\epsilon_{c,2}-\epsilon_{f}\right|$.
We then focus on the conduction band in the projected model in the
next subsection when adding a magnetic field.
\begin{figure}
    \centering
    \includegraphics[width=0.8\linewidth]{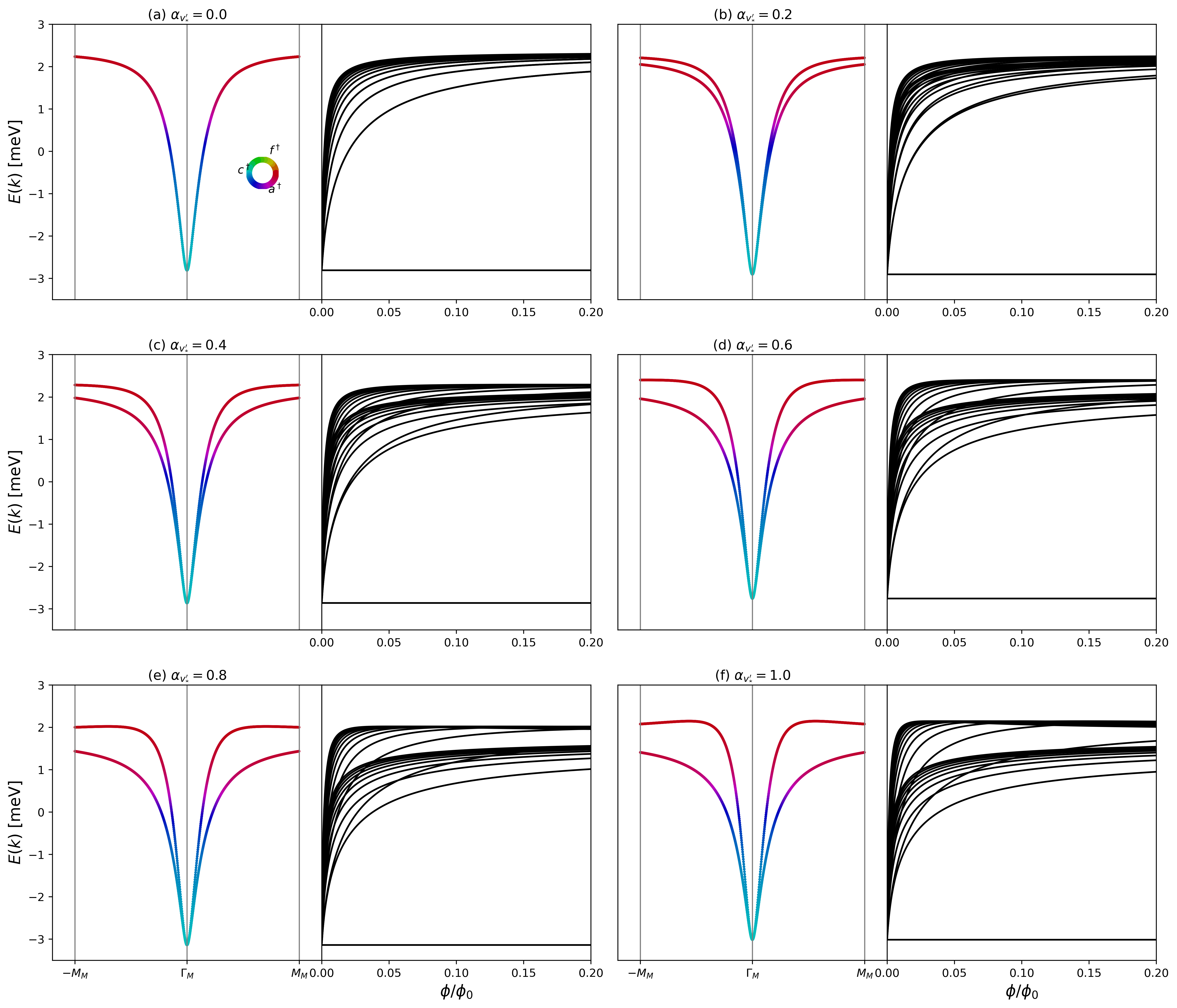}
    \caption{\textbf{Mott semimetal:}  Dispersion and LL spectra of the THF model with relaxation for $M=0$, and $\lambda=0$. We show the spectra as a function of an artificial prefactor $\alpha_{v_\ast^\prime}\in[0,1]$ that multiplies $v_\ast^\prime$. The color wheel in (a) denotes the fermion character at each k-point, obtained from the expectation value of $Z_{\rm char} = e^{i\pi}\sigma_0 \oplus e^{i\pi}\sigma_0 \oplus e^{i\pi/3}\sigma_0 \oplus e^{-i\pi/3}\sigma_0$ such that $c-$electrons are colored cyan, $f$-electrons are colored yellow, and $a-$electrons are colored magenta.} 
    \label{fig:projdispcomporbs}
\end{figure}

\subsection{Poles of the Green's function with \texorpdfstring{$M=0$}{M=0} at finite \texorpdfstring{$B$}{B}}\label{subsec:projGreen'sM0}

To understand the effect of deviations from the Chiral limit ($v_{\ast}^{\prime}=0$)
we first solve for the poles of the Green's function with increasing
$v_{\ast}^{\prime}=0$. For this we introduce a parameter $\alpha_{v_{\ast}^{\prime}}\in[0,1]$
to tune the Chiral velocity to its physical value in Tab.(\ref{tab:model_params})
when $\alpha_{v_{\ast}^{\prime}}=1$. Throughout this subsection we again
set $M=0$. The relaxed Hubbard I Hamiltonian at finite field with
these definitions is 
\begin{equation}
\tilde{h}_{i,i',\eta=+}^{\text{MF}}\left(B\right)=\begin{pmatrix}\mu_{1} & 0 & -i\Delta_{B}a & 0 & \gamma & i\alpha_{v_{\ast}^{\prime}}\Delta_{B}^{\prime}a^{\dagger} & 0 & 0\\
0 & \mu_{1} & 0 & i\Delta_{B}a^{\dagger} & -i\alpha_{v_{\ast}^{\prime}}\Delta_{B}^{\prime}a & \gamma & 0 & 0\\
i\Delta_{B}a^{\dagger} & 0 & \mu_{2} & 0 & 0 & 0 & 0 & 0\\
0 & -i\Delta_{B}a & 0 & \mu_{2} & 0 & 0 & 0 & 0\\
\gamma & i\alpha_{v_{\ast}^{\prime}}\Delta_{B}^{\prime}a^{\dagger} & 0 & 0 & \varepsilon_{f} & 0 & \frac{U_1}{2} & 0\\
-i\alpha_{v_{\ast}^{\prime}}\Delta_{B}^{\prime}a & \gamma & 0 & 0 & 0 & \varepsilon_{f} & 0 & \frac{U_1}{2}\\
0 & 0 & 0 & 0 & \frac{U_1}{2} & 0 & \varepsilon_{f} & 0\\
0 & 0 & 0 & 0 & 0 & \frac{U_1}{2} & 0 & \varepsilon_{f}
\end{pmatrix}\label{eq:mean_field_neut_chiral_mdiag}
\end{equation}
where we introduced $\Delta_{B}^{\prime}=\sqrt{2}v_{\ast}^{\prime}/\ell.$
Because this Hamiltonian preserves $SO\left(2\right)$ symmetry, it
is block diagonalized by acting on the following ansatz spinor 
\begin{equation}
|\psi_{m}\rangle=\left(\begin{array}{cccccccc}
\psi_{\Gamma_{3},m}^{1}|m\rangle, & \psi_{\Gamma_{3},m}^{2}|m-1\rangle, & \psi_{\Gamma_{1,2},m}^{1}|m-2\rangle, & \psi_{\Gamma_{1,2},m}^{2}|m+1\rangle, & \psi_{f,m}^{1}|m\rangle, & \psi_{f,m}^{2}|m-1\rangle, & \psi_{a,m}^{1}|m\rangle, & \psi_{a,m}^{2}|m-1\rangle\end{array}\right),
\end{equation}
The resulting spectra is shown in Fig.(\ref{fig:projdispcomporbs}).
At $\alpha_{v_{\ast}^{\prime}}=0$ each LL is doubly degenerate but any
finite $\alpha_{v_{\ast}^{\prime}}$ breaks this degeneracy. The splitting
naturally increases continuously with increasing magnetic field and
is only noticeable in Fig.(\ref{fig:projdispcomporbs}) beyond $\phi/\phi_{0}\approx0.15$
for $\alpha_{v_{\ast}^{\prime}}=0.2.$ For larger $\alpha_{v_{\ast}^{\prime}}$,
however, the LL splitting is significant event at small flux and we
find several level crossings between LLs that occur at different magnetic
fields depending on the specific LL index $m$, with larger $m$ having
crossings at lower flux. 

\begin{figure}
    \centering
    \includegraphics[width=0.8\linewidth]{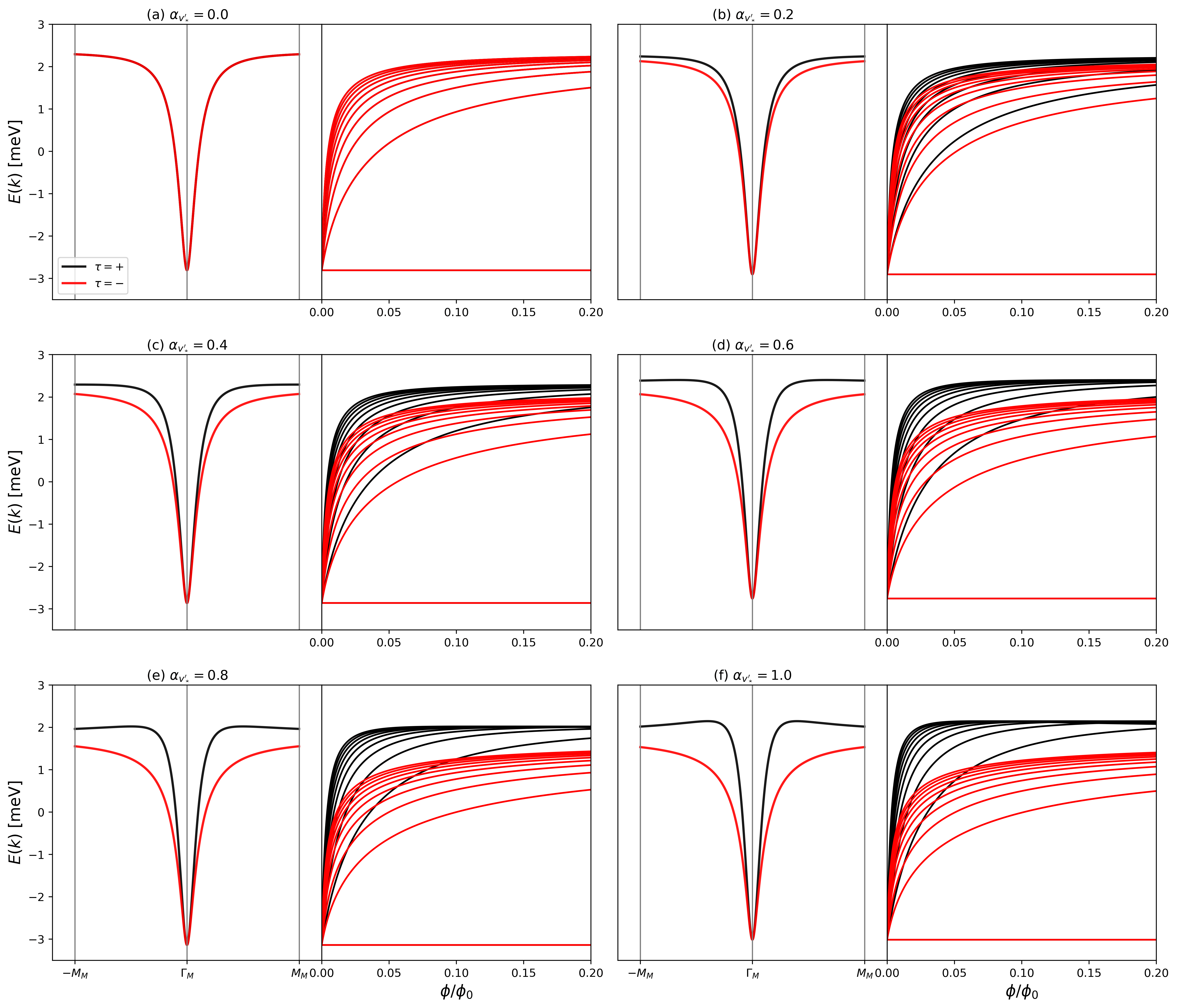}
    \caption{\textbf{Mott semimetal:} Dispersion and semiclassical spectra of the projected THF model with relaxation for $M=0$, and $\lambda=0$. We show the spectra as a function of an artificial prefactor $\alpha_{v_\ast^\prime}\in[0,1]$ that multiplies $v_\ast^\prime$. Red lines denote the dispersion and Onsager orbits in the $\zeta=+$ sector and black lines denote the dispersion and Onsager orbits in the $\zeta=-$ sector. Crossings between the orbits that emanate from each $\zeta$ sector explain the gap closings and reopenings for different Chern gaps.} 
    \label{fig:projdispcompons}
\end{figure}

We can understand this behavior from the projected model via semiclassical
quantization. In the $SO\left(2\right)$ symmetric limit, the cyclotron
orbits are circular and we can determine the semiclassical spectrum
at finite $B$-field by the simple replacement $\left|\text{\ensuremath{\mathbf{k}}}\right|\rightarrow k_m=\frac{\sqrt{2}}{\ell}\sqrt{m}$
on Eq.(\ref{eq:projected_spec}). The results for the semiclassical
LL spectra are shown in Fig.(\ref{fig:projdispcompons}). We see the
same features as in the full diagonalization of the auxiliary fermion
Hamiltonian at finite field, with LLs being degenerate at $\alpha_{v_{\ast}^{\prime}}=0$
and splitting for $\alpha_{v_{\ast}^{\prime}}\neq0$. However, in the
present setting we can clearly identify the split levels as cyclotron
orbits from bands on each separate $\zeta$ sector of the projected
THF auxiliary Hamiltonian in Eq.(\ref{eq:auxiliaryHam_projected}).
Orbits from different sectors are represented by the red ($\zeta=+$)
and black ($\zeta=-$) in Fig.(\ref{fig:projdispcompons}). The presence
of a finite $v_{\ast}^{\prime}$ reduces the energy for the $\zeta=-$
sector relative to the $\zeta=+$ sector such that the cyclotron orbits
from the $\zeta=+$ sector grow more rapidly with increasing magnetic
field. The crossings we found in the full spectra in Fig.(\ref{fig:projdispcompons})
correspond to the $m$th LL in the $\zeta=+$ sector crossing an LL
with larger $m$ in the $\zeta=-$ sector.

For fixed Hubbard-I branch $s=\pm$, the $\zeta=\pm$ dependence enters only through
\begin{equation}
    x_\zeta(k)=
\frac{v_\ast^2 k^2}
{(\gamma+\zeta  v_\ast' k)^2+v_\ast^2 k^2},
\end{equation}
and the pole energy is a one-to-one function of $x_\zeta$, except at the
fine-tuned point $U_1=2|\epsilon_{c,2}-\epsilon_f|$. Therefore a 
gap closure for either $C=+8$ or $C=-8$, would require
\begin{equation}
E_s^+(k_m)=E_s^-(k_m)
\quad\Rightarrow\quad
x_+(k_m)=x_-(k_m).
\end{equation}
Since the numerators are identical, this requires
\begin{equation}
(\gamma+ v_\ast' k_m)^2
=
(\gamma- v_\ast' k_m)^2,
\end{equation}
hence
\begin{equation}
4\gamma  v_\ast' k_m=0.
\end{equation}
For $\gamma\neq0$, finite $v_\ast'$, and
$\alpha_{v_\ast'}\neq0$, this implies $k_m=0$, i.e. $B=0$. Thus, neither
the $C=+8$ nor the $C=-8$ same $m$-index semiclassical gap closes at any
finite magnetic field in this projected $M=0$, $SO(2)$-symmetric model. Note that the previous proof only shows that this intra-doublet gap cannot close at
finite $B$. The $C=12$ gap closing is not of this type. Instead, it is an
inter-index crossing between different semiclassical cyclotron orbits,
\begin{equation}
E_s^{+}(k_m)=E_s^{-}(k_n),
\qquad m\neq n .
\end{equation}
For this case, the crossing condition is
\begin{equation}
x_+(k_m)=x_-(k_n),
\end{equation}
which gives
\begin{equation}
\gamma(k_m-k_n) =2 v_\ast' k_m k_n .
\end{equation}
Using $k_m=\sqrt{2m}/\ell_B$, this gives the finite-field crossing
\begin{equation}
B_{m,n}
=\frac{\hbar\gamma^2}
{8e(v_\ast')^2}
\frac{(\sqrt m-\sqrt n)^2}{mn},
\end{equation}
provided the corresponding sign condition is satisfied. We thus identify 
\begin{equation}
    f_{nm}=\frac{1}
{8}
\frac{(\sqrt m-\sqrt n)^2}{mn},
\end{equation}
in the main text. Thus the $C=12$
gap closure is an $m\neq n$ crossing between different cyclotron orbits,
not a same-$m$ split-doublet closure. Note that in the presence of a finite $M$ the two sectors  involved in an LL crossing  are allowed to mix and the LL crossings will become avoided crossings. We now turn to the effect of a finite $M$, which we explore in the following section.

%----------------Section Change----------------------------

\section{\texorpdfstring{$k\cdot p$}{k.p} model with relaxation effects at CNP within Hubbard-I}\label{sec:kdotprelaxationneut}

\begin{figure}
    \centering
    \includegraphics[width=0.5\linewidth]{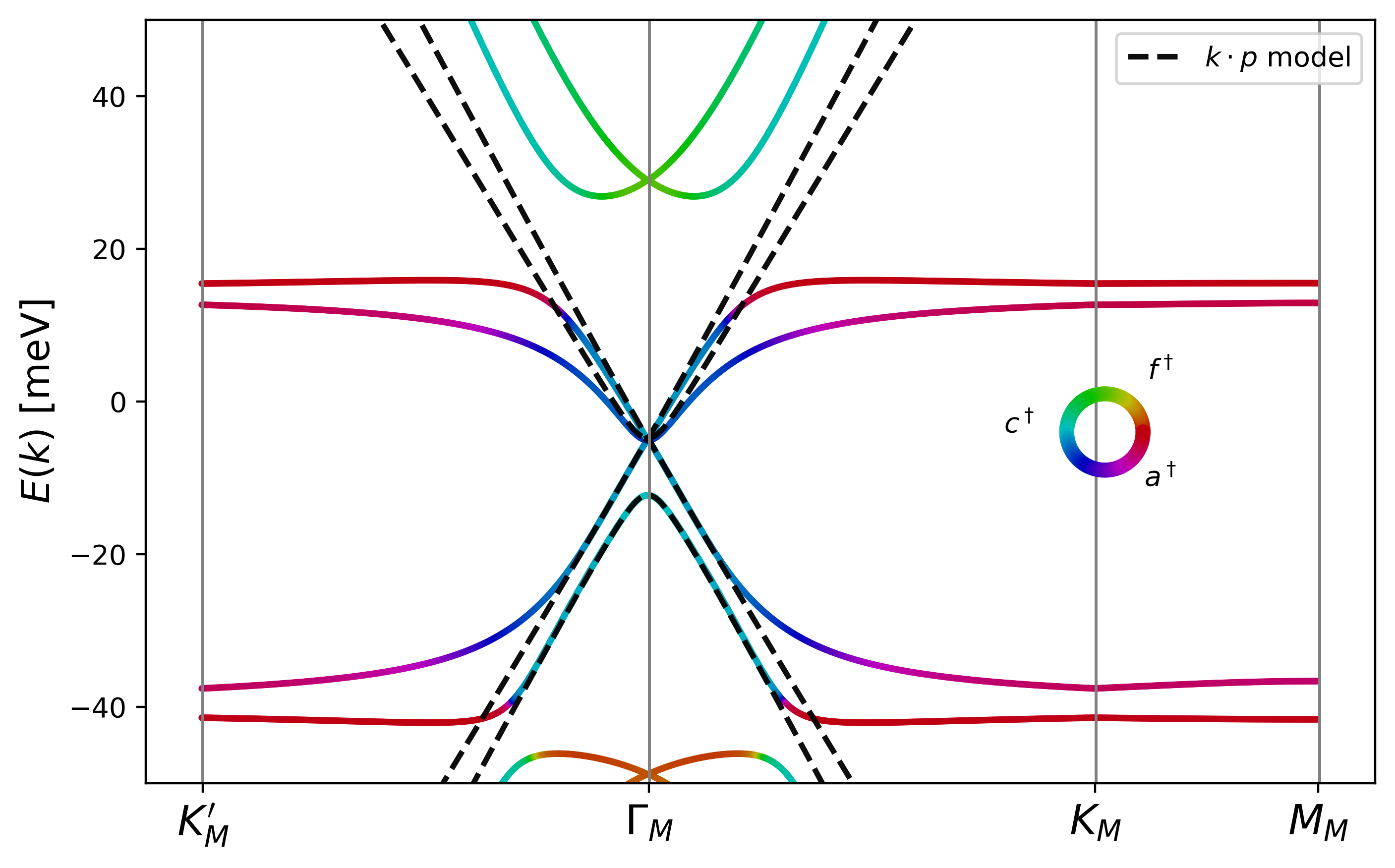}
    \caption{\textbf{Mott semimetal:} Comparison of the Hubbard-I single particle spectra obtained from the topological heavy fermion (THF) model (colored solid lines) with a $k\cdot p$ expansion around the $\Gamma_M$ point (black dashed lines) in Eq.(\ref{eq:kdotp_full_leading}).  The color wheel denotes the fermion character at each k-point, obtained from the expectation value of $Z_{\rm char} = e^{i\pi}\sigma_0 \oplus e^{i\pi}\sigma_0 \oplus e^{i\pi/3}\sigma_0 \oplus e^{-i\pi/3}\sigma_0$ such that $c-$electrons are colored cyan, $f$-electrons are colored yellow, and $a-$electrons are colored magenta.}
    \label{fig:vpmassdepkdotp}
\end{figure}

In this section we derive a $\bk\cdot\bp$ model in vicinity of $\Gamma$ in presence of relaxation and determine whether it can induce a gap. To keep the derivations general,
we first derive the effective Hamiltonian near $\Gamma_{M}$ for arbitrary
filling, but then we specialize to the case of charge neutrality.
To simplify notation, we define 
\begin{equation}
\varepsilon_{f}=\epsilon_{f}+\frac{U_{1}\left(N_f-1\right)}{N_f}\nu_{f}\ \ \ \text{and}\ \ \ U_{1,af}=\frac{U_{1}\sqrt{\left(N_f/2+\nu_{f}\right)\left(N_f/2-\nu_{f}\right)}}{2N_{f}}
\end{equation}
 The auxiliary fermion Hamiltonian has a matrix representation 
\begin{equation}
\tilde{h}_{i,i',\eta}^{\text{MF}}=\begin{pmatrix}\epsilon_{c,1} & 0 & v_{\ast}\eta\left|\mathbf{k}\right|e^{+i\eta\theta} & 0 & \gamma & v'_{\star}\eta\left|\mathbf{k}\right|e^{-i\eta\theta} & 0 & 0\\
0 & \epsilon_{c,1} & 0 & v_{\ast}\eta\left|\mathbf{k}\right|e^{+i\eta\theta} & v'_{\star}\eta\left|\mathbf{k}\right|e^{+i\eta\theta} & \gamma & 0 & 0\\
v_{\ast}\eta\left|\mathbf{k}\right|e^{-i\eta\theta} & 0 & \epsilon_{c,2} & M & 0 & 0 & 0 & 0\\
0 & v_{\ast}\eta\left|\mathbf{k}\right|e^{+i\eta\theta} & M & \epsilon_{c,2} & 0 & 0 & 0 & 0\\
\gamma & v'_{\star}\eta\left|\mathbf{k}\right|e^{-i\eta\theta} & 0 & 0 & \varepsilon_{f} & 0 & U_{1,af} & 0\\
v'_{\star}\eta\left|\mathbf{k}\right|e^{+i\eta\theta} & \gamma & 0 & 0 & 0 & \varepsilon_{f} & 0 & U_{1,af}\\
0 & 0 & 0 & 0 & U_{1,af} & 0 & \varepsilon_{f} & 0\\
0 & 0 & 0 & 0 & 0 & U_{1,af} & 0 & \varepsilon_{f}
\end{pmatrix}_{i,i'}
\end{equation}
We now perform a $k\cdot p$ expansion near the $\Gamma$ point. The
Hamiltonian at the $\Gamma$ point is given by
\begin{equation}
\tilde{h}_{i,i',\eta}^{\text{MF}}=\begin{pmatrix}\epsilon_{c,1} & 0 & 0 & 0 & \gamma & 0 & 0 & 0\\
0 & \epsilon_{c,1} & 0 & 0 & 0 & \gamma & 0 & 0\\
0 & 0 & \epsilon_{c,2} & M & 0 & 0 & 0 & 0\\
0 & 0 & M & \epsilon_{c,2} & 0 & 0 & 0 & 0\\
\gamma & 0 & 0 & 0 & \varepsilon_{f} & 0 & U_{1,af} & 0\\
0 & \gamma & 0 & 0 & 0 & \varepsilon_{f} & 0 & U_{1,af}\\
0 & 0 & 0 & 0 & U_{1,af} & 0 & \varepsilon_{f} & 0\\
0 & 0 & 0 & 0 & 0 & U_{1,af} & 0 & \varepsilon_{f}
\end{pmatrix}_{i,i'}.
\end{equation}
First we consider the states with $\varepsilon_{f}=\epsilon_{c,1}=\epsilon_{c,2}=M=0$.
The flat band states are then 
\begin{align}
\phi_{0} & =\frac{1}{\sqrt{\frac{U_{1,af}^{2}}{\gamma^{2}}+1}}\left(0,-\frac{U_{1,af}}{\gamma},0,0,0,0,0,1\right)^{T},\\
\phi_{1} & =\frac{1}{\sqrt{\frac{U_{1,af}^{2}}{\gamma^{2}}+1}}\left(-\frac{U_{1,af}}{\gamma},0,0,0,0,0,1,0\right)^{T},\\
\phi_{2} & =\left(0,0,0,1,0,0,0,0\right)^{T},\\
\phi_{3} & =\left(0,0,1,0,0,0,0,0\right)^{T}.
\end{align}
Now we consider the first order correction to these states by the
inclusion of the perturbation 
\begin{equation}
    \mathcal{E}_{i,i'}=\begin{pmatrix}\epsilon_{c,1} & 0 & 0 & 0 & 0 & 0 & 0 & 0\\
0 & \epsilon_{c,1} & 0 & 0 & 0 & 0 & 0 & 0\\
0 & 0 & \epsilon_{c,2} & M & 0 & 0 & 0 & 0\\
0 & 0 & M & \epsilon_{c,2} & 0 & 0 & 0 & 0\\
0 & 0 & 0 & 0 & \varepsilon_{f} & 0 & 0 & 0\\
0 & 0 & 0 & 0 & 0 & \varepsilon_{f} & 0 & 0\\
0 & 0 & 0 & 0 & 0 & 0 & \varepsilon_{f} & 0\\
0 & 0 & 0 & 0 & 0 & 0 & 0 & \varepsilon_{f}
\end{pmatrix}_{i,i'}.
\end{equation}
To first order in the diagonal energies, the states at the $\Gamma$ point become
\begin{align}
\tilde{\phi}_{0} & =\frac{1}{\sqrt{\frac{U_{1,af}^{2}}{\gamma^{2}}+1}}\left(0,-\frac{U_{1,af}}{\gamma},0,0,0,-\frac{U_{1,af}\left(\varepsilon_{f}-\epsilon_{c,1}\right)}{\gamma^{2}+U_{1,af}^{2}},0,1\right)^{T},\\
\tilde{\phi}_{1} & =\frac{1}{\sqrt{\frac{U_{1,af}^{2}}{\gamma^{2}}+1}}\left(-\frac{U_{1,af}}{\gamma},0,0,0,-\frac{U_{1,af}\left(\varepsilon_{f}-\epsilon_{c,1}\right)}{\gamma^{2}+U_{1,af}^{2}},0,1,0\right)^{T},\\
\phi_{2} & =\left(0,0,0,1,0,0,0,0\right)^{T},\\
\phi_{3} & =\left(0,0,1,0,0,0,0,0\right)^{T}.
\end{align}
Projecting the full Hamiltonian onto the active bands near neutrality
we then obtain 
\begin{equation}
h_{\text{kp},ii'}^{\eta}\left(\mathbf{k}\right)=\left(\begin{array}{cccc}
\frac{\varepsilon_{f}\gamma^{2}+\epsilon_{c,1}U_{1,af}^{2}}{U_{1,af}^{2}+\gamma^{2}} & \frac{2\gamma U_{1,af}^{2}\left(\varepsilon_{f}-\epsilon_{c,1}\right)}{\left(\gamma^{2}+U_{1,af}^{2}\right)^{2}}\eta v'_{\star}\left|\mathbf{k}\right|e^{+i\eta\theta} & -\frac{U_{1,af}}{\gamma}\frac{\eta v_{\ast}}{\sqrt{\frac{U_{1,af}^{2}}{\gamma^{2}}+1}}\left|\mathbf{k}\right|e^{-i\eta\theta} & 0\\
\frac{2\gamma U_{1,af}^{2}\left(\varepsilon_{f}-\epsilon_{c,1}\right)}{\left(\gamma^{2}+U_{1,af}^{2}\right)^{2}}v'_{\star}\eta\left|\mathbf{k}\right|e^{-i\eta\theta} & \frac{\varepsilon_{f}\gamma^{2}+\epsilon_{c,1}U_{1,af}^{2}}{U_{1,af}^{2}+\gamma^{2}} & 0 & -\frac{U_{1,af}}{\gamma}\frac{\eta v_{\ast}}{\sqrt{\frac{U_{1,af}^{2}}{\gamma^{2}}+1}}\left|\mathbf{k}\right|e^{+i\eta\theta}\\
-\frac{U_{1,af}}{\gamma}\frac{v_{\ast}}{\sqrt{\frac{U_{1,af}^{2}}{\gamma^{2}}+1}}\eta\left|\mathbf{k}\right|e^{+i\eta\theta} & 0 & \epsilon_{c,2} & M\\
0 & -\frac{U_{1,af}}{\gamma}\frac{\eta v_{\ast}}{\sqrt{\frac{U_{1,af}^{2}}{\gamma^{2}}+1}}\left|\mathbf{k}\right|e^{-i\eta\theta} & M & \epsilon_{c,2}
\end{array}\right)_{ii'}\label{eq:kdophamgen}
\end{equation}
We can then identify the parameters 
\begin{equation}
\tilde{\mu}_{1}=\frac{\varepsilon_{f}\gamma^{2}+\epsilon_{c,1}U_{1,af}^{2}}{U_{1,af}^{2}+\gamma^{2}},\ \ \tilde{s}_{a}=\frac{2\gamma U_{1,af}^{2}\left(\varepsilon_{f}-\epsilon_{c,1}\right)}{\left(\gamma^{2}+U_{1,af}^{2}\right)^{2}},\ \ \tilde{s}_{b}=-\frac{U_{1,af}}{\gamma}\frac{1}{\sqrt{\frac{U_{1,af}^{2}}{\gamma^{2}}+1}}.
\end{equation}
Here, we label the basis in which \ref{eq:kdophamgen} is written
as $\left\{ |1,+\rangle,|1,-\rangle,|2,+\rangle,|2,-\rangle\right\} $
such that the Hamiltonian can be expressed as a sum of tensor products
of two two-level systems. The Pauli matrices $\tilde{\tau}_{i}$ act
on the index $\left(1,2\right)$ while the $\tilde{\sigma}_{i}$ on
the Dirac pseudospin $\left(+,-\right)$. We also define the Pauli
vector $\tilde{\boldsymbol{\sigma}}_{\eta}=\left(\tilde{\sigma}_{x},\eta\tilde{\sigma}_{y}\right)$.
With these definitions, the Hamiltonian becomes:
\begin{equation}
h_{\text{kp}}^{\eta}\left(\mathbf{k}\right)=\left[\frac{\tilde{\mu}_{1}+\epsilon_{c,2}}{2}\tilde{\tau}_{0}+\frac{\tilde{\mu}_{1}-\epsilon_{c,2}}{2}\tilde{\tau}_{z}\right]\otimes\tilde{\sigma}_{0}+\frac{1+\tilde{\tau}_{z}}{2}\otimes\tilde{s}_{a}v_{\ast}'\eta\mathbf{k}\cdot\tilde{\boldsymbol{\sigma}}_{\eta}+\frac{1-\tilde{\tau}_{z}}{2}\otimes M\tilde{\sigma}_{x}+v_{\ast}\tilde{s}_{b}\eta(\tilde{\tau}_{x}\otimes\mathbf{k}\cdot\tilde{\boldsymbol{\sigma}}_{\eta}^{*})\label{eq:kdotp_full_leading}
\end{equation}
Note that at neutrality in the absence of relaxation, we have $\varepsilon_{f}=\epsilon_{c,1}=\epsilon_{c,2}=0$
and the system becomes gapless with a double degeneracy at $\Gamma$
and two additional bands split by $\pm M$ as expected. In the presence
of relaxation, particle-hole is broken and $\epsilon_{c,1}=-\mu+\frac{V^{(0)}}{\Omega_{0}}\nu_{c}+W_{1}\nu_{f}+\mu_{1},$
$\epsilon_{c,2}=-\mu+\frac{V^{(0)}}{\Omega_{0}}\nu_{c}+W_{3}\nu_{f}+\mu_{2}$
and $\varepsilon_{f}=-\mu+W_{1}(\nu_{c,1}+\nu_{c,2})+W_{3}(\nu_{c,2}+\nu_{c,3})+\frac{U_{1}\left(N_f-1\right)}{N_f}\nu_{f}$.
We compare the spectrum from Eq.(\ref{eq:kdotp_full_leading}) with
the full dispersion in Fig.(\ref{fig:vpmassdepkdotp}). 
When either $v_{\ast}^{\prime}=0$ or $M=0$ the Hamiltonian above
has an $SO\left(2\right)$ rotation symmetry. We first focus on the
$v'_{\star}=0$ case, in which the Hamiltonian in Eq.(\ref{eq:kdotp_full_leading}) can be put in block-diagonal
form using the following unitary transformation:
\begin{equation}
U\left(\theta\right)=\left(\begin{array}{cccc}
\frac{e^{-i\theta}}{\sqrt{2}} & 0 & \frac{e^{-i\theta}}{\sqrt{2}} & 0\\
\frac{e^{i\theta}}{\sqrt{2}} & 0 & -\frac{e^{i\theta}}{\sqrt{2}} & 0\\
0 & \frac{1}{\sqrt{2}} & 0 & \frac{1}{\sqrt{2}}\\
0 & \frac{1}{\sqrt{2}} & 0 & -\frac{1}{\sqrt{2}}
\end{array}\right).
\end{equation}
The Hamiltonian then becomes:
\begin{equation}
U\left(\theta\right)h_{\text{kp},ii'}^{\eta}\left(\mathbf{k}\right)U^{\dagger}\left(\theta\right)=\left(\begin{array}{cccc}
\tilde{\mu}_{1} & \eta v_{\ast}\tilde{s}_{b}\left|\mathbf{k}\right| & 0 & 0\\
\eta v_{\ast}\tilde{s}_{b}\left|\mathbf{k}\right| & \epsilon_{c,2}+M & 0 & 0\\
0 & 0 & \tilde{\mu}_{1} & \eta v_{\ast}\tilde{s}_{b}\left|\mathbf{k}\right|\\
0 & 0 & \eta v_{\ast}\tilde{s}_{b}\left|\mathbf{k}\right| & \epsilon_{c,2}-M
\end{array}\right)
\label{eqn:blockDiagkdotp}
\end{equation}
Such that the spectrum decouples into two independent sectors with
energies:
\begin{align}
E_{kp}^{\left(1\right)}\left(\mathbf{k}\right) & =\frac{1}{2}\left(-\sqrt{\left(\tilde{\mu}_{1}-\epsilon_{c,2}-M\right)^{2}+4v_{\ast}^{2}\tilde{s}_{b}^{2}\left|\mathbf{k}\right|^{2}}+\right.\nonumber \\
 & \left.\tilde{\mu}_{1}+\epsilon_{c,2}+M\right) \label{eqn:eigkdotpEM1}\\
E_{kp}^{\left(2\right)}\left(\mathbf{k}\right) & =\frac{1}{2}\left(\sqrt{\left(\tilde{\mu}_{1}-\epsilon_{c,2}-M\right)^{2}+4v_{\ast}^{2}\tilde{s}_{b}^{2}\left|\mathbf{k}\right|^{2}}+\right.\nonumber \\
 & \left.\tilde{\mu}_{1}+\epsilon_{c,2}+M\right)\label{eqn:eigkdotpEM2}\\
E_{kp}^{\left(3\right)}\left(\mathbf{k}\right) & =\frac{1}{2}\left(-\sqrt{\left(\tilde{\mu}_{1}-\epsilon_{c,2}+M\right)^{2}+4v_{\ast}^{2}\tilde{s}_{b}^{2}\left|\mathbf{k}\right|^{2}}+\right.\nonumber \\
 & \left.\tilde{\mu}_{1}+\epsilon_{c,2}-M\right)\label{eqn:eigkdotpEM3}\\
E_{kp}^{\left(4\right)}\left(\mathbf{k}\right) & =\frac{1}{2}\left(\sqrt{\left(\tilde{\mu}_{1}-\epsilon_{c,2}+M\right)^{2}+4v_{\ast}^{2}\tilde{s}_{b}^{2}\left|\mathbf{k}\right|^{2}}+\right.\nonumber \\
 & \left.\tilde{\mu}_{1}+\epsilon_{c,2}-M\right)\label{eqn:eigkdotpEM4}
\end{align}

\subsection{Conditions for the existence of a gap at \texorpdfstring{$\Gamma_{M}$}{GammaM}}\label{sec:existencegapgamma}

To determine whether the  spectrum of the Hamiltonian in Eq.(\ref{eqn:blockDiagkdotp}) is gapped or gapless at the $\Gamma_{M}$
point, we inspect the $\left|\mathbf{k}\right|=0$ eigenvalues in Eq.(\ref{eqn:eigkdotpEM1})-Eq.(\ref{eqn:eigkdotpEM4}). Focusing
on the case of $\nu_{f}=0$, we note that including relaxation will
induce deviations from $\nu_{c}=\nu_{f}=0$. For simplicity, consider
the case where $\nu_{c}=\nu_{f}=0$ but $\left|\nu_{c,a}\right|\neq0$,
which is strictly valid in the zero-hybridization limit, but remains
a good approximation even when $\gamma\neq0$ for small $\mu_{1},\mu_{2}$
compared to $U_{1},\gamma$. At neutrality, we also have $U_{1,af}=U_{1}/2$.
In this case, we have different potentials for $f$, $\Gamma_{1}\oplus\Gamma_{2}$
$c$-fermions, and $\Gamma_{3}$ $c$ fermions and the $\mathbf{k}=0$
energies become:
\begin{align}
E_{kp}^{\left(1\right)}\left(\mathbf{k}=0\right) & =\frac{1}{2}\left(-\left|\frac{1}{1+\frac{4\gamma^{2}}{U_{1}^{2}}}\mu_{1}-\mu_{2}-M\right|+\frac{1}{1+\frac{4\gamma^{2}}{U_{1}^{2}}}\mu_{1}+\mu_{2}+M\right)-\mu\\
E_{kp}^{\left(2\right)}\left(\mathbf{k}=0\right) & =\frac{1}{2}\left(\left|\frac{1}{1+\frac{4\gamma^{2}}{U_{1}^{2}}}\mu_{1}-\mu_{2}-M\right|+\frac{1}{1+\frac{4\gamma^{2}}{U_{1}^{2}}}\mu_{1}+\mu_{2}+M\right)-\mu\\
E_{kp}^{\left(3\right)}\left(\mathbf{k}=0\right) & =\frac{1}{2}\left(-\left|\frac{1}{1+\frac{4\gamma^{2}}{U_{1}^{2}}}\mu_{1}-\mu_{2}+M\right|+\frac{1}{1+\frac{4\gamma^{2}}{U_{1}^{2}}}\mu_{1}+\mu_{2}-M\right)-\mu,\\
E_{kp}^{\left(4\right)}\left(\mathbf{k}=0\right) & =\frac{1}{2}\left(\left|\frac{1}{1+\frac{4\gamma^{2}}{U_{1}^{2}}}\mu_{1}-\mu_{2}+M\right|+\frac{1}{1+\frac{4\gamma^{2}}{U_{1}^{2}}}\mu_{1}+\mu_{2}-M\right)-\mu.
\end{align}
To assess whether the system is gapped or gapless at the $\Gamma_{M}$
point, we then have to consider 4 different cases which depend on
the relative magnitude of $\frac{1}{1+\frac{4\gamma^{2}}{U_{1}^{2}}}\mu_{1}-\mu_{2}$
and $M$, and their relative signs. For simplicity we consider $M>0$,
which is valid near $\theta\approx1.05$ although for larger twist
angles around $\theta\approx1.1$ $M<0$. With this simplification
we only have three cases to consider:

Case I: $\left|\frac{1}{1+\frac{4\gamma^{2}}{U_{1}^{2}}}\mu_{1}-\mu_{2}\right|<M$ , in this case we have
\begin{align}
E_{kp}^{\left(1\right)}\left(\mathbf{k}=0\right) & =\frac{1}{1+\frac{4\gamma^{2}}{U_{1}^{2}}}\mu_{1}-\mu & E_{kp}^{\left(2\right)}\left(\mathbf{k}=0\right) & =M+\mu_{2}-\mu,\\
E_{kp}^{\left(3\right)}\left(\mathbf{k}=0\right) & =-M+\mu_{2}-\mu & E_{kp}^{\left(4\right)}\left(\mathbf{k}=0\right) & =\frac{1}{1+\frac{4\gamma^{2}}{U_{1}^{2}}}\mu_{1}-\mu,
\end{align}
In this case the system is gapless with bands $1$ and $4$ forming
a node at the $\Gamma_{M}$ point and bands $2$ and $3$ being gapped. 

Case II: $\left|\frac{1}{1+\frac{4\gamma^{2}}{U_{1}^{2}}}\mu_{1}-\mu_{2}\right|>M$
and , $\frac{1}{1+\frac{4\gamma^{2}}{U_{1}^{2}}}\mu_{1}-\mu_{2}<0$,
in this case we have
\begin{align}
E_{kp}^{\left(1\right)}\left(\mathbf{k}=0\right) & =\frac{1}{1+\frac{4\gamma^{2}}{U_{1}^{2}}}\mu_{1}-\mu & E_{kp}^{\left(2\right)}\left(\mathbf{k}=0\right) & =M+\mu_{2}-\mu,\\
E_{kp}^{\left(3\right)}\left(\mathbf{k}=0\right) & =\frac{1}{1+\frac{4\gamma^{2}}{U_{1}^{2}}}\mu_{1}-\mu & E_{kp}^{\left(4\right)}\left(\mathbf{k}=0\right) & =-M+\mu_{2}-\mu,
\end{align}
In this case bands $1$ and $3$ form a node and the system overall
is gapped.

\begin{figure}
    \centering
    \includegraphics[width=0.9\linewidth]{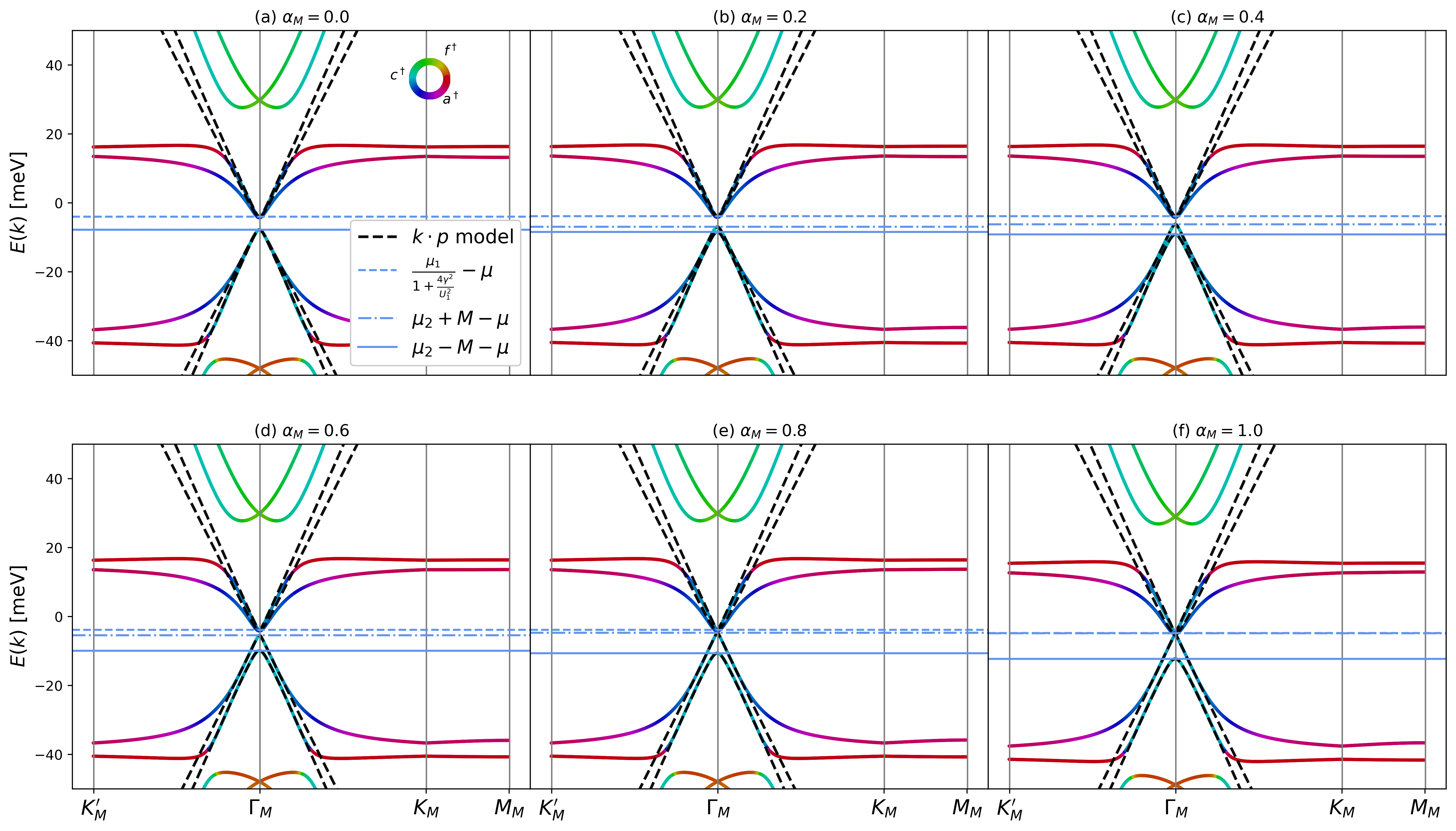}
    \caption{\textbf{Mott semimetal:} Comparison of the Hubbard-I single particle spectra obtained from the topological heavy fermion (THF) model (colored solid lines) with a $k\cdot p$ expansion around the $\Gamma_M$ point (black dashed lines). The panels (a)–(f) display the band dispersion $E(k)$ along the high-symmetry path $K_M^\prime - \Gamma_M - K_M - M_M$ for varying values of the artificial scaling parameter $\alpha_M\in[0,1]$ which multiplies $M$, the splitting of the $\Gamma_1\oplus\Gamma_2$ $c-$fermions. The color of the solid lines represents the orbital character of the bands, as indicated by the inset color wheel. The color wheel denotes the fermion character at each k-point, obtained from the expectation value of $Z_{\rm char} = e^{i\pi}\sigma_0 \oplus e^{i\pi}\sigma_0 \oplus e^{i\pi/3}\sigma_0 \oplus e^{-i\pi/3}\sigma_0$ such that $c-$electrons are colored cyan, $f$-electrons are colored yellow, and $a-$electrons are colored magenta. The $k\cdot p$ model effectively captures the low-energy physics and band crossings near the $\Gamma_M$ point. }

        \label{fig:massdepkdotp}
\end{figure}

Case III: $\left|\frac{1}{1+\frac{4\gamma^{2}}{U_{1}^{2}}}\mu_{1}-\mu_{2}\right|>M$
and , $\frac{1}{1+\frac{4\gamma^{2}}{U_{1}^{2}}}\mu_{1}-\mu_{2}>0$,
we have 
\begin{align}
E_{kp}^{\left(1\right)}\left(\mathbf{k}=0\right) & =M+\mu_{2}-\mu & E_{kp}^{\left(2\right)}\left(\mathbf{k}=0\right) & =\frac{1}{1+\frac{4\gamma^{2}}{U_{1}^{2}}}\mu_{1}-\mu,\\
E_{kp}^{\left(3\right)}\left(\mathbf{k}=0\right) & =-M+\mu_{2}-\mu & E_{kp}^{\left(4\right)}\left(\mathbf{k}=0\right) & =\frac{1}{1+\frac{4\gamma^{2}}{U_{1}^{2}}}\mu_{1}-\mu,
\end{align}
In this case bands $2$ and $4$ form a node and the system overall
is gapped. We introduce a parameter $\alpha_{M}\in[0,1]$ to tune
the splitting of the $\Gamma_{1}\oplus\Gamma_{2}$ $c$-fermion to
its physical value in Tab.(\ref{tab:model_params}) when $\alpha_{M}=1$.
We now compare these expressions with the full numerical results in
the Hubbard I approximation, these are shown in Fig.(\ref{fig:massdepkdotp}).
When $\alpha_{M}=0$ we have case III above and the system is gapped
at the $\Gamma_{M}$ point. As $M$ increases the system goes to case
$I$ and bands 1 and 4 form a node. Additionally, for $\alpha_{M}=1$
we find $E_{kp}^{\left(4\right)}\left(\mathbf{k}=0\right)\approx E_{kp}^{\left(3\right)}\left(\mathbf{k}=0\right)$
giving rise to an approximate three-fold degenerate node at $\Gamma_{M}$
(strictly the node is twofold degenerate). 

\begin{figure}
    \centering
    \includegraphics[width=0.9\linewidth]{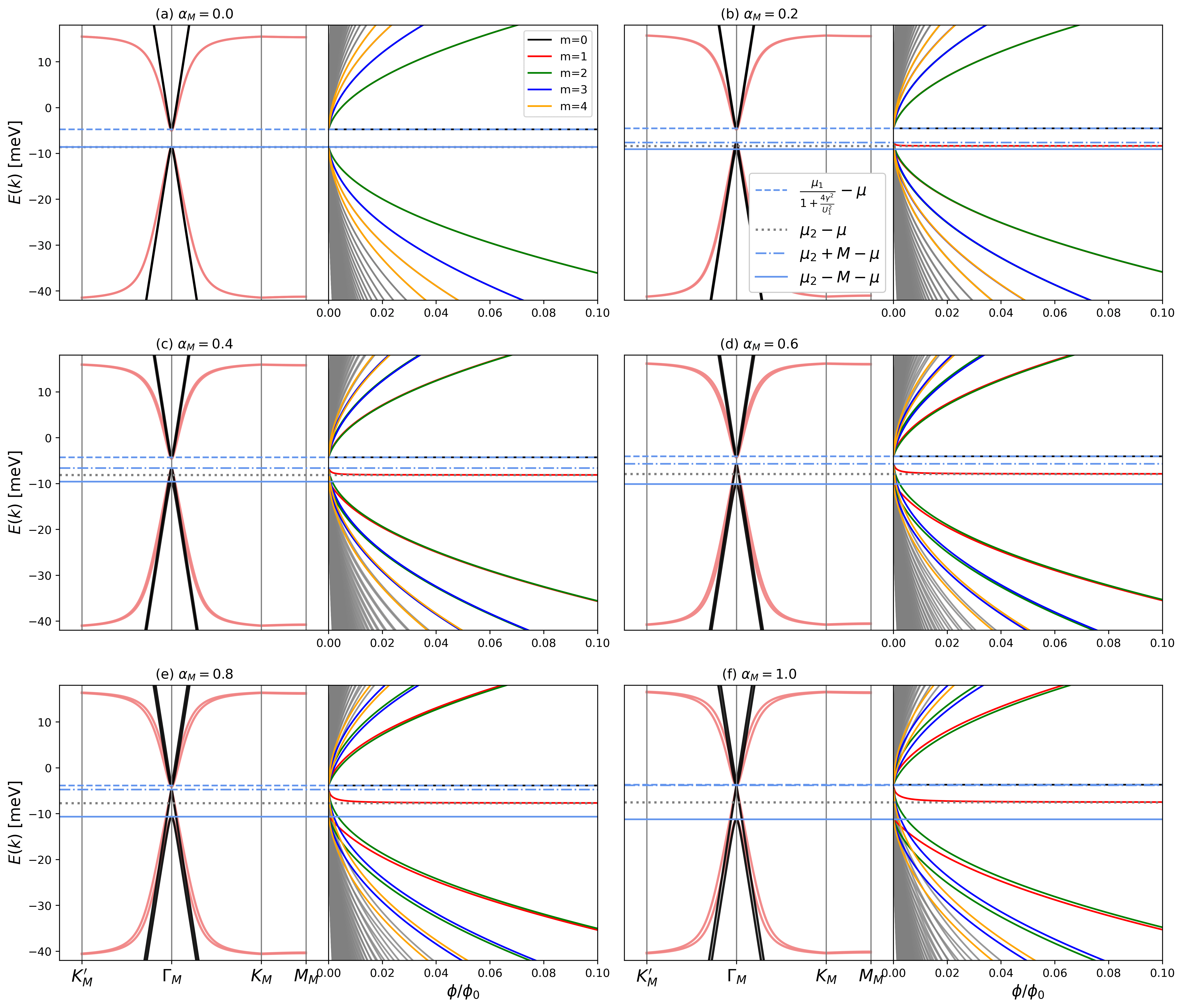}
    \caption{\textbf{Mott semimetal:}Landau level spectrum derived from the $\Gamma_M$-point $k\cdot p$ effective Hamiltonian via canonical substitution. Panels (a)–(f) display the spectral evolution for varying coupling strengths $\alpha_{M}$ ranging from 0.0 to 1.0. Each panel contains two sub-plots: the left sub-plot shows the zero-field band dispersion near the $\Gamma_M$ point, while the right sub-plot presents the Landau level energies as a function of magnetic flux $\phi/\phi_0$.  The angular momentum is shown explicitly for LLs with total angular momentum $m\leq4$. Here, black corresponds to $m=0$, red to $m=1$, green to $m=2$, blue to $m=3$, and yellow to $m=4$. All higher angular momenta are shown in gray. }
    \label{fig:massdepkdotpcanon}
\end{figure}

\subsection{Canonical substitution for the \texorpdfstring{$k\cdot p$}{k.p} Hamiltonian \texorpdfstring{$M\protect\neq0$}{M!=0},
\texorpdfstring{$v_{\ast}^{\prime}=0$}{vstarprime=0}}
 \label{sec:mneq0vprimezero}

Upon confirming the excellent agreement between the $k\cdot p$ Hamiltonian,
we now want to leverage this insight to capture the effect of
$M$ on the Landau level spectrum. To do this, we perform canonical
substitution on Eq.(\ref{eq:kdophamgen}), yielding:
\begin{equation}
h_{kp}^{\eta=+}\left(B\right)=\left(\begin{array}{cccc}
\tilde{\mu}_{1} & 0 & \frac{\sqrt{2}}{\ell}\tilde{s}_{b}v_{\ast}ia^{\dagger} & 0\\
0 & \tilde{\mu}_{1} & 0 & \frac{\sqrt{2}}{\ell}\tilde{s}_{b}v_{\ast}(-ia)\\
\frac{\sqrt{2}}{\ell}\tilde{s}_{b}v_{\ast}(-ia) & 0 & \epsilon_{c,2} & M\\
0 & \frac{\sqrt{2}}{\ell}\tilde{s}_{b}v_{\ast}ia^{\dagger} & M & \epsilon_{c,2}
\end{array}\right)
\end{equation}
Since the $k\cdot p$ Hamiltonian retains full $SO(2)$ symmetry,
we can block diagonalize the spectrum each total angular momentum
sector by choosing the ansatz:
\begin{equation}
\psi_{kp,m}=\left(\begin{array}{cccc}
\psi_{m}^{1}|m+2\rangle, & \psi_{m}^{2}|m\rangle, & \psi_{m}^{3}|m+1\rangle, & \psi_{m}^{4}|m+1\rangle\end{array}\right)^{T}
\end{equation}
Then we get the eigenvalue problem for the matrix:
\begin{align}
h_{kp,m}^{\eta=+}\left(B\right) & =\Delta_{B}^{\star}\left(\begin{array}{cccc}
\tilde{\varepsilon}_{1} & 0 & i\sqrt{m+2} & 0\\
0 & \tilde{\varepsilon}_{1} & 0 & -i\sqrt{m+1}\\
-i\sqrt{m+2} & 0 & \tilde{\varepsilon}_{2} & \tilde{M}\\
0 & i\sqrt{m+1} & \tilde{M} & \tilde{\varepsilon}_{2}
\end{array}\right)
\end{align}
Where we defined the cyclotron energy for the $k\cdot p$ Hamiltonian:
\begin{equation}
\Delta_{B}^{\star}=\frac{\sqrt{2}\tilde{s}_{b}v_{\ast}}{\ell}
\end{equation}
and the renormalized energies for the $c$ and $f$ electrons:
\begin{equation}
\tilde{\varepsilon}_{1}=\frac{\tilde{\mu}_{1}}{\Delta_{B}^{\star}},\ \ \tilde{\varepsilon}_{2}=\frac{\epsilon_{c,2}}{\Delta_{B}^{\star}}.
\end{equation}
The numerical diagonalization of this Hamiltonian is shown in Fig.(\ref{fig:massdepkdotpcanon}).
At $\alpha_{M}=0$ the spectrum near $\Gamma_{M}$ is well captured
by two degenerate massive Dirac cones and so the LLs that emerge from
this dispersion are doubly degenerate on the scale where the $k\cdot p$
model is accurate. As $\alpha_{M}$ increases the degeneracy between
the two massive Dirac cones is split and so is the degeneracy between
LL's. Around $\alpha_{M}=1.0$, we noted that for the parameters in
Tab.(\ref{tab:model_params}) we have a node coexisting with a
quadratic Dirac, which upon minimal coupling, leads to two sets of
LLs that at finite $B$ field. These levels cross at finite $B$ due
to the approximate $SO(2)$ rotational symmetry around the $\Gamma_{M}$
point which block-diagonalizes the LL Hamiltonian. In the full THF
we thus expect these crossing points to become avoided crossings.
We now turn to the anomalous modes in the following section. 

\subsection{Canonical substitution for the \texorpdfstring{$k\cdot p$}{k.p} Hamiltonian \texorpdfstring{$M=0$}{M=0}, \texorpdfstring{$v_{\ast}^{\prime}\protect\neq0$}{vstarprime!=0}} \label{sec:mzerovprimeneq0}

\begin{figure}
    \centering
    \includegraphics[width=0.9\linewidth]{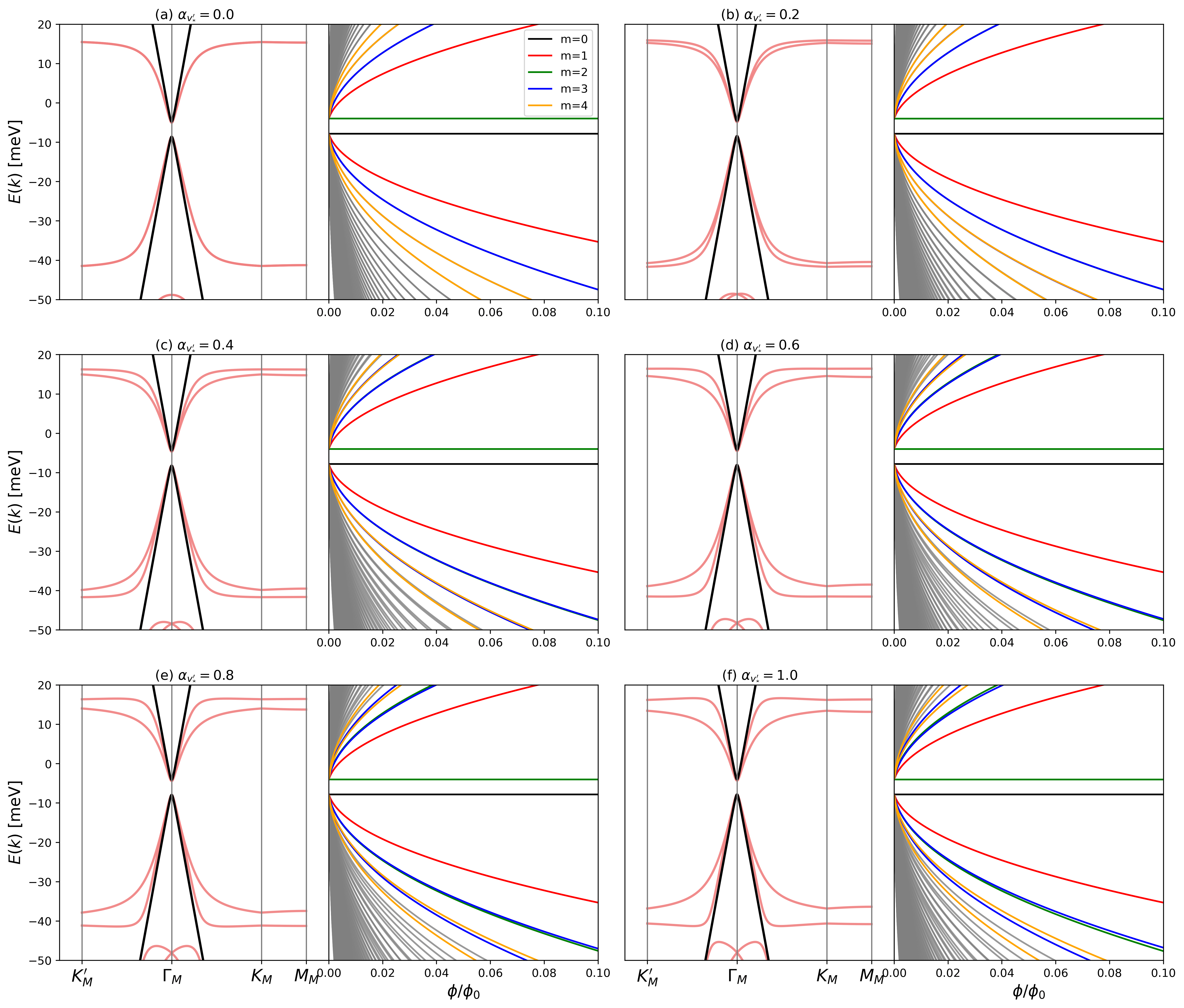}
    \caption{\textbf{Mott semimetal:} Landau level spectrum derived from the $\Gamma_M$-point $k\cdot p$ effective Hamiltonian via canonical substitution. Panels (a)–(f) display the spectral evolution for varying coupling strengths $\alpha_{v^\prime_\ast}$ ranging from 0.0 to 1.0. Each panel contains two sub-plots: the left sub-plot shows the zero-field band dispersion near the $\Gamma_M$ point, while the right sub-plot presents the Landau level energies as a function of magnetic flux $\phi/\phi_0$. The angular momentum is shown explicitly for LLs with total angular momentum $m\leq4$. Here, black corresponds to $m=0$, red to $m=1$, green to $m=2$, blue to $m=3$, and yellow to $m=4$. All higher angular momenta are shown in gray. }
    \label{fig:massdepkdotpcanon2}
\end{figure}

In this case we once again recover an $SO\left(2\right)$ symmetric
limit in which the Hamiltonian in $B$ field is block diagonal. We
perform canonical substitution on Eq.(\ref{eq:kdophamgen}), which
yields:
\begin{equation}
h_{kp}^{\eta=+}\left(B\right)=\left(\begin{array}{cccc}
\tilde{\mu}_{1} & \frac{\sqrt{2}}{\ell}\tilde{s}_{a}v_{\ast}^{\prime}(-ia) & \frac{\sqrt{2}}{\ell}\tilde{s}_{b}v_{\ast}ia^{\dagger} & 0\\
\frac{\sqrt{2}}{\ell}\tilde{s}_{a}v_{\ast}^{\prime}ia^{\dagger} & \tilde{\mu}_{1} & 0 & \frac{\sqrt{2}}{\ell}\tilde{s}_{b}v_{\ast}(-ia)\\
\frac{\sqrt{2}}{\ell}\tilde{s}_{b}v_{\ast}(-ia) & 0 & \epsilon_{c,2} & 0\\
0 & \frac{\sqrt{2}}{\ell}\tilde{s}_{b}v_{\ast}ia^{\dagger} & 0 & \epsilon_{c,2}
\end{array}\right)
\end{equation}
Since the $k\cdot p$ Hamiltonian retains full $SO(2)$ symmetry,
we can block diagonalize the spectrum each total angular momentum
sector by choosing the ansatz:
\begin{equation}
\psi_{kp,m}=\left(\begin{array}{cccc}
\psi_{m}^{1}|m+1\rangle, & \psi_{m}^{2}|m+2\rangle, & \psi_{m}^{3}|m\rangle, & \psi_{m}^{4}|m+3\rangle\end{array}\right)^{T}
\end{equation}
Then we get the eigenvalue problem for the matrix:
\begin{align}
h_{kp,m}^{\eta=+}\left(B\right) & =\Delta_{B}^{\star}\left(\begin{array}{cccc}
\tilde{\varepsilon}_{1} & -i\frac{\tilde{s}_{a}v_{\ast}^{\prime}}{\tilde{s}_{b}v_{\ast}}\sqrt{m+1} & i\sqrt{m+2} & 0\\
i\frac{\tilde{s}_{a}v_{\ast}^{\prime}}{\tilde{s}_{b}v_{\ast}}\sqrt{m+1} & \tilde{\varepsilon}_{1} & 0 & -i\sqrt{m+1}\\
-i\sqrt{m+2} & 0 & \tilde{\varepsilon}_{2} & 0\\
0 & i\sqrt{m+1} & 0 & \tilde{\varepsilon}_{2}
\end{array}\right)
\end{align}
The numerical diagonalization of this Hamiltonian is shown in Fig.(\ref{fig:massdepkdotpcanon2}).
At $\alpha_{v_{\ast}^{\prime}}=0$ the spectrum near $\Gamma_{M}$
is well captured by two degenerate massive Dirac cones and so the
LLs that emerge from this dispersion are doubly degenerate on the
scale where the $k\cdot p$ model is accurate. However, the angular
momentum labeling at finite field differs from that in Fig.(\ref{fig:massdepkdotpcanon}).
This stems from the fact that in the $v_{\ast}^{\prime}=0,M=0$ limit,
the angular momentum labeling of the two independent Dirac cones has
a gauge freedom as they can be rotated independently by separate $SO(2)$
rotations. We also see the absence of a crossing that closes the $C=8$
gap in this limit. As such only in the limit where both $v_{\ast}^{\prime}\neq0,M\neq0$
we find a crossing that closes the $C=8$ gap as we show in the next
section. 

\subsection{Canonical substitution for the \texorpdfstring{$k\cdot p$}{k.p} Hamiltonian \texorpdfstring{$M\protect\neq0$}{M!-0},
\texorpdfstring{$v_{\ast}^{\prime}\protect\neq0$}{vstarprime!=0}}

\begin{figure}
    \centering
    \includegraphics[width=0.5\linewidth]{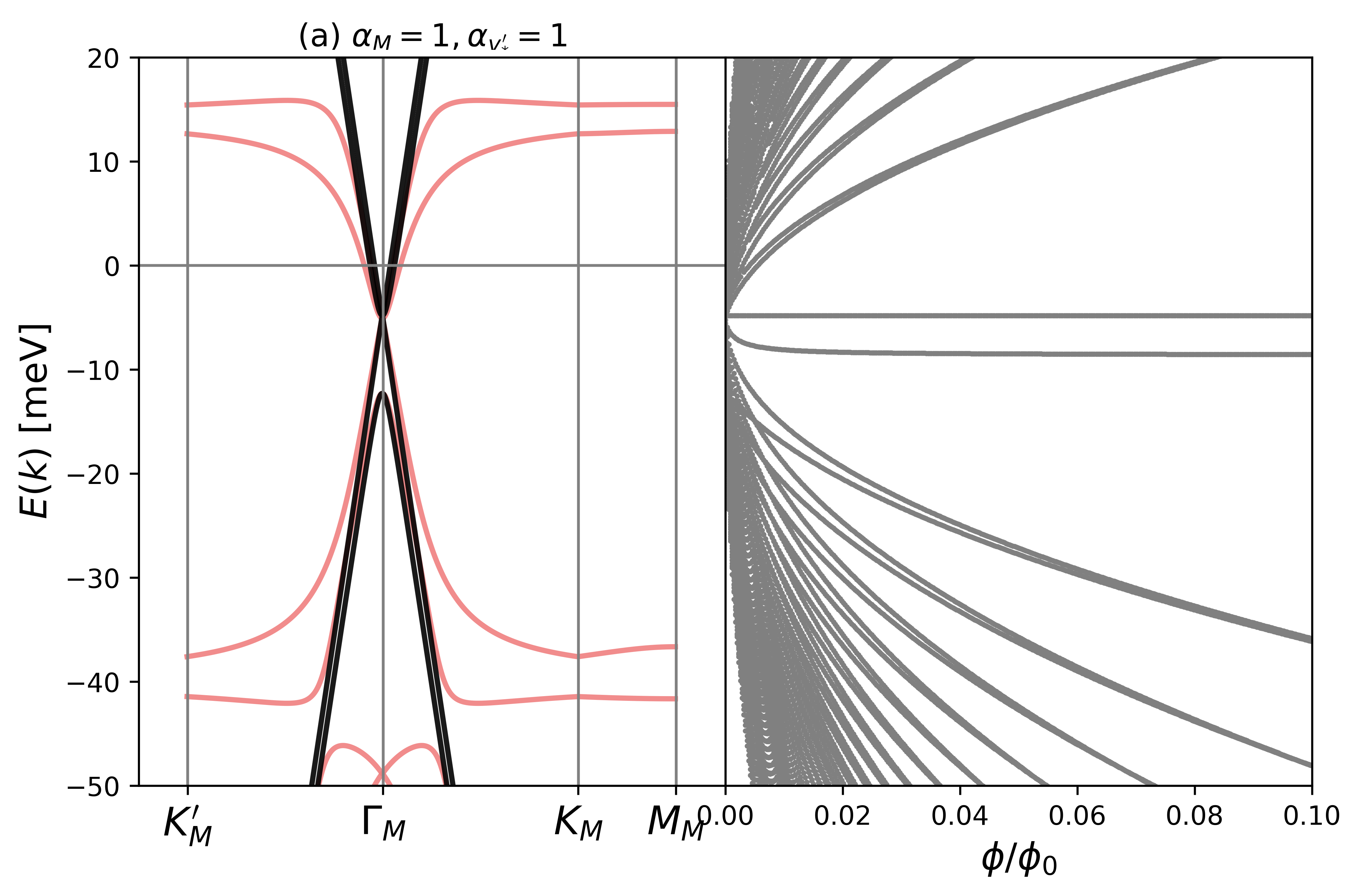}
    \caption{\textbf{Mott semimetal:} Landau level spectrum derived from the $\Gamma_M$-point $k\cdot p$ effective Hamiltonian via canonical substitution. Panels (a)–(f) display the spectral evolution for varying coupling strengths $\alpha_M$ ranging from 0.0 to 1.0. Each panel contains two sub-plots: the left sub-plot shows the zero-field band dispersion near the $\Gamma_M$ point, while the right sub-plot presents the Landau level energies as a function of magnetic flux $\phi/\phi_0$. }
    \label{fig:massdepkdotpcanon3}
\end{figure}

In this limit the $SO(2)$ symmetry is completely broken, the Hamiltonian
within valley $\eta=+$ is 
\begin{equation}
h_{kp}^{\eta=+}\left(B\right)=\left(\begin{array}{cccc}
\tilde{\mu}_{1} & \frac{\sqrt{2}}{\ell}\tilde{s}_{a}v_{\ast}^{\prime}(-ia) & \frac{\sqrt{2}}{\ell}\tilde{s}_{b}v_{\ast}ia^{\dagger} & 0\\
\frac{\sqrt{2}}{\ell}\tilde{s}_{a}v_{\ast}^{\prime}ia^{\dagger} & \tilde{\mu}_{1} & 0 & \frac{\sqrt{2}}{\ell}\tilde{s}_{b}v_{\ast}(-ia)\\
\frac{\sqrt{2}}{\ell}\tilde{s}_{b}v_{\ast}(-ia) & 0 & \epsilon_{c,2} & M\\
0 & \frac{\sqrt{2}}{\ell}\tilde{s}_{b}v_{\ast}ia^{\dagger} & M & \epsilon_{c,2}
\end{array}\right)
\end{equation}
We also carry out a full numerical diagonalization of this Hamiltonian, and the result is shown in Fig.(\ref{fig:massdepkdotpcanon3}). Note that the value of the gap beyond the crossing is much smaller than in the full calculation.

\section{Spectral functions of the anomalous modes within Hubbard-I with relaxation in absence of Zeeman}\label{sec:anomalous}

We can now calculate the effect of the relaxation and chiral symmetry
breaking from the operator:
\begin{equation}
V_{ii'\eta=+}\left(B\right)=\begin{pmatrix}\mu_{1} & i\Delta_{B,2}a^{\dagger} & 0 & -i\Delta_{B,1}a & 0 & 0 & 0 & 0\\
-i\Delta_{B,2}a & \mu_{1} & i\Delta_{B,1}a^{\dagger} & 0 & 0 & 0 & 0 & 0\\
0 & -i\Delta_{B,1}a & \mu_{2} & 0 & 0 & 0 & 0 & 0\\
i\Delta_{B,1}a^{\dagger} & 0 & 0 & \mu_{2} & 0 & 0 & 0 & 0\\
0 & 0 & 0 & 0 & \varepsilon_{f} & 0 & 0 & 0\\
0 & 0 & 0 & 0 & 0 & \varepsilon_{f} & 0 & 0\\
0 & 0 & 0 & 0 & 0 & 0 & \varepsilon_{f} & 0\\
0 & 0 & 0 & 0 & 0 & 0 & 0 & \varepsilon_{f}
\end{pmatrix}_{ii'}
\end{equation}

where $\Delta_{B,i}=\sqrt{2}v_{i}\ell^{-1}$.
For simplicity, first consider the $\varepsilon_{f}=0$ limit which
holds at charge neutrality considering the zero-shot mean field density
matrix at neutrality. The leading order corrections to the energy
are then given by 

\begin{equation}
E_{0,1}^{(1)}=\frac{\mu_{1}}{1+\left(\frac{2\gamma\Sigma\left(0\right)}{U_{1}}\right)^{2}+\left(\frac{2\Delta_{B}^{\prime}\Sigma\left(1\right)}{U_{1}}\right)^{2}},\ \ \ E_{0,2}^{(1)}=\frac{\mu_{1}+\left(\frac{\Delta_{B}}{M}\right)^{2}\mu_{2}-\frac{2\Delta_{B}\Delta_{B,1}}{M}}{1+\left(\frac{\Delta_{B}}{M}\right)^{2}+\left(\frac{2\gamma\Sigma\left(1\right)}{U_{1}}\right)^{2}+2\left(\frac{2\Delta_{B}^{\prime}\Sigma\left(2\right)}{U_{1}}\right)^{2}}.\label{eq:pertheoanom}
\end{equation}
\begin{figure}
    \centering
    \includegraphics[width=0.7\linewidth]{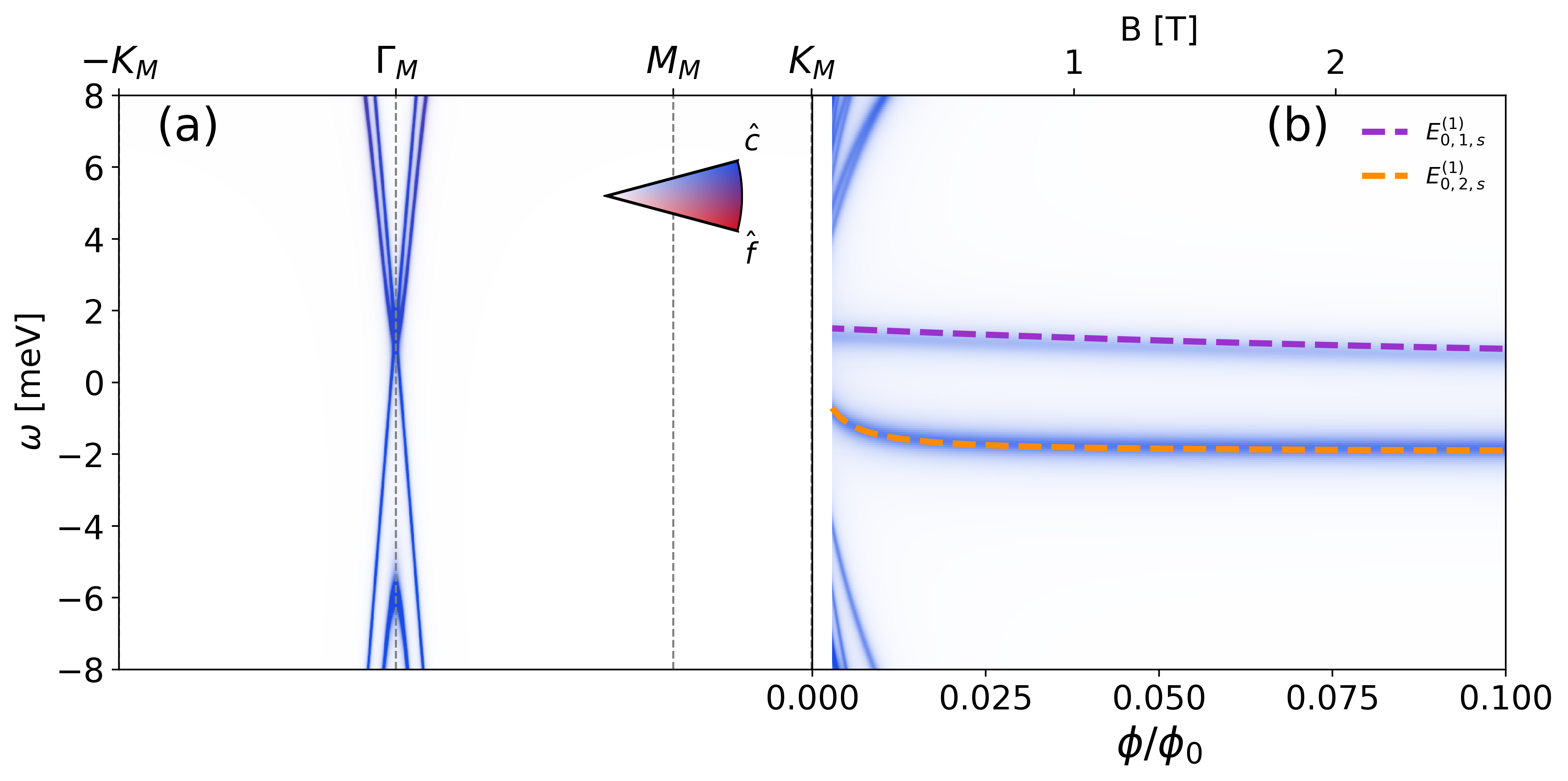}
    \caption{ \textbf{Mott semimetal:}
    (a) Zero-field spectral function along the high-symmetry path $K_M-\Gamma_M-M_M-K_M$, showing the low-energy Mott-semimetal cones near $\Gamma_M$. The color intensity denotes spectral weight, and the inset indicates the relative $\hat c$- and $\hat f$- character of the bands. (b) Finite-field spectrum as a function of flux ($\phi/\phi_0$), with the corresponding magnetic field ($B$) shown on the upper axis. The blue intensity shows the numerical spectral weight, while the dashed curves are the analytical anomalous-mode energies $E^{(1)}_{0,1,s}$ and $E^{(1)}_{0,2,s}$ obtained from the exact chiral-limit zero modes, with perturbative corrections from the relaxation parameters. The close agreement confirms that the low-energy anomalous modes are accurately captured by the reduced analytical description.
    }
    \label{fig:spectral_anom_comp}
\end{figure}
Importantly, we note that $\Delta_{B,2}$ does not affect the energies
of the zero modes to leading order in perturbation theory. For simplicity
we set $\Sigma\left(0\right)=\Sigma\left(1\right)=\Sigma\left(2\right)=1$
and we now calculate the slope of these anomalous modes around the
$B=0$ limit. We now define, 
\begin{equation}
A_{\star}=\frac{2ev_{\ast}^{2}}{\hbar M^{2}},\qquad A_{\star}^{\prime}=\frac{8ev_{\ast}^{\prime2}}{\hbar U_{1}^{2}}\qquad A_{\star1}=\frac{4ev_{\ast}v_{1}}{\hbar M}\qquad D_{\star}=1+\left(\frac{2\gamma}{U_{1}}\right)^{2}
\end{equation}

\begin{figure}
    \centering
    \includegraphics[width=0.5\linewidth]{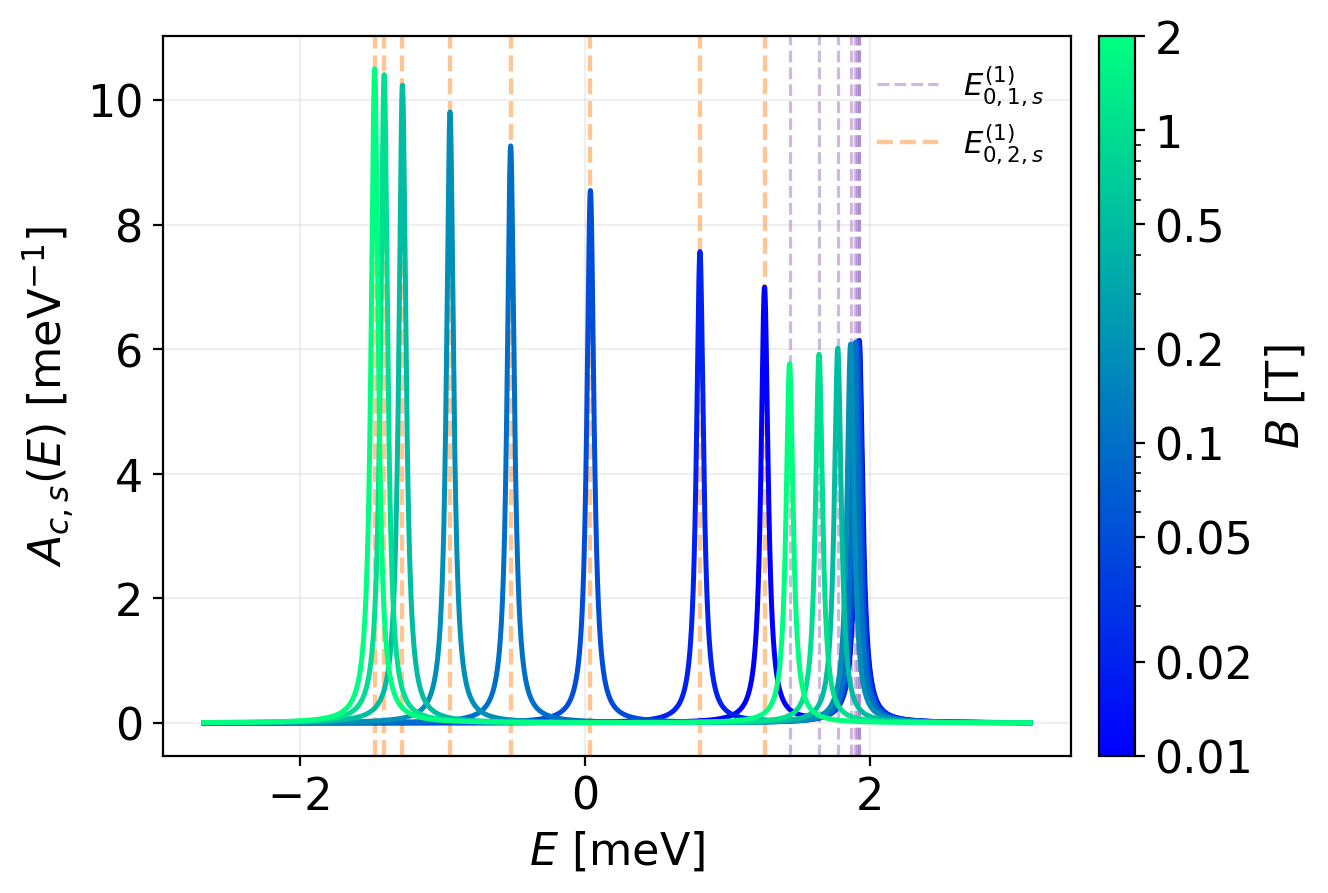}
    \caption{\textbf{Mott semimetal:} Local $c$-electron spectral function $A_{c,s}(E)$ for a fixed spin sector $s$, plotted for magnetic fields $B=0.01-2,\mathrm{T}$. The color of each solid curve denotes $B$, on the logarithmic scale shown by the colorbar. The vertical dashed lines mark the analytical anomalous-mode energies $E^{(1)}_{0,1,s}$ and $E^{(1)}_{0,2,s}$ at the corresponding fields. The agreement between the spectral peaks and the dashed lines shows that the low-energy finite-field spectral weight is controlled by the two analytically obtained anomalous modes.}
    \label{fig:spectral_anom}
\end{figure}

In terms of these definitions, the anomalous Landau levels have the
following form 
\begin{equation}
E_{0,1}^{(1)}=\frac{\mu_{1}}{D_{\star}+A_{\star}^{\prime}B},\ \ \ E_{0,2}^{(1)}=\frac{\mu_{1}+\left(A_{\star}\mu_{2}-A_{\star1}\right)B}{D_{\star}+\left(A_{\star}+2A_{\star}^{\prime}\right)B}.
\end{equation}
As such, near $B=0$ we can expand to linear order in the magnetic
field:
\begin{align}
E_{0,1}^{(1)} & \approx\frac{\mu_{1}}{D_{\star}}-\mu_{1}A_{\star}^{\prime}\frac{B}{D_{\star}^{2}}\\
E_{0,2}^{(1)} & \approx\frac{\mu_{1}}{D_{\star}}+\left(D_{\star}\left(A_{\star}\mu_{2}-A_{\star1}\right)-\left(A_{\star}+2A_{\star}^{\prime}\right)\mu_{1}\right)\frac{B}{D_{\star}^{2}}.
\end{align}
First, note that as $B\rightarrow0$ the energies of these two modes
coincide and is pinned to the energy of the node derived in Sec.(\ref{sec:kdotprelaxationneut}).
Since $D_{\star}>0$ and $A_{\star}^{\prime}>0$ we conclude that
$E_{0,1}^{(1)}$ will disperse downward with magnetic field as long
as $\mu_{1}>0$. Note that in the chiral limit $v_{\ast}^{\prime}=0$
$E_{0,1}^{(1)}$ becomes field independent. Similarly, $E_{0,2}^{(1)}$
will disperse downward with magnetic field as long as $\left(A_{\star}+2A_{\star}^{\prime}\right)\mu_{1}>D_{\star}\left(A_{\star}\mu_{2}-A_{\star1}\right)$.
In the $v_{\ast}^{\prime}=0$ chiral limit and using the simplified
relaxation model where $v_{1}=0$ this condition for $E_{0,2}^{(1)}$
dispersing downward with $B$ greatly simplifies to $\mu_{1}>\left(1+\left(\frac{2\gamma\Sigma\left(0\right)}{U_{1}}\right)^{2}\right)\mu_{2}$.
In Fig.(\ref{fig:spectral_anom_comp}) we show the comparison of the analytical expressions
from perturbation theory in Eq.(\ref{eq:pertheoanom}) and the
numerically obtained Landau level spectrum in the Hubbard I approximation.
We can then calculate the spectral function of these two anomalous
modes if we ignoring mixing with higher Landau levels from $v_{1},v_{2},\mu_{1},\mu_{2}$.
To leading order in the relaxation parameters, $v_{1},v_{2},\mu_{1},\mu_{2}$,
the spectral function is given by 
\begin{equation}
A_{c}\left(E\right)=\frac{1}{1+\left(\frac{2\gamma\Sigma\left(0\right)}{U_{1}}\right)^{2}+\left(\frac{2\Delta_{B}^{\prime}\Sigma\left(1\right)}{U_{1}}\right)^{2}}\delta\left(E-E_{0,1}^{(1)}\right)+\frac{1+\left(\frac{\Delta_{B}}{M}\right)^{2}}{1+\left(\frac{\Delta_{B}}{M}\right)^{2}+\left(\frac{2\gamma\Sigma\left(1\right)}{U_{1}}\right)^{2}+2\left(\frac{2\Delta_{B}^{\prime}\Sigma\left(2\right)}{U_{1}}\right)^{2}}\delta\left(E-E_{0,2}^{(1)}\right)
\end{equation}
where the spectral weights are obtained directly from the wavefunctions in Eqs.(\ref{eq:zeromodenonchiral}) and (\ref{eq:zeromodenonchiral2}).
The behavior of these two flat modes is shown in Fig.(\ref{fig:spectral_anom}).
The mode that disperses with magnetic field ($E_{0,1}^{(1)}$ when
$\Sigma(0)=1$) becomes sharper with increasing magnetic field while
the non-dispersing mode has a quasi-particle weight that remains constant
with magnetic field. However, note that the Luttinger count is not
preserved by the Hubbard-I approximation, which is why the spectral
weight in these two Landau levels does not remain fixed as a function
of $B$. Deviations from $\Sigma(0)=1$ will induce a finite $B$
dependence on both modes and to leading order in $B$ both modes decrease
in energy with increasing magnetic field. 

%------------_Section Change----------------------------

\section{Landau level spectrum of the unstrained Mott Semimetal including the Zeeman
coupling}\label{sec:LLzeemann}

We now turn to the problem of determining the Landau-level spectrum
within the Hubbard I approximation when the Zeeman coupling, $-g\mu_Bs/2$ with s=$\{\uparrow,\downarrow\}=\{+1,-\}$, is included.
The spin diagonal atomic self-energy in presence of a Zeeman field for the THF model derived in \cite{aux_map_B2026}
is given by:
\begin{equation}
\Sigma_{s}^{\text{At}}\left(i\omega_{n}\right)=\begin{cases}
\tilde{\Sigma}_{s}^{\text{At}}\left(i\omega_{n}\right)\delta_{ii'}\delta_{\eta\eta'}\delta_{ss;} & \text{if }i=5,6\\
0 & \text{otherwise}
\end{cases}
\end{equation}
for spin $s$. The scalar self energy for integer filling of $f$ electrons
in the ground state is given by

\begin{equation}
\tilde{\Sigma}_{s}^{\text{At}}\left(i\omega_{n}\right)=\frac{U_{1}^{2}n_{s}(1-n_{s})}{i\omega_{n}+\mu+hs+\frac{U_{1}}{2}(1-2n_{s})},\ \ 
\end{equation}
 where $U_{1}$ is the strength of the Hubbard interaction, $N_{f}=8$
is the number of $f$-electron flavors, $h=\frac{g\mu_{B}B}{2}$ is the Zeeman coupling and $n_{s}$ is the filling
fraction of the $f$-electrons. Additionally, the mean-field Hamiltonian
is given by 
\begin{equation}
h_{i\eta s;i'\eta's'}^{\text{MF}}(\mathbf{k})=\delta_{\eta,\eta'}\delta_{s,s'}\tilde{h}_{i,i',\eta,s}^{\text{MF}}(\mathbf{k})
\end{equation}

with the orbital blocks of the Hamiltonian given by

\begin{align}
 & h_{i,i',\eta=+,s}^{\text{MF}}=\\
 & \begin{pmatrix}\epsilon_{c,1}-sh & 0 & v_{\ast}k & 0 & \gamma e^{-\lambda^{2}\frac{\left|\mathbf{k}\right|^{2}}{2}} & v'_{\star}e^{-\lambda^{2}\frac{\left|\mathbf{k}\right|^{2}}{2}}\bar{k}\\
0 & \epsilon_{c,1}-sh & 0 & v_{\ast}\bar{k} & v'_{\star}e^{-\lambda^{2}\frac{\left|\mathbf{k}\right|^{2}}{2}}k & \gamma e^{-\lambda^{2}\frac{\left|\mathbf{k}\right|^{2}}{2}}\\
v_{\ast}\bar{k} & 0 & \epsilon_{c,2}-J\left(n_{s}-\frac{1}{2}\right)-sh & M & 0 & 0\\
0 & v_{\ast}k & M & \epsilon_{c,2}-J\left(n_{s}-\frac{1}{2}\right)-sh & 0 & 0\\
\gamma e^{-\lambda^{2}\frac{\left|\mathbf{k}\right|^{2}}{2}} & v'_{\star}e^{-\lambda^{2}\frac{\left|\mathbf{k}\right|^{2}}{2}}\bar{k} & 0 & 0 & \epsilon_{f}+\frac{U_{1}}{2}(1-2n_{s})-sh & 0\\
v'_{\star}e^{-\lambda^{2}\frac{\left|\mathbf{k}\right|^{2}}{2}}k & \gamma e^{-\lambda^{2}\frac{\left|\mathbf{k}\right|^{2}}{2}} & 0 & 0 & 0 & \epsilon_{f}+\frac{U_{1}}{2}(1-2n_{s})-sh
\end{pmatrix}.\nonumber 
\end{align}

In the above, we only consider the Hartree potentials $\epsilon_{c,1}=-\mu+\frac{V^{(0)}}{\Omega_{0}}\nu_{c}+W_{1}\nu_{f},$
$\epsilon_{c,2}=-\mu+\frac{V^{(0)}}{\Omega_{0}}\nu_{c}+W_{3}\nu_{f}$
and $\epsilon_{f}=-\mu+W_{1}(\nu_{c,1}+\nu_{c,2})+W_{3}(\nu_{c,2}+\nu_{c,3})$,
with $\nu_{c,a}$ denote the filling of the $c$ electrons with orbital
index $a$ and $\nu_{c}$ is the total filling fraction of the $c$
electrons and $h\left(B\right)=\frac{g\mu_{B}B}{2}$. One can reproduce
the pole structure of the Green's function by introducing an additional
coupling to auxiliary fermions $\hat{a}_{\mathbf{k}\alpha\eta s}$.
The coupled $f-a$ block of the Hamiltonian in momentum space is given
by 

\begin{align}
\hat{H}^{\text{Aux},fa} & =\sum_{\mathbf{k}\alpha\eta s}\left[\epsilon_{f}+\frac{U_{1}}{2}(1-2n_{s})-sh\left(B\right)\right]\hat{f}_{\mathbf{k}\alpha\eta s}^{\dagger}\hat{f}_{\mathbf{k}\alpha\eta s}\nonumber \\
 & +\sum_{\mathbf{k}\alpha\eta s}\left[\epsilon_{f}+\frac{U_{1}}{2}(2n_{s}-1)-sh\left(B\right)\right]\hat{a}_{\mathbf{k}\alpha\eta s}^{\dagger}\hat{a}_{\mathbf{k}\alpha\eta s}\nonumber \\
 & +\sum_{\mathbf{k}\alpha\eta s}U_{1}\sqrt{n_{s}(1-n_{s})}\left(\hat{a}_{\mathbf{k}\alpha\eta s}^{\dagger}\hat{f}_{\mathbf{k}\alpha\eta s}+\text{h.c}\right),
\end{align}

where $\hat{f}_{\mathbf{k}\alpha\eta s}^{\dagger}$ and $\hat{a}_{\mathbf{k}\alpha\eta s}^{\dagger}$
are the creation operators for the $f$ and auxiliary fermions at
momentum $\mathbf{k}$, valley $\eta$ and spin $s$, respectively.
Consider the case where the Hartree corrections vanish setting $\nu_{c}=,\nu_{f}=0$.
Furthermore, we introduce the following definitions 
\begin{equation}
V_{s}\left(B\right)=U_{1}\sqrt{n_{s}(1-n_{s})}\qquad U_{s}\left(B\right)=U_{1}\left(n_{s}-\frac{1}{2}\right)\qquad J_{s}\left(B\right)=-J\left(n_{s}-\frac{1}{2}\right).\label{eq:Defszeeman}
\end{equation}

To determine the spectrum at finite $B-$field, we perform minimal
substitution making the replacement $(k_{x}+ik_{y})\rightarrow-i\frac{\sqrt{2}}{\ell}\hat{a}$.
Following the same conventions as in Sec.(\ref{sec:Landau-Level-spectrum}), the auxiliary Hubbard-I
Hamiltonian takes the following form:

\begin{align}
 & \tilde{h}_{i,i',\eta=+,s}^{\text{MF}}\left(B\right)=\nonumber \\
 & \begin{pmatrix}-sh & 0 & -i\Delta_{B}a & 0 & \gamma\Sigma\left(a^{\dagger}a\right) & i\Delta_{B}^{\prime}a^{\dagger}\Sigma\left(a^{\dagger}a\right) & 0 & 0\\
0 & -sh & 0 & i\Delta_{B}a^{\dagger} & -i\Delta_{B}^{\prime}a\Sigma\left(a^{\dagger}a\right) & \gamma\Sigma\left(a^{\dagger}a\right) & 0 & 0\\
i\Delta_{B}a^{\dagger} & 0 & J_{s}\left(B\right)-sh & M & 0 & 0 & 0 & 0\\
0 & -i\Delta_{B}a & M & J_{s}\left(B\right)-sh & 0 & 0 & 0 & 0\\
\gamma\Sigma\left(a^{\dagger}a\right) & i\Delta_{B}^{\prime}a^{\dagger}\Sigma\left(a^{\dagger}a\right) & 0 & 0 & -U_{s}\left(B\right)-sh & 0 & V_{s}\left(B\right) & 0\\
-i\Delta_{B}^{\prime}a\Sigma\left(a^{\dagger}a\right) & \gamma\Sigma\left(a^{\dagger}a\right) & 0 & 0 & 0 & -U_{s}\left(B\right)-sh & 0 & V_{s}\left(B\right)\\
0 & 0 & 0 & 0 & V_{s}\left(B\right) & 0 & U_{s}\left(B\right)-sh & 0\\
0 & 0 & 0 & 0 & 0 & V_{s}\left(B\right) & 0 & U_{s}\left(B\right)-sh
\end{pmatrix}.\label{eq:the_MF_ham_Zee}
\end{align}

\begin{figure}
    \centering
    \includegraphics[width=1.0\linewidth]{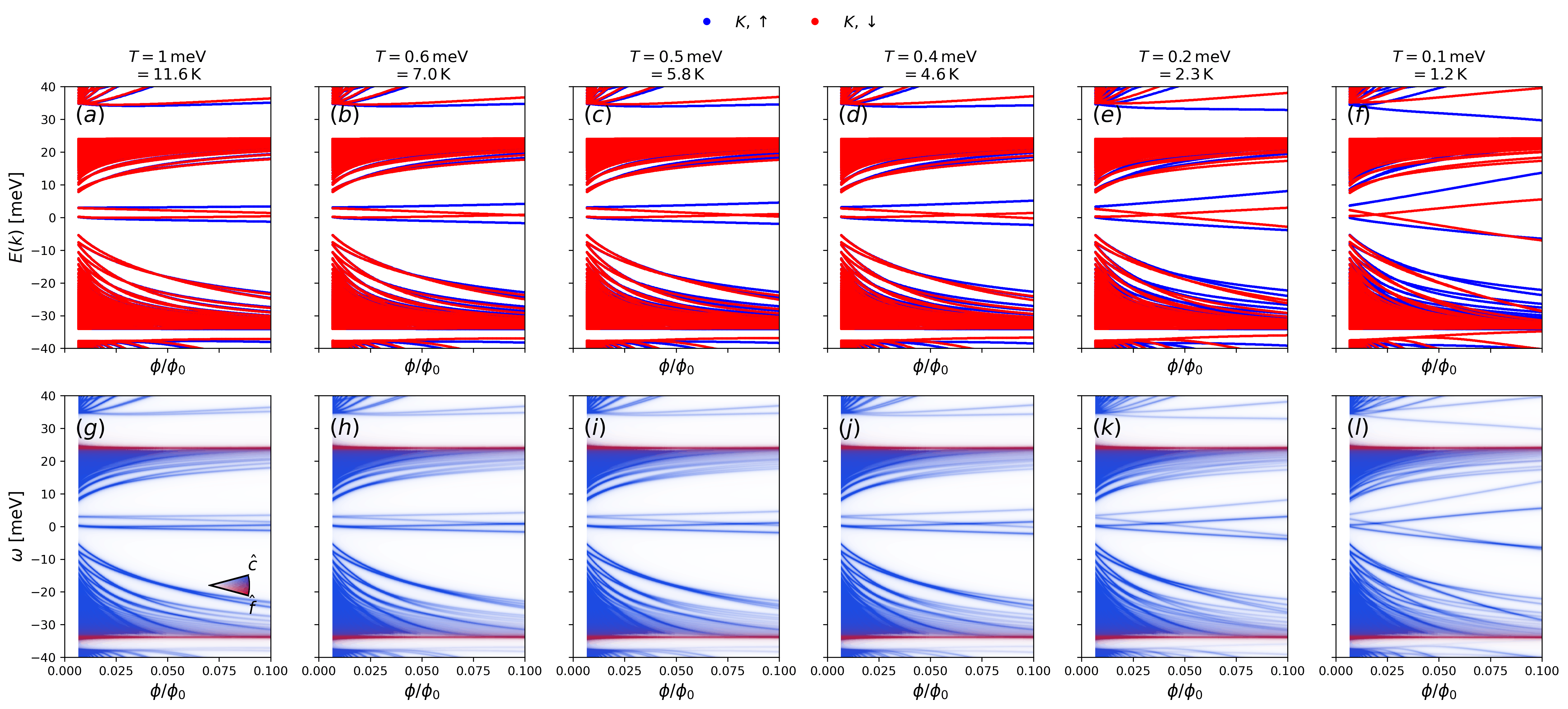}
    \caption{\textbf{Mott semimetal:} Landau-level spectrum of the THF model at charge neutrality including relaxation and the Zeeman-corrected self-energy at different temperatures as a function of flux $\phi/\phi_0$. The upper panels show the spin-resolved spectrum in the (K) valley, with blue and red denoting the two spin sectors. The lower panels show the corresponding low-energy spectral weight, with the color intensity indicating the relative orbital character as illustrated by the inset. Upon lowering temperature, the anomalous modes acquire a stronger field dependence and a larger spin splitting. This reflects the Curie enhancement of the local-moment susceptibility, which amplifies the Hartree-Fock contribution from the exchange interaction and controls the dispersion of the anomalous modes at small magnetic field. }
    \label{fig:tempevolrel}
\end{figure}

We now further add the relaxation terms in Eq.(\ref{Eq:Relaxation_for_Hubb1}), the resulting
spin-resolved spectrum is shown in Fig.(\ref{fig:tempevolrel}) as a function of temperature.
As temperature decreases, the fluctuations of the local moments are
quenched and the spectrum is strongly modified since $n_{s}\rightarrow\delta_{s,\uparrow}$. To quantify the onset of this high-temperature
limit, note that the characteristic Zeeman energy scale is $h=\frac{g\mu_{B}B}{2}$
such that at $B=1$T and assuming a $g-$factor of $g=2$, we find
$h\approx0.1$meV. Consequently, we anticipate that the effect of the
Zeeman splitting is negligible at temperatures comparable to previous
spectroscopic measurements (5-6K) Ref.\cite{2020Natur.582..198W}. At low temperatures when
$\beta h\gg1$, the system becomes spin polarized deviating strongly
from the fluctuating moment regime analyzed in Sec.(\ref{sec:Landau-Level-spectrum}) which is
reflected in the fact that the hybridization between the $f$ and
auxiliary fermions vanishes $V_{s}\left(B\right)\rightarrow0$ as
$\beta h\rightarrow\infty$. Nevertheless, we find that at low magnetic
fields we recover a spectra that resembles the gap hierarchy of the
Hamiltonian in Sec.(\ref{sec:Landau-Level-spectrum}) where the $C=\pm8$ gap is negligible compared
to the $C=\pm12$ gap. 

\begin{figure}
    \centering
    \includegraphics[width=1.0\linewidth]{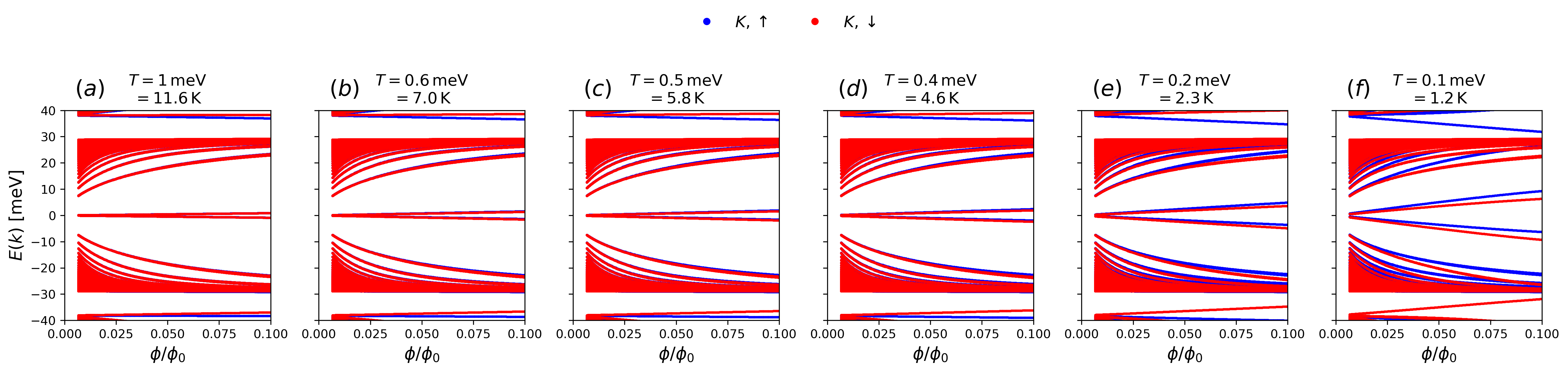}
    \caption{\textbf{Mott semimetal:} Landau-level spectrum of the THF model at charge neutrality in the absence of relaxation and including the Zeeman-corrected self-energy at different temperatures as a function of flux $\phi/\phi_0$ in the chiral-flat limit $M=v^\prime_\ast=0$. Each subplot shows spin-resolved spectrum in the (K) valley, with blue and red denoting the two spin sectors. Upon lowering temperature, the anomalous modes acquire a stronger field dependence and a larger spin splitting. However, at higher temperatures above $5$K the spin-splitting becomes negligible compared with }
    \label{fig:tempevo}
\end{figure}

To simplify the analysis of the Landau level spectra, we show the
location of the poles of the Green's function obtained by diagonalizing
the Hamiltonian in Eq.(\ref{eq:the_MF_ham_Zee}) in the $M=v_{\ast}^{\prime}=0$
limit in Fig.(\ref{fig:tempevo}). Note that at $T=1$meV the spin-splitting does
not lead to a qualitative change from the analysis in Sec.(\ref{sec:Landau-Level-spectrum}). However,
note that even in the absence of relaxation, the anomalous modes near
zero energy are more dramatically affected by the change in temperature
than the rest of the spectrum. To understand this analytically, we
study the experimentally relevant intermediate-temperature/small-$B$
limit where $\beta h\ll1$ and $\beta U_{1}\gg1$. In this limit,
we can make analytic progress using perturbation theory. Keeping the
linear order terms in $\beta h$ we approximate
\begin{equation}
n_{s}=\frac{1}{2}+\frac{1}{2}s\chi_{0}B+\mathcal{O}\left(B^{2}\right),
\end{equation}

where we define the Curie susceptibility of the $f$ local moments
from the derivative of the magnetization:
\begin{equation}
\chi_{0}=\left.\frac{\partial\left(n_{\uparrow}-n_{\downarrow}\right)}{\partial B}\right|_{h=0}=\frac{\beta g\mu_{B}N_{f}}{4(N_{f}-1)}.
\end{equation}

where the last equality is derived in Ref\cite{aux_map_B2026}. Consequently,
we have the following approximations 
\begin{equation}
U_{s}\left(B\right)=s\frac{U_{1}}{2}\chi_{0}B+\mathcal{O}\left(\left(\beta h\right)^{2}\right),\qquad J_{s}\left(B\right)=-s\frac{J}{2}\chi_{0}B+\mathcal{O}\left(\left(\beta h\right)^{2}\right),\quad\text{and}\quad V_{s}\left(B\right)=\frac{U_{1}}{2}+\mathcal{O}\left(\left(\beta h\right)^{2}\right).\label{eq:approxZeeman}
\end{equation}
Consequently, in the limit where $\beta h\ll1$ and $\beta U_{1}\gg1$,
we can consider $U_{s}\left(B\right)$ and $J_{s}\left(B\right)$
as small perturbations on top of the Hamiltonian in Eq.(\ref{Appx:Eq:Hubbar_I_magnetic_after_aux}). The
starting point for perturbation theory is once again the exact zero
modes in Eq.(\ref{eq:pertheoanom-zee}) and Eq.(\ref{eq:pertheoanom-zee2}).
We can now calculate the energy-shift caused by the relaxation and
the Hartree-Fock energies at finite temperature from the following
operator:
\begin{align}
 & V_{ii',s,\eta=+}\left(B\right)=\\
 & \begin{pmatrix}\mu_{1}-sh & i\Delta_{B,2}a^{\dagger} & 0 & -i\Delta_{B,1}a & 0 & 0 & 0 & 0\\
-i\Delta_{B,2}a & \mu_{1}-sh & i\Delta_{B,1}a^{\dagger} & 0 & 0 & 0 & 0 & 0\\
0 & -i\Delta_{B,1}a & \mu_{2}+J_{s}\left(B\right)-sh & 0 & 0 & 0 & 0 & 0\\
i\Delta_{B,1}a^{\dagger} & 0 & 0 & \mu_{2}+J_{s}\left(B\right)-sh & 0 & 0 & 0 & 0\\
0 & 0 & 0 & 0 & -U_{s}\left(B\right)-sh & 0 & 0 & 0\\
0 & 0 & 0 & 0 & 0 & -U_{s}\left(B\right)-sh & 0 & 0\\
0 & 0 & 0 & 0 & 0 & 0 & +U_{s}\left(B\right)-sh & 0\\
0 & 0 & 0 & 0 & 0 & 0 & 0 & +U_{s}\left(B\right)-sh
\end{pmatrix}_{ii'}\label{Eq:Perturbation_Zeema_LL}
\end{align}
The leading order, corrections to the energy are then given by 
\begin{align}
E_{0,1,s}^{(1)} & =\frac{\mu_{1}+\left(\left(2\gamma\Sigma\left(0\right)\right)^{2}+\left(2\Delta_{B}^{\prime}\Sigma\left(1\right)\right)^{2}\right)\frac{U_{s}\left(B\right)}{U_{1}^{2}}}{1+\left(\frac{2\gamma\Sigma\left(0\right)}{U_{1}}\right)^{2}+\left(\frac{2\Delta_{B}^{\prime}\Sigma\left(1\right)}{U_{1}}\right)^{2}}-sh,\label{eq:pertheoanom-zee}\\
E_{0,2,s}^{(1)} & =\frac{\mu_{1}+\left(\frac{\Delta_{B}}{M}\right)^{2}\left(\mu_{2}+J_{s}\left(B\right)\right)-\frac{2\Delta_{B}\Delta_{B,1}}{M}+\left(\left(2\gamma\Sigma\left(1\right)\right)^{2}+2\left(2\Delta_{B}^{\prime}\Sigma\left(2\right)\right)^{2}\right)\frac{U_{s}\left(B\right)}{U_{1}^{2}}}{1+\left(\frac{\Delta_{B}}{M}\right)^{2}+\left(\frac{2\gamma\Sigma\left(1\right)}{U_{1}}\right)^{2}+2\left(\frac{2\Delta_{B}^{\prime}\Sigma\left(2\right)}{U_{1}}\right)^{2}}-sh\label{eq:pertheoanom-zee2}
\end{align}
\begin{figure}
    \centering
    \includegraphics[width=1.0\linewidth]{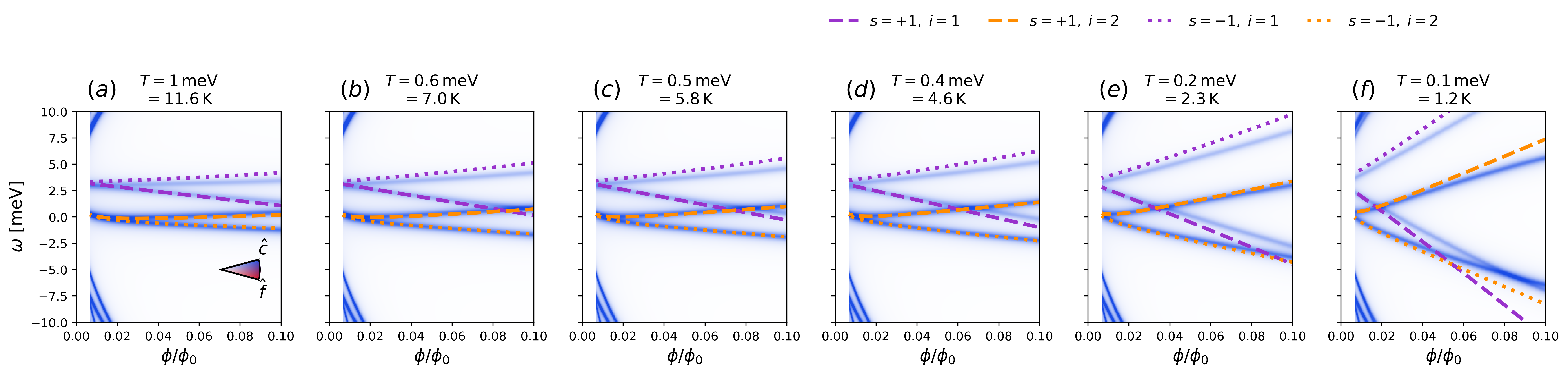}
    \caption{\textbf{Mott semimetal:} Finite-field spectral function focusing on the energy range where the anomalous modes of the Mott-Semimetal reside as a function of flux $\phi/\phi_0$ for several temperatures including relaxation and away from the chiral-flat limit at $\nu=0$. The blue intensity denotes the numerical spectral weight, with the relative $\hat c$- and $\hat f$-character indicated by the inset. The dashed and dotted curves show the analytical anomalous-mode energies $E^{(1)}_{0,i,s}$, with $i=1,2$ labeling the two anomalous branches and $s=\pm1$ the spin sector. Lowering the temperature strongly enhances the field dependence and spin splitting of the anomalous modes. This behavior follows from the Curie enhancement of the local-moment susceptibility, which amplifies the Hartree-Fock contribution generated by the exchange interaction and produces the rapid temperature dependence of the low-field slopes. }
    \label{fig:Aspectempevolrel}
\end{figure}
We can understand the behavior of the energies of the anomalous modes
by referring to the analysis in Sec.(\ref{sec:anomalous}).
Note that with the inclusion of the Zeeman coupling there are two
types of contributions. On the one hand we have the simple Zeeman
shift of the anomalous states which, as we have argued above is parametrically
small, leading to $0.1$meV energy differences at $B=1$T. More importantly,
the contributions from $J_{s}\left(B\right)$ and $U_{s}\left(B\right)$
arise from the Hartree-Fock energy gained by filling one spin sector
favored by the Zeeman coupling. To analyze these interaction-induced
contributions, we introduce the definitions 
\begin{equation}
A_{U,1}=\frac{2}{U_{1}}\chi_{0}\gamma^{2}\qquad A_{U,2}^{\prime}=\frac{8ev_{\ast}^{\prime2}}{\hbar}\frac{\chi_{0}}{2U_{1}}\qquad A_{J,2}=\frac{2ev_{\ast}^{2}}{\hbar}\frac{\chi_{0}J}{2M^{2}}\qquad M_{g}=\frac{g\mu_{B}}{2}
\end{equation}
In the $\Sigma\left(0\right)=\Sigma\left(1\right)=\Sigma\left(2\right)=1$
limit, we then obtain 
\begin{align}
E_{0,1,s}^{(1)} & =\frac{\mu_{1}+sA_{U,1}B+sA_{U,2}^{\prime}B^{2}}{D_{\star}+A_{\star}^{\prime}B}-sM_{g}B\\
E_{0,2,s}^{(1)} & =\frac{\mu_{1}+\left(A_{\star}\mu_{2}-A_{\star1}+sA_{U,1}\right)B+s\left(2A_{U,2}^{\prime}-A_{J,2}\right)B^{2}}{D_{\star}+\left(A_{\star}+2A_{\star}^{\prime}\right)B}-sM_{g}B
\end{align}
Note that for realistic THF parameters we obtain:
\begin{equation}
\frac{A_{J,2}}{2A_{U,2}^{\prime}}=\frac{v_{\ast}^{2}}{8v_{\ast}^{\prime2}}\frac{JU_{1}}{M^{2}}\gg1.
\end{equation}
Such that the exchange term, $A_{J,2}$, has a dominant effect over
the on-site Hubbard term $A_{U,2}^{\prime}$. From this estimate we
conclude that at large $B$ where the $B^{2}$ terms dominate in the
numerator, $E_{0,1s}^{(1)}$ and $E_{0,2s}^{(1)}$ disperse in opposite
directions in energy for the same spin polarization $s$. This behavior
of the anomalous modes is shown in Fig.(\ref{fig:Aspectempevolrel}). The opposite dispersion of the two anomalous modes is shown in Fig.~\ref{fig:Aspectempevolrel}. This demonstrates that exchange controls their relative energetics: if $J$ were set to zero, the quadratic-in-`B` corrections to $E_{0,1s}^{(1)}$ and $E_{0,2s}^{(1)}$ would have the same sign. At large flux and low temperature, however, the assumptions used in deriving Eq.~(\ref{eq:approxZeeman}) are no longer valid when the Zeeman splitting becomes comparable with $T$, so the perturbative energies deviate from the numerically computed spectral function. We can explicitly calculate
the spin-resolved spectral function neglecting LL mixing:
\begin{equation}
A_{c,s}\left(E\right)=\frac{1}{1+\left(\frac{2\gamma\Sigma\left(0\right)}{U_{1}}\right)^{2}+\left(\frac{2\Delta_{B}^{\prime}\Sigma\left(1\right)}{U_{1}}\right)^{2}}\delta\left(E-E_{0,1,s}^{(1)}\right)+\frac{1+\left(\frac{\Delta_{B}}{M}\right)^{2}}{1+\left(\frac{\Delta_{B}}{M}\right)^{2}+\left(\frac{2\gamma\Sigma\left(1\right)}{U_{1}}\right)^{2}+2\left(\frac{2\Delta_{B}^{\prime}\Sigma\left(2\right)}{U_{1}}\right)^{2}}\delta\left(E-E_{0,2,s}^{(1)}\right)
\end{equation}
Once again, we find that the top mode given by $E_{0,1,s}$ has a smaller quasiparticle weight compared to $E_{0,2,s}$ due to the presence of the additional factor of $\left(\frac{\Delta_{B}}{M}\right)^{2}$ which derives from the weight of the second anomalous mode in Eq.(\ref{eq:zeromodenonchiral2}) on the $\Gamma_1\oplus\Gamma_2$ $c-$ fermion sector.

\section{Spectral maps for main-text Figures}
\label{Appx:Spectral_map_color}

In this section we elaborate the color scheme summarized briefly in the caption of Fig.(\ref{fig:main:fig2}), and used in Fig.(\ref{fig:main:fig2}a,b).
\subsection{Orbital-resolved spectral weight}
\subsubsection{Mott Semimetal}
As discussed in the previous sections, the Green's function for physical fermions can be obtained in the Hubbard-I approximation through the auxiliary fermion mapping. We now discuss how to obtain the spectral function for the physical fermions from the auxiliary fermion construction. At a given $\bk$ (at $\bB=0$) or $\phi/\phi_0$ (at $\bB\neq0$), the spectral function for the physical fermions is given by 
\begin{eqnarray}
A(\omega)=\sum_{n}P_{\rm{phy}}|u_n\rangle\langle u_n|P_{\rm{phy}} \delta(\omega-\epsilon_n)
\end{eqnarray}
where $n$ labels the normalized eigenvectors $|u_n\rangle$ with eigenvalue $\epsilon_n$ of the auxiliary fermion mapped non-interacting Hamiltonian at the given $\bk$ or $\phi/\phi_0$. Here $P_{\rm{phy}}$ denotes the projection operator onto the physical (non-auxiliary) fermions. The respective $c$ and $f$ orbital character for the spectral function is obtained by calculating 
\begin{eqnarray}
    A_{c(f)}(\omega)=\rm{Tr}[P_{c(f)}A(\omega)]
\end{eqnarray}
where $P_{c(f)}$ denotes the projector onto $c(f)$ fermion orbitals. In practice, to build a smooth spectral map, we replace the delta functions by Lorentzian functions, i.e.
\begin{eqnarray}
    \delta(\omega-\epsilon_n)\rightarrow\frac{1}{\pi}\frac{w}{(\omega-\epsilon_n)^2+w^2}
\end{eqnarray}
with width $w=0.23$ meV.

\subsubsection{Strained Anisotropic Semimetal}
We use a similar approach as above to obtain the orbital-resolved spectral function for this $T=0$ state. At every $\bk$ (at $\bB=0$) or $\phi/\phi_0$(at $\bB\neq 0$), we diagonalize the corresponding Hamiltonian and obtain eigen-energies $E_n$ together with a eigen-mode resolved `$f$-weight' $w^{(f)}_{n}\in[0,1]$ (and thus $c$-weight $w^{(c)}_{n}=1-w^{(f)}_{n}$). To build a smooth spectral map, we broaden each discrete level by a Lorentzian of fixed width $0.23$ meV and sum over eigenmodes obtained above. Concretely, on an energy grid $\omega\in[\omega_{\min},\omega_{\max}]$, we compute two partial spectral intensities,
\begin{align}
A_f(\omega) &= \sum_{n} w^{(f)}_{n}\, L_w\!\left(\omega-E_n\right),\\
A_c(\omega) &= \sum_{n} w^{(c)}_{n}\, L_w\!\left(\omega-E_n\right),
\end{align}
with $w=0.23$ meV and 
\begin{equation}
L_w(x)=\frac{1}{\pi}\frac{w}{x^2+w^2}.
\end{equation}
At $\bB\neq0$, we linearly interpolate $A_{f,c}(\omega)$ between the sampled $\phi/\phi_0$ points, i.e. the $1/q$ sequence. In other words, the spectral weight $y$, at a given $\omega$, between sampled $\phi/\phi_0$'s $x_j$ and $x_{j+1}$ is calculated by 
\begin{eqnarray}
    y(x) = y(x_j) + \frac{x-x_j}{x_{j+1}-x_j}(y(x_{j+1})-y(x_j))
\end{eqnarray}

\subsection{Color mixing}
 Rather than plotting a single scalar colormap, we encode the relative $c$ and $f$ spectral weight directly in pixel color. We start from a white background, and subtract two `color shifts' proportional to the (normalized) partial intensities.
We first normalize each channel by its own global maximum over the plotted window,
\begin{equation}
n_c(\omega)=\alpha\,\frac{A_c(\omega)}{\max(A_c)+\varepsilon}~~~,\qquad
n_f(\omega)=\alpha\,\frac{A_f(\omega)}{\max(A_f)+\varepsilon}
\end{equation}
where $\alpha$ is a fixed contrast factor (in our implementation $\alpha=1$) and $\varepsilon$ is a tiny constant to avoid division by zero. We then define two anchor colors, blue and red for $c$ and $f$, respectively. In RGB format 
\begin{eqnarray}
\mathbf{c}&=&(0.10,\,0.31,\,0.90)\\
\mathbf{f}&=&(0.85,\,0.02,\,0.10)
\end{eqnarray}
and the white vector $\mathbf{w}=(1,1,1)$. The RGB color at each pixel is then obtained by
\begin{eqnarray}
    \mathbf{C}(\omega) =  \frac{n_c(\omega)\mathbf{c} + n_f(\omega)\mathbf{f}}{n_c(\omega)+n_f(\omega)}
\end{eqnarray}
We further clip the total spectral function by introducing
\begin{equation}
\alpha_{cf}\left(\omega\right)=\left(n_{c}\left(\omega\right)+n_{f}\left(\omega\right)\right)\theta\left(1-\left|n_{c}\left(\omega\right)+n_{f}\left(\omega\right)\right|\right)+\left(1-\theta\left(1-\left|n_{c}\left(\omega\right)+n_{f}\left(\omega\right)\right|\right)\right),
\end{equation}
where $\theta(x)$ is the Heaviside theta function. Finally, we obtain the RGB color at each pixel by computing
\begin{equation}
    \mathbf{RGB}(\omega) = (1-\alpha_{cf}\left(\omega\right))\mathbf{w}+\alpha_{cf}\left(\omega\right)\mathbf{C}(\omega).
\end{equation}
In this scheme, large $A_c$ shifts the pixel color toward blue, large $A_f$ shifts it toward red, and simultaneous weight in both channels produces intermediate hues, while regions with negligible spectral weight remain close to white.

\section{Numerical band structure and gap hierarchy results with canonical substitution}
\label{sec:numerical_results}
In this appendix, we present comprehensive numerical results for the THF model under the influence of variable heterostrain, relaxation, and on-site f-electron-electron interactions. The calculations are divided into two primary approaches: self-consistent Hartree-Fock (SCHF) theory for the strained semimetal and the Hubbard I approximation for the Mott semimetal. All finite-magnetic-field calculations below are performed using the canonical substitution \emph{without} projecting to the hybrid-Wannier basis.

\subsection{Numerical results for strained Topological Heavy Fermions with relaxation}
\label{subsec:schf_results}

\begin{figure}
    \centering
    \includegraphics[width=0.8\linewidth]{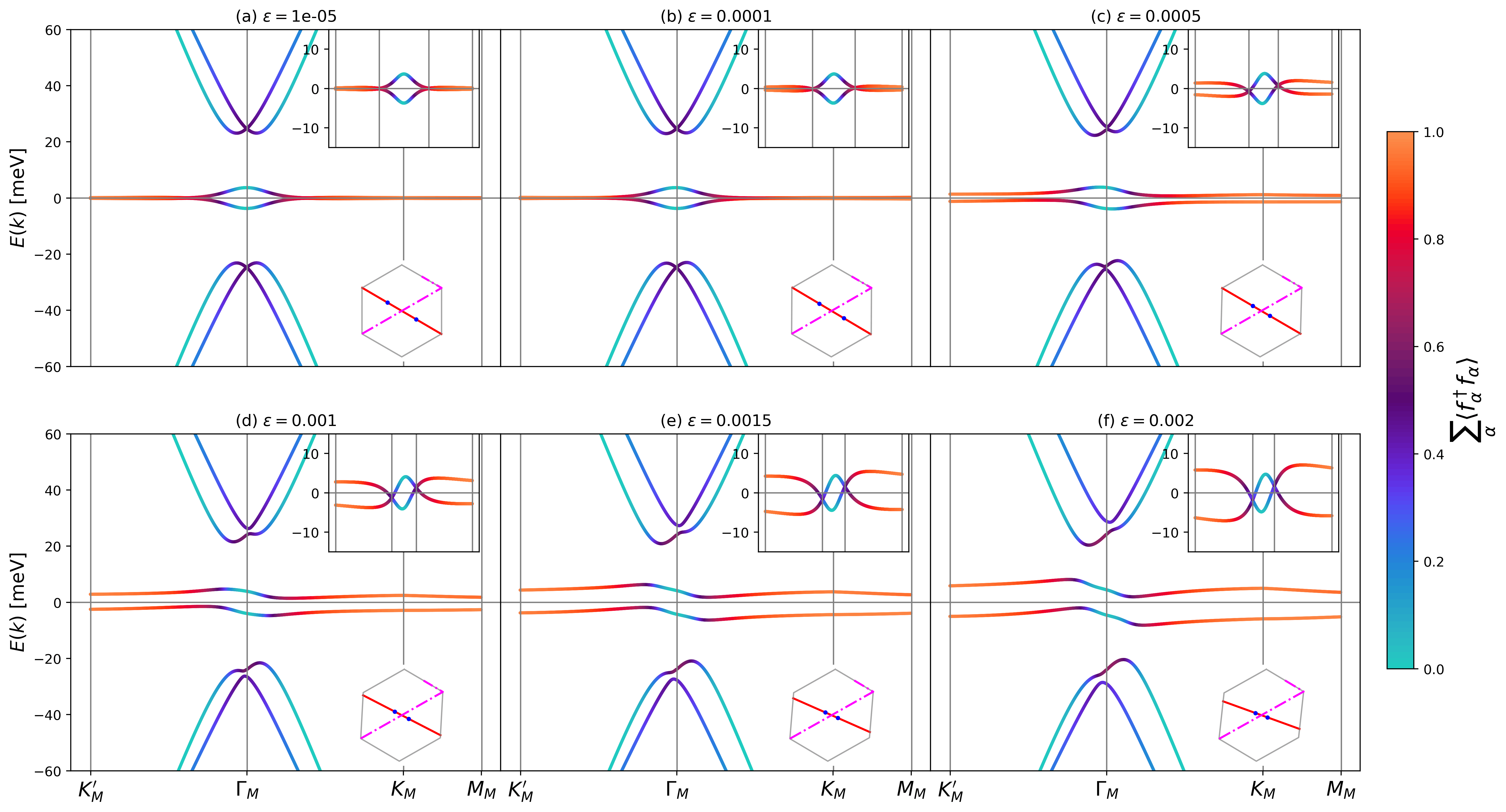}
    \caption{\textbf{Non-interacting strained THF:} Dispersion of the topological heavy fermion model including the full strain correction as a function of $\varepsilon$. The insets show the high symmetry path on the strained BZ and the dispersion along a cut that includes the Dirac crossing. } 
    \label{fig:strainsweep}
\end{figure}

\begin{figure}
    \centering
    \includegraphics[width=0.8\linewidth]{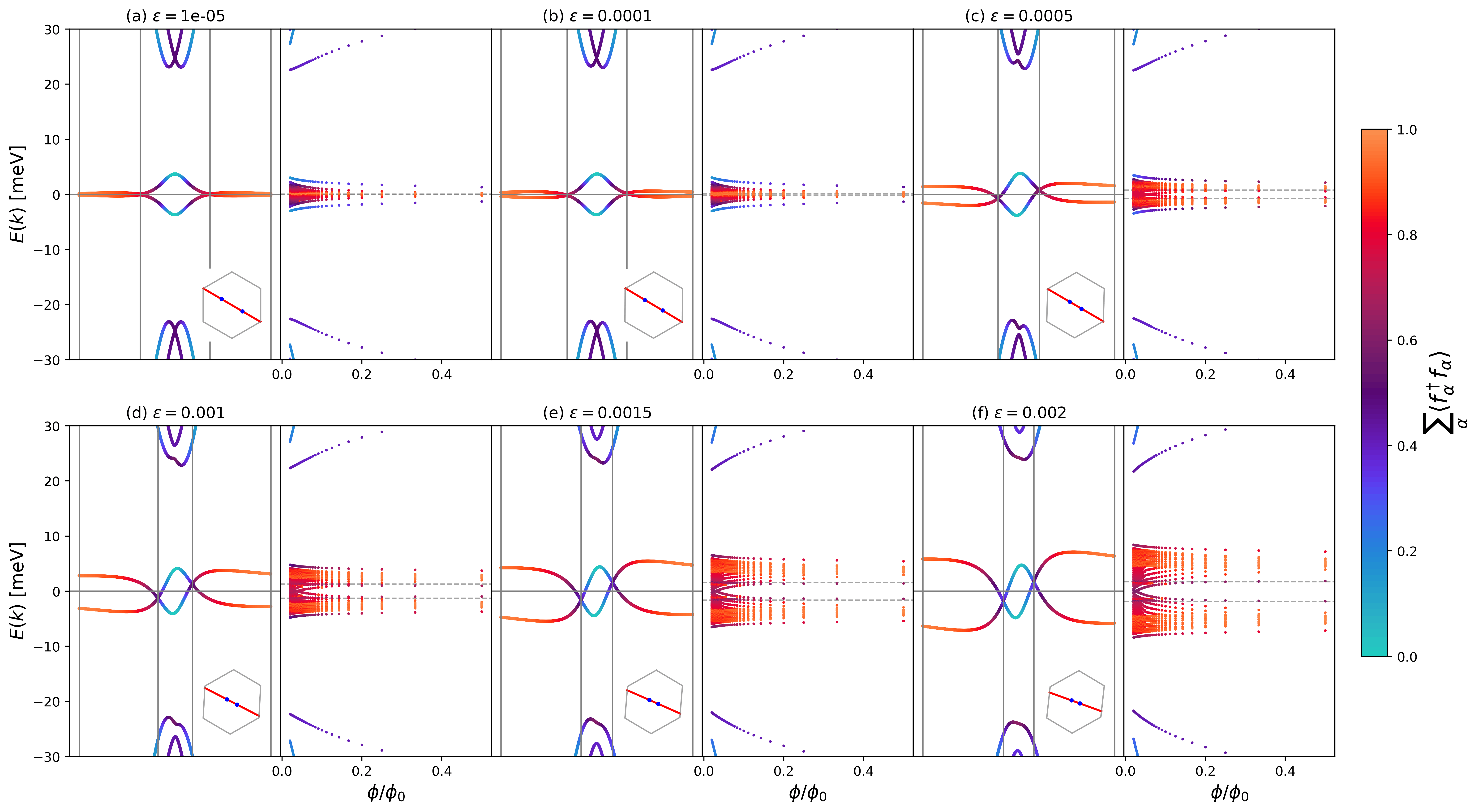}
    \caption{ \textbf{Non-interacting strained THF:}  Dispersion and Landau level spectrum of the topological heavy fermion model including the full strain correction as a function of $\varepsilon$.  The insets show a path in the strained BZ that includes the Dirac crossing. The anomalous behavior of the $f$-electrons lead to a divergence in the number of low energy states as the LL cutoff increases obscuring the low-energy Dirac-like spectrum near the shifted Dirac points.   } 
    \label{fig:strainsweep_LL}
\end{figure}

\begin{figure}
    \centering
    \includegraphics[width=0.8\linewidth]{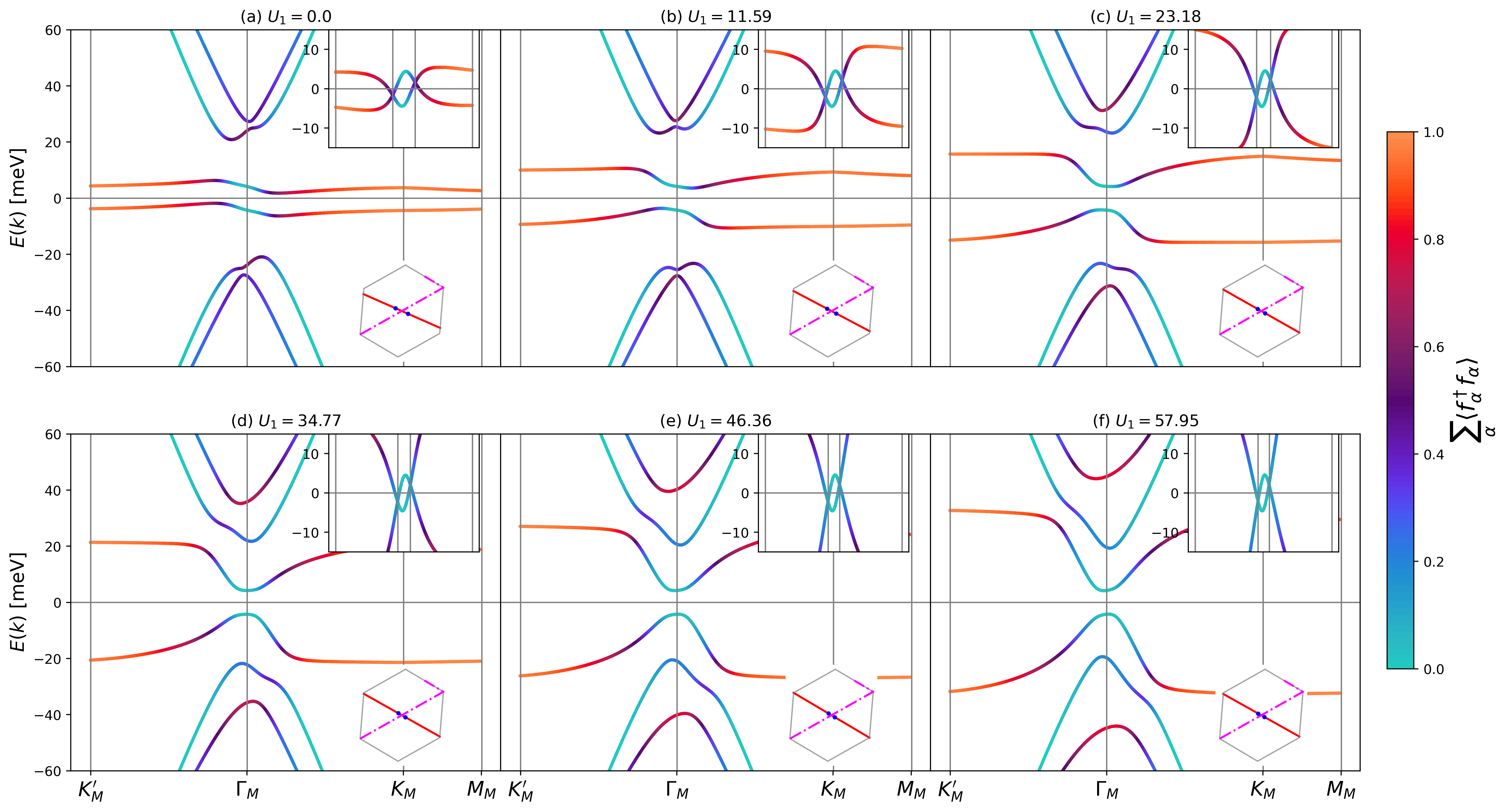}
    \caption{ \textbf{Strain-induced anisotropic semimetal:}  Dispersion of the topological heavy fermion model including the full strain correction as a function of $U_1$ including the self-consistent Hartree-Fock correction. The insets show the high symmetry path on the strained BZ and the dispersion along a cut that includes the Dirac crossing. } 
    \label{fig:intsweep}
\end{figure}

\begin{figure}
    \centering
    \includegraphics[width=0.8\linewidth]{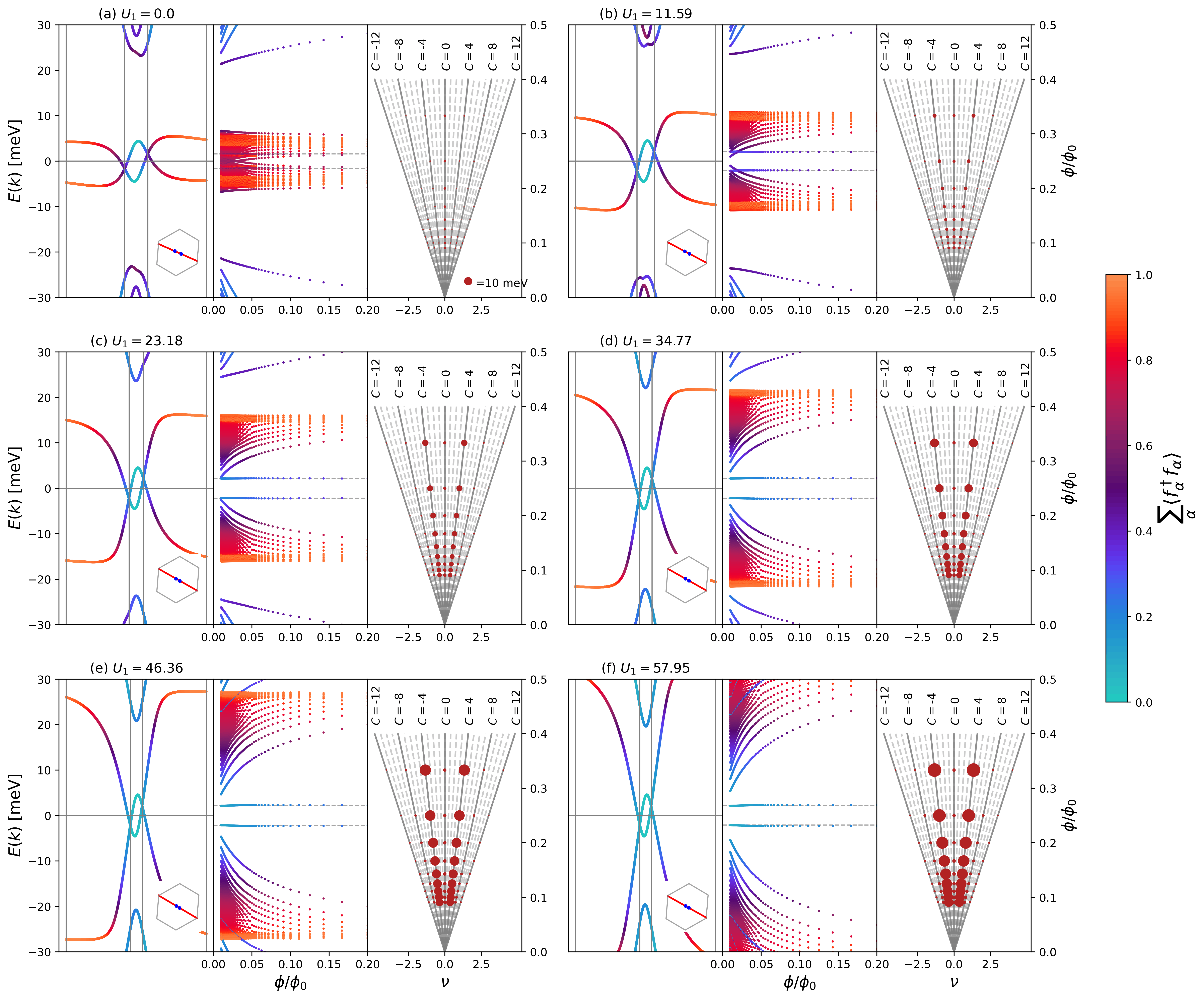}
    \caption{\textbf{Strain-induced anisotropic semimetal:}  Dispersion, Landau level spectrum, and Landau fan diagram of the topological heavy fermion model including the full strain correction as a function of $U_1$ for $\varepsilon=0.15\%$. At low interaction strength the Landau level spectrum includes anomalous $f$-electron levels. At larger interaction strength the spectrum is dominated by the $c-$electrons. Strain shifts the Dirac cones towards the center of the BZ and away from zero energy opening up a $C=0$ gap. The observed Chern sequence would be  $\nu=0,\pm4,\pm8,\pm12...$ in the absence of quantum-hall ferromagnetism.  } 
    \label{fig:intsweep_LL}
\end{figure}

\begin{figure}
    \centering
    \includegraphics[width=0.8\linewidth]{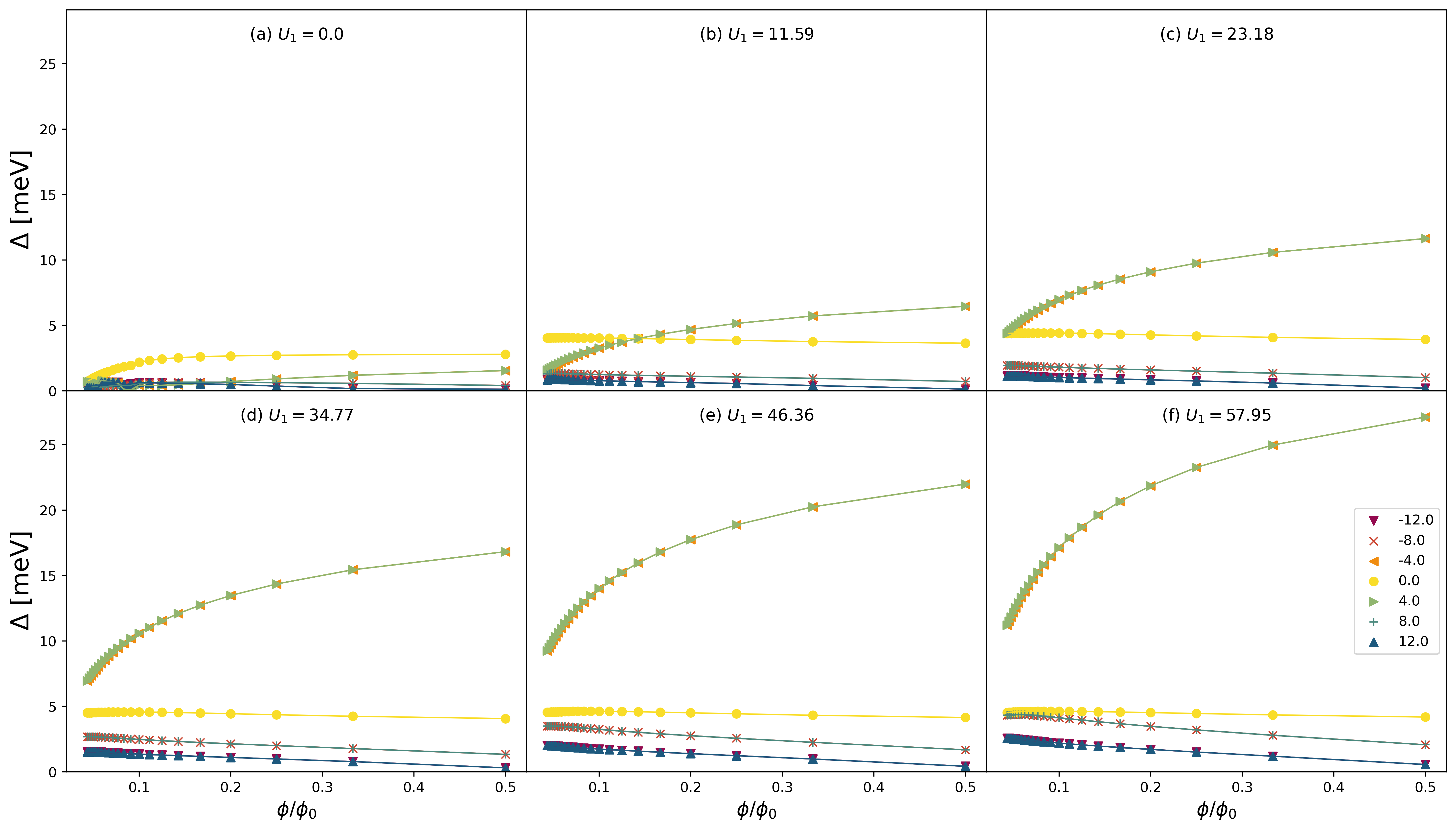}
    \caption{\textbf{Strain-induced anisotropic semimetal:}   Evolution of the energy gap ($\Delta$) as a function of magnetic flux ($\phi/\phi_0$) for the topological heavy fermion model including heterostrain. The panels (a–f) illustrate the gap dependence for increasing interaction strengths, $U_1$ , ranging from 0.0 to 57.95. Different symbols and colors correspond to different Chern gaps as indicated in the legend in panel (f), spanning from -12 to 12. At low interaction strengths (a–b), the gaps remain relatively small. As $U_1$ increases (c–f), a significant enhancement of the $C=\pm4$ gap, where $\Delta$ scales non-linearly with flux, exceeding 25 meV in the strongest interaction regime is attributed to the Hartree-Fock renormalization of the bands increasing the separation in energy of the f-electron states, while the anomalous zero modes remain strongly $U_1$ independent} 
    \label{fig:intsweep_LL_gap}
\end{figure}

\begin{figure}
    \centering
    \includegraphics[width=0.8\linewidth]{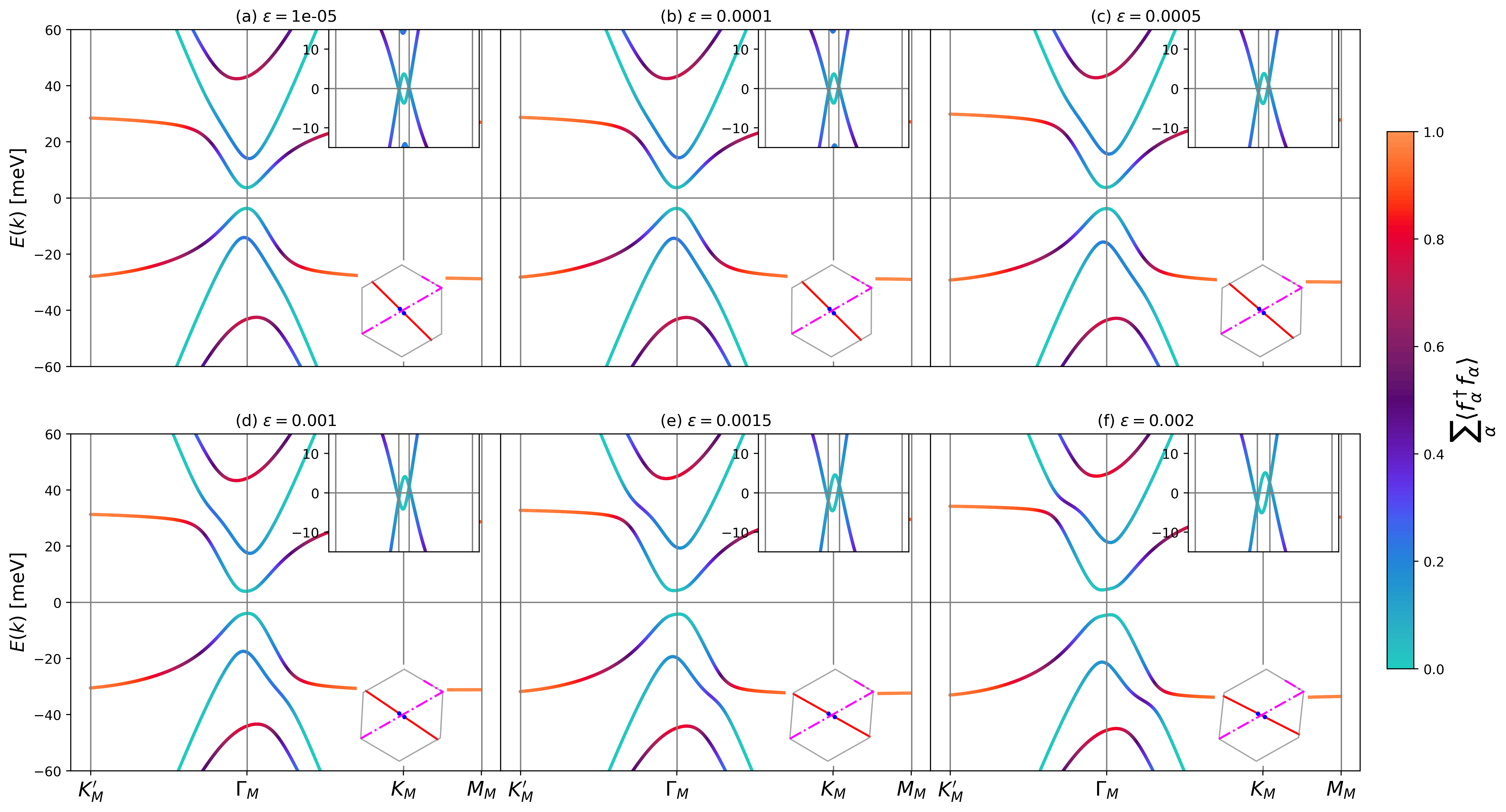}
    \caption{ \textbf{Strain-induced anisotropic semimetal:} Dispersion of the topological heavy fermion model including the full strain correction as a function of $\varepsilon$ including the self-consistent Hartree-Fock correction for $U_1=57.95$meV. The insets show the high symmetry path on the strained BZ and the dispersion along a cut that includes the Dirac crossing. } 
    \label{fig:int_strainsweep}
\end{figure}

\begin{figure}
    \centering
    \includegraphics[width=0.8\linewidth]{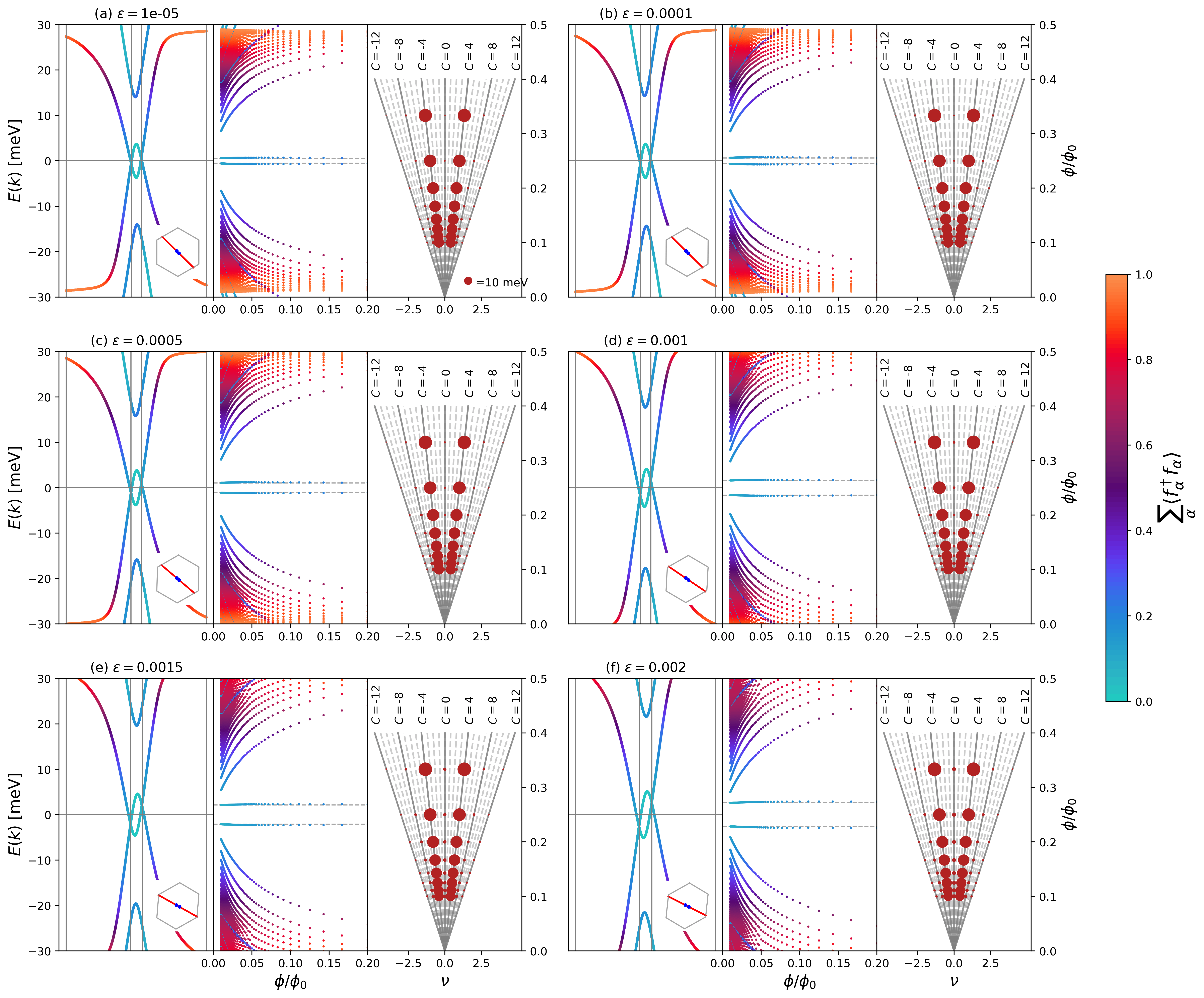}
    \caption{\textbf{Strain-induced anisotropic semimetal:}  Dispersion, Landau level spectrum, and Landau fan diagram of the topological heavy fermion model including the full strain correction as a function of $\varepsilon$ including the self-consistent Hartree-Fock correction for $U_1=57.95$meV. At low strain the $C=0$ gap is negligible and the $C=\pm4$ are the most prominent ones in the Landau fan diagram. The observed Chern sequence would be  $\nu=0,\pm4,\pm8,\pm12...$ in the absence of quantum-hall ferromagnetism.} 
    \label{fig:int_strainsweep_LL}
\end{figure}

\begin{figure}
    \centering
    \includegraphics[width=0.8\linewidth]{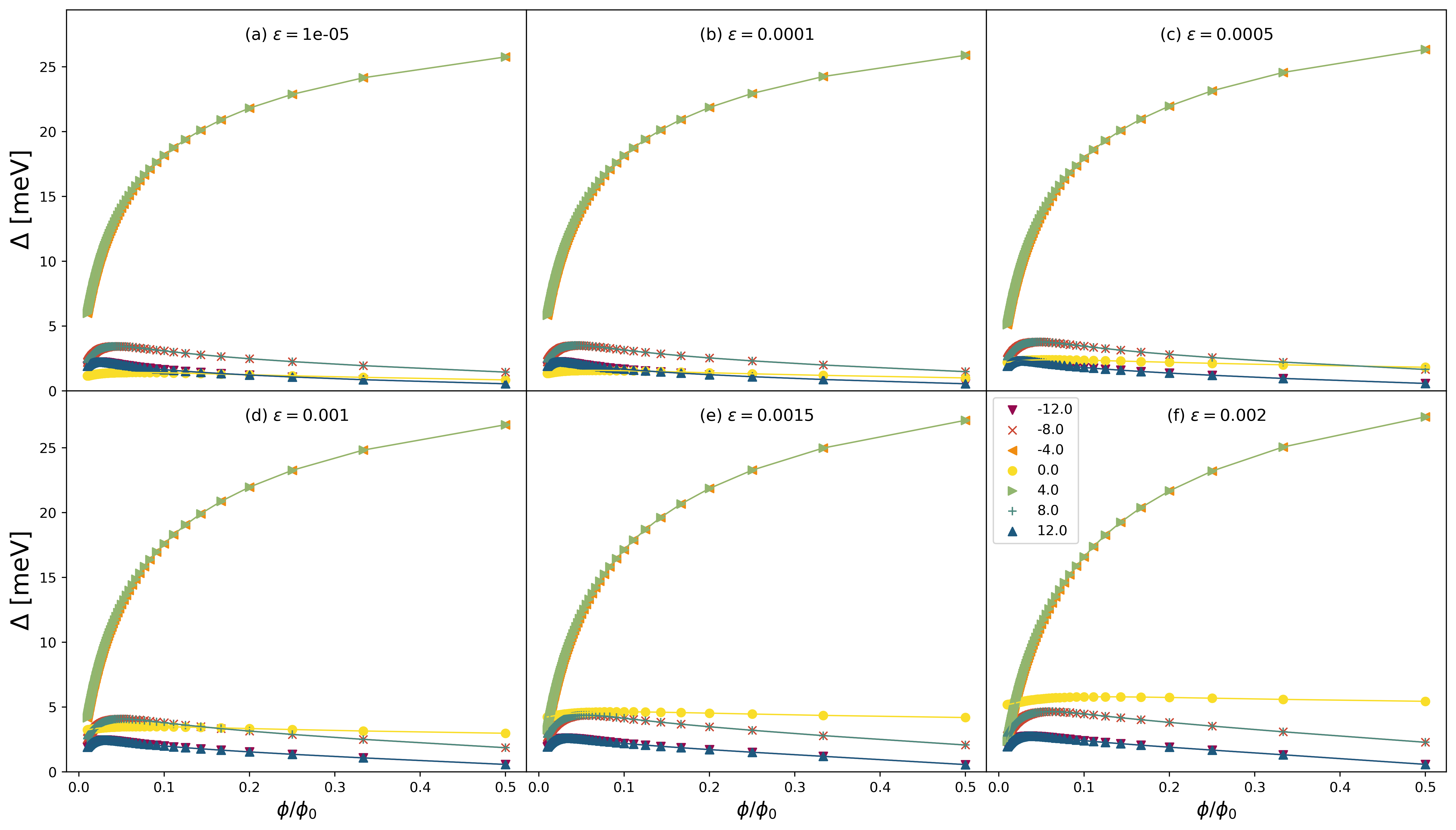}
    \caption{\textbf{Strain-induced anisotropic semimetal:}  Evolution of the energy gap ($\Delta$) as a function of magnetic flux ($\phi/\phi_0$) for the topological heavy fermion model including heterostrain. The panels (a–f) illustrate the gap dependence for increasing heterostrain, $\varepsilon$ , ranging from $10^{-5}$ to $0.002$. Different symbols and colors correspond to different Chern gaps as indicated in the legend in panel (f), spanning from -12 to 12. At low strain (a–c), the gap hierarchy is such that the $C=\pm4$ gap is the largest followed by the $C=\pm8$. As strain increases, so does the splitting of the anomlous modes which leads to a larger $C=0$ gap which eventually becomes larger than the $C=\pm8$ gap (d-f) at large flux. } 
    \label{fig:int_strainsweep_LL_gap}
\end{figure}

\begin{figure}
    \centering
    \includegraphics[width=0.8\linewidth]{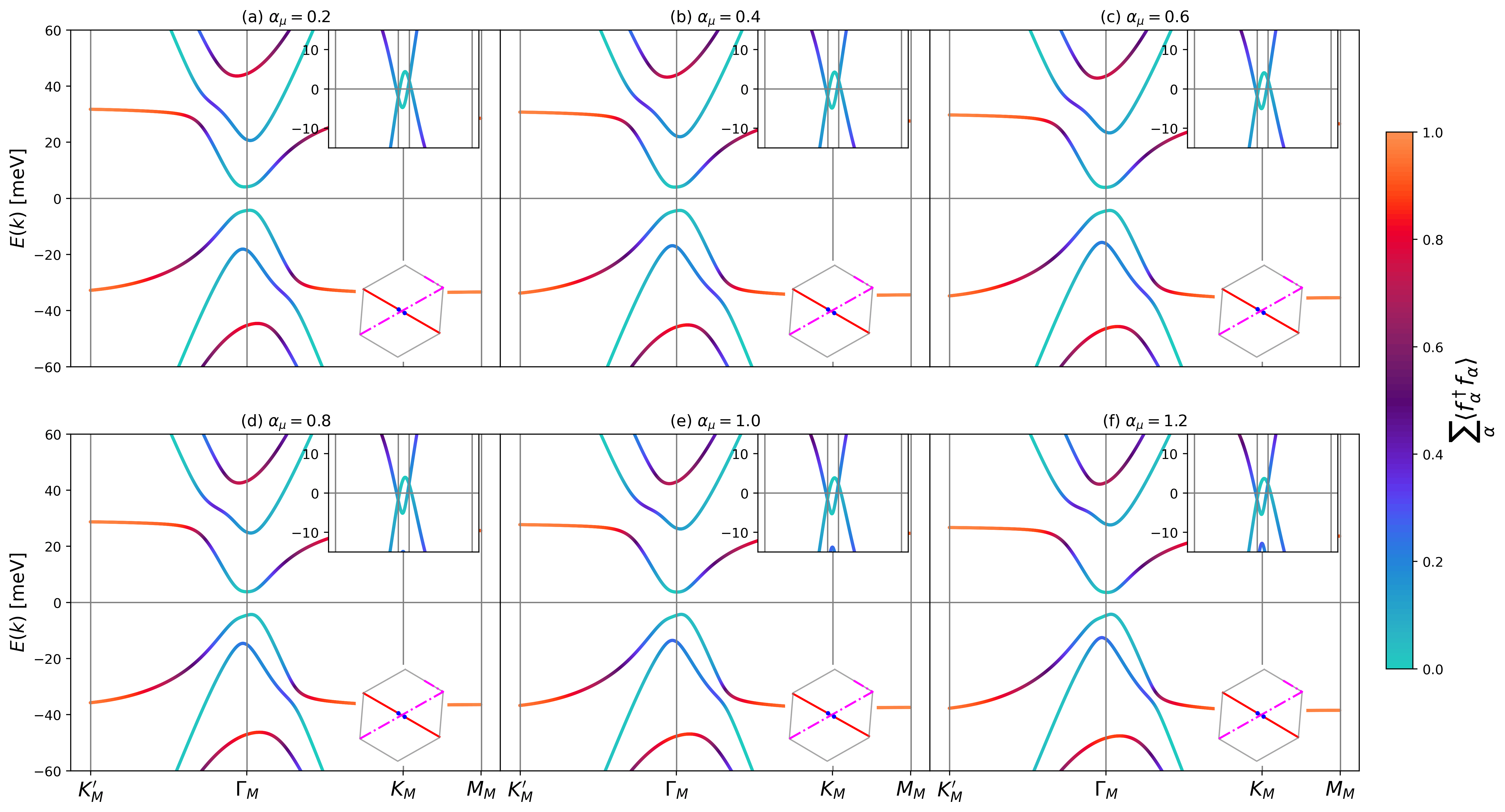}
    \caption{\textbf{Strain-induced anisotropic semimetal:}   Dispersion of the topological heavy fermion model including the full strain correction and the simplified relaxation correction. We multiply the relaxation correction by $\alpha_\mu$, where $\alpha_\mu=1$ corresponds to the realistic parameters derived in \cite{herzog2025topological}. We include the self-consistent Hartree-Fock correction for $U_1=57.95$meV. The insets show the high symmetry path on the strained BZ and the dispersion along a cut that includes the Dirac crossing.} 
    \label{fig:int_strain_relsweep}
\end{figure}

\begin{figure}
    \centering
    \includegraphics[width=0.8\linewidth]{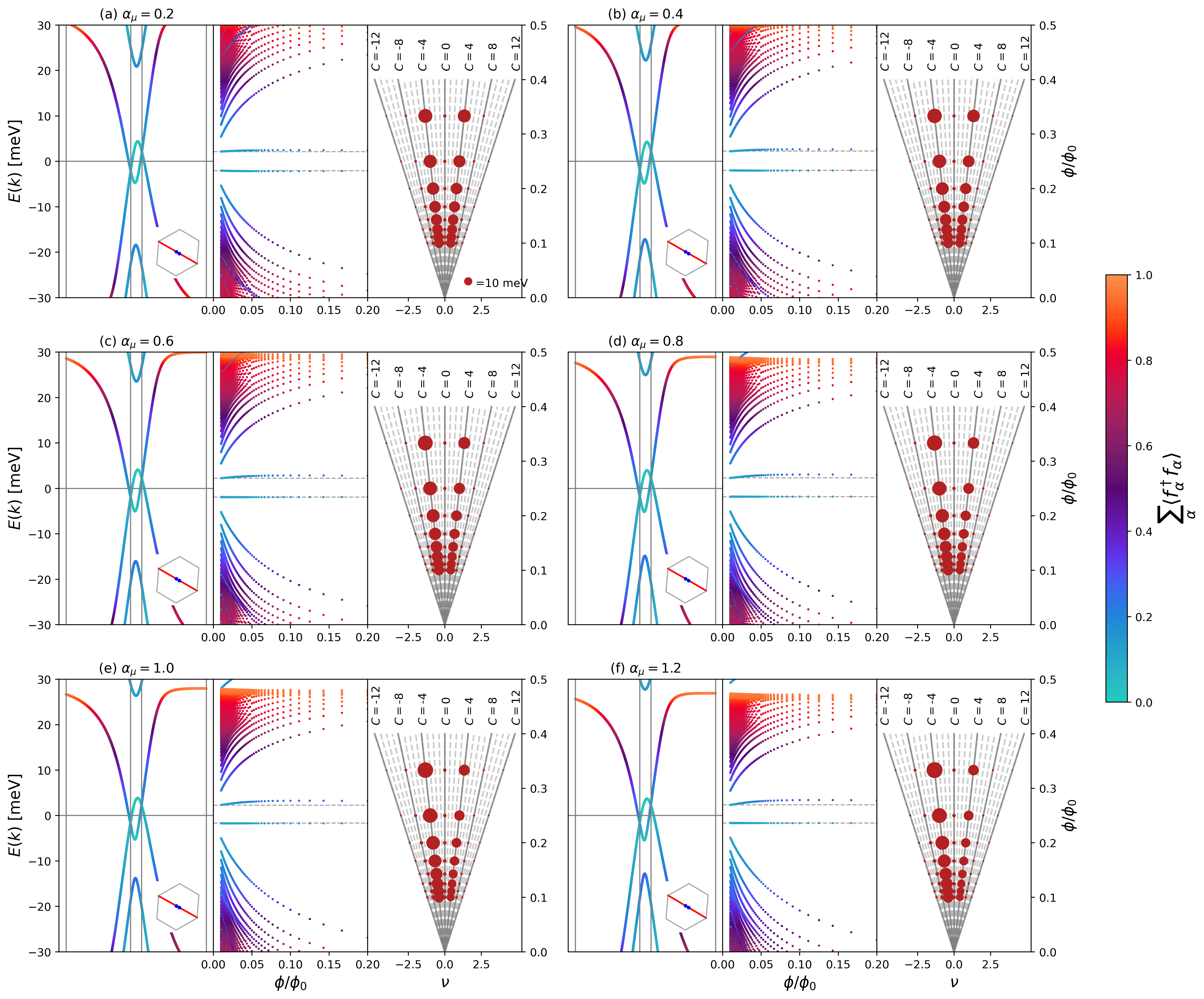}
    \caption{\textbf{Strain-induced anisotropic semimetal:}  Dispersion, Landau level spectrum, and Landau fan diagram of the topological heavy fermion model including the full strain correction and the simplified relaxation correction from \cite{herzog2025topological}. We multiply the relaxation correction by $\alpha_\mu$, where $\alpha_\mu=1$ corresponds to the realistic parameters derived in \cite{herzog2025topological}. For realistic values of the relaxation correction, the most prominent gap in the Landau fan diagram is the $C=-4$ gap. The observed Chern sequence would be  $\nu=0,\pm4,\pm8,\pm12...$ in the absence of quantum-hall ferromagnetism.} 
    \label{fig:int_strain_relsweep_LL}
\end{figure}

\begin{figure}
    \centering
    \includegraphics[width=0.8\linewidth]{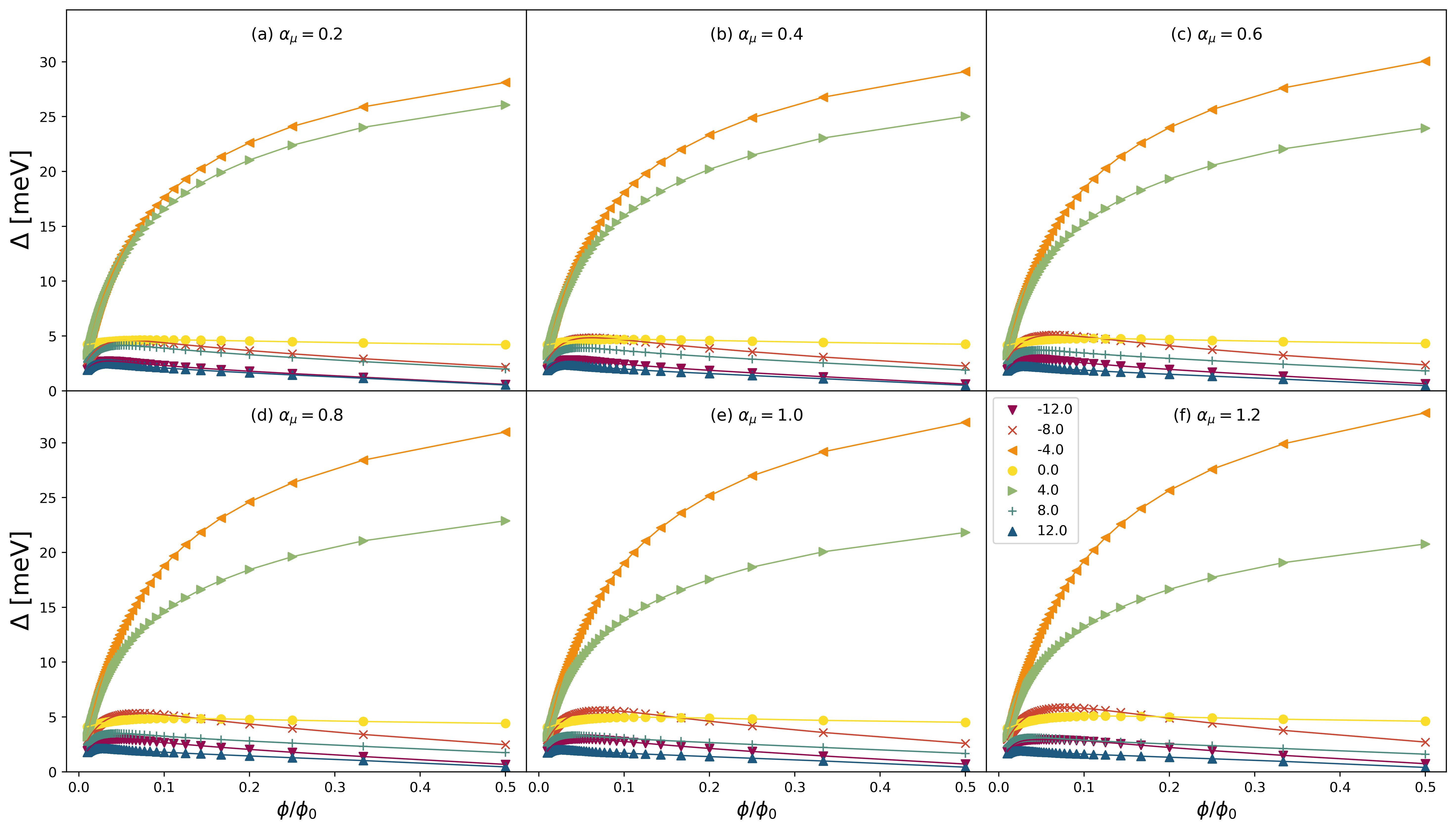}
    \caption{\textbf{Strain-induced anisotropic semimetal:}  Evolution of the energy gap ($\Delta$) as a function of magnetic flux ($\phi/\phi_0$) for the topological heavy fermion model including heterostrain and relaxation. The panels (a–f) illustrate the gap dependence for increasing relaxation, obtained by increasing the artificial pre-factor $\alpha_r$, ranging from $0.2$ to the physical value of $1.0$. Different symbols and colors correspond to different Chern gaps as indicated in the legend in panel (f), spanning from -12 to 12. At finite relaxation strength (a), the $C>0$ and $C<0$ split with $C<0$ gaps becoming larger. Physically, this asymmetry arises from the relaxation-induced downward shift of f-electron energy levels relative to conduction (c) electrons; this energy renormalization enhances the interacting dispersion for occupied states ($E<0$), thereby increasing the dispersion of the associated Landau levels with the applied magnetic field.} 
    \label{fig:int_strain_relsweep_LL_gap}
\end{figure}

\begin{figure}
    \centering
    \includegraphics[width=0.8\linewidth]{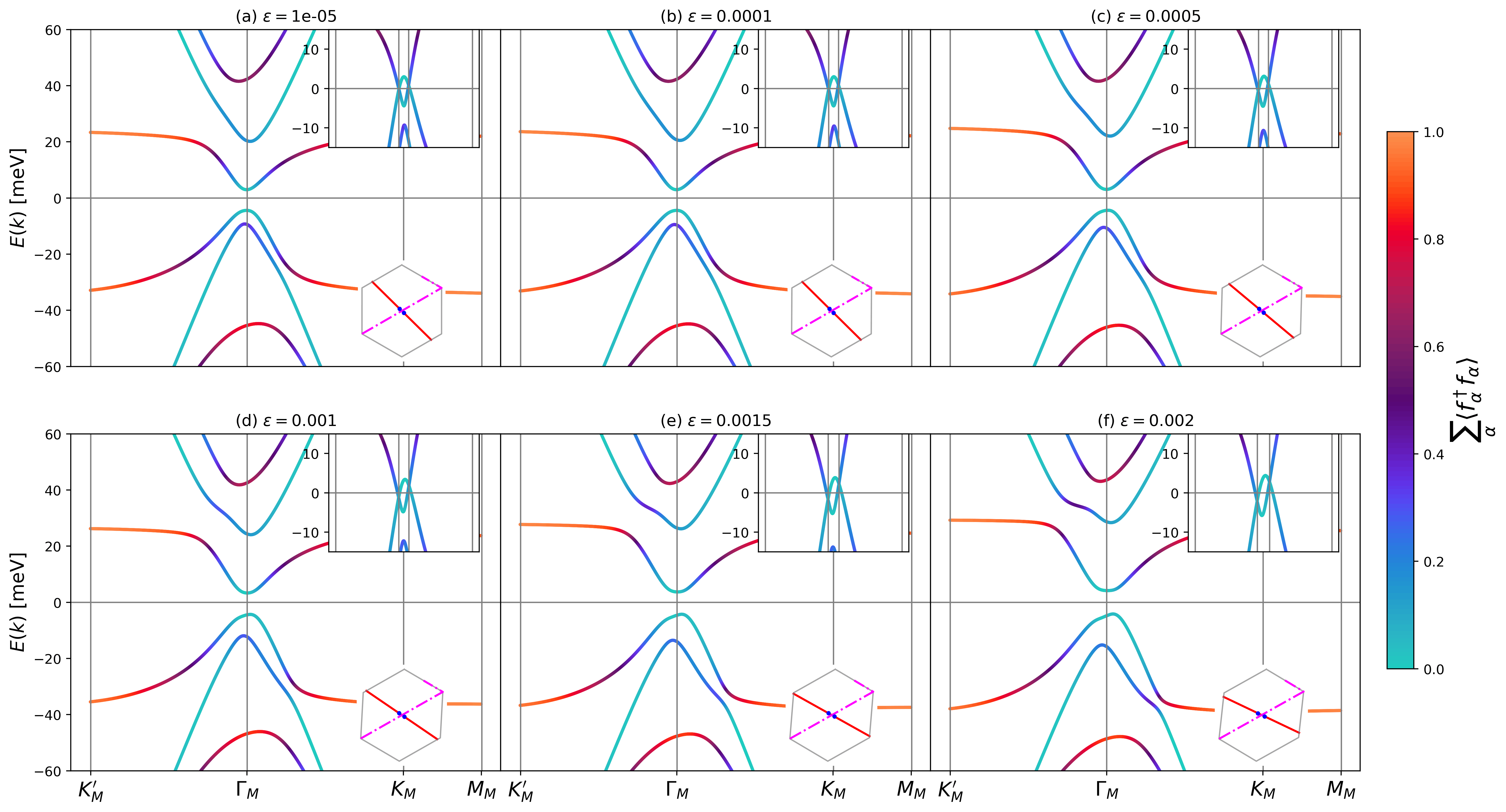}
    \caption{ Dispersion of the topological heavy fermion model including the full strain correction and the simplified full relaxation correction as a function of heterostrain $\varepsilon$. Although the primary $f$-electron splitting is dominated by the interaction term $U_1$, the impact of heterostrain is most significant near the $\Gamma_M$ point. In this region, the energy splitting scales proportionally with $\varepsilon$, lifting the degeneracy characteristic of the unstrained lattice.} 
    \label{fig:int_strainsweep_rel}
\end{figure}

\begin{figure}
    \centering
    \includegraphics[width=0.8\linewidth]{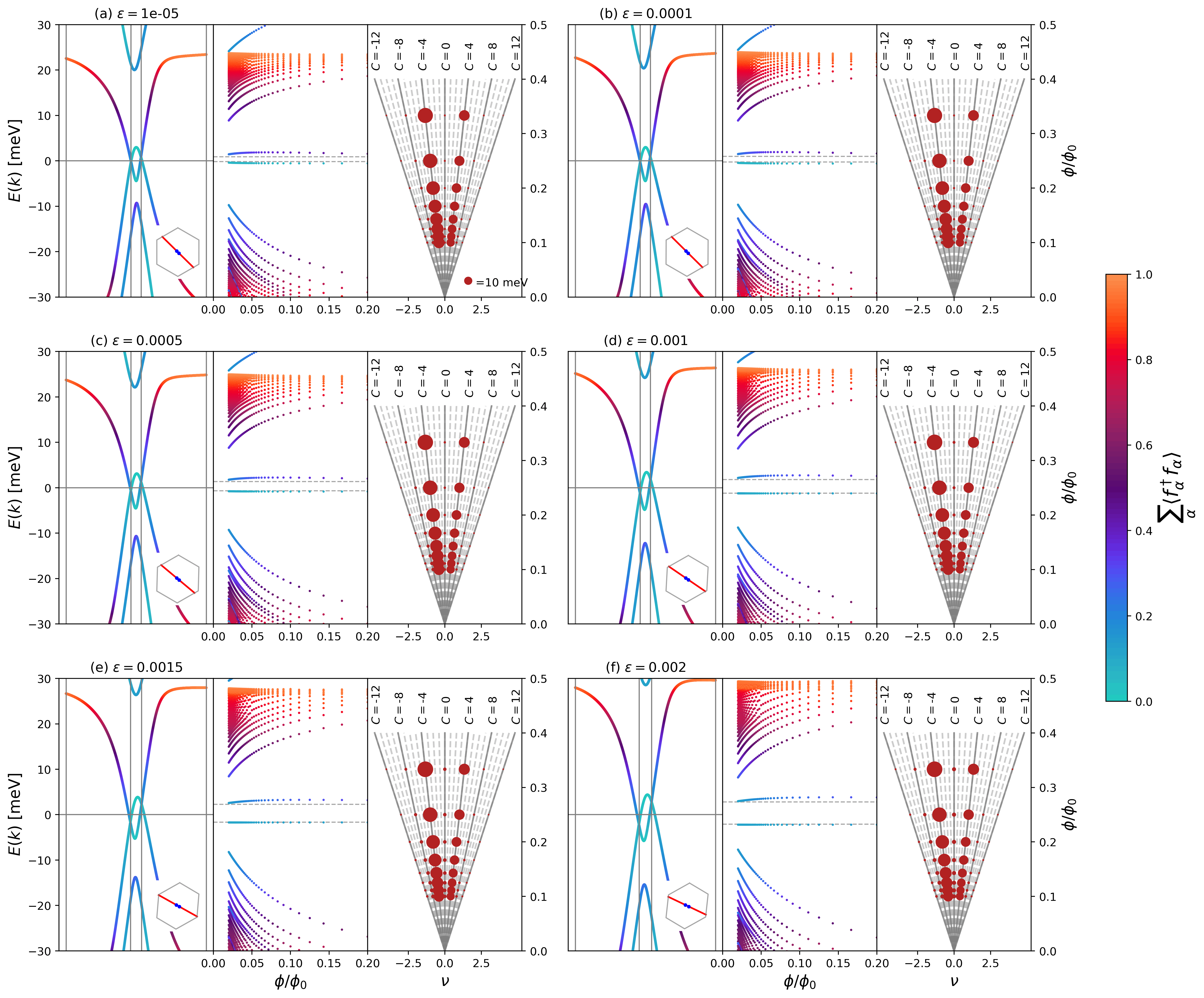}
    \caption{\textbf{Strain-induced anisotropic semimetal:} Dispersion, Landau level spectrum, and Landau fan diagram of the topological heavy fermion model including the full strain correction and the simplified full relaxation correction as a function of heterostrain $\epsilon$. Similar to the dispersion in Fig.(\ref{fig:int_strainsweep_rel}), increasing $\epsilon$ primarily impacts the region near the $\Gamma_M$ point, shifting the Dirac crossings and lifting degeneracies. In the Landau fan diagrams (right subpanels), the evolution of the Chern sequence is tracked as the strain scales, demonstrating the robustness of the $C=0$ gap opening and the sequence $\nu =0,\pm4, \pm8, ...$ across the sampled strain values. } 
    \label{fig:int_strainsweep_rel_LL}
\end{figure}

\begin{figure}
    \centering
    \includegraphics[width=0.8\linewidth]{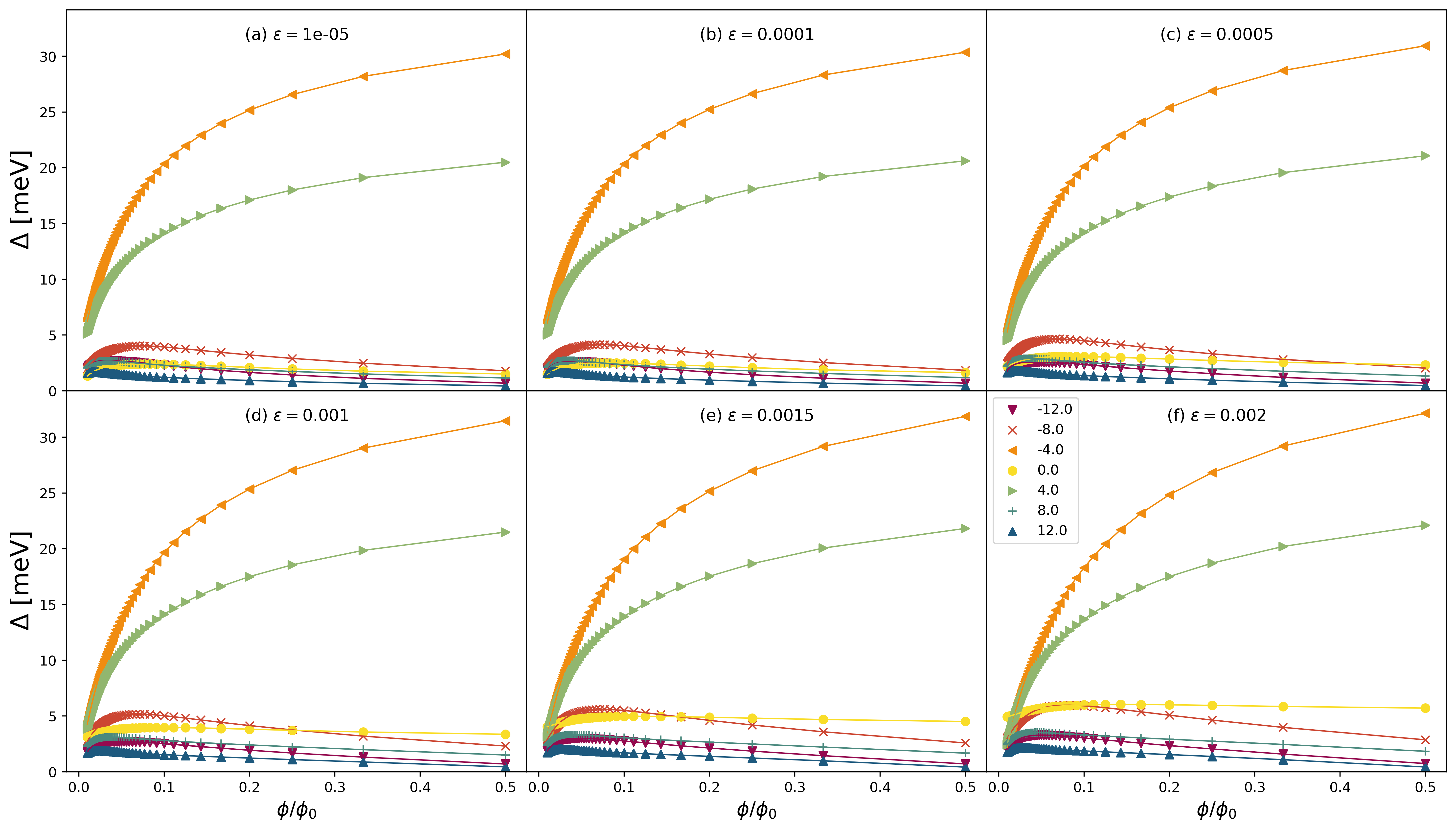}
    \caption{\textbf{Strain-induced anisotropic semimetal:} Evolution of the energy gap ($\Delta$) as a function of magnetic flux ($\phi/\phi_0$) for the topological heavy fermion model including the full strain correction and simplified relaxation as a function of heterostrain $\epsilon$. The panels (a–f) track the gap dependence for increasing strain values from $\epsilon = 10^{-5}$ to $0.002$. A key feature of this evolution is the shifting hierarchy between the $C = 8$ and $C = 0$ gaps; at lower strain, the $C = 8$ gap is dominant, but as strain increases, the $C = 0$ gap becomes larger, eventually overtaking it in the high-strain regime. Additionally, the inclusion of relaxation effects introduces a visible splitting between the negative and positive Chern gaps, particularly noticeable in the divergence of the $C = \pm 4$ and $C = \pm 8$ sequences as $\phi/\phi_0$ increases. } 
    \label{fig:int_strainsweep_rel_LL_gap}
\end{figure}

We first investigate the interplay between strain and interactions within a self-consistent Hartree-Fock framework. This subsection details the evolution of the electronic structure through several parameter sweeps:
\begin{itemize}
    \item \textbf{Variable Strain without Interactions or relaxation:} We analyze the dispersion and LL spectra as a function of the magnitude of heterostrain, $\varepsilon$ (see Figs.~\ref{fig:strainsweep},\ref{fig:strainsweep_LL}).
    \item \textbf{Variable Interaction with fixed strain:} We analyze the dispersion and LL spectra as a function of the Hubbard $U_1$ interaction under the influence of a $\varepsilon=0.2\%$ strain (see Figs.~\ref{fig:intsweep}, \ref{fig:intsweep_LL}, \ref{fig:intsweep_LL_gap}).
    \item \textbf{Fixed Interaction with variable strain:} We analyze the dispersion and LL spectra as a function of variable magnitude of heterostrain, $\varepsilon$ at fixed the Hubbard interaction $U_1=57.95$ meV (see Figs.~\ref{fig:int_strainsweep}, \ref{fig:int_strainsweep_LL}, \ref{fig:int_strainsweep_LL_gap}).
    \item \textbf{Variable Relaxation at fixed strain:} We examine the impact of variable relaxation at fixed strain and interaction, exploring how the breaking of particle-hole symmetry modifies the topological gaps at fixed heterostrain, $\varepsilon=0.2\%$.(see Figs.~\ref{fig:int_strain_relsweep}, \ref{fig:int_strain_relsweep_LL}, \ref{fig:int_strain_relsweep_LL_gap}).
    \item \textbf{Variable strain at fixed Relaxation:} We examine the impact of variable strain at fixed relaxation and interaction, exploring how the breaking of particle-hole symmetry modifies the topological gaps.(see Figs.~\ref{fig:int_strainsweep_rel}, \ref{fig:int_strainsweep_rel_LL}, \ref{fig:int_strainsweep_rel_LL_gap}).
\end{itemize}
The SCHF results for the strained semimetal reveal that the Chern gap sequence is generally monotonically decreasing with increasing $|C|$ for $|C| \geq 4$. The $C=0$ gap is a persistent feature and, at large strain values (approaching $0.2\%$), can become comparable to or larger than the $|C|=8$ gaps. Notably, the $|C|=4$ gaps remain the largest across the sampled parameter space. The introduction of relaxation significantly breaks the symmetry between the positive and negative Chern sectors; we find that $C<0$ gaps are generally larger than $C>0$ gaps. This is attributed to the $f$-electron states being pushed lower in energy relative to the $c$-fermion $\Gamma_1$ and $\Gamma_2$ states, thereby increasing the energy separation between the Dirac cones and the occupied $f$-character portions of the bands.

\subsection{Hubbard I approximation with relaxation}
\label{subsec:hubbard_i_results}

\begin{figure}
    \centering
    \includegraphics[width=0.8\linewidth]{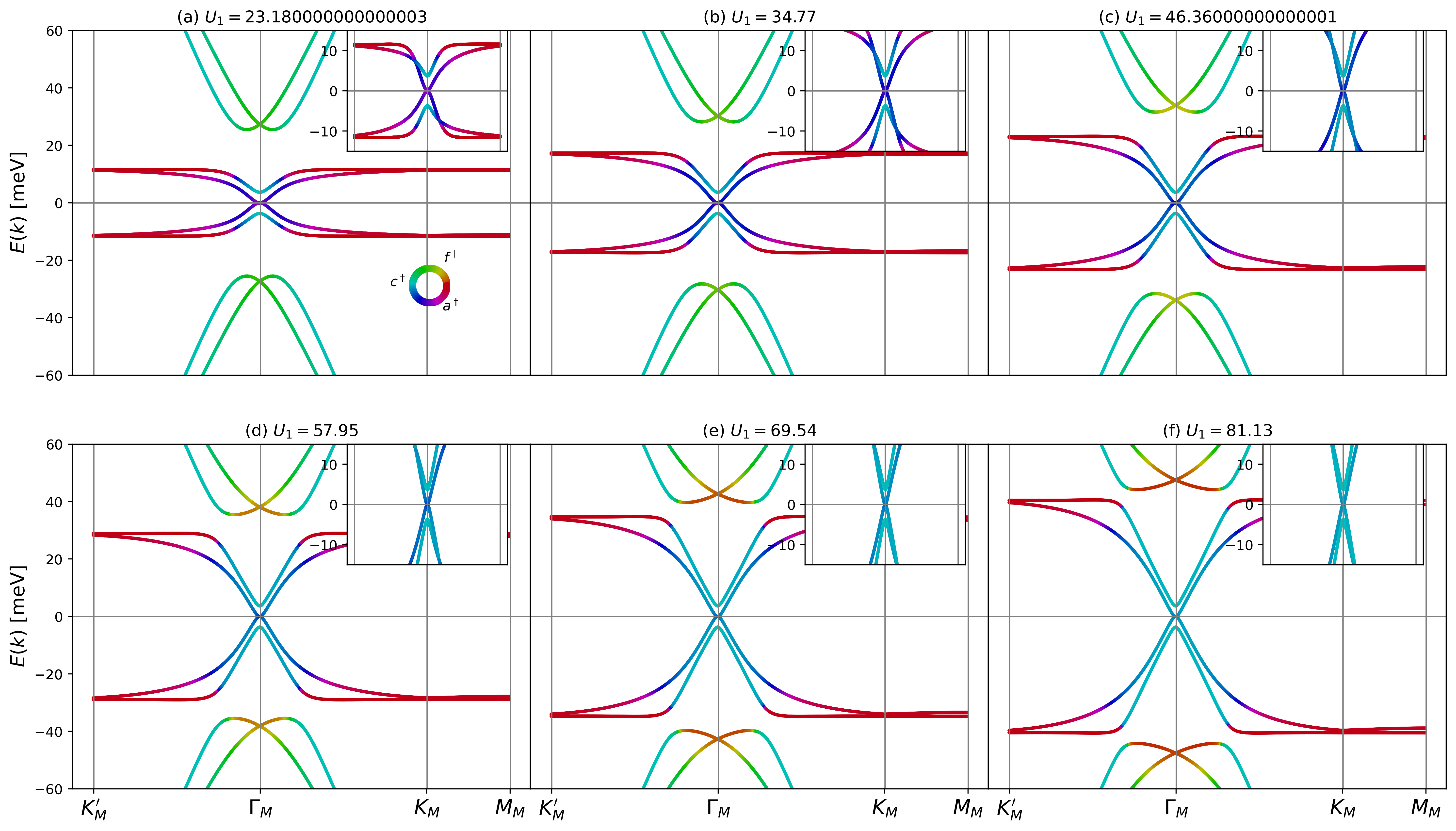}
    \caption{\textbf{Mott semimetal:}  Dispersion of the auxiliary fermion Hamiltonian within the Hubbard I approximation as a function of the on-site Hubbard interaction $U_1$. The panels illustrate the evolution of the poles, where the color mapping indicates the character of the states (mixture of auxiliary $a$, $c$, and $f$ fermions). For small interaction strengths $U_1$ (panels a--c), the bands closest to the chemical potential near the $\Gamma_M$ point exhibit a significant mixture of $a$ and $c$ character. As $U_1$ increases (panels d--f), the character of these states undergoes a clear transition, becoming predominantly $c$-like in the large $U$ regime. The insets provide a zoomed-in view of the Dirac crossing regions, highlighting the renormalization of the bandwidth and the shifting energy of the $f$-electron poles as the interaction strength scales.} 
    \label{fig:intsweep_nostrain}
\end{figure}

\begin{figure}
    \centering
    \includegraphics[width=0.8\linewidth]{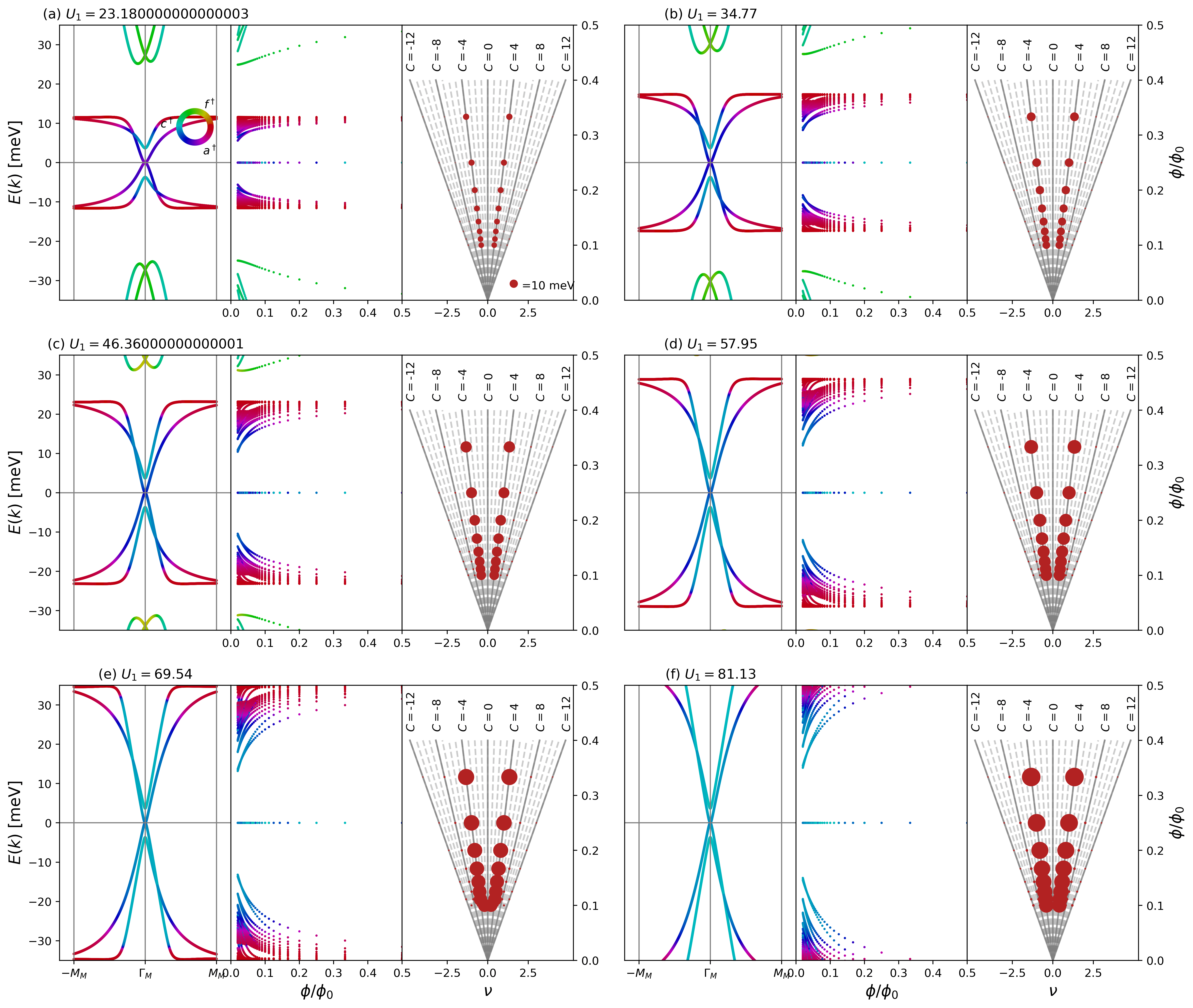}
    \caption{\textbf{Mott semimetal:} Dispersion, Landau level spectrum, and Landau fan diagram of the auxiliary fermion Hamiltonian within the Hubbard I approximation as a function of the on-site Hubbard interaction $U_1$ in the absence of relaxation. Because of the inherent Particle-Hole (PH) symmetry of the model in this approximation, the positive and negative Chern gaps are identical in magnitude. Notably, no $C=0$ gap appears in the Landau fan diagrams. } 
    \label{fig:intsweep_nostrain_LL}
\end{figure}

\begin{figure}
    \centering
    \includegraphics[width=0.8\linewidth]{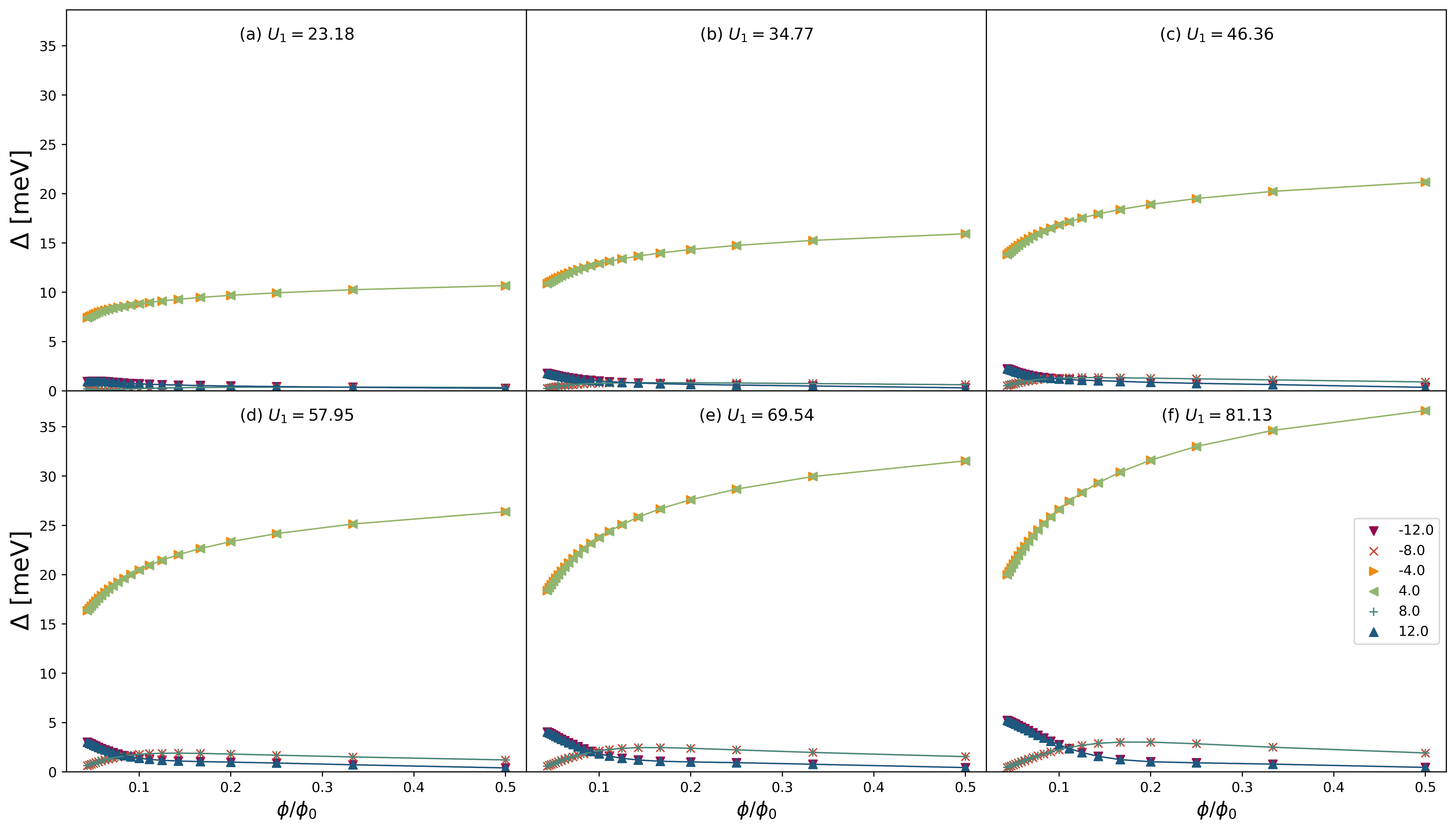}
    \caption{\textbf{Mott semimetal:} Evolution of the energy gap ($\Delta$) as a function of magnetic flux ($\phi/\phi_0$) within the Hubbard I approximation for increasing interaction strengths $U_1$. In the absence of relaxation effects, the model preserves Particle-Hole (PH) symmetry, resulting in identical magnitudes for the positive and negative Chern gaps ($C = \pm 4, \pm 8, \pm 12$). A notable feature at low magnetic flux is that the gap sizes do not decrease monotonically with increasing Chern number; for instance, the $C = \pm 12$ gaps can appear larger or comparable to the $C = \pm 8$. As $U_1$ increases. } 
    \label{fig:intsweep_nostrain_LL_gap}
\end{figure}

\begin{figure}
    \centering
    \includegraphics[width=0.8\linewidth]{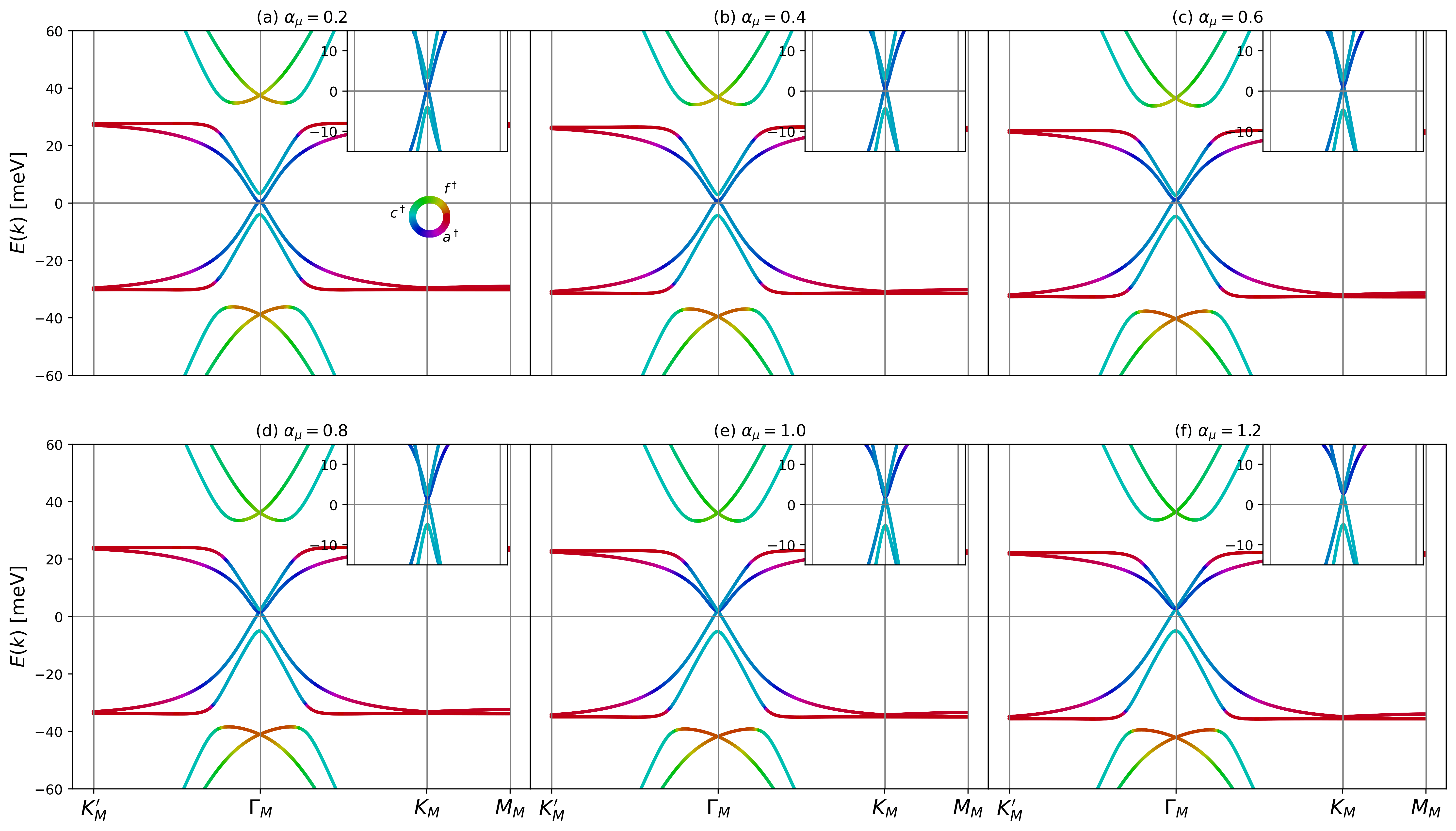}
    \caption{\textbf{Mott semimetal:} Dispersion of the auxiliary fermion Hamiltonian within the Hubbard I approximation as a function of the relaxation strength $\alpha_\mu$ for a fixed interaction $U_1$. The introduction of relaxation explicitly breaks the Particle-Hole (PH) symmetry characteristic of the unrelaxed model. For realistic values of the relaxation parameter (panels c--e), the nodal crossing at the $\Gamma_M$ point shifts in energy, becoming nearly degenerate with the local minimum of the quadratic-like band immediately above it. } 
    \label{fig:int_simp_relsweep}
\end{figure}

\begin{figure}
    \centering
    \includegraphics[width=0.8\linewidth]{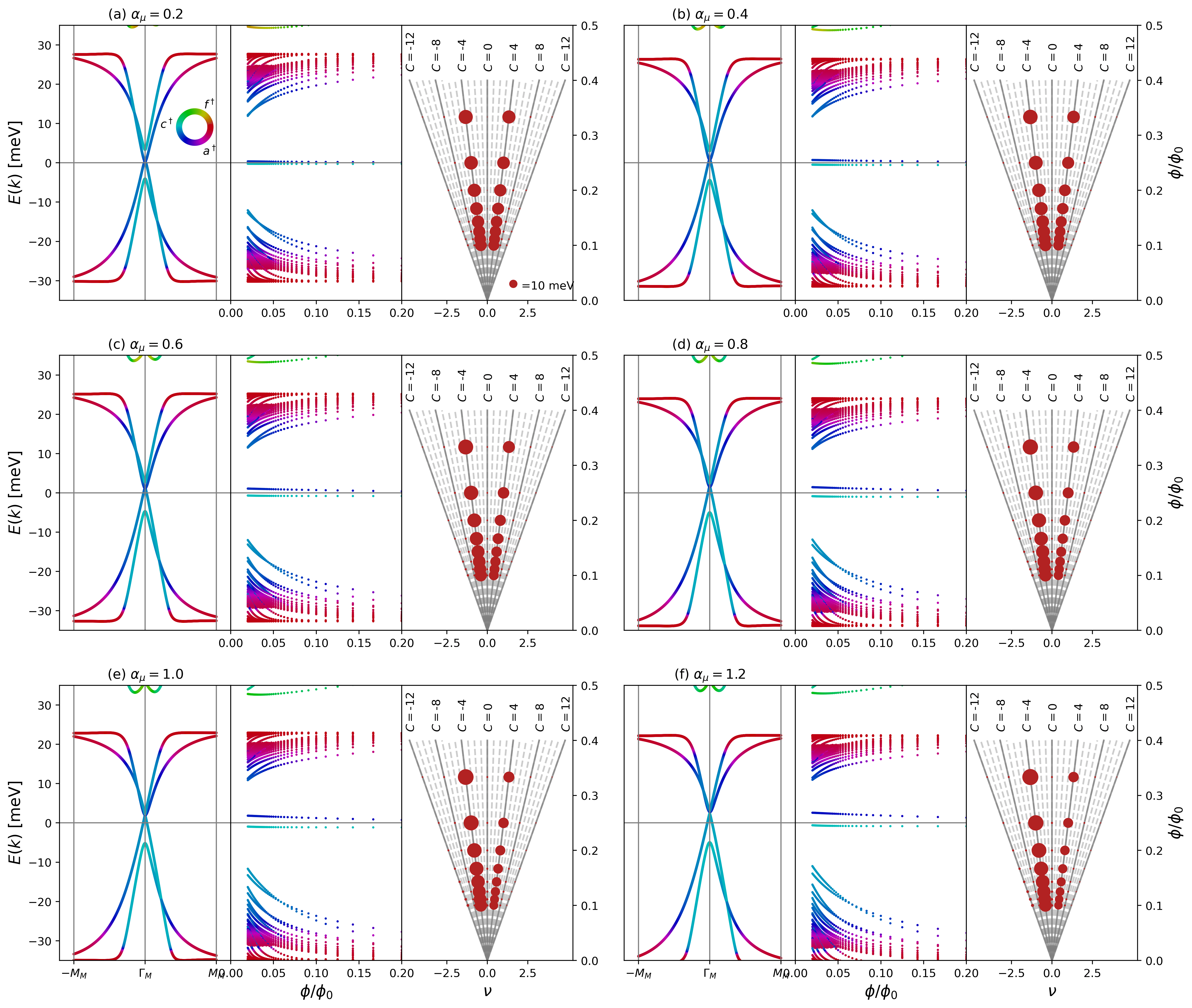}
    \caption{\textbf{Mott semimetal:} Dispersion, Landau level spectrum, and Landau fan diagram of the auxiliary fermion Hamiltonian within the Hubbard I approximation as a function of the relaxation strength $\alpha_\mu$. The explicit breaking of Particle-Hole (PH) symmetry by relaxation leads to the opening of a $C=0$ gap in the Landau fan, a feature absent in the symmetric case. The fan diagrams (right subpanels) reveal a complex topological landscape characterized by multiple gap closings and reopenings as the flux and relaxation parameters evolve. This behavior underscores the sensitivity of the Chern sequence to the specific relaxation-induced shifting of the poles of the Green's function near the chemical potential. } 
    \label{fig:int_simp_relsweep_LL}
\end{figure}

\begin{figure}
    \centering
    \includegraphics[width=0.8\linewidth]{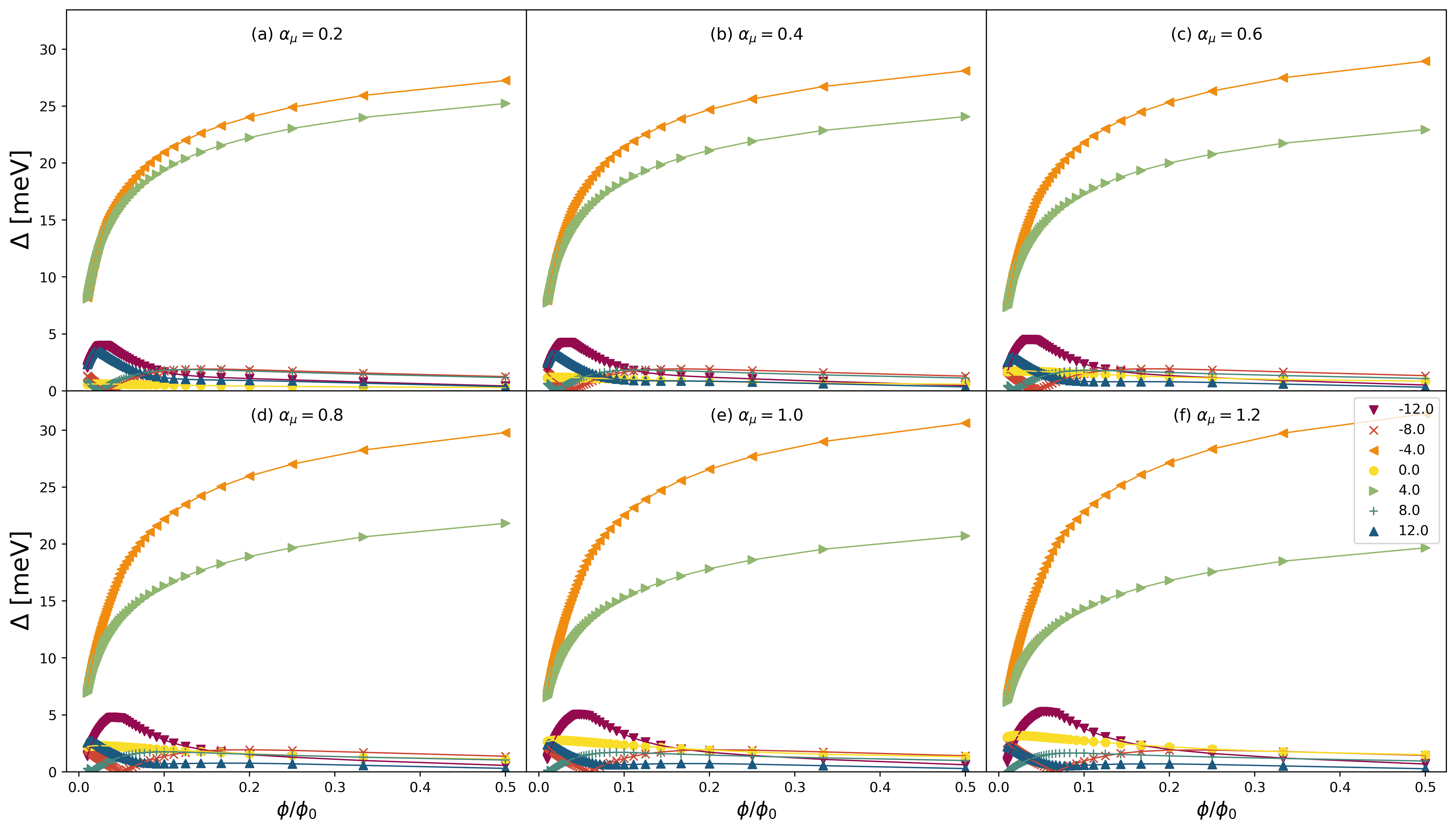}
    \caption{\textbf{Mott semimetal:} Evolution of the energy gap ($\Delta$) as a function of magnetic flux ($\phi/\phi_0$) within the Hubbard I approximation for increasing relaxation strengths $\alpha_\mu$. The explicit breaking of Particle-Hole (PH) symmetry is evidenced by the distinct evolution and magnitude differences between the positive and negative Chern gaps (e.g., $C = +4$ vs. $C = -4$). The profiles exhibit complex behavior characterized by multiple gap closings and reopenings as the flux varies. Notably, for certain relaxation and flux regimes, the high-order Chern gaps ($C = \pm 12$) can exceed the magnitude of the $C = 0$ gap induced by relaxation.} 
    \label{fig:int_simp_relsweep_LL_gap}
\end{figure}

\begin{figure}
    \centering
    \includegraphics[width=0.8\linewidth]{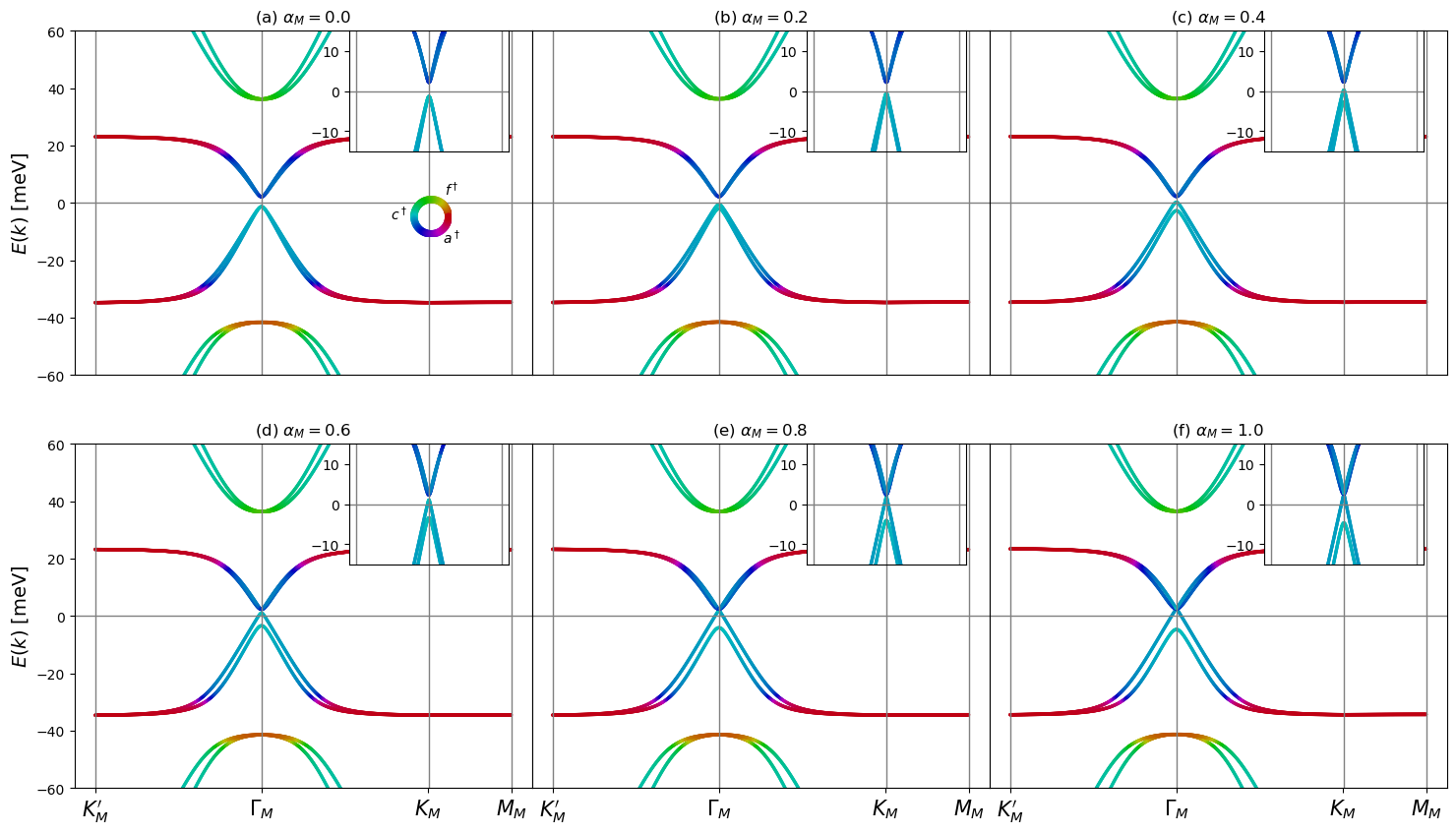}
    \caption{\textbf{Mott semimetal:} Dispersion of the auxiliary fermion Hamiltonian within the Hubbard I approximation as a function of the artificial scaling factor $\alpha_M$ for the mass term $M$ in the chiral limit. This term explicitly controls the energy splitting between the $\Gamma_1$ and $\Gamma_2$ $c$-fermion states. While the system is fully gapped at charge neutrality for $\alpha_M = 0$ (panel a), increasing the mass term leads to a progressive reduction of the charge gap. For the realistic relaxation parameters specified in Tab.~(\ref{tab:model_params}), the gap eventually closes, forming a nodal crossing that overlaps closely in energy with the local minimum of the dispersive quadratic-like band immediately above it (panels d--f).}
    \label{fig:int_rel_massweep_chiral}
\end{figure}

\begin{figure}
    \centering
    \includegraphics[width=0.8\linewidth]{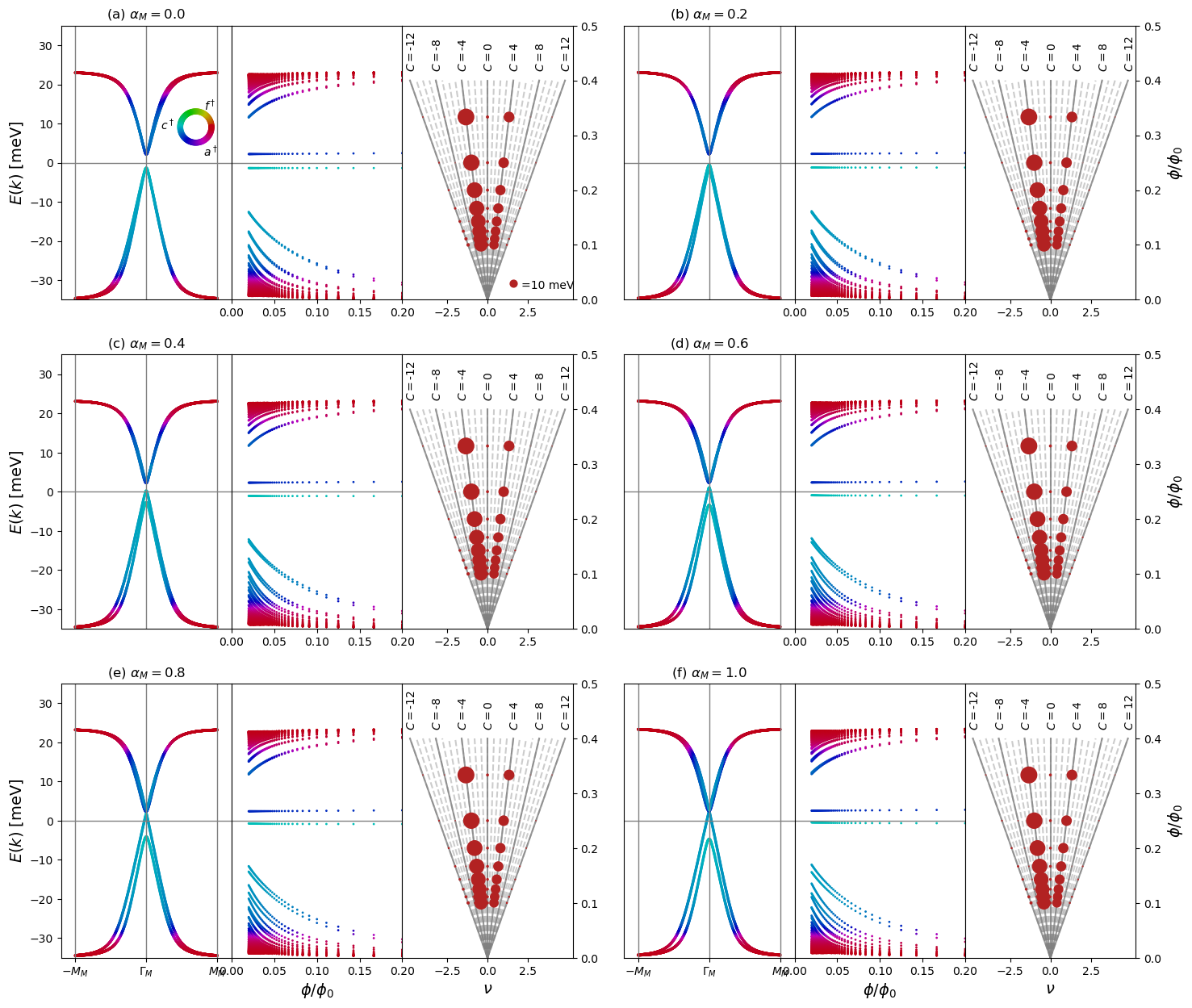}
    \caption{\textbf{Mott semimetal:} Dispersion, Landau level spectrum, and Landau fan diagram of the auxiliary fermion Hamiltonian within the Hubbard I approximation as a function of the artificial scaling factor $\alpha_M$, which controls the energy splitting between the $\Gamma_1$ and $\Gamma_2$ $c$-fermion states in the chiral limit. At zero mass ($\alpha_M = 0$) and low magnetic flux, the $C = \pm 8$ and $C = \pm 16$ Landau levels converge toward zero energy as $\phi/\phi_0 \to 0$, a consequence of the bands being described by degenerate massive Dirac fermions near the $\Gamma_M$ point. As finite $M$ is introduced ($\alpha_M > 0$), the Dirac degeneracy is lifted and the bands split. In the Landau fan diagrams (right subpanels), this splitting manifests as a series of near-closings and reopenings of the $C = \pm 8$ and $C = \pm 16$ Chern gaps, reflecting the sensitive restructuring of the topological states near the chemical potential.}
    \label{fig:int_rel_massweep_LL_chiral}
\end{figure}

\begin{figure}
    \centering
    \includegraphics[width=0.75\linewidth]{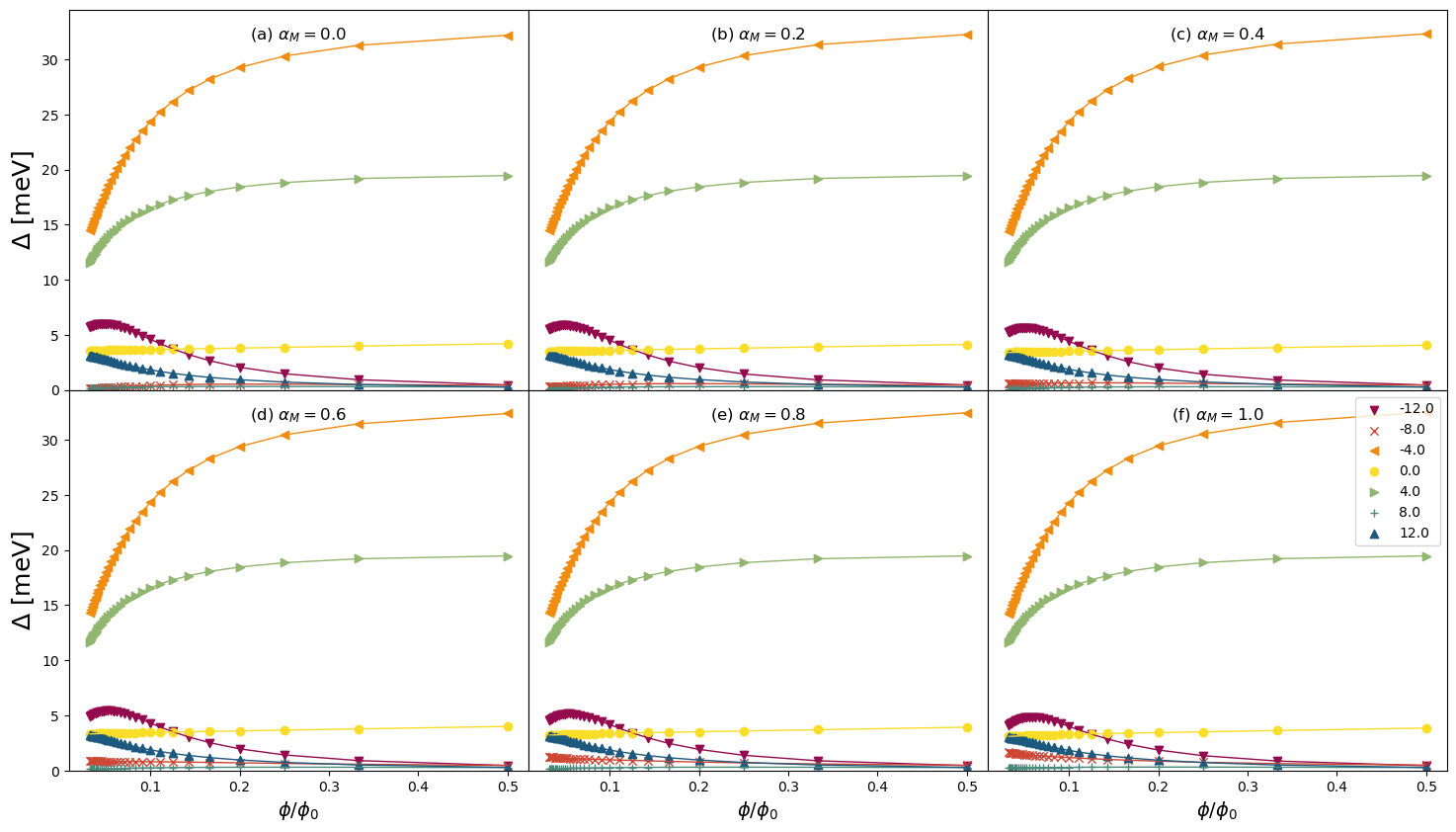}
    \caption{\textbf{Mott semimetal:} Evolution of the energy gap ($\Delta$) as a function of magnetic flux ($\phi/\phi_0$) within the Hubbard I approximation for increasing artificial mass scaling factor $\alpha_M$ in the chiral limit. The parameter $\alpha_M$ modulates the splitting between the $\Gamma_1$ and $\Gamma_2$ $c$-fermion states. Notably, at low flux, the $C = \pm 8$ gaps exhibit prominent closings and reopenings as $\alpha_M$ increases, corresponding to the avoided crossings and Dirac splittings observed in the Landau level spectra of Fig.(\ref{fig:int_rel_massweep_LL_chiral}).  }
    \label{fig:int_rel_massweep_LL_chiralgap}
\end{figure}

\begin{figure}
    \centering
    \includegraphics[width=0.8\linewidth]{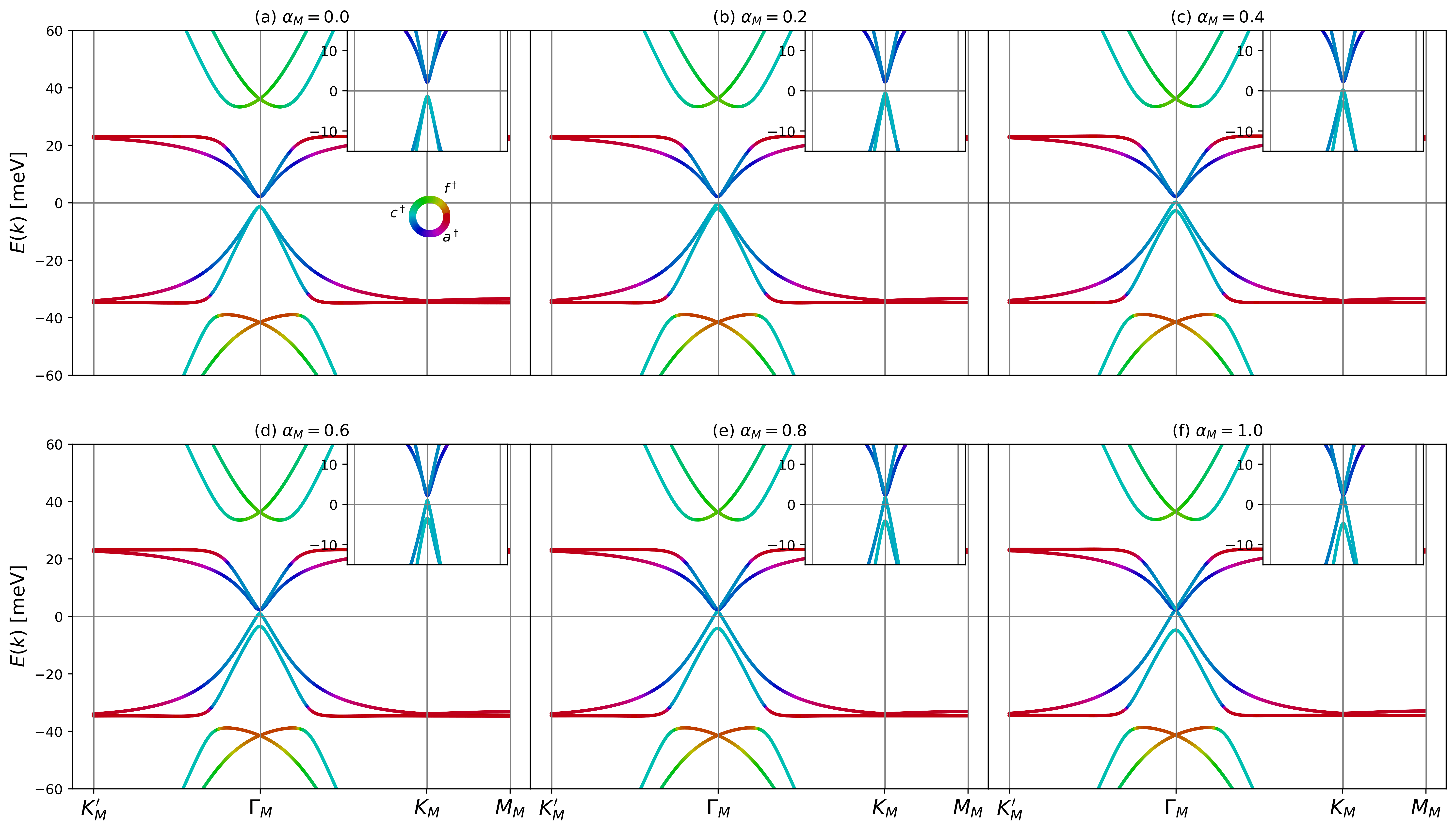}
    \caption{\textbf{Mott semimetal:}  Dispersion of the auxiliary fermion Hamiltonian within the Hubbard I approximation as a function of the artificial scaling factor $\alpha_M$ for the mass term $M$. This term explicitly controls the energy splitting between the $\Gamma_1$ and $\Gamma_2$ $c$-fermion states. While the system is fully gapped at charge neutrality for $\alpha_M = 0$ (panel a), increasing the mass term leads to a progressive reduction of the charge gap. For the realistic relaxation parameters specified in Tab.~(\ref{tab:model_params}), the gap eventually closes, forming a nodal crossing that overlaps closely in energy with the local minimum of the dispersive quadratic-like band immediately above it (panels d--f).} 
    \label{fig:int_rel_massweep}
\end{figure}

\begin{figure}
    \centering
    \includegraphics[width=0.8\linewidth]{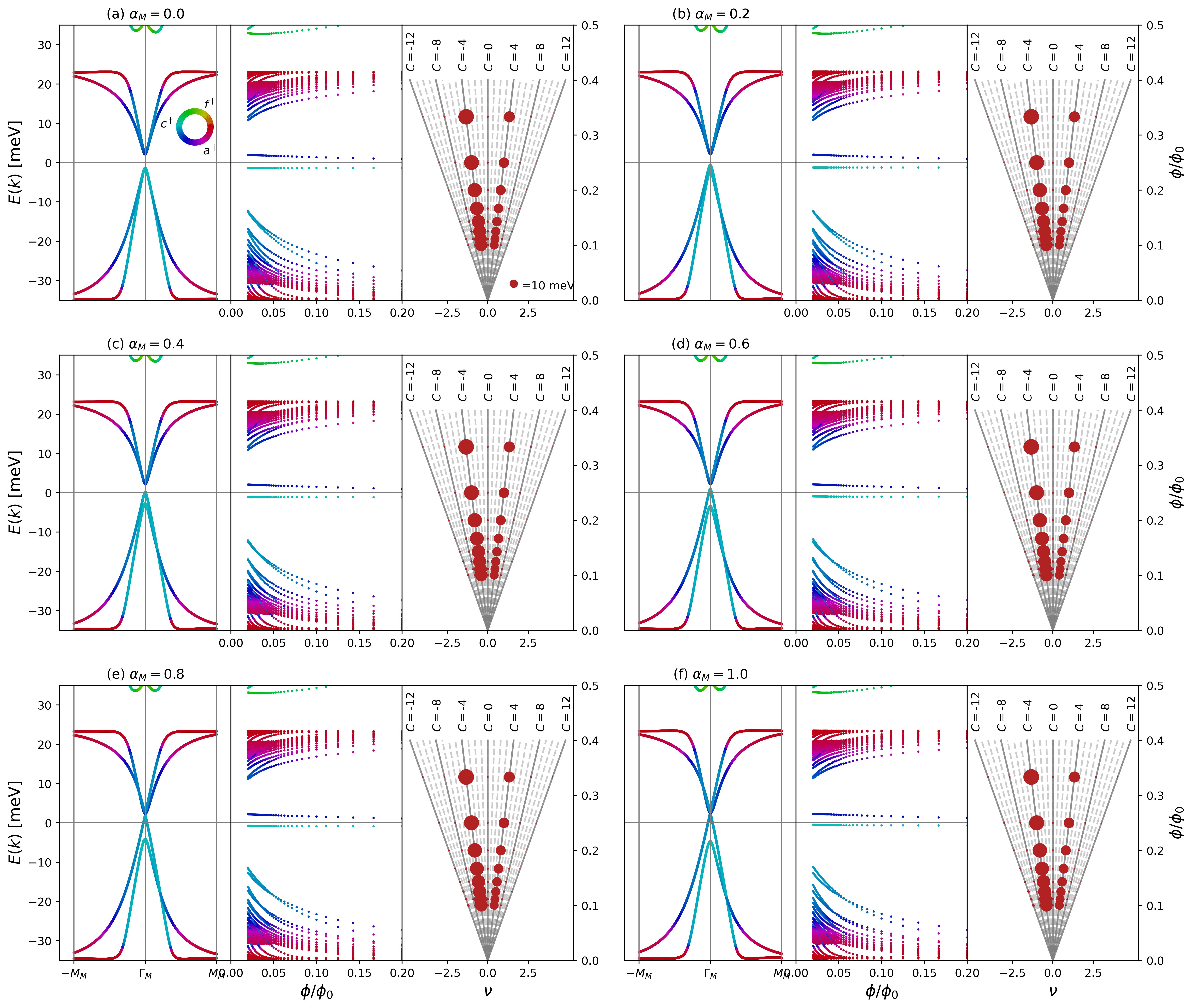}
    \caption{\textbf{Mott semimetal:}  Dispersion, Landau level spectrum, and Landau fan diagram of the auxiliary fermion Hamiltonian within the Hubbard I approximation as a function of the artificial scaling factor $\alpha_M$, which controls the energy splitting between the $\Gamma_1$ and $\Gamma_2$ $c$-fermion states. At zero mass ($\alpha_M = 0$) and low magnetic flux, the $C = \pm 8$ and $C = \pm 16$ Landau levels converge toward zero energy as $\phi/\phi_0 \to 0$, a consequence of the bands being described by degenerate massive Dirac fermions near the $\Gamma_M$ point. As finite $M$ is introduced ($\alpha_M > 0$), the Dirac degeneracy is lifted and the bands split. In the Landau fan diagrams (right subpanels), this splitting manifest as a series of near-closings and reopenings of the $C = \pm 8$ and $C = \pm 16$ Chern gaps, reflecting the sensitive restructuring of the topological states near the chemical potential.} 
    \label{fig:int_rel_massweep_LL}
\end{figure}

\begin{figure}
    \centering
    \includegraphics[width=0.8\linewidth]{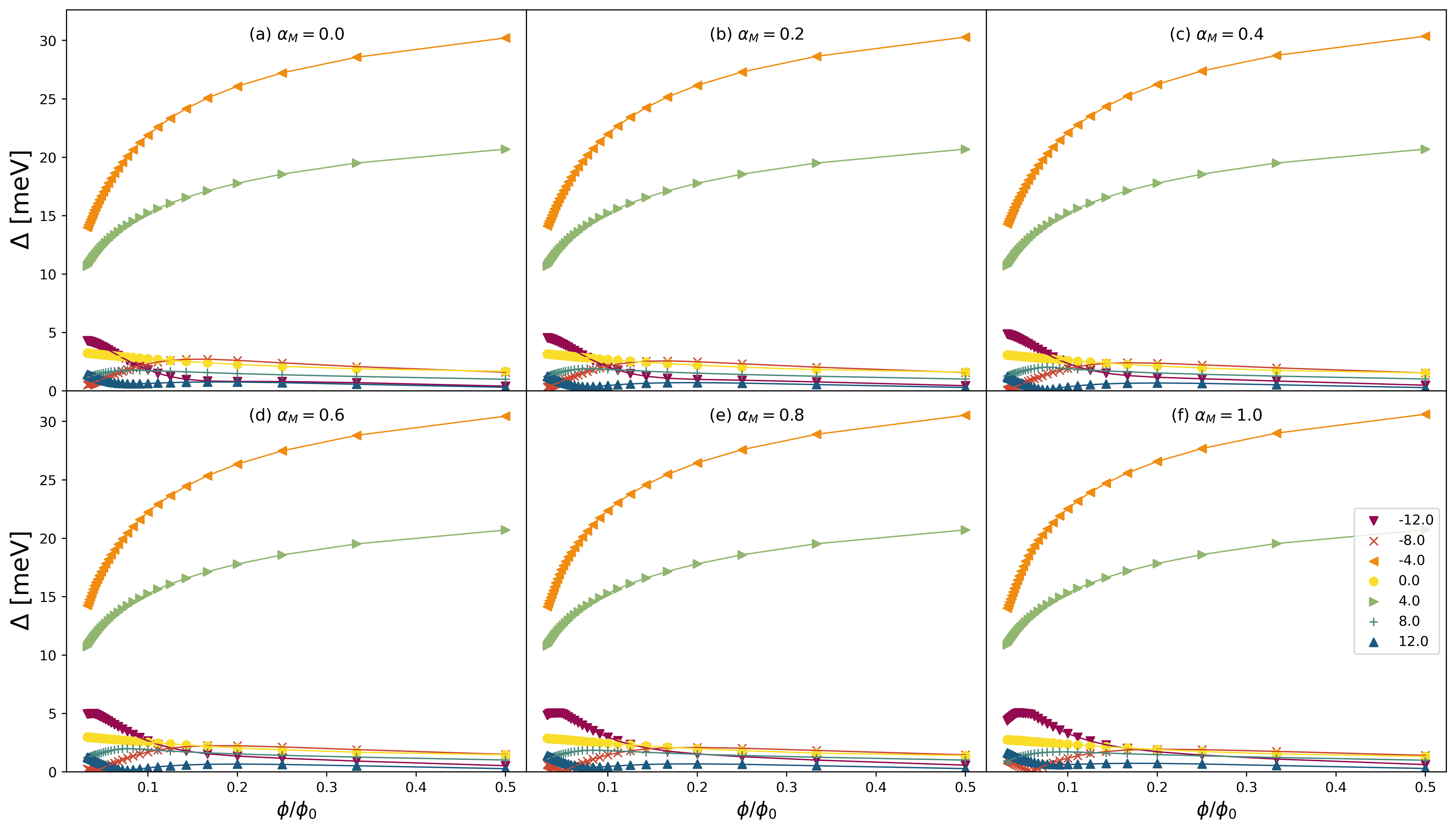}
    \caption{\textbf{Mott semimetal:}  Evolution of the energy gap ($\Delta$) as a function of magnetic flux ($\phi/\phi_0$) within the Hubbard I approximation for increasing artificial mass scaling factor $\alpha_M$. The parameter $\alpha_M$ modulates the splitting between the $\Gamma_1$ and $\Gamma_2$ $c$-fermion states. Notably, at low flux, the $C = \pm 8$ gaps exhibit prominent closings and reopenings as $\alpha_M$ increases, corresponding to the avoided crossings and Dirac splittings observed in the Landau level spectra of Fig.(\ref{fig:int_rel_massweep_LL}). } 
    \label{fig:int_rel_massweep_LL_gap}
\end{figure}

To further understand the role of local correlations, we employ the Hubbard I approximation to calculate the poles of the Green's function. This subsection covers:
\begin{itemize}
    \item \textbf{Interaction Sweeps:} The evolution of the auxiliary fermion poles and Landau fans as a function of $U_1$ (see Figs.~\ref{fig:intsweep_nostrain}, \ref{fig:intsweep_nostrain_LL}, \ref{fig:intsweep_nostrain_LL_gap}).
    \item \textbf{Relaxation and Particle Hole Symmetry Breaking:} The effect of variable relaxation $\alpha_\mu$ at fixed $U_1$ (see Figs.~\ref{fig:int_simp_relsweep}, \ref{fig:int_simp_relsweep_LL}, \ref{fig:int_simp_relsweep_LL_gap}).
    \item \textbf{$M$ Scaling in the chiral limit:} The impact of the artificial mass parameter $\alpha_M$, which modulates the $\Gamma_1 / \Gamma_2$ splitting (see Figs.~\ref{fig:int_rel_massweep_chiral}, \ref{fig:int_rel_massweep_LL_chiral}, \ref{fig:int_rel_massweep_LL_chiralgap}).
    \item \textbf{$M$ Scaling:} The impact of the artificial mass parameter $\alpha_M$, which modulates the $\Gamma_1 / \Gamma_2$ splitting (see Figs.~\ref{fig:int_rel_massweep}, \ref{fig:int_rel_massweep_LL}, \ref{fig:int_rel_massweep_LL_gap}).
\end{itemize}

In the absence of relaxation, the Hubbard I results exhibit no $C=0$ gap, and the gap hierarchy is generically non-monotonic with respect to $|C|$, depending strongly on the magnetic flux. The introduction of relaxation opens a $C=0$ gap and triggers a complex landscape of gap closings and reopenings. Intriguingly, in the Hubbard I regime at low magnetic flux, we identify parameter ranges where the $|C|=12$ gaps exceed the magnitude of both the $|C|=8$ and the relaxation-induced $C=0$ gaps.

%-------------------Section Change-----------------

\section{Orbital coupling of the Wannier states to an out-of-plane magnetic field}
In this section we show that the orbital coupling of the $\bB=0$ $p_x\pm i p_y$  Wannier states to an out-of-plane magnetic field is absent up to linear order in $\bB$. We start by noting that
\begin{eqnarray}
    H_{BM}^\bK(\bp-\frac{e}{c}\bA(\br)) = H_{BM}^\bK(\bp) - \frac{e}{c}\rho_0\sigma\cdot\bA(\br)
\end{eqnarray}
at valley $\bK$ (valley $\bK'$ can be obtained by minimal substitution into the time-reversed BM model). The Pauli matrices $\rho$ and $\sigma$ act in layer and orbital spaces, respectively. The linear orbital coupling can be estimated by calculating
\begin{eqnarray}
  g_{ab} =   -\frac{e}{c}\int_\br W_a^\dagger(\br) \rho_0 \sigma\cdot\bA(\br)W_b(\br)
\end{eqnarray}
where $W_1(\br)=W_{\mathbf{0},1 \bK}$ and $W_2(\br)=W_{\mathbf{0},2 \bK}$ are the $\bB=0$ $p_x+ip_y$ and $p_x-ip_y$ Wannier states, respectively. Before further simplifying the orbital coupling $g_{ab}$, we start by noting that under $C_{3z}$, $C_{2z}\mathcal{T}$ and (anti-commuting) particle hole symmetry $P$ (of the BM model), these states transform as \cite{song2022magic}
\begin{eqnarray}
    U_{C_{3z}} W_1(C^{-1}_{3z}\br) &=& e^{+i\frac{2\pi}{3}}W_1(\br)\\
    U_{C_{3z}} W_2(C^{-1}_{3z}\br)&=& e^{-i\frac{2\pi}{3}}W_2(\br)\\
    U_{C_{2z}\mathcal{T}} K W_1(C^{-1}_{2z}\br) &=& W_2(\br)\\
   U_{C_{2z}\mathcal{T}}K W_2(C^{-1}_{2z}\br) &=&  W_1(\br) \\U_P W_1(-\br) &=&+i W_1(\br)\\
    U_P W_2(-\br) &=&-iW_2(\br) \\
\end{eqnarray}
where the representation (at $\bK$) read \cite{song2022magic}
\begin{eqnarray}
    U_{C_{3z}} &=&\rho_0e^{i\frac{2\pi}{3}\sigma_z}\\
U_{C_{2z}\mathcal{T}} &=&\rho_0\sigma_x\\
    U_P &=& -i\rho_y\sigma_0
\end{eqnarray}
Now see that Landau gauge $\sigma\cdot\bA$ can be decomposed as
\begin{eqnarray}
   \sigma\cdot\bA(\br) = B\left(A_1(\br) + A_2(\br) + A_3(\br)\right) 
\end{eqnarray}
where
\begin{eqnarray}
   A_1(\br)&=&\frac{x\sigma_y-y\sigma_x}{2}\\
   A_2(\br)&=&+\frac{i}{4}(x-iy)(\sigma_x-i\sigma_y)\\
   A_3(\br)&=&-\frac{i}{4}(x+iy)(\sigma_x+i\sigma_y)
\end{eqnarray}
with the property that $U_{C_{3z}}  \rho_0A_1(C^{-1}_{3z}(\br))U^\dagger_{C_{3z}}=A_1(\br)$, $U_{C_{3z}}  \rho_0A_2(C^{-1}_{3z}(\br))U^\dagger_{C_{3z}} =\omega^\ast A_2(\br)$ and $U_{C_{3z}}  \rho_0A_3(C^{-1}_{3z}(\br))U^\dagger_{C_{3z}} =\omega A_3(\br)$, where $\omega=e^{i2\pi/3}$.
Now consider the diagonal term \begin{eqnarray}
    g_{aa} &=&  -\frac{e}{c}\int_\br W_a^\dagger(\br) \rho_0 \sigma\cdot\bA(\br)W_a(\br) = -\frac{eB}{c}\int_\br W^\dagger_a(\br) \rho_0 A_1(\br)W_a(\br)\\
    &=& -\frac{eB}{c}\int_\br W^\dagger_a(-\br) U^\dagger_P \left( U_P \rho_0 A_1(-\br)U^\dagger_P\right)U_P W_a(-\br)\label{Eq:Diagonal_orbital_coup_1}\\
    &=&+\frac{eB}{c}\int_\br W^\dagger_a(\br)   \rho_0A_1(\br)W_a(\br) = 0\label{Eq:Diagonal_orbital_coup_2}
\end{eqnarray}
where we used the $C_{3z}$ and $P$ symmetries. For the off-diagonal term, we have
\begin{eqnarray}
    g_{12} = g^\ast_{21} &=&  -\frac{e}{c}\int_\br W^\dagger_1(\br) \rho_0 \sigma\cdot\bA(\br)W_2(\br)\\
    &=& -\frac{eB}{c}\int_\br \left(U_{C_{2z}\mathcal{T}}K[(-\rho_0 x\sigma_y) W_2(-\br)]\right)^\dagger \left(U_{C_{2z}\mathcal{T}}KW_1(-\br) \right)\\
    &=& -\frac{eB}{c}\int_\br \left([U_{C_{2z}\mathcal{T}}(-\rho_0 x\sigma_y)^\ast U^\dagger_{C_{2z}\mathcal{T}}] U_{C_{2z}\mathcal{T}}W^\ast_2(-\br)\right)^\dagger \left(U_{C_{2z}\mathcal{T}}W^\ast_1(-\br) \right)\\
    &=& -\frac{eB}{c}\int_\br \left((-\rho_0 x\sigma_y)W_1(\br)\right)^\dagger W_2(\br) = +\frac{eB}{c}\int_\br W_1^\dagger(\br)\rho_0  x\sigma_y W_2(\br)\\
    &=& +\frac{e}{c}\int_\br W_1^\dagger(\br) \rho_0 \sigma\cdot\bA(\br)W_2(\br)\\
    &=&0
\end{eqnarray}
where we used the fact that $C_{2z}\mathcal{T}$ is anti-unitary. Thus $g_{ab}=0$ in Landau gauge. This can straightforwardly be extended to symmetric gauge by noting that
\begin{eqnarray}
    \sigma\cdot\bA = BA_1(\br)
\end{eqnarray}
in symmetric gauge and thus off-diagonal terms are trivially zero. The diagonal term was shown to be zero in Eq.(\ref{Eq:Diagonal_orbital_coup_2}).
We thus ignored orbital coupling of the $p_x\pm ip_y$ Wannier states to magnetic field in our calculation so far.

\end{document}